# The Sources of Statistical Bias Series: Simulated Demonstrations to Illustrate the Causes and Effects of Biases in Statistical Estimates

Ian A. Silver Ph.D.[1]

**Acknowledgements:**
I would like to take a moment to thank the people who have shaped this endeavor. I truly can not state this with any more clarity, I do not believe I would understand statistics nearly as well as I do today without John Wooldredge. John stressed the importance of learning the math when most others stress the importance of learning software. Additionally, without the encouragement of J.C. Barnes and Joseph L. Nedelec my interest in statistics would not have developed. I can never thank these individual enough for the time they have spent – and will spend – talking to me about statistics.

---

[1] For Questions Please Email: silveria@rowan.edu
**Primary Affiliation:**
Law and Justice Department, Rowan University, Glassboro, New Jersey 08028
**Secondary Affiliation:**
Corrections Institute, University of Cincinnati, Cincinnati, OH 45221-0389

# Table of Contents

# Introduction

When teaching and discussing statistical biases, our focus is oftentimes placed on how to test and address potential issues rather than the sources and effects of statistical biases on the estimates produced by our statistical models. The latter represents a potential avenue to help us better understand the impact of researcher degrees of freedom on the statistical estimates we produce. The ***Sources of Statistical Bias Series*** is an endeavor I have undertaken to demonstrate the sources of statistical bias on the estimates produced across various statistical models. The series will review assumptions associated with estimating causal associations, as well as more complicated statistical models including, but not limited to, multilevel models, path models, structural equation models, and Bayesian models. In addition to the primary goal, the series of entries are designed to illustrate how simulations can be used to develop a comprehensive understanding of applied statistics. I personally believe that simulations can provide knowledge of statistics without relying on the formal calculations that often scare individuals away. You, however, be the judge of that. If you are interested in the effects of violating a specific assumption please feel free to reach out to me at silveria@rowan.edu. This document will be updated with each new entry.

**Upcoming Sections and Entries:**
1. ***Measurement and Statistical Bias***
    1. Measurement Creation: Dichotomies, Ordinal Measures, Aggregate Events, & Variety Scores
    2. Measurement Creation: Aggregate Scale, Average Scale, Standardized Scale, Weighted Scale & CFA
    3. Measurement Error
    4. Distributional Assumptions for the Dependent Variable (Linear Regression Models)
    5. Distributional Assumptions for the Dependent Variable (Mixed-Effects Linear Models)
    6. Distributional Assumptions for the Dependent Variable (Non-Parametric Regression Models)
    7. Distributional Assumptions for the Lagged-Endogenous Variables (Structural Equation Modeling)
    8. Distributional Assumptions for the Endogenous Variable (Structural Equation Modeling)
    9. Longitudinal Clustering
    10. Spatial Clustering

2. ***Statistical Biases when Examining Causal Associations (Revisited)***
3. ***Missing Data & Imputation***
4. ***Fixed-Effects Models***
5. ***Random-Effects Models***
6. ***Measurement Models***
7. ***Path Models***
8. ***Structural Equation Models***
9. ***Advanced SEM Techniques***
10. ***Behavioral Genetic Methodologies***

**Manuscripts Aligned with the Sources of Statistical Biases Series:**

**Silver, Ian. A.,** D'Amato, Christopher, and Wooldredge, John.
2022 "The heterogeneous effects of confounder bias: Evaluating the implications for social behavioral research." Preprint available at: 10.31234/osf.io/kc9b4

**Silver, Ian. A.,** and Wooldredge, John.
2022 "Colliders: A potential biasing factor when evaluating the causal structure of an unobserved construct in a SEM?" Preprint available at: 10.31234/osf.io/5tkq9

**Silver, Ian. A.,** Longergan, Holly, and Nedelec, Joseph L.
2022 "On the selection of variables in criminology: Adjusting for the descendants of unobserved confounders." *Journal of Criminal Justice.* https://doi.org/10.1016/j.jcrimjus.2022.101924

**Silver, Ian A.,** Liu, Hexuan, and Nedelec, Joseph L.
2022 "Genetically adjusted propensity scores: A methodological proposal and simulated comparison to discordant MZ twin models." *Twin Research and Human Genetics.* Published Online. https//doi.org/10.1017/thg.2022.2

**Silver, Ian A**. and Kelsay, James D.
2021 "The moderating effects of population characteristics: A potential biasing factor when employing non-random samples to conduct experimental research." *Journal of Experimental Criminology*. Published Online. https://doi.org/10.1007/s11292-021-09478-7

**Silver, Ian A**., Nedelec, Joseph L., Segal, Nancy L., and Lonergan, Holly.
2020 "Heteropaternal siblings misclassified as dizygotic twins: A potential biasing factor for heritability estimates" *Behavior Genetics*. Published Online. https://doi.org/10.1007/s10519-020-10039-3.

**Silver, Ian A**., Wooldredge, John., Sullivan, Christopher J., and Nedelec, Joseph L.
2020 "Longitudinal propensity score matching: A demonstration of counterfactual conditions adjusted for longitudinal clustering." *Journal of Quantitative Criminology*. Published Online. https://doi.org/10.1007/s10940-020-09455-9.

**Silver, Ian A**.
2019 "Linear *and* non-linear: A exploration of the variation in the functional form of IQ and antisocial behavior as adolescents age into adulthood." *Intelligence*. *76*, 1-14. https://doi.org/10.1016/j.intell.2019.101375.

# Violating the Fundamental Assumptions of Linear Regression Models

Let us begin by reviewing how violations of fundamental assumptions influence the estimates produced by ordinary least squares (OLS) regression models. The first four entries focus on: (1) the linearity assumption, (2) the homoscedasticity assumption, (3) the collinearity assumption, and (4) the normality assumption: outliers.[2] Please enjoy!

---

[2] Individual PDFs and the R-code associated with each entry is available on ianasilver.com.

**Entry 1: The Linearity Assumption**

**Introduction**
Satisfying the assumption of linearity in an Ordinary Least Squares (OLS) regression model is vital to the development of unbiased slope coefficients, standardized coefficients, standard errors, and the model $R^2$. Simply put, if a non-linear relationship exists, the estimates produced from specifying a linear association between two variables will be biased. But how biased will the slope coefficients, standardized coefficients, standard errors, and model $R^2$ be when we violate the linearity assumption in OLS regression model? Moreover, what would violating the linearity assumption do to our interpretation of an association between two variables?

Let's start with a basic review of the linearity assumption. The linearity assumption is the belief that the expected value of a dependent variable will change at a constant rate across values of an independent variable (i.e., a linear function). To provide an example of the linearity assumption, if we increase the independent variable by 1-point and observe a 1-point increase in the dependent variable, we would assume that any subsequent 1-point increase in the independent variable would result in a 1-point increase in the dependent variable. Although mathematically logical, non-linear associations often exist between variables. A non-linear association is simply a relationship where the *direction and rate of change* in the dependent variable will differ as we increase the score on the independent variable. Considering the example above, if the relationship is non-linear we would **not** assume that any subsequent 1-point increase in the independent variable would result in a 1-point increase in the dependent variable.

Determining if an association is linear or non-linear is important as it guides how we specify OLS regression models. For a linear association (the most common assumption) we would regress the dependent variable on the independent variable, and for a non-linear association with a single curve we would regress the dependent variable on the independent variable and the independent variable squared. The differences between the specifications simply inform the model that we either expect a linear association or expect a threshold (i.e., curve) to exist in the association of interest. If our expectations and specifications do not match the observed data, we would violate our assumptions in the estimated model. More often than not, we assume that a linear association exists between the dependent variable and the independent variable because theory and prior research rarely guide us to the alternative. However, if the observed data violates this assumption (the linearity assumption), the results of our models could be biased.

To determine the degree of bias that exists after violating the linearity assumption, we will conduct directed equation simulations. Simulations are a common analytical technique used to explore how the coefficients produced by statistical models deviate from reality (the simulated relationship) when certain assumptions are violated. The current analysis focuses on violations of the linearity assumption.

**Satisfying the Linearity Assumption: Linear Association**
We will first begin by simulating a linear relationship between a dependent variable (identified as *Y* in the code) and an independent variable (identified as *X* in the code). All of the examples reviewed from herein are based on 100 cases (identified as *n* in the code). The dependent variable (*X*) is specified as a normally distributed construct with a mean of 5 and a standard deviation of 1.

If we did not specify *X* as a normally distributed construct we would run the risk of violating other assumptions and create an ambiguous examination of the degree of bias that exists after violating the linearity assumption. After simulating *X*, we specify that *Y* is equal to .25×*X* plus .025×normally distributed error. In this example, .25 is the **true slope coefficient** for the relationship between *X* and *Y*, where a 1-point increase in *X* corresponds to a .25 increase in *Y*. Including normally distributed error ensures we don't perfectly predict the dependent variable when estimating the regression model. Importantly, the set.seed portion of the code ensures that the random variables are equal each time we run a simulation.

```
> n <- 100
> set.seed(1001)
> X<-rnorm(n,5,1)   # Specification of the independent variable
> set.seed(32)
> Y<-(.25*X)+.025*rnorm(n,5,1) # Specification of the dependent variable
```

Now that we have our data, let's estimate an OLS regression model. Following the *R*-code for estimating a regression model, we regress *Y* on *X*. The results suggest that the slope coefficient for the association is .250, the standard error is .002, and the standardized coefficient is .997. This is exactly what we would expect to find given the specification of the data. In addition to just estimating the model, let's plot this relationship using ggplot2. Ggplot2 is the best package ever. As demonstrated, the specification of the relationship between *X* and *Y* in our first simulated dataset is perfectly linear (a 1-point increase in *X* corresponds to a .25-point increase in *Y*).

```
> M<-lm(Y~X)  # Estimating regression model (Linear relationship between X and Y
assumed)
>
> # Results
> summary(M)  # Summary of model results

Call:
lm(formula = Y ~ X)

Residuals:
          Min              1Q          Median              3Q             Max
-0.06579416011  -0.01478787853   0.00240413934   0.01523569058   0.04045840087

Coefficients:
                  Estimate    Std. Error    t value             Pr(>|t|)
(Intercept) 0.12328940395 0.00975526777   12.63824 < 0.000000000000000222 ***
X           0.25002056505 0.00190264638  131.40674 < 0.000000000000000222 ***
---
Signif. codes:  0 '***' 0.001 '**' 0.01 '*' 0.05 '.' 0.1 ' ' 1

Residual standard error: 0.0216664405 on 98 degrees of freedom
Multiple R-squared:  0.994356702,    Adjusted R-squared:  0.994299117
F-statistic: 17267.7323 on 1 and 98 DF,  p-value: < 0.00000000000000002220446

>
> library(lm.beta)
> lm.beta(M) # Standardized coefficients

Call:
lm(formula = Y ~ X)
```

```
Standardized Coefficients::
    (Intercept)                X
0.000000000000 0.997174358939
```

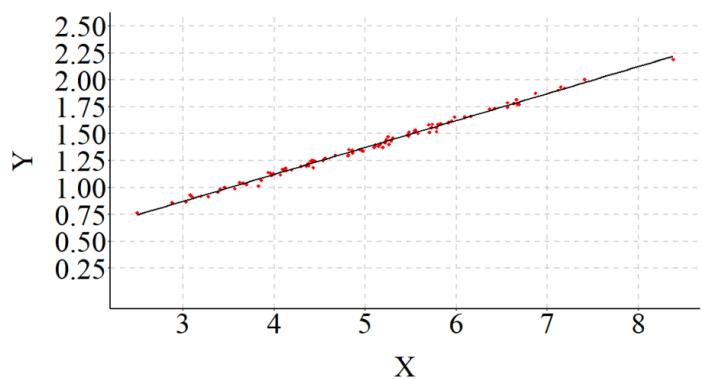

**Violating the Linearity Assumption**
We are presented with a unique challenge when simulating a curvilinear association between our dependent and independent variables. Similar to the preceding example, the independent variable (*X*) is specified as a normally distributed variable with a mean of 5 and a standard deviation of 1. To simulate a curvilinear association, however, we need to include another specification of *X* that is $X^2$ (labeled as X2 in the model). We can create *X2* by multiplying *X* by *X* or specifying X^2. For simplicity we rely on the former. Next, *Y* is defined to have a positive .25 slope coefficient with *X*, a negative .025 slope coefficient with *X2*, and a .025 slope coefficient with the normally distributed error. Again, the normally distributed error only serves to ensure that our model does not perfectly predict *Y*.

```
> set.seed(1001)
> X<-rnorm(n,5,1) # Specification of the independent variable
> X2<-X*X # Specification of the independent variable squared
> set.seed(32)
> Y<-(.25*X)-.025*X2+.025*rnorm(n,5,1) # Specification of the dependent variable
```

After simulating a curvilinear association in the data, we estimate a regression model that assumes a linear association between *Y* and *X* (we are knowingly violating the linearity assumption). The findings of the misspecified model suggest that a 1-point increase in *X* is associated with a .007 decrease in *Y* (SE = .004; *β* = -.173). Consistent with the misspecification, the estimated slope coefficient deviates from the specified slope coefficients between *X* and *Y*, and *X2* and *Y*. Additionally, the $R^2$ value suggests that the linear specification of the association only explains .02 percent of the variation in *Y*, which is a substantial departure from reality.

```
> # Misspecified model (Linear relationship between X and Y assumed)
> Mis<-lm(Y~X)
>
> # Results
> summary(Mis)
```

```
Call:
lm(formula = Y ~ X)

Residuals:
          Min            1Q        Median            3Q           Max
-0.2597571656 -0.0206712212  0.0113012471  0.0283234287  0.0682318921

Coefficients:
                Estimate     Std. Error  t value             Pr(>|t|)
(Intercept)  0.75134794607  0.02089728531 35.95433 < 0.0000000000000002 ***
X           -0.00707614052  0.00407576144 -1.73615              0.08568 .
---
Signif. codes:  0 '***' 0.001 '**' 0.01 '*' 0.05 '.' 0.1 ' ' 1

Residual standard error: 0.0464128509 on 98 degrees of freedom
Multiple R-squared:  0.0298395903,  Adjusted R-squared:  0.0199399943
F-statistic: 3.01422303 on 1 and 98 DF,  p-value: 0.0856797297

> lm.beta(Mis)

Call:
lm(formula = Y ~ X)

Standardized Coefficients::
    (Intercept)                X
 0.000000000000 -0.172741397271
```

Moreover, as demonstrated in the figure below, the specification of a linear relationship – for our curvilinear data – creates a relatively horizontal OLS regression line.

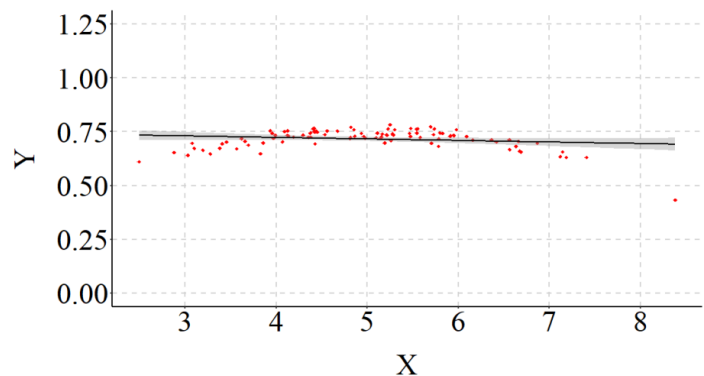

To make sure the example was completed correctly, however, we should estimate a model that properly specifies the relationship (i.e., the inclusion of $X2$ in the regression model) and compare the results. As expected, when we estimate the model assuming a curvilinear association we observe coefficients that match our specification. Additionally, the $R^2$ jumps from .02 to .79, indicating that the explanatory power of our model substantively increases when the relationship is properly specified.

```
> # Properly Specified Model (Curvilinear relationship between X and Y assumed)
```

```
> PS<-lm(Y~X+I(X^2))
> 
> # Results
> summary(PS)

Call:
lm(formula = Y ~ X + I(X^2))

Residuals:
         Min              1Q         Median              3Q            Max
-0.06629449249 -0.01476256564  0.00216027121  0.01361241435  0.03760637947

Coefficients:
                  Estimate     Std. Error    t value                  Pr(>|t|)
(Intercept)   0.15397661671  0.03328538416    4.62595                0.00001153 ***
X             0.23745870851  0.01316497818   18.03715 < 0.000000000000000222 ***
I(X^2)       -0.02377848916  0.00126670865  -18.77187 < 0.000000000000000222 ***
---
Signif. codes:  0 '***' 0.001 '**' 0.01 '*' 0.05 '.' 0.1 ' ' 1

Residual standard error: 0.0216741925 on 97 degrees of freedom
Multiple R-squared:  0.790589444,    Adjusted R-squared:  0.786271701
F-statistic: 183.102461 on 2 and 97 DF,  p-value: < 0.00000000000000002220446

> lm.beta(PS)

Call:
lm(formula = Y ~ X + I(X^2))

Standardized Coefficients::
   (Intercept)                X            I(X^2)
 0.00000000000   5.79679685375  -6.03292108222
```

The differences between the estimates for the misspecified model and the properly specified model provide an illustration of how our interpretations can change when we violate the linearity assumption. The standardized coefficients from the misspecified model suggest that *X* has a negligible linear effect on *Y*. Nevertheless, this negligible linear effect does not represent reality. Our specification – reviewed above – dictates that *X* has a substantive and statistically significant curvilinear association with *Y*, where *X* has a positive influence on *Y* until the threshold, and then has a negative influence on Y.

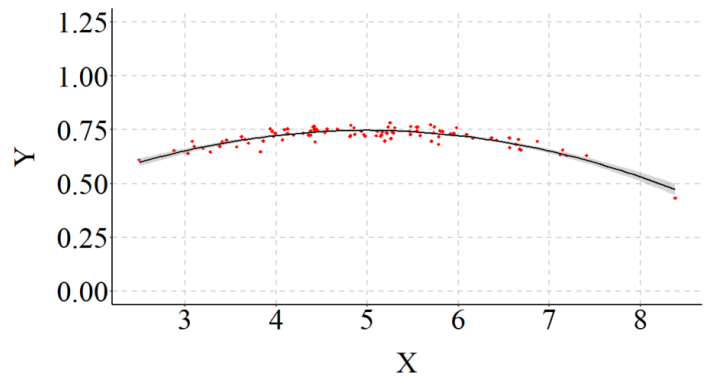

Taking this example to an extreme (specification in the code), we can see that the magnitude – or strength – of a linear regression line fit on curvilinear data is substantially weaker than reality (i.e., the specifications of the data). Of particular importance is the observation that *X* has minimal influence on *Y* in the misspecified model.

```
> set.seed(1001)
> X<-rnorm(n,5,1)
> X2<-X*X
> set.seed(32)
> Y<-5.00*X-.50*X2+.025*rnorm(n,5,1)
> # Misspecified model (Linear relationship between X and Y assumed)
> Mis<-lm(Y~X)
>
> summary(Mis)

Call:
lm(formula = Y ~ X)

Residuals:
        Min          1Q      Median          3Q         Max
-4.631367315 -0.337941262  0.360400075  0.582827465  0.688120764

Coefficients:
              Estimate   Std. Error  t value              Pr(>|t|)
(Intercept) 12.6844602464 0.3882829355 32.66809 < 0.0000000000000002 ***
X           -0.1419135464 0.0757298661 -1.87394              0.063917 .
---
Signif. codes:  0 '***' 0.001 '**' 0.01 '*' 0.05 '.' 0.1 ' ' 1

Residual standard error: 0.862376032 on 98 degrees of freedom
Multiple R-squared:  0.0345937282,  Adjusted R-squared:  0.0247426438
F-statistic: 3.51166702 on 1 and 98 DF,  p-value: 0.0639172523

> lm.beta(Mis)

Call:
lm(formula = Y ~ X)

Standardized Coefficients::
    (Intercept)                  X
```

```
 0.000000000000 -0.185993892912
```

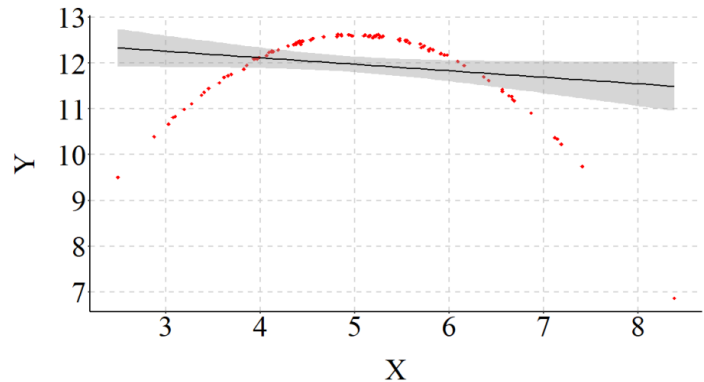

However, when we correctly assume the relationship is curvilinear and properly specify our model, the results align with reality.

```
> # Properly Specified Model (Curvilinear relationship between X and Y assumed)
> PS<-lm(Y~X+I(X^2))
>
> summary(PS)

Call:
lm(formula = Y ~ X + I(X^2))

Residuals:
         Min            1Q         Median            3Q           Max
-0.06629449249 -0.01476256564  0.00216027121  0.01361241435  0.03760637947

Coefficients:
                Estimate     Std. Error   t value              Pr(>|t|)
(Intercept)  0.15397661671  0.03328538416    4.62595            0.00001153 ***
X            4.98745870851  0.01316497818  378.84291 < 0.000000000000000222 ***
I(X^2)      -0.49877848916  0.00126670865 -393.75944 < 0.000000000000000222 ***
---
Signif. codes:  0 '***' 0.001 '**' 0.01 '*' 0.05 '.' 0.1 ' ' 1

Residual standard error: 0.0216741925 on 97 degrees of freedom
Multiple R-squared:  0.999396401,    Adjusted R-squared:  0.999383956
F-statistic: 80302.9033 on 2 and 97 DF,  p-value: < 0.00000000000000002220446

> lm.beta(PS)

Call:
lm(formula = Y ~ X + I(X^2))

Standardized Coefficients::
    (Intercept)                 X           I(X^2)
 0.00000000000    6.53663363645   -6.79400644477
```

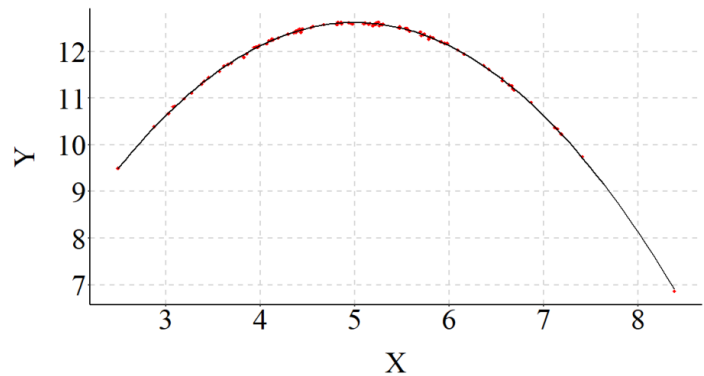

*Exceptions*

There are conditions where the shape of the association is curvilinear, but the misspecification of the relationship as linear only results in negligible reductions in the variation explained by the model. The example below provides an illustration of this situation. The relationship is primarily negative and it is quite difficult to visually evaluate where the curve occurs. Although only negligible differences exist in the variation explained by the dependent variable, the unstandardized coefficient and standard error is considerably different than the specification.

```
> set.seed(1001)
> X<-rnorm(n,5,1)
> X2<-X*X
> set.seed(32)
> Y<-.50*X-.50*X2+.025*rnorm(n,5,1)
> # Misspecified model (Linear relationship between X and Y assumed)
> Mis<-lm(Y~X)
>
> summary(Mis)

Call:
lm(formula = Y ~ X)

Residuals:
        Min          1Q      Median          3Q         Max
-4.631367315 -0.337941262  0.360400075  0.582827465  0.688120764

Coefficients:
                Estimate    Std. Error   t value             Pr(>|t|)
(Intercept) 12.6844602464  0.3882829355   32.66809 < 0.000000000000000222 ***
X           -4.6419135464  0.0757298661  -61.29568 < 0.000000000000000222 ***
---
Signif. codes:  0 '***' 0.001 '**' 0.01 '*' 0.05 '.' 0.1 ' ' 1

Residual standard error: 0.862376032 on 98 degrees of freedom
Multiple R-squared:  0.974579526,   Adjusted R-squared:  0.974320134
F-statistic: 3757.16024 on 1 and 98 DF,  p-value: < 0.00000000000000002220446

> lm.beta(Mis)

Call:
lm(formula = Y ~ X)

Standardized Coefficients::
   (Intercept)                X
```

```
 0.00000000000 -0.98720794474
```

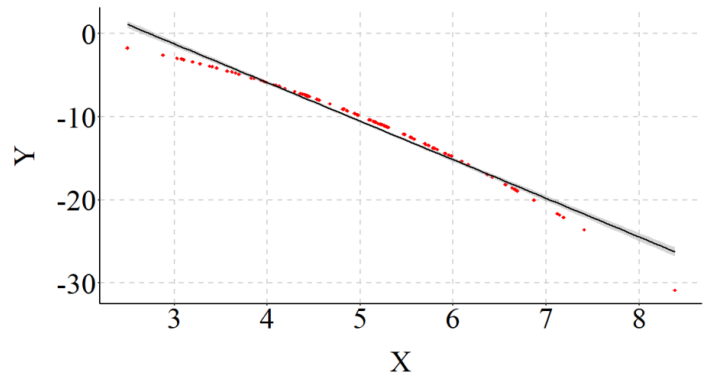

The results from a model assuming a curvilinear relationship between *X* and *Y* are presented below.

```
> # Properly Specified Model (Curvilinear relationship between X and Y assumed)
> PS<-lm(Y~X+I(X^2))
> 
> summary(PS)

Call:
lm(formula = Y ~ X + I(X^2))

Residuals:
         Min              1Q          Median              3Q             Max
-0.06629449249 -0.01476256564  0.00216027121  0.01361241435  0.03760637947

Coefficients:
                  Estimate     Std. Error    t value                    Pr(>|t|)
(Intercept)    0.15397661671  0.03328538416    4.62595                 0.00001153 ***
X              0.48745870851  0.01316497818   37.02693 < 0.000000000000000222 ***
I(X^2)        -0.49877848916  0.00126670865 -393.75944 < 0.000000000000000222 ***
---
Signif. codes:  0 '***' 0.001 '**' 0.01 '*' 0.05 '.' 0.1 ' ' 1

Residual standard error: 0.0216741925 on 97 degrees of freedom
Multiple R-squared:  0.999984106,   Adjusted R-squared:  0.999983779
F-statistic: 3051497.63 on 2 and 97 DF,  p-value: < 0.00000000000000002220446

> lm.beta(PS)

Call:
lm(formula = Y ~ X + I(X^2))

Standardized Coefficients::
    (Intercept)               X            I(X^2)
 0.000000000000   0.103669123728  -1.102459685779
```

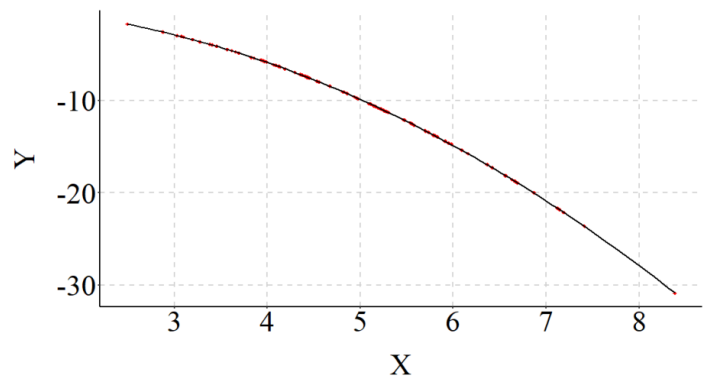

*Implications*
Considering all of the examples reviewed thus far, it appears that the magnitude of the curve for the association determines the degree to which the misspecification of a linear relationship influences the slope coefficient, standard error, standardized effect and $R^2$ value. Considering the potential effects, we should test if the association between the dependent variable and the independent variable of interest is linear or curvilinear before estimating our OLS regression models. This added step could help ensure that we don't misspecify our model.

**The Linearity Assumption in Multivariable Models**
In and of itself, it is relatively easy to test and address the misspecification of a linear association between two variables. The misspecification of linear associations, however, become substantially more difficult to test and address in multivariable models. When dealing with a large number of covariates, conducting bivariate tests of the structure of the association between each covariate and the dependent variable could take a large amount of time. Moreover, shared variation between constructs could invalidate the classifications determined by bivariate tests. This is all to say that we have an increased likelihood of misspefiying our model by assuming linear relationships. Though, why does it matter?

It matters because when we misspecify the structure of the association between a covariate and the dependent variable, the findings associated with the relationship of interest can be altered. The direction and magnitude of the bias introduced into the relationship, however, depends on the shape (i.e., standard bell curve or inverted bell curve) of the curvilinear association between the covariate and the dependent variable. To illustrate, let's go back to the drawing board (or the simulation code).

*Diminishing and Amplifying the Magnitude of the Relationship of Interest*
Again, we will start off by simulating our data. This time, however, we will call our first variable $C$ – for confounder – which is a normally distributed variable with a mean of 5 and a standard deviation of 1. Similar to the previous examples, we also multiple $C$ by $C$ to create our $C^2$ term. For the purpose of this example, we want a linear association between our confounding variable ($C$) and our independent variable ($X$). As such, $C^2$ is not included when simulating the data for $X$. The slope of the association between $C$ and $X$ is specified as .25, where a 1-point increase in $C$ corresponds to a .25-point increase in $X$. Importantly, to ensure that the confounder does not completely nullify the association between *X* and *Y*, we specify that a large proportion of *X* is normally distributed and not influenced by *C*.

To simulate the dependent variable (*Y*), we now designate a linear relationship between *X* and *Y*, where a 1-point increase in *X* is associated with a .25 increase in *Y*. In addition to specifying a linear relationship between *X* and *Y*, we inform the computer that the confounder (or *C*) has a curvilinear effect on *Y* ($Y = -4.0*C + .50*C^2$). Again, to ensure we don't estimate a perfectly specified model, we include some normally distributed error in the simulation of *Y*.

```
> set.seed(1001)
> C<-rnorm(n,5,1)
> C2<-C*C
> set.seed(884680)
> X<-.25*C+1*rnorm(n,5,1)
> set.seed(32)
> Y<-(.25*X)+4.00*C+(-.50*C2)+.025*rnorm(n,5,1)
```

To compare the slope coefficients for the association between *X* and *Y*, we estimate one model assuming a linear association between *C* and *Y* (the misspecified model) and one model assuming a curvilinear association between *C* and *Y* (the properly specified model). For the misspecified model, the slope coefficient for *Y* on *X* is closer to zero (*b* = .173) than reality (*b* = .250). When compared to the properly specified model, the standard error also appears to be quite large (Misspecified *SE* = .084) and the magnitude of the standardized effects is reduced (Misspecified *β* = .127). In this example, the magnitude of the association between *X* and *Y* was attenuated when we assumed that a linear relationship existed between *C* and *Y*.

```
> # Misspecified model (Linear relationship between C and Y assumed)
> Mis<-lm(Y~X+C)
>
> summary(Mis)

Call:
lm(formula = Y ~ X + C)

Residuals:
        Min          1Q      Median          3Q         Max
-4.624276409 -0.351528084  0.323115155  0.565400618  0.797768249

Coefficients:
                Estimate    Std. Error    t value             Pr(>|t|)
(Intercept) 13.0397899244  0.5461585065   23.87547 < 0.0000000000000002 ***
X            0.1726791657  0.0835168910    2.06760             0.041341 *
C           -1.1173989631  0.0802778032  -13.91915 < 0.0000000000000002 ***
---
Signif. codes:  0 '***' 0.001 '**' 0.01 '*' 0.05 '.' 0.1 ' ' 1

Residual standard error: 0.863005351 on 97 degrees of freedom
Multiple R-squared:  0.674374746,	Adjusted R-squared:  0.667660824
F-statistic: 100.444222 on 2 and 97 DF,  p-value: < 0.00000000000000002220446

> lm.beta(Mis)

Call:
lm(formula = Y ~ X + C)

Standardized Coefficients::
```

```
           (Intercept)                      X                         C
   0.000000000000    0.126896684596   -0.854274488166
```

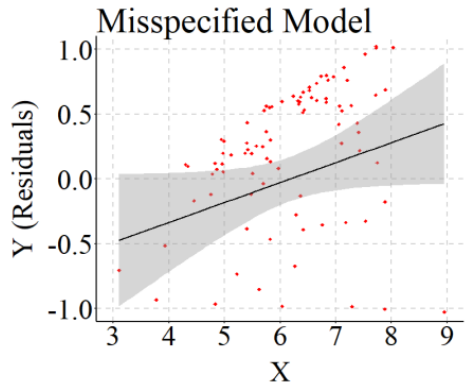
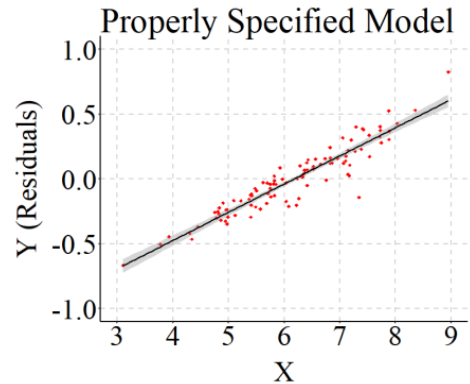

To reverse the effects, we will reverse the specification of the curvilinear association between *C* and *Y* ($Y = -4.00*C + .50*C^2$). Every other specification remains the same.

```
> set.seed(1001)
> C<-rnorm(n,5,1)
> C2<-C*C
> set.seed(884680)
> X<-.25*C+1*rnorm(n,5,1)
> set.seed(32)
> Y<-(.25*X)+(-4.00*C)+(.50*C2)+.025*rnorm(n,5,1)
```

When the same comparison is conducted, it can be observed that the slope coefficient and the standard error for the association between *X* and *Y* is higher than reality ($b = .321$; $SE = .084$; $p < .001$).

```
> # Misspecified model (Linear relationship between C and Y assumed)
> Mis<-lm(Y~X+C)
>
> summary(Mis)

Call:
lm(formula = Y ~ X + C)

Residuals:
         Min            1Q        Median            3Q           Max
-0.834434950  -0.561855959  -0.340962810   0.329700840   4.565545886

Coefficients:
                  Estimate    Std. Error    t value               Pr(>|t|)
(Intercept) -12.7624272229   0.5492533751  -23.23596 < 0.000000000000000222 ***
X             0.3206221640   0.0839901488    3.81738              0.00023763 ***
C             1.1195639077   0.0807327064   13.86754 < 0.000000000000000222 ***
---
Signif. codes:  0 '***' 0.001 '**' 0.01 '*' 0.05 '.' 0.1 ' ' 1

Residual standard error: 0.867895668 on 97 degrees of freedom
Multiple R-squared:  0.736643463,   Adjusted R-squared:  0.731213431
```

```
F-statistic: 135.660988 on 2 and 97 DF,  p-value: < 0.00000000000000002220446

> lm.beta(Mis)

Call:
lm(formula = Y ~ X + C)

Standardized Coefficients::
   (Intercept)              X              C
0.000000000000 0.210699184493 0.765415128876
```

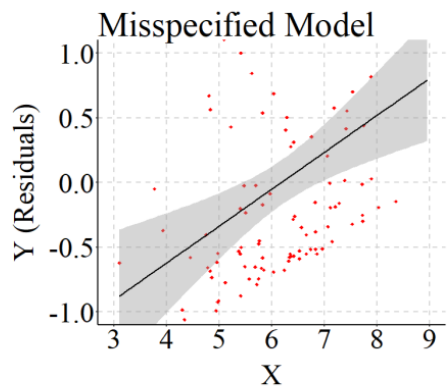
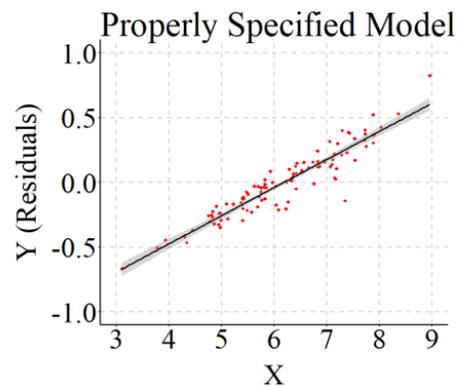

As both examples illustrated, the findings associated with *Y* on *X* is altered by misspecifying the relationship between *C* and *Y*. More broadly, by misspecifying the structure of the associations between confounding variables and the dependent variable we can nullify or amplify the relationship between the dependent variable and the independent variable of interest. These misspecifications can result in an increased probability of type 1 (when the relationship is amplified) or type 2 error (when the relationship is nullified) and lead to conclusions not supported by reality. This highlights how cautious we should be when assuming linear relationships between our covariates and the dependent variable.

**Conclusions**
Overall, this demonstration is intended to illustrate how violating the linearity assumption in an OLS regression model effects the results of the model. Unlike other assumptions (which will be reviewed later), we can influence the slope coefficients, standard errors, and standardized effects in our model by accidentally violating the linearity assumption. To avoid the potential pitfalls associated with violating the linearity assumption, we should do our due diligence and examine if the dependent variable is linearly related to all of the constructs included in our models. If we suspect that the specification of a linear association between two constructs is not supported by the data, we can readily remedy this issue by specifying a curvilinear association or relying on alternative modeling strategies.

# Entry 2: The Homoscedasticity Assumption

**Introduction**

Homoscedasticity and heteroscedasticity are not just difficult words to pronounce (homo·sce·das·tic·i·ty & hetero·sce·das·tic·i·ty), but also terms used to describe a key assumption about the distribution of error in regression models. In statistics we regularly make assumptions about the structure of error. For an OLS regression model, one of the primary assumptions is that the distribution of error – or scores – in the dependent variable will remain consistent across the distribution of scores on the independent variable. Using the proper term, we assume that the relationship between the dependent variable and independent variable are homoscedastic. If the relationship is heteroscedastic – the distribution of error in the dependent variable is not consistent across scores on the independent variable – we have violated the homoscedasticity assumption and biased the standard errors and the estimates conditional upon the standard errors. Using directed equation simulations, we will explore how violations of the homoscedasticity assumption alter the results of an OLS regression model.

**A Homoscedastic Relationship**

Simulating a homoscedastic relationship between two variables in *R* is relatively straightforward using traditional specifications. In the simplest of forms, specifying that *Y* is equal to the slope of *X* plus normally distributed values – see code – will create an association that satisfies the homoscedasticity assumption.

```
n<-200 # N-Size
set.seed(12) # Seed
X<-rnorm(n,10,1) # Normal Distribution of 200 cases with M = 10 & SD = 1
Y<-2*X+1*rnorm(n,10,2)
```

For illustrative purposes, however, we will rely on a more complicated specification for the simulation. The simulation begins by creating a categorical independent variable for 1000 cases. Scores on the categorical variable are specified to range between 1 and 5 with 200 respondents receiving each score (*n* in code). We begin the simulation of the dependent variable (*Y*) by specifying distinct constructs for each score on the independent variable. Simply, *Y1* is to *X* = 1 as *Y5* is to *X* = 5. Each of the five distinct *Y* constructs are simulated to equal an intercept (the value) plus normally distributed values with a mean of 10 and a standard deviation of 1. The only distinction between the five constructs is that the intercept increases by 2 for each subsequent *Y* specification. This 2-point increase in the intercept corresponds to the slope of the association. After specifying the five distinct *Y* constructs, we create an empty *Y* vector (column; *Y* = NA) and include *Y* alongside *X* in a data frame (df1 in the code). The empty *Y* vector (column; *Y* = NA) is then filled in with the values of the five distinct *Y* constructs when *X* equals the corresponding value. The resulting association between *X* and *Y* will be homoscedastic with a positive slope coefficient of 2.00.[i]

```
n<-200 # N-Size
set.seed(12) # Seed
X<-rep(c(1:5), each = n) # Repeat Each Value 200 times

set.seed(15)
Y1<-2+rnorm(n,10,1) # Y1 Specification
set.seed(15)
```

```
Y2<-4+rnorm(n,10,1)  # Y2 Specification
set.seed(15)
Y3<-6+rnorm(n,10,1)  # Y3 Specification
set.seed(15)
Y4<-8+rnorm(n,10,1)  # Y4 Specification
set.seed(15)
Y5<-10+rnorm(n,10,1) # Y5 Specification

Y<-rep(NA, each = 1000) # Empty Vector for Y with 1000 cells

df1<-data.frame(X,Y) # Dataframe containing X and Y (Empty Vector)
df1$Y[df1$X==1]<-Y1  # When X = 1, Y = Y1
df1$Y[df1$X==2]<-Y2  # When X = 2, Y = Y2
df1$Y[df1$X==3]<-Y3  # When X = 3, Y = Y3
df1$Y[df1$X==4]<-Y4  # When X = 4, Y = Y4
df1$Y[df1$X==5]<-Y5  # When X = 5, Y = Y5
```

A density plot and plot mapping the linear relationship of the association between *X* and *Y* were then created using ggplot2.

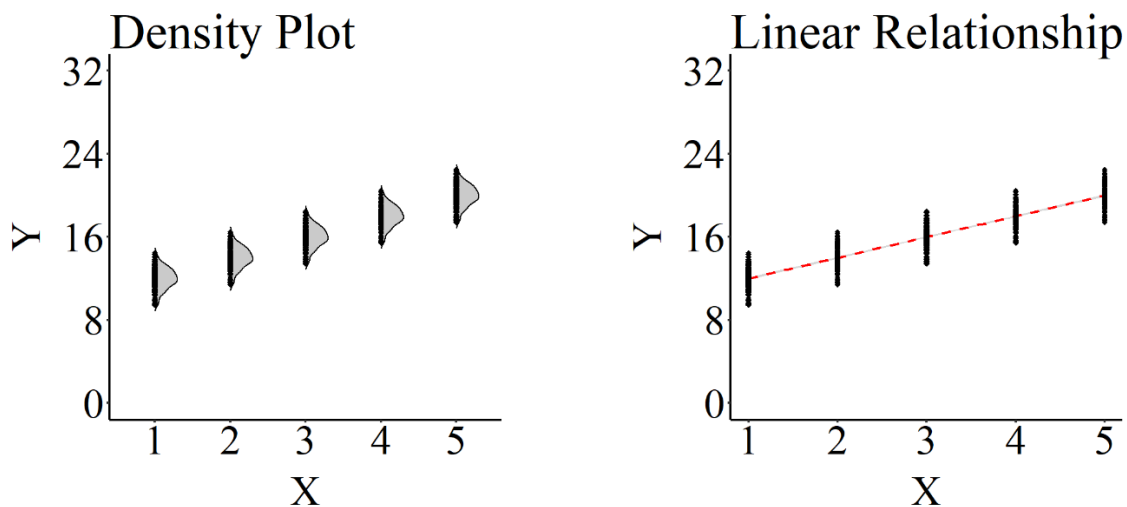

Furthermore, an OLS regression model was estimated where *Y* was regressed on *X* for the 1,000 cases. The estimates produced by the OLS regression model indicate that the slope coefficient was 2.000, the standard error of the association was .023, the *p*-value < .001, and the standardized slope coefficient was .937.

```
M<-lm(Y~X,data = df1) # Model of Y regressed on X

summary(M) # Summary of Results (estimate = b)

Call:
lm(formula = Y ~ X, data = df1)

Residuals:
         Min              1Q          Median              3Q             Max
```

```
-2.5823123488 -0.6183026704  0.0184348923  0.7294629685  2.4902740119

Coefficients:
              Estimate    Std. Error  t value               Pr(>|t|)
(Intercept) 9.9949045000 0.0778204590 128.43543 < 0.000000000000000222 ***
X           2.0000000000 0.0234637512  85.23786 < 0.000000000000000222 ***
---
Signif. codes:  0 '***' 0.001 '**' 0.01 '*' 0.05 '.' 0.1 ' ' 1

Residual standard error: 1.04933086 on 998 degrees of freedom
Multiple R-squared:  0.879227831,   Adjusted R-squared:  0.879106816
F-statistic: 7265.49319 on 1 and 998 DF,  p-value: < 0.0000000000000002220446

lm.beta(M) # Standardized Coefficient for Association

Call:
lm(formula = Y ~ X, data = df1)

Standardized Coefficients::
   (Intercept)              X
0.000000000000 0.937671493949
```

**Violating the Homoscedasticity Assumption**

Now that we have simulated a homoscedastic relationship and estimated the coefficients using an OLS regression model, we must violate the model assumptions by specifying a heteroscedastic relationship between the variables. The relationship between *X* and *Y* can be simulated as heteroscedastic by simply replicating the specification but varying the standard deviation value for the five distinct *Y* constructs (reminder: *Y1* is to *X* = 1 as *Y5* is to *X* = 5). Varying the standard deviation specification will ensure that the distribution of scores on the dependent variable is not consistent across scores of the independent variable. For illustrative purposes, three simulations with distinct specification patterns for the standard deviation were constructed. These patterns of specifying the standard deviations were: (1) an inconsistent specification, (2) an expanding specification, and (3) a tightening specification.

The inconsistent specification – as alluded to by the name – were just values selected by myself when specifying the standard deviation of the five distinct *Y* constructs. I, with no reason in mind, selected the values .2, 3, 1.4, 4, .4.[ii] After specifying the five distinct *Y* constructs, the remaining steps of the specification were repeated, the graphs were produced, and an OLS regression model was estimated.

```
n<-200 # N-Size
set.seed(12) # Seed
X<-rep(c(1:5), each = n) # Repeat Each Value 200 times

set.seed(15)
Y1<-2+rnorm(n,10,.2) # Y1 Specification
set.seed(15)
Y2<-4+rnorm(n,10,3) # Y2 Specification
set.seed(15)
Y3<-6+rnorm(n,10,1.4) # Y3 Specification
set.seed(15)
Y4<-8+rnorm(n,10,4) # Y4 Specification
set.seed(15)
Y5<-10+rnorm(n,10,.4) # Y5 Specification
```

```
Y<-rep(NA, each = 1000) # Empty Vector for Y with 1000 cells

df1<-data.frame(X,Y) # Dataframe containing X and Y (Empty Vector)
df1$Y[df1$X==1]<-Y1 # When X = 1, Y = Y1
df1$Y[df1$X==2]<-Y2 # When X = 2, Y = Y2
df1$Y[df1$X==3]<-Y3 # When X = 3, Y = Y3
df1$Y[df1$X==4]<-Y4 # When X = 4, Y = Y4
df1$Y[df1$X==5]<-Y5 # When X = 5, Y = Y5
```

As demonstrated in the plot below, and consistent with our specifications, the distribution of scores on *Y* across values of *X* is not consistent. Specifically, the distribution of scores on *Y* when *X* equals 1, 3, or 5 is substantively tighter than the distribution of scores on *Y* when *X* equals 2 or 4. The differences between the distributions for *Y* at different scores on *X* means that we properly specified a heteroscedastic association. The density plot provides the perfect visual illustration of heteroscedasticity. Excluding the visual differences in the plotted values, the slope of the regression line for the heteroscedastic relationship appears to be – and is almost – identical to the slope of the regression for the homoscedastic specification. The key difference between the *linear relationship* plots is the estimated standard error represented by the gray area.

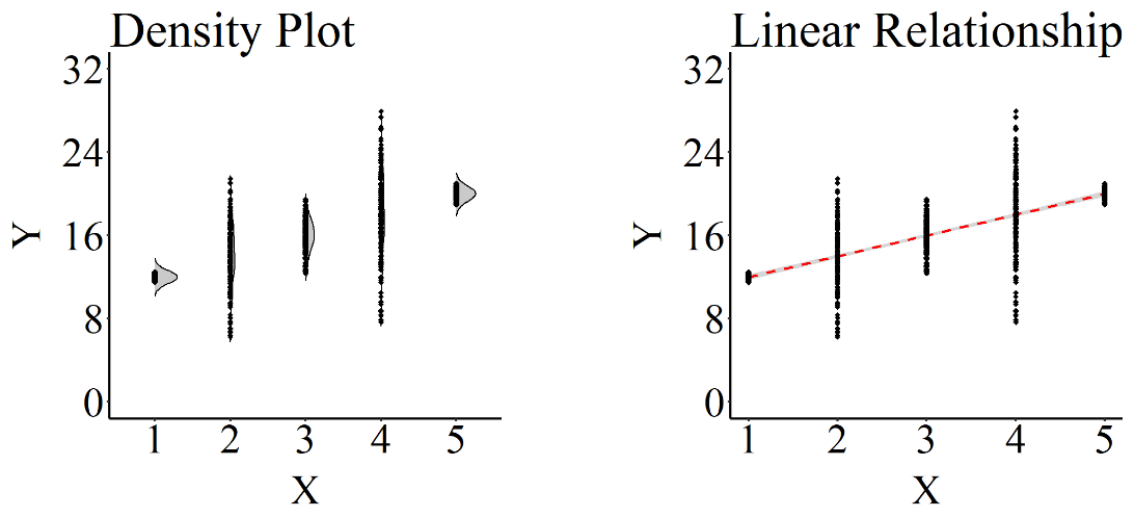

Although visually similar, the estimated standard error of the association between *X* and *Y* almost doubles to .054 (Non-violated *SE* = .023) and the standardized slope coefficient is reduced to .757 (Non-violated $\beta$ = .937).

```
M<-lm(Y~X,data = df1)

summary(M)

Call:
lm(formula = Y ~ X, data = df1)

Residuals:
         Min              1Q          Median              3Q             Max
```

```
-10.339746125  -0.597435168   0.015822992   0.652115189   9.950599318

Coefficients:
              Estimate    Std. Error  t value             Pr(>|t|)
(Intercept) 9.9929682100 0.1813742803 55.09584 < 0.000000000000000222 ***
X           1.9992866300 0.0546864031 36.55912 < 0.000000000000000222 ***
---
Signif. codes:  0 '***' 0.001 '**' 0.01 '*' 0.05 '.' 0.1 ' ' 1

Residual standard error: 2.4456503 on 998 degrees of freedom
Multiple R-squared:  0.572512111,   Adjusted R-squared:  0.572083767
F-statistic: 1336.56906 on 1 and 998 DF,  p-value: < 0.00000000000000002220446

lm.beta(M)

Call:
lm(formula = Y ~ X, data = df1)

Standardized Coefficients::
   (Intercept)                X
0.000000000000 0.756645300717
```

The complete results of the OLS regression model can be observed using the corresponding syntax file. The exact same pattern of results can be observed when the distribution of scores on Y expands as values on *X* increase and when the distribution of scores on Y tightens as values on *X* increase.

*Expanding Distribution*
The specification for this association and results of the OLS regression model can be observed in the syntax below.

```
n<-200 # N-Size
set.seed(12) # Seed
X<-rep(c(1:5), each = n) # Repeat Each Value 200 times

set.seed(15)
Y1<-2+rnorm(n,10,1) # Y1 Specification
set.seed(15)
Y2<-4+rnorm(n,10,1.7) # Y2 Specification
set.seed(15)
Y3<-6+rnorm(n,10,2.4) # Y3 Specification
set.seed(15)
Y4<-8+rnorm(n,10,3.1) # Y4 Specification
set.seed(15)
Y5<-10+rnorm(n,10,3.8) # Y5 Specification

Y<-rep(NA, each = 1000) # Empty Vector for Y with 1000 cells

df1<-data.frame(X,Y) # Dataframe containing X and Y (Empty Vector)
df1$Y[df1$X==1]<-Y1 # When X = 1, Y = Y1
df1$Y[df1$X==2]<-Y2 # When X = 2, Y = Y2
df1$Y[df1$X==3]<-Y3 # When X = 3, Y = Y3
df1$Y[df1$X==4]<-Y4 # When X = 4, Y = Y4
df1$Y[df1$X==5]<-Y5 # When X = 5, Y = Y5
```

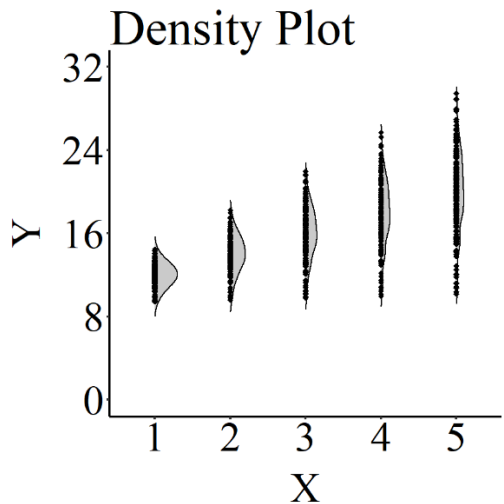
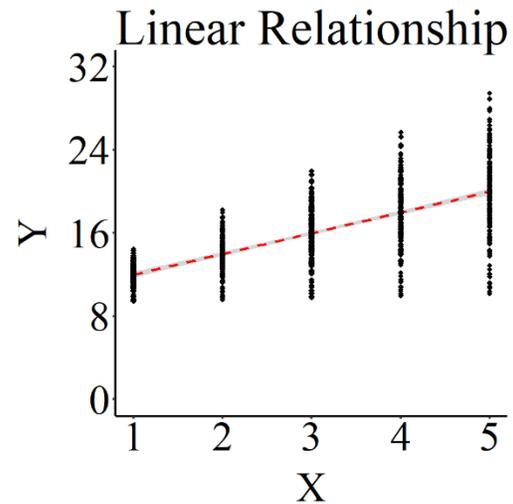

```
M<-lm(Y~X,data = df1)

summary(M)

Call:
lm(formula = Y ~ X, data = df1)

Residuals:
        Min          1Q      Median          3Q         Max
-9.812786926 -1.359686355  0.030912321  1.529959707  9.463041245

Coefficients:
              Estimate   Std. Error  t value             Pr(>|t|)
(Intercept) 9.9984713500 0.2020336623 49.48914 < 0.0000000000000000222 ***
X           1.9964331500 0.0609154412 32.77384 < 0.0000000000000000222 ***
---
Signif. codes:  0 '***' 0.001 '**' 0.01 '*' 0.05 '.' 0.1 ' ' 1

Residual standard error: 2.72422135 on 998 degrees of freedom
Multiple R-squared:  0.518368786,   Adjusted R-squared:  0.517886189
F-statistic: 1074.12483 on 1 and 998 DF,  p-value: < 0.00000000000000002220446

lm.beta(M)

Call:
lm(formula = Y ~ X, data = df1)

Standardized Coefficients::
   (Intercept)                    X
0.000000000000  0.719978322972
```

### *Tightening Distribution*
The specification for this association and results of the OLS regression model can be observed in the syntax below.

```
n<-200 # N-Size
```

```
set.seed(12) # Seed
X<-rep(c(1:5), each = n) # Repeat Each Value 200 times

set.seed(15)
Y1<--2+rnorm(n,10,3.8) # Y1 Specification
set.seed(15)
Y2<--4+rnorm(n,10,3.1) # Y2 Specification
set.seed(15)
Y3<--6+rnorm(n,10,2.4) # Y3 Specification
set.seed(15)
Y4<--8+rnorm(n,10,1.7) # Y4 Specification
set.seed(15)
Y5<--10+rnorm(n,10,1) # Y5 Specification

Y<-rep(NA, each = 1000) # Empty Vector for Y with 1000 cells

df1<-data.frame(X,Y) # Dataframe containing X and Y (Empty Vector)
df1$Y[df1$X==1]<--Y1 # When X = 1, Y = Y1
df1$Y[df1$X==2]<--Y2 # When X = 2, Y = Y2
df1$Y[df1$X==3]<--Y3 # When X = 3, Y = Y3
df1$Y[df1$X==4]<--Y4 # When X = 4, Y = Y4
df1$Y[df1$X==5]<--Y5 # When X = 5, Y = Y5
```

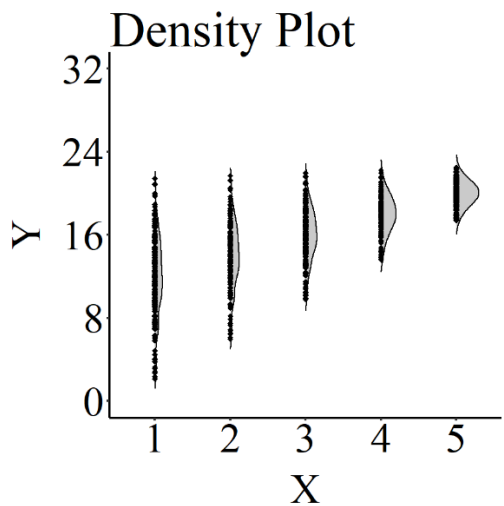
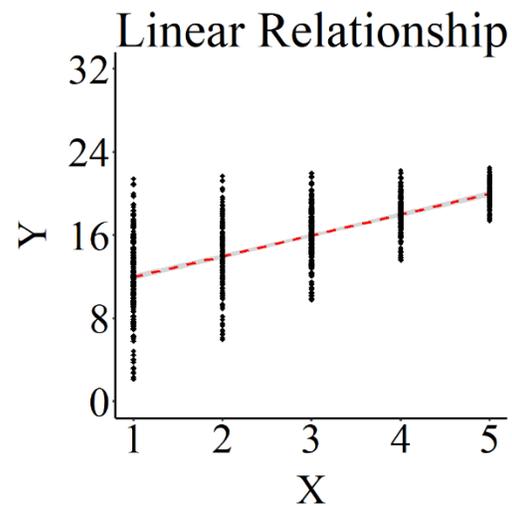

```
M<-lm(Y~X,data = df1)

> summary(M)

Call:
lm(formula = Y ~ X, data = df1)

Residuals:
        Min          1Q      Median          3Q         Max
-9.812786926 -1.359686355  0.030912321  1.529959707  9.463041245

Coefficients:
              Estimate    Std. Error   t value              Pr(>|t|)
(Intercept) 9.9770702501 0.2020336623  49.38321 < 0.0000000000000000222 ***
X           2.0035668500 0.0609154412  32.89095 < 0.0000000000000000222 ***
---
```

```
Signif. codes:  0 '***' 0.001 '**' 0.01 '*' 0.05 '.' 0.1 ' ' 1

Residual standard error: 2.72422135 on 998 degrees of freedom
Multiple R-squared:  0.520149565,   Adjusted R-squared:  0.519668752
F-statistic: 1081.81472 on 1 and 998 DF,  p-value: < 0.00000000000000002220446

> lm.beta(M)

Call:
lm(formula = Y ~ X, data = df1)

Standardized Coefficients::
   (Intercept)                X
0.000000000000 0.721213951981
```

Distinct from other assumptions, violating the homoscedasticity assumption will not alter the regression line estimated in an OLS regression model. Standard errors, however, will be upwardly biased, while hypothesis tests and the standardized slope coefficients will be downwardly biased. The same holds true in multivariable models.

**Violating the Homoscedasticity Assumption: Multivariable Models**
The simulation code provided below relies on the simpler method of simulating homoscedastic and heteroscedastic associations.[iii] The simulation begins by specifying that *C* (representing "confounder") is a random sample of 1,000 values ranging between 1 and 5 and *X* is equal to 1*C plus a random sample of 1,000 values ranging between 1 and 5. This specification ensures that *C* confounds the association between *X* and *Y*. The simulation is then specified to set scores on the dependent variable (*Y*) equal to 2*X plus 1*C plus 2* a normal distribution of values. The normal distribution of values is set to have a mean of 15 and a standard deviation of 1 for the homoscedastic specification, while the normal distribution of values is set to have a mean of 15 and a standard deviation equal to the square root of X*C raised to the 1.5 power for the heteroscedastic association. The results of violating the homoscedasticity assumption in a multivariable OLS regression model directly emulates the bivariate results reviewed above.

*Multivariable Homoscedastic Association*
The specification for this association and results of the OLS regression model can be observed in the syntax below.

```
n<-1000
set.seed(12)
C<-sample(c(1:5),size=n, replace = T)
set.seed(43)
X<-.1*C+2*sample(c(1:5),size=n, replace = T)
set.seed(29)
Y<-2*X+1*C+5*rnorm(n,15,1)
df1<-data.frame(X,Y,C)
```

To plot the bivariate association between X and Y, the Y-residuals were calculated. The y-axis on the density plot and linear relationship plot are the Y residuals.

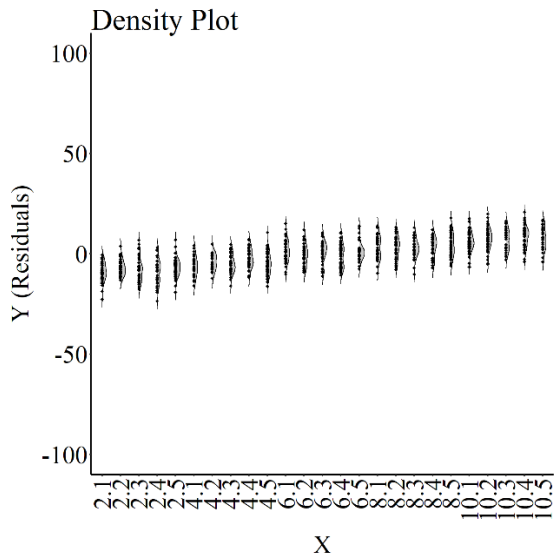
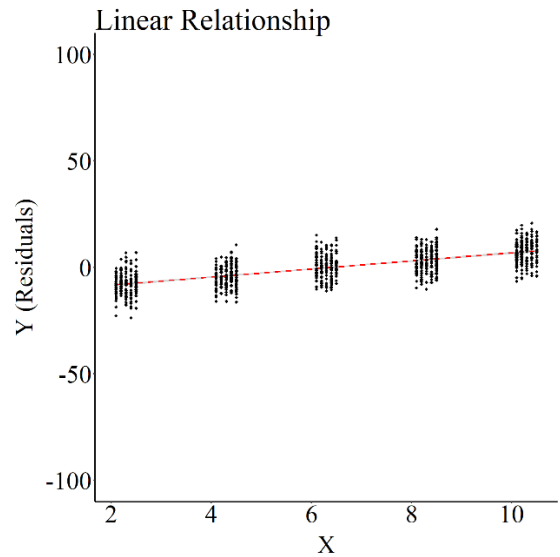

```
M<-lm(Y~X+C,data = df1)

summary(M)

Call:
lm(formula = Y ~ X + C, data = df1)

Residuals:
         Min             1Q         Median             3Q            Max
-16.069812664   -3.585479819   -0.038827947    3.816366395    15.775871868

Coefficients:
               Estimate     Std. Error   t value              Pr(>|t|)
(Intercept) 75.5558677681   0.5565820207  135.74975 < 0.0000000000000000222 ***
X            1.8937163747   0.0603399595   31.38412 < 0.0000000000000000222 ***
C            0.9686775249   0.1190453249    8.13705   0.0000000000000011985 ***
---
Signif. codes:  0 '***' 0.001 '**' 0.01 '*' 0.05 '.' 0.1 ' ' 1

Residual standard error: 5.35416372 on 997 degrees of freedom
Multiple R-squared:  0.512197284,   Adjusted R-squared:  0.511218743
F-statistic:   523.42953 on 2 and 997 DF,  p-value: < 0.00000000000000002220446

> lm.beta(M)

Call:
lm(formula = Y ~ X + C, data = df1)

Standardized Coefficients::
    (Intercept)                  X                  C
0.000000000000 0.694225418791 0.179993768445
```

### *Multivariable Heteroscedastic Association*
The specification for this association and results of the OLS regression model can be observed in the syntax below.

```
n<-1000
set.seed(12)
C<-sample(c(1:5),size=n, replace = T)
set.seed(43)
X<-.1*C+2*sample(c(1:5),size=n, replace = T)
set.seed(29)
Y<-2*X+1*C+2*rnorm(n,15,sqrt((X*C)^1.5))
df1<-data.frame(X,Y,C)
```

To plot the bivariate association between X and Y, the Y-residuals were calculated. The y-axis on the density plot and linear relationship plot are the Y residuals.

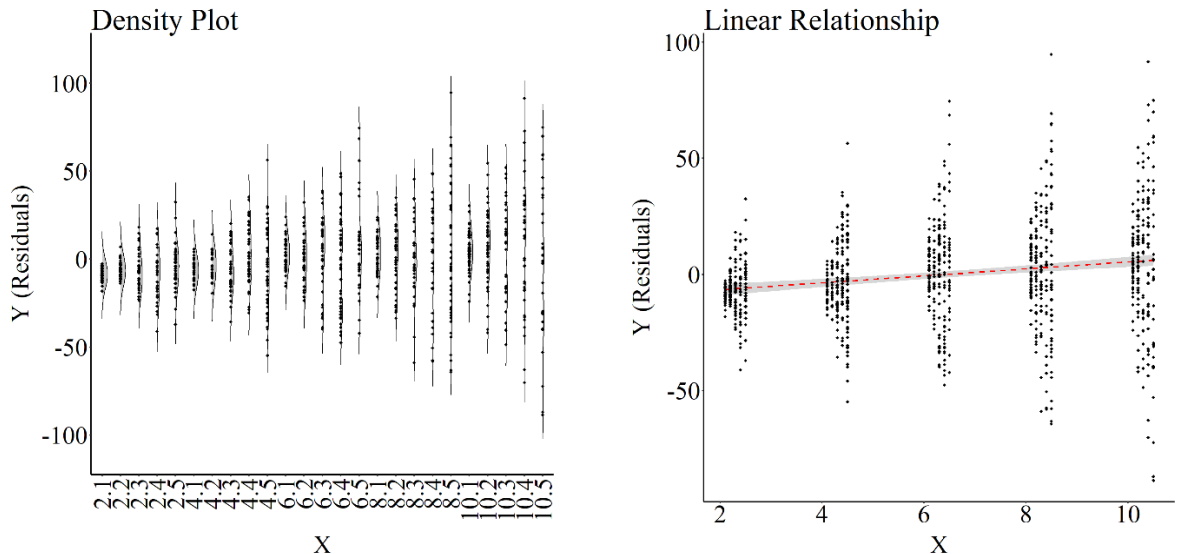

```
M<-lm(Y~X+C,data = df1)

summary(M)

Call:
lm(formula = Y ~ X + C, data = df1)

Residuals:
        Min          1Q      Median          3Q         Max
-94.82062676  -9.77572616  -0.25253024  10.81786931  91.49539700

Coefficients:
              Estimate    Std. Error   t value              Pr(>|t|)
(Intercept) 33.198290533  2.244998933  14.78766 < 0.000000000000000222 ***
X            1.497595236  0.243383975   6.15322       0.0000000010981 ***
C            0.614402524  0.480174741   1.27954                 0.201
---
Signif. codes:  0 '***' 0.001 '**' 0.01 '*' 0.05 '.' 0.1 ' ' 1

Residual standard error: 21.5962632 on 997 degrees of freedom
Multiple R-squared:  0.0379854149,  Adjusted R-squared:  0.0360555963
F-statistic: 19.6834119 on 2 and 997 DF,  p-value: 0.00000000413100503

> lm.beta(M)

Call:
lm(formula = Y ~ X + C, data = df1)
```

```
Standardized Coefficients::
    (Intercept)              X              C
0.0000000000000 0.1911444894215 0.0397477905866
```

**Conclusions**
Violations of the homoscedasticity assumption do bias estimates produced by OLS regression models. The estimates influenced by violations of the homoscedasticity assumption are the standard error, *t*-value, *p*-value, and the standardized slope coefficients. Although the degree of bias can vary dependent upon the inconsistencies in the width of the *Y* distribution across scores on *X*, the direction of the bias will remain consistent across all heteroscedastic associations. Specifically, the standard error and the *p*-value will be upwardly biased, while the *t*-value and the standardized slope coefficients will be downwardly biased.[iv] This means that estimating a hypothesis test on a heteroscedastic association will increase the likelihood of committing type 2 errors (failing to reject the null hypothesis when we should've rejected the null hypothesis) but not increase our likelihood of committing type 1 errors (rejecting the null hypothesis when we should've failed to reject the null hypothesis). In conclusion, although we should make efforts to reduce our likelihood of violating model assumptions, violating the homoscedasticity assumption can only serve to reduce the likelihood of rejecting the null hypothesis.

---

[i] Although more complicated, simulating the relationship in this manner provides the ability to alter the structure of the heteroscedasticity in an inconsistent manner, which is key to one of the subsequent examples.
[ii] Now looking back upon it, the values I *randomly* chose do create a pattern…
[iii] The specification process used for the bivariate associations cannot be easily translated to multivariable models.
[iv] The slope coefficient estimated when homoscedasticity is violated could slightly vary but not enough to fall outside of the confidence intervals of the slope coefficient when homoscedasticity is satisfied

# Entry 3: The Collinearity Assumption

**Introduction**

Regression based techniques are one of the most frequently used statistical approaches for hypothesis testing. The primary benefit of regression techniques is the ability to adjust estimates for the variation across multiple independent variables – otherwise known as the statistical control approach. This is extremely valuable when we are interested in the relationship between an independent variable and a dependent variable, but are unable to conduct a randomized controlled trial (i.e., true experiment). However, when introducing numerous statistical controls into a regression model we make assumptions about the structure of the variation across the independent variables. One assumption is that the independent variables are only highly correlated with the dependent variable. To state differently, we assume that multicollinearity – high levels of intercorrelation – does not exist across the independent variables. The current entry will illustrate how violating this assumption, and the degree of the violation, affect our estimates produced by OLS regression models.

**Defining Statistical Controls and Multicollinearity**

Before defining multicollinearity, it is important to review how the statistical control method works. In a traditional bivariate regression model, we are interested in estimating if variation in $X$ predicts variation in $Y$ and the strength of that prediction. Ordinary least squares regression – the most frequently used linear regression model – estimates an association by calculating a linear regression line where the predicted $x,y$ point is as close as possible to every observed $x,y$ point in the scatterplot. When a statistical control is introduced into a model, the linear regression line is altered by removing variation in $Y$ predicted by variation in $Z$ or variation in $X$ predicted by $Z$. This means that observed $x,y$ points that correspond with $y,z$ points or $x,z$ points are no longer included in the calculation of the linear regression line. To state differently, all of the cases where values of $Y$ correlate with values of $Z$ or values of $X$ correlate with values of $Z$ are removed from the calculation of the linear regression line. By removing these points – or cases – we can adjust our linear regression line for the influence of $Z$ and increase our ability to estimate the true association between $X$ and $Y$. Through this simplistic description of statistical controls, we can now define multicollinearity in the context of statistical controls. Multicollinearity is defined as the observation of a noticeable degree of correspondence between the variation in two or more independent variables. In this sense, multicollinearity is a special condition of a correlation, where $Z_1$ through $Z_k$ (or $X_1$ through $X_k$) are highly intercorrelated.

When using the statistical control method, we assume that variation in a control variable predicts variation in the dependent variable ($Y$), which by adjusting for increases our ability to estimate the true association between $X$ and $Y$. When including multiple control variables, we assume that the variation in $Z_1$ through $Z_k$ predicts variation in $Y$, but is not highly correlated with variation in $X$ or variation in $Z_1$ through $Z_k$. This assumption, however, can be violated when $Z_1$ through $Z_k$ are highly correlated with $X$ and highly correlated with each other. As demonstrated in this entry, violating this assumption means that we can not estimate the true association between $X$ and $Y$ when $Z_1$ through $Z_k$ are included as statistical controls in the regression model.

**Simulation 1: No Intercorrelation**

Let's start by discussing a simulation where variation in the control variables **only** predict variation in *Y*. Due to the need to simulate intercorrelated data, we will rely on the matrix-based simulation technique. Matrix-based simulations will create intercorrelated data that is consistent with a correlation matrix provided by the user. Specifically, we will simply provide *R* with a correlation matrix and *R* will simulate data that produces the specified correlation matrix. For the current simulation, our correlation matrix will be specified where the first row and column correspond to the dependent variable (*Y*), the second row and column correspond to the independent variable of interest (*X*), and the remaining four rows and columns will correspond to the control variables ($Z_1$-$Z_4$). Consistent with a correlation matrix, each cell in the diagonal of the matrix is specified as 1 and all corresponding cells above and below the diagonal are identical.

To create an association between *X* and *Y*, the cells corresponding to row 1 column 2 and row 2 column 1 will be specified to have a correlation of .50. The remaining cells in row 1 and column 1 are specified to have a correlation of .10, which means that $Z_1$-$Z_4$ will each possess a correlation of .10 with *Y*. The remaining cells receive a 0 to ensure that none of the independent variables – *X* and $Z_1$-$Z_4$ – are correlated with each other. After specifying the matrix, we simulate the data using the mvrnorm command (from the MASS package), where n is the number of cases (n = 1000), mu is the mean of the variables (mu = 0 for each variable), Sigma is the specified correlation matrix (Sigma = CM), and empirical informs R to not deviate from the specified correlation matrix (empirical = TRUE). After simulating the data, we use the tidyverse package to rename the columns to our desired specifications.

```
CM<-matrix(c(1,.50,.10,.10,.10,.10,
             .50,1,.0,.0,.0,.0,
             .10,.0,1,0,0,0,
             .10,.0,0,1,0,0,
             .10,.0,0,0,1,0,
             .10,.0,0,0,0,1), nrow = 6)

set.seed(56)
DF<-data.frame(mvrnorm(n = 1000, mu = rep(0, 6), Sigma = CM, empirical = TRUE))

DF<-DF %>%
  rename(
    Y = X1,
    X = X2,
    Z1 = X3,
    Z2 = X4,
    Z3 = X5,
    Z4 = X6
  )
```

For illustrative purposes, we then use GGally in combination with ggplot2 to create a visual correlation matrix. As observed, the simulated data has a correlation matrix identical to our specification of the data.

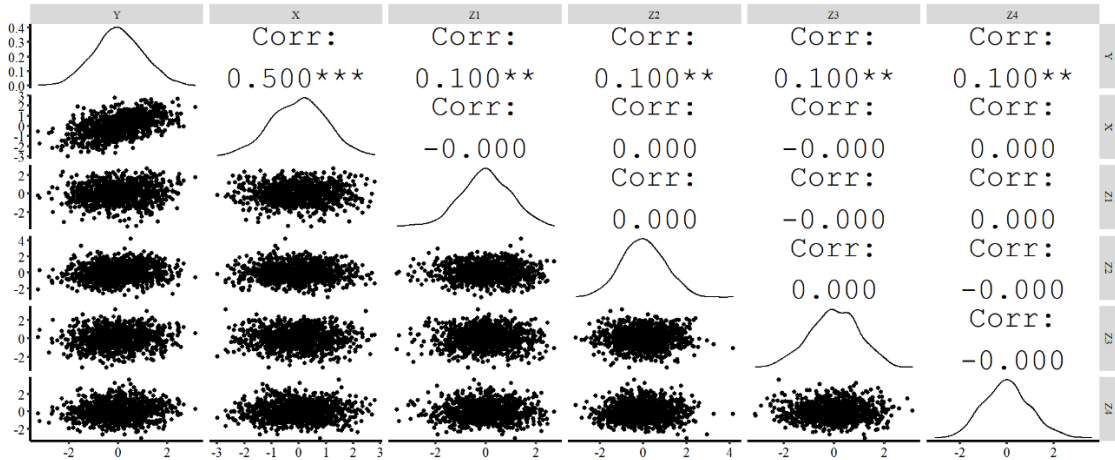

After producing the correlation matrix, we will estimate an ordinary least squares (OLS) regression model where $Y$ is regressed on $X$ and $Z_1$-$Z_4$. The results of the model are consistent with our specification, where the estimated slope of the association between $X$ and $Y$ is .50 and the estimated slope of the associations between $Z_1$-$Z_4$ and $Y$ are .10. The resulting slope coefficients are consistent with the specification of the simulation because the mean of every construct is 0 and the standard deviation is 1.

```
> M<-lm(Y~X+Z1+Z2+Z3+Z4, data = DF)
> 
> summary(M)

Call:
lm(formula = Y ~ X + Z1 + Z2 + Z3 + Z4, data = DF)

Residuals:
   Min     1Q Median     3Q    Max
-2.743 -0.569  0.010  0.522  2.264

Coefficients:
                          Estimate            Std. Error t value              Pr(>|t|)
(Intercept) -0.0000000000000000214  0.0267127577867098606    0.00               1.00000
X            0.5000000000000003331  0.0267261241912424390   18.71 < 0.0000000000000002 ***
Z1           0.0999999999999996170  0.0267261241912424286    3.74               0.00019 ***
Z2           0.1000000000000000611  0.0267261241912424320    3.74               0.00019 ***
Z3           0.1000000000000002137  0.0267261241912424494    3.74               0.00019 ***
Z4           0.0999999999999996586  0.0267261241912424112    3.74               0.00019 ***
---
Signif. codes:  0 '***' 0.001 '**' 0.01 '*' 0.05 '.' 0.1 ' ' 1

Residual standard error: 0.845 on 994 degrees of freedom
Multiple R-squared:  0.29,  Adjusted R-squared:  0.286
F-statistic: 81.2 on 5 and 994 DF,  p-value: <0.0000000000000002

> lm.beta(M)

Call:
lm(formula = Y ~ X + Z1 + Z2 + Z3 + Z4, data = DF)

Standardized Coefficients::
(Intercept)           X          Z1          Z2          Z3          Z4
        0.0         0.5         0.1         0.1         0.1         0.1
```

```
> ols_coll_diag(M)
Tolerance and Variance Inflation Factor
---------------------------------------
  Variables Tolerance VIF
1         X         1   1
2        Z1         1   1
3        Z2         1   1
4        Z3         1   1
5        Z4         1   1

Eigenvalue and Condition Index
------------------------------
  Eigenvalue Condition Index intercept        X       Z1        Z2       Z3     Z4
1          1               1         0 0.002163 0.942058 0.0009771 0.038598 0.0162
2          1               1         0 0.141103 0.040010 0.0202728 0.455664 0.3429
3          1               1         0 0.029184 0.000000 0.5818457 0.197989 0.1910
4          1               1         1 0.000000 0.000000 0.0000000 0.000000 0.0000
5          1               1         0 0.575826 0.009721 0.1730437 0.003303 0.2381
6          1               1         0 0.251724 0.008211 0.2238607 0.304446 0.2118
```

We then created a scatterplot of the association, including the linear regression line, between $X$ and $Y$ (Residuals) after adjusting for $Z_1$ through $Z_4$.[i]

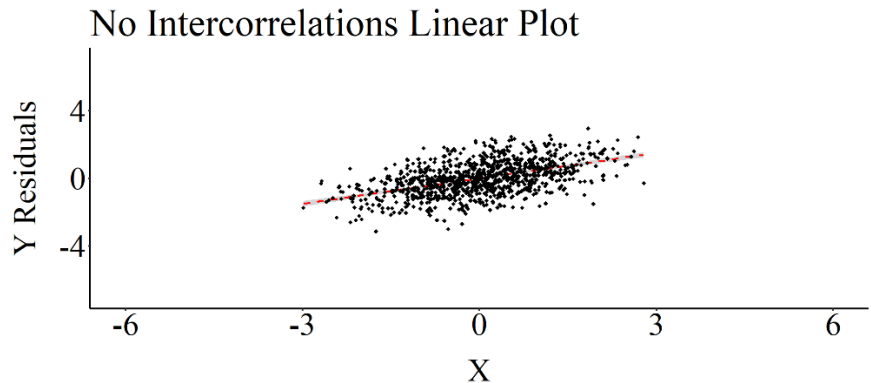

As demonstrated in the above simulation, the association between $X$ and $Y$ is not biased when $Z_1$-$Z_4$ are included in the regression model. This is because the independent variables – X and $Z_1$-$Z_4$ – satisfy the assumption that variation in the control variables **only** predict variation in $Y$.

**Simulation 2: Small Intercorrelation**

For our second simulation, we will replicate the first simulation except slightly increase the correlation among the independent variables (X and $Z_1$-$Z_4$). Specifically, in the first simulation we specified that X and $Z_1$-$Z_4$ were uncorrelated, while in the second simulation we indicate that X and $Z_1$-$Z_4$ are intercorrelated at .10 (i.e., every cell containing a zero in the previous simulation was specified to be .10 in the current simulation). Importantly, the relationship between $X$ and $Y$ was specified to be .50.

```
CM<-matrix(c(1,.50,.10,.10,.10,.10,
             .50,1,.10,.10,.10,.10,
             .10,.10,1,.10,.10,.10,
             .10,.10,.10,1,.10,.10,
             .10,.10,.10,.10,1,.10,
             .10,.10,.10,.10,.10,1), nrow = 6)

set.seed(56)
DF<-data.frame(mvrnorm(n = 1000, mu = rep(0, 6), Sigma = CM, empirical = TRUE))

DF<-DF %>%
  rename(
    Y = X1,
    X = X2,
    Z1 = X3,
    Z2 = X4,
    Z3 = X5,
    Z4 = X6
  )
```

The visual correlation matrix was then recreated. The correlation matrix of the simulated data perfectly matched our specifications.

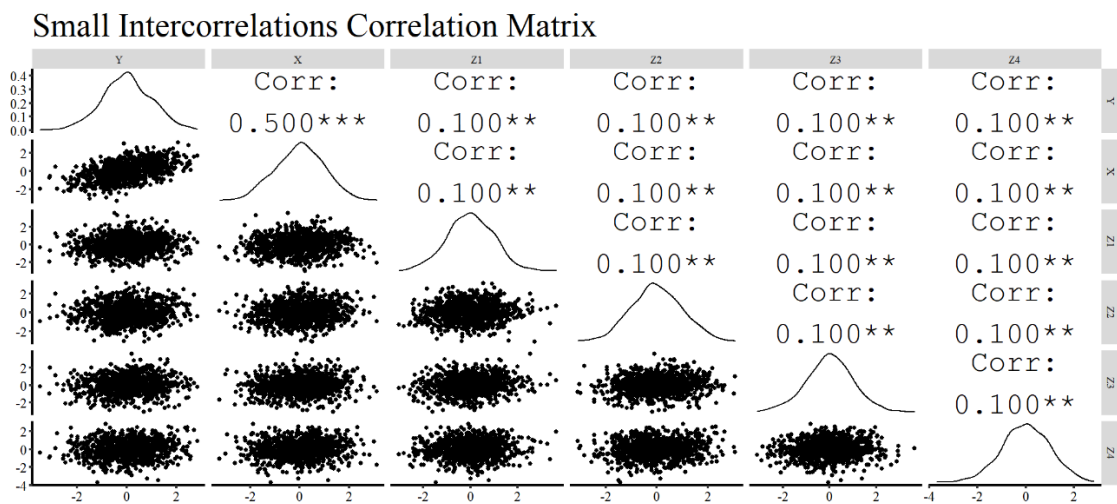

The OLS regression model was then reestimated and we recreated a scatterplot of the association, including the linear regression line, between *X* and *Y* (Residuals) after adjusting for $Z_1$ through $Z_4$. Although not a noticeable deviation from the specification, the estimated association between *X* and *Y* was .484 (not .50 like the specification).

```
> M<-lm(Y~X+Z1+Z2+Z3+Z4, data = DF)
>
> summary(M)

Call:
lm(formula = Y ~ X + Z1 + Z2 + Z3 + Z4, data = DF)

Residuals:
    Min      1Q  Median      3Q     Max
 -3.254  -0.593  -0.011   0.570   2.594
```

```
Coefficients:
                      Estimate              Std. Error t value             Pr(>|t|)
(Intercept) -0.0000000000000000471 0.0273092695948878271    0.00                 1.00
X            0.4841269841269854601 0.0277532433843363582   17.44 <0.0000000000000002 ***
Z1           0.0396825396825397497 0.0277532433843363478    1.43                 0.15
Z2           0.0396825396825404228 0.0277532433843363860    1.43                 0.15
Z3           0.0396825396825400828 0.0277532433843363513    1.43                 0.15
Z4           0.0396825396825398885 0.0277532433843363929    1.43                 0.15
---
Signif. codes:  0 '***' 0.001 '**' 0.01 '*' 0.05 '.' 0.1 ' ' 1

Residual standard error: 0.864 on 994 degrees of freedom
Multiple R-squared:  0.258, Adjusted R-squared:  0.254
F-statistic: 69.1 on 5 and 994 DF,  p-value: <0.0000000000000002

> lm.beta(M)

Call:
lm(formula = Y ~ X + Z1 + Z2 + Z3 + Z4, data = DF)

Standardized Coefficients::
(Intercept)           X          Z1          Z2          Z3          Z4
    0.00000     0.48413     0.03968     0.03968     0.03968     0.03968

> ols_coll_diag(M)
Tolerance and Variance Inflation Factor
---------------------------------------
  Variables Tolerance   VIF
1         X    0.9692 1.032
2        Z1    0.9692 1.032
3        Z2    0.9692 1.032
4        Z3    0.9692 1.032
5        Z4    0.9692 1.032

Eigenvalue and Condition Index
------------------------------
  Eigenvalue Condition Index intercept        X       Z1       Z2       Z3       Z4
1        1.4            1.000         0 0.138462 0.13846 0.13846 0.138462 0.13846154
2        1.0            1.183         1 0.000000 0.00000 0.00000 0.000000 0.00000000
3        0.9            1.247         0 0.734130 0.25952 0.03559 0.002818 0.04486554
4        0.9            1.247         0 0.114162 0.45025 0.18589 0.326582 0.00003915
5        0.9            1.247         0 0.011382 0.03619 0.52447 0.500425 0.00445671
6        0.9            1.247         0 0.001864 0.11558 0.11558 0.031714 0.81217706
```

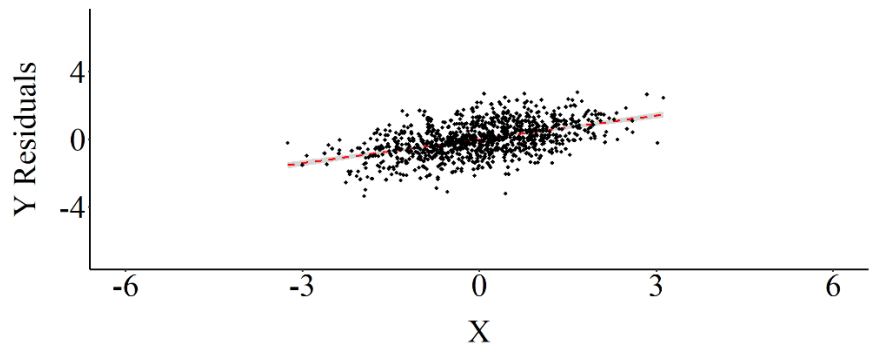

Small Intercorrelations Linear Plot

The effects, however, become more noticeable as the intercorrelation between the independent variables increase.

## Simulations 3-5: Moderately Small, Moderate, and High Intercorrelation

For the remaining simulations – moderately small, moderate, and high –, we increased the correlations amongst the independent variables to .25, .50, and .75 (respectively). The code for the remaining simulations is identical to the previous simulations except for altering the correlation matrix specification amongst the independent variables.

### *Moderately Small Intercorrelation Simulation Code*

```
CM<-matrix(c(1,.50,.10,.10,.10,.10,
             .50,1,.25,.25,.25,.25,
             .10,.25,1,.25,.25,.25,
             .10,.25,.25,1,.25,.25,
             .10,.25,.25,.25,1,.25,
             .10,.25,.25,.25,.25,1), nrow = 6)

set.seed(56)
DF<-data.frame(mvrnorm(n = 1000, mu = rep(0, 6), Sigma = CM, empirical = TRUE))

DF<-DF %>%
  rename(
    Y = X1,
    X = X2,
    Z1 = X3,
    Z2 = X4,
    Z3 = X5,
    Z4 = X6
  )
```

### *Moderate Intercorrelation Simulation Code*

```
CM<-matrix(c(1,.50,.10,.10,.10,.10,
             .50,1,.50,.50,.50,.50,
             .10,.50,1,.50,.50,.50,
             .10,.50,.50,1,.50,.50,
             .10,.50,.50,.50,1,.50,
             .10,.50,.50,.50,.50,1), nrow = 6)

set.seed(56)
DF<-data.frame(mvrnorm(n = 1000, mu = rep(0, 6), Sigma = CM, empirical = TRUE))

DF<-DF %>%
  rename(
    Y = X1,
    X = X2,
    Z1 = X3,
    Z2 = X4,
    Z3 = X5,
    Z4 = X6
  )
```

### *High Intercorrelation Simulation Code*

```
CM<-matrix(c(1,.50,.10,.10,.10,.10,
             .50,1,.75,.75,.75,.75,
             .10,.75,1,.75,.75,.75,
             .10,.75,.75,1,.75,.75,
             .10,.75,.75,.75,1,.75,
             .10,.75,.75,.75,.75,1), nrow = 6)

set.seed(56)
```

```
DF<-data.frame(mvrnorm(n = 1000, mu = rep(0, 6), Sigma = CM, empirical = TRUE))

DF<-DF %>%
  rename(
    Y = X1,
    X = X2,
    Z1 = X3,
    Z2 = X4,
    Z3 = X5,
    Z4 = X6
  )
```

The correlation matrices were then produced, the OLS regression models were estimated, and scatterplots of the association between *X* and *Y* (Residuals) after adjusting for $Z_1$ through $Z_4$ were created. The results of the regression model estimated on the simulated data with moderately small intercorrelation demonstrated only a slight upward bias (Estimated *b* = .52; True *b* = .50). Nevertheless, the regression model estimated using the simulated data with moderate intercorrelation demonstrated a .20 upward departure from the true slope coefficient (Estimated *b* = .70; True *b* = .50) and the regression model estimated using the simulated data with high intercorrelation demonstrated a .82 upward departure from the true slope coefficient (Estimated *b* = 1.32; True *b* = .50).

*Moderately Small Intercorrelation Results*

```
> M<-lm(Y~X+Z1+Z2+Z3+Z4, data = DF)
>
> summary(M)

Call:
lm(formula = Y ~ X + Z1 + Z2 + Z3 + Z4, data = DF)

Residuals:
   Min     1Q Median     3Q    Max
-3.612 -0.588 -0.009  0.562  2.712

Coefficients:
                        Estimate            Std. Error t value           Pr(>|t|)
(Intercept) -0.0000000000000000249  0.0274243976326386460    0.00               1.00
X            0.5166666666666698271  0.0296365569627647200   17.43 <0.0000000000000002 ***
Z1          -0.0166666666666668468  0.0296365569627647027   -0.56               0.57
Z2          -0.0166666666666665277  0.0296365569627646611   -0.56               0.57
Z3          -0.0166666666666659656  0.0296365569627647166   -0.56               0.57
Z4          -0.0166666666666670966  0.0296365569627646680   -0.56               0.57
---
Signif. codes:  0 '***' 0.001 '**' 0.01 '*' 0.05 '.' 0.1 ' ' 1

Residual standard error: 0.867 on 994 degrees of freedom
Multiple R-squared:  0.252, Adjusted R-squared:  0.248
F-statistic: 66.9 on 5 and 994 DF,  p-value: <0.0000000000000002

> lm.beta(M)

Call:
lm(formula = Y ~ X + Z1 + Z2 + Z3 + Z4, data = DF)

Standardized Coefficients::
(Intercept)           X          Z1          Z2          Z3          Z4
    0.00000     0.51667    -0.01667    -0.01667    -0.01667    -0.01667

> ols_coll_diag(M)
Tolerance and Variance Inflation Factor
---------------------------------------
  Variables Tolerance    VIF
```

```
1        X    0.8571 1.167
2       Z1    0.8571 1.167
3       Z2    0.8571 1.167
4       Z3    0.8571 1.167
5       Z4    0.8571 1.167

Eigenvalue and Condition Index
------------------------------
  Eigenvalue Condition Index intercept         X        Z1       Z2       Z3        Z4
1      2.00             1.000         0 0.0857143 0.08571  0.08571  0.08571 0.0857143
2      1.00             1.414         1 0.0000000 0.00000  0.00000  0.00000 0.0000000
3      0.75             1.633         0 0.0003314 0.10090  0.45241  0.10440 0.4848087
4      0.75             1.633         0 0.6203864 0.36620  0.13077  0.02507 0.0004327
5      0.75             1.633         0 0.2201670 0.41516  0.04601  0.07849 0.3830299
6      0.75             1.633         0 0.0734010 0.03202  0.28509  0.70633 0.0460144
```

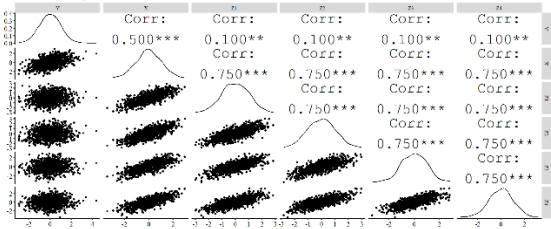
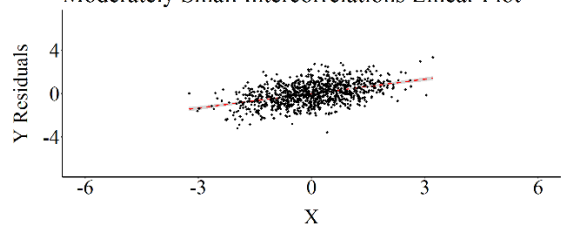

*Moderate Intercorrelation Results*

```
> M<-lm(Y~X+Z1+Z2+Z3+Z4, data = DF)
>
> summary(M)

Call:
lm(formula = Y ~ X + Z1 + Z2 + Z3 + Z4, data = DF)

Residuals:
    Min      1Q  Median      3Q     Max
-2.9677 -0.5623 -0.0094  0.5669  3.0278

Coefficients:
                      Estimate            Std. Error t value          Pr(>|t|)
(Intercept) -0.000000000000000153  0.026333834224239020    0.00            1.0000
X            0.699999999999995182  0.034013844973766146   20.58 <0.0000000000000002 ***
Z1          -0.099999999999999326  0.034013844973766215   -2.94            0.0034 **
Z2          -0.099999999999998465  0.034013844973766263   -2.94            0.0034 **
Z3          -0.099999999999998618  0.034013844973766250   -2.94            0.0034 **
Z4          -0.099999999999997577  0.034013844973766284   -2.94            0.0034 **
---
Signif. codes:  0 '***' 0.001 '**' 0.01 '*' 0.05 '.' 0.1 ' ' 1

Residual standard error: 0.833 on 994 degrees of freedom
Multiple R-squared:  0.31,   Adjusted R-squared:  0.307
F-statistic: 89.3 on 5 and 994 DF,  p-value: <0.0000000000000002

> lm.beta(M)

Call:
lm(formula = Y ~ X + Z1 + Z2 + Z3 + Z4, data = DF)

Standardized Coefficients::
(Intercept)           X          Z1          Z2          Z3          Z4
        0.0         0.7        -0.1        -0.1        -0.1        -0.1
```

```
> ols_coll_diag(M)
Tolerance and Variance Inflation Factor
---------------------------------------
  Variables Tolerance   VIF
1         X       0.6 1.667
2        Z1       0.6 1.667
3        Z2       0.6 1.667
4        Z3       0.6 1.667
5        Z4       0.6 1.667

Eigenvalue and Condition Index
------------------------------
  Eigenvalue Condition Index intercept        X       Z1      Z2      Z3         Z4
1        3.0             1.000         0 0.040000 0.04000 0.04000 0.04000 0.04000000
2        1.0             1.732         1 0.000000 0.00000 0.00000 0.00000 0.00000000
3        0.5             2.449         0 0.472281 0.59491 0.09748 0.03323 0.00210347
4        0.5             2.449         0 0.008781 0.12671 0.34900 0.71545 0.00005152
5        0.5             2.449         0 0.403519 0.13903 0.48748 0.16151 0.00846069
6        0.5             2.449         0 0.075418 0.09935 0.02604 0.04981 0.94938432
```

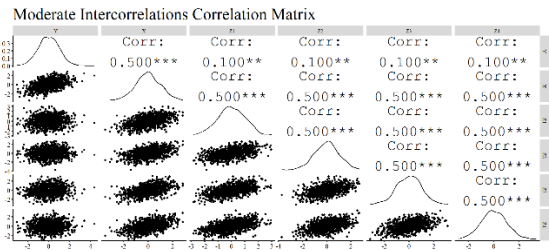
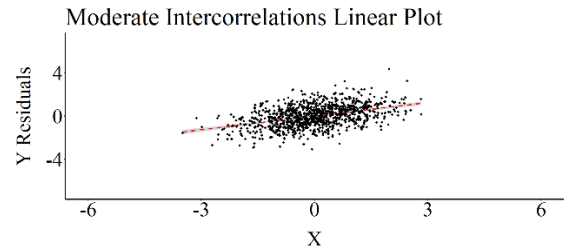

## *High Intercorrelation Results*

```
> M<-lm(Y~X+Z1+Z2+Z3+Z4, data = DF)
>
> summary(M)

Call:
lm(formula = Y ~ X + Z1 + Z2 + Z3 + Z4, data = DF)

Residuals:
    Min      1Q  Median      3Q     Max
-2.3608 -0.4411 -0.0101  0.4459  2.2218

Coefficients:
                        Estimate            Std. Error t value            Pr(>|t|)
(Intercept)  0.0000000000000835  0.0212073337795262128    0.00                   1
X            1.3249999999999657607  0.0382511950603019180   34.64 < 0.0000000000000002 ***
Z1          -0.2749999999999924172  0.0382511950603023065   -7.19       0.0000000000013 ***
Z2          -0.2749999999999905853  0.0382511950603023343   -7.19       0.0000000000013 ***
Z3          -0.2749999999999913625  0.0382511950603023482   -7.19       0.0000000000013 ***
Z4          -0.2749999999999901967  0.0382511950603023204   -7.19       0.0000000000013 ***
---
Signif. codes:  0 '***' 0.001 '**' 0.01 '*' 0.05 '.' 0.1 ' ' 1

Residual standard error: 0.671 on 994 degrees of freedom
Multiple R-squared:  0.552, Adjusted R-squared:  0.55
F-statistic:  245 on 5 and 994 DF,  p-value: <0.0000000000000002

> lm.beta(M)

Call:
lm(formula = Y ~ X + Z1 + Z2 + Z3 + Z4, data = DF)
```

```
Standardized Coefficients::
(Intercept)           X          Z1          Z2          Z3          Z4
      0.000       1.325      -0.275      -0.275      -0.275      -0.275

> ols_coll_diag(M)
Tolerance and Variance Inflation Factor
---------------------------------------
  Variables Tolerance  VIF
1         X    0.3077 3.25
2        Z1    0.3077 3.25
3        Z2    0.3077 3.25
4        Z3    0.3077 3.25
5        Z4    0.3077 3.25

Eigenvalue and Condition Index
------------------------------
  Eigenvalue Condition Index intercept         X        Z1       Z2        Z3        Z4
1       4.00               1         0 0.0153846   0.01538  0.01538  0.015385  0.015385
2       1.00               2         1 0.0000000   0.00000  0.00000  0.000000  0.000000
3       0.25               4         0 0.4223173   0.44481  0.01270  0.229181  0.121764
4       0.25               4         0 0.5613878   0.05807  0.03687  0.572729  0.001715
5       0.25               4         0 0.0002569   0.19251  0.18745  0.004139  0.846421
6       0.25               4         0 0.0006533   0.28924  0.74759  0.178568  0.014715
```

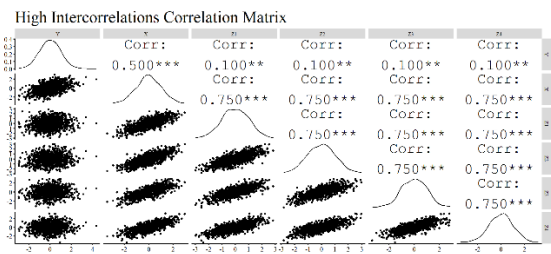

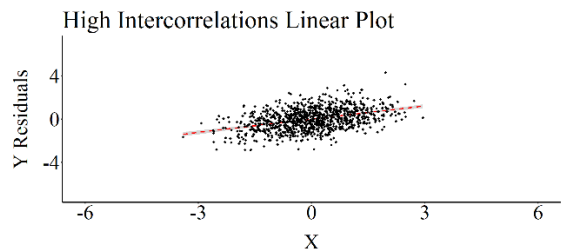

**Conclusion**

As demonstrated by the five simulations, multicollinearity amongst the independent variables can upwardly bias the estimated association between *X* and *Y*. This upward bias occurred because high intercorrelation between the independent variables creates a condition where the variation in *Y* predicted by *X* is difficult to discern from the variation in *Y* predicted by $Z_1$-$Z_4$. Downwardly biased estimates are also possible when we violate this assumption. For instance, while our focus was placed primarily on *X,* the association between *Y* and $Z_1$-$Z_4$ were downwardly biased in the regression model estimated using the high intercorrelation data (*b* = -.27 for all constructs). As demonstrated, however, the magnitude of the intercorrelations directly influence the amount of bias that will be introduced into the estimates. The regression models estimated on the datasets with intercorrelations below .25 produced slope coefficients that slightly deviated from the true slope coefficient, while the regression models estimated on the datasets with intercorrelations above .25 produced slope coefficients with more substantial departures from the true slope coefficients. Overall, multicollinearity can bias the slope coefficients from a regression model.

*Important Information*

No simulated example is perfect and the examples selected to demonstrate the effects of multicollinearity in the current entry do not represent real life data. As such, the examples leave readers desiring how intercorrelated independent variables could introduce bias into their regression models. Fret not! The simulation technique – matrix-based simulations – used throughout the current entry can simulate data based on any correlation matrix including, but not limited to, real correlation matrices from your data. Simply, (1) set your *X* and *Y* correlation, (2) plug in the observed correlation matrix, (3) set your means and standard deviations, and (4) simulate your data. After simulating your data, you can observe the degree of bias in the slope coefficients with the real data intercorrelations and compare it to the true slope correlation. This can be used to demonstrate the amount of bias in the slope coefficients when multicollinearity becomes a problem or to illustrate that the intercorrelations amongst the independent variables is not biasing the slope coefficients. I hope you find value in this extra information.

---

[i] This plot was created by regressing *Y* on $Z_1$-$Z_4$ and calculating the residuals of that regression model. The residuals of *Y* were then used as the dependent variable in the scatterplot.

# Entry 4: The Normality Assumption: Outliers

**Introduction**
An outlier is a case, datapoint, or score meaningfully removed from the mass of the distribution as to be recognizably different from the remainder of cases, datapoints, or scores. Consistent with this definition, outliers are conditioned upon the observed data and can vary between samples. For example, an individual with 10 armed robberies might be an outlier in a nationally representative sample, but will likely be well within the distribution in a sample of high-risk incarcerated individuals. Consistent with the inability to explicitly define an outlier, we make no assumptions about the existence of outliers when estimating a regression model. We do, however, assume that the residual error – the difference between the predicted value ($Y_{hat}$) and the observed value – of the dependent variable ($Y$) will be normally distributed (i.e., the normality assumption). The inclusion of particular outliers in a model could violate the normality assumption and introduce bias into the estimates derived from a linear regression model. The current entry reviews the effects of outliers on the estimates derived from an ordinary least squares (OLS) regression model.

**Normality Assumption**
Before defining the normality assumption, it is important to discuss what residuals refer to in the context of a regression model. A regression model – with different estimation procedures depending upon the model – estimates the intercept (the value of $Y$ when $X$ is zero) and the slope coefficient (the change in $Y$ corresponding to a change in $X$) that best fits the data. Fundamentally, this is a single line used to illustrate the **predicted value** of the dependent variable ($Y_{hat}$) at any score on the independent variable ($X$). This single regression line, however, is unlikely to make the appropriate prediction for the dependent variable ($Y$) at every value on the independent variable ($X$). Explicitly, $Y_{hat}$ (the predicted value of $Y$) for any one X-value is likely not going to equal the value of $Y$ observed in the data at the same X-value. The difference between the value of $Y$ observed in the data and $Y_{hat}$ at a specified value on the independent variable ($X$) is a residual. A residual is simply a method of quantifying the amount of unpredicted variation – or error – remaining after the regression model is estimated. Using this calculation, we can find the residual for each case, at each value of $X$, and build a distribution of residuals.

We have satisfied the normality assumption when the residuals for the dependent variable are normally distributed across the values on the independent variable (i.e., the traditional bell-curve distribution). When the residuals for the dependent variable are not normally distributed across the values on the independent variable we have violated the normality assumption. Although the normality assumption can be violated by various data peculiarities, the inclusion of outliers in our regression models represents one way we can violate the normality assumption.

The existence of outliers in a dataset – observations defined as substantively different than every other observation – can create a condition where the residuals of the dependent variable are not normally distributed across the values on the independent variable. Only certain outliers, however, can violate the normality assumption and alter the results of the regression model. In a bivariate regression model three types of outliers can exist: outlier on $X$ only, outlier on $Y$ only, and an outlier on $X$ and $Y$. The remainder of the entry will explore how these three types of outliers can affect the slope coefficients, standard errors, and hypothesis tests conducted in an OLS regression model.

**No Outliers**

As we begin with every simulated exploration, let us start out by specifying a directed equation simulation that does not contain any outliers. For this example, we simply specify that we want a sample of 100 cases (N in code) with unique identifiers (ID in code). Additionally, we specify that *Y* is equal to .60 multiplied by *X* (a normally distributed variable with a mean of 10 and a SD of 1) plus .50 multiplied by a normally distributed variable with a mean of 10 and a SD of 1 (representing additional random error).[i] After specifying the data, we than create a data frame, estimate a linear OLS regression model, and plot the bivariate linear association between *X* and *Y*. The slope of the association was equal to .5934, the standard error was .0582, the standardized slope coefficient was .7174, and the association was significant at $p < .001$.

```
> N<-100
> ID<-c(1:100)
> set.seed(32)
> X<-rnorm(N,10,1)
> set.seed(15)
> Y<-.60*X+.50*rnorm(N,10,1)
>
> Data<-data.frame(ID,Y,X)
>
> M<-lm(Y~X, data = Data)
> summary(M)

Call:
lm(formula = Y ~ X, data = Data)

Residuals:
    Min      1Q  Median      3Q     Max
-1.2695 -0.3101 -0.0388  0.3855  1.1858

Coefficients:
            Estimate Std. Error t value            Pr(>|t|)
(Intercept)   5.1165     0.5806    8.81   0.000000000000045 ***
X             0.5934     0.0582   10.19 < 0.0000000000000002 ***
---
Signif. codes:  0 '***' 0.001 '**' 0.01 '*' 0.05 '.' 0.1 ' ' 1

Residual standard error: 0.499 on 98 degrees of freedom
Multiple R-squared:  0.515, Adjusted R-squared:  0.51
F-statistic:   104 on 1 and 98 DF,  p-value: <0.0000000000000002

> lm.beta(M)

Call:
lm(formula = Y ~ X, data = Data)

Standardized Coefficients::
(Intercept)           X
     0.0000      0.7174
```

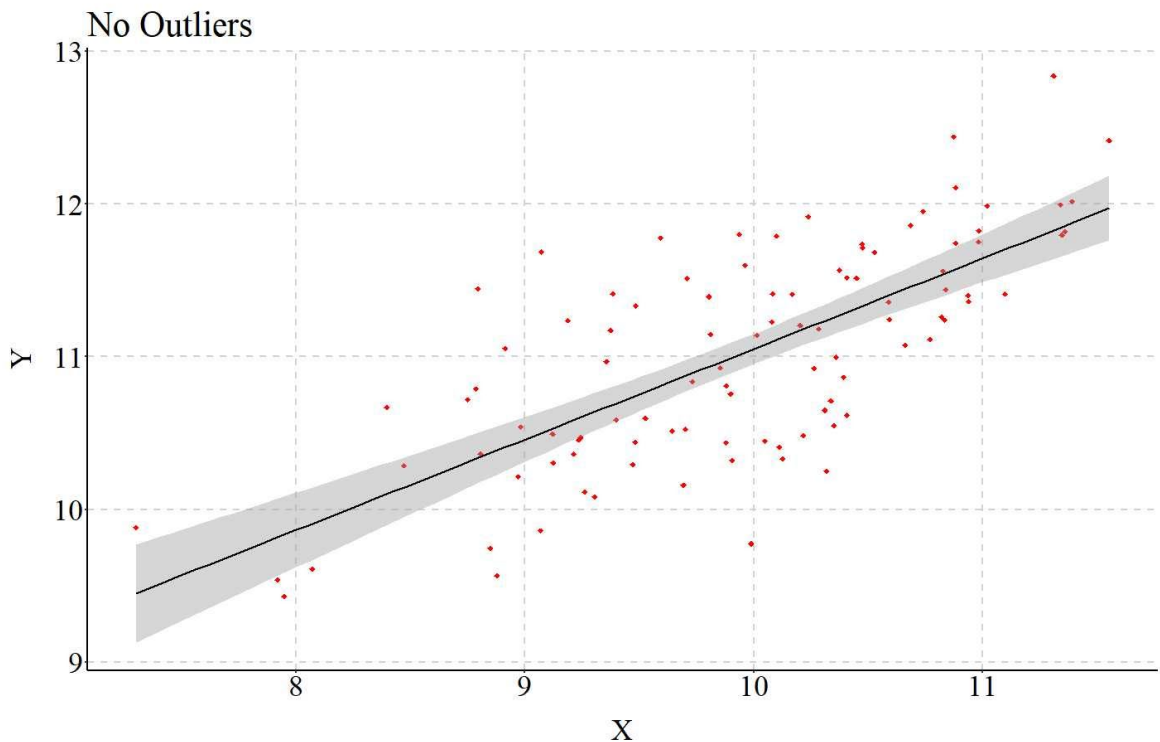

**Outliers on One Variable (*X* & *Y*) But Not the Other (N = 101)** [ii]
*Example 1: Outliers on X but not Y (X = 16)*
To simulate an outlier on *X* but not *Y* we will replicate the process for no outliers, except add a single case in the second step. Specifically, we specify an additional datapoint (ID = 101) with our desired *X* and *Y* values. For our first example, we wanted to simulate a case that is a small outlier on *X* but not an outlier on *Y*. As such, the outlying case was specified to have a value of 16 on *X* and a *Y* value with the same relationship with *X* as the remainder of the *Y* datapoints. After creating the outlier, the model is estimated and the plot was recreated (the outlier is the blue point in the scatterplot). The results of the linear OLS regression model suggested that the outlier has a slight biasing effect on the slope coefficient, downwardly biased (i.e., closer to zero) the standard error, upwardly biased (i.e., further from zero) the standardized slope coefficient ($b$ = .5999; SE = .0474; $β$ = .7862), and no effect on the *p*-value.

```
> ## Data
> N<-100
> ID<-c(1:100)
> set.seed(32)
> X<-rnorm(N,10,1)
> set.seed(15)
> Y<-.60*X+.50*rnorm(N,10,1)
> Data<-data.frame(ID,Y,X)
>
> ## Outlier
> ID<-101
> X<-16
> set.seed(15)
> Y<-.60*X+.50*rnorm(1,10,1)
> O<-data.frame(ID,Y,X)
>
> D_O<-rbind(Data,O)
```

```
> 
> M<-lm(Y~X, data = D_O)
> summary(M)

Call:
lm(formula = Y ~ X, data = D_O)

Residuals:
    Min      1Q  Median      3Q     Max
-1.2707 -0.3116 -0.0405  0.3757  1.1906

Coefficients:
            Estimate Std. Error t value             Pr(>|t|)
(Intercept)   5.0532     0.4762    10.6 <0.0000000000000002 ***
X             0.5999     0.0474    12.7 <0.0000000000000002 ***
---
Signif. codes:  0 '***' 0.001 '**' 0.01 '*' 0.05 '.' 0.1 ' ' 1

Residual standard error: 0.497 on 99 degrees of freedom
Multiple R-squared:  0.618, Adjusted R-squared:  0.614
F-statistic:   160 on 1 and 99 DF,  p-value: <0.0000000000000002

> lm.beta(M)

Call:
lm(formula = Y ~ X, data = D_O)

Standardized Coefficients::
(Intercept)           X
     0.0000      0.7862
```

### Example 2: Outliers on X but not Y (X = 50)

For the second example, our outlying case was specified to have a value of 50 on *X* and a *Y* value with the same relationship with *X* as the remainder of the *Y* datapoints (N = 101). Presented below are the results corresponding to the estimated linear OLS regression model. The results illustrated, similar to the previous example, that the outlier on only *X* had a slight biasing effect on the slope coefficient, downwardly biased the standard error, upwardly biased the standardized slope coefficient ($b$ = .6016; SE = .0122; $\beta$ = .9803), and no effect on the *p*-value.

```
> ## Data
> N<-100
> ID<-c(1:100)
> set.seed(32)
> X<-rnorm(N,10,1)
> set.seed(15)
> Y<-.60*X+.50*rnorm(N,10,1)
> Data<-data.frame(ID,Y,X)
> 
> ## Outlier
> ID<-101
> X<-50
> set.seed(15)
> Y<-.60*X+.50*rnorm(1,10,1)
> O<-data.frame(ID,Y,X)
> D_O<-rbind(Data,O)
> 
> M<-lm(Y~X, data = D_O)
> summary(M)

Call:
lm(formula = Y ~ X, data = D_O)

Residuals:
    Min      1Q  Median      3Q     Max
-1.2701 -0.3125 -0.0404  0.3754  1.1927
```

```
Coefficients:
            Estimate Std. Error t value        Pr(>|t|)
(Intercept)   5.0356     0.1353    37.2 <0.0000000000000002 ***
X             0.6016     0.0122    49.4 <0.0000000000000002 ***
---
Signif. codes:  0 '***' 0.001 '**' 0.01 '*' 0.05 '.' 0.1 ' ' 1

Residual standard error: 0.497 on 99 degrees of freedom
Multiple R-squared:  0.961, Adjusted R-squared:  0.961
F-statistic: 2.44e+03 on 1 and 99 DF,  p-value: <0.0000000000000002

> lm.beta(M)

Call:
lm(formula = Y ~ X, data = D_O)

Standardized Coefficients::
(Intercept)           X
     0.0000       0.9803
```

## *Example 3: Outliers on Y but not X (Y = 17; X = 10 [mean of X])*

After exploring how outliers on *X* but not *Y* influenced our estimated coefficients, we must explore how outliers on *Y* but not *X* influence the estimated coefficients. For the third example, we replicated the above specification except the *X* and *Y* values for the outlier were specified to be 10 (the mean of *X*) and 17, respectively. The results of the OLS regression model, again, illustrated that the outlier on *Y* but not *X* only slightly biased the slope coefficient, upwardly biased the standard error, downwardly biased the standardized slope coefficient (*b* = .5986; SE = .0904; *β* = .5542), and no effect on the *p*-value.

```
> ## Data
> N<-100
> ID<-c(1:100)
> set.seed(32)
> X<-rnorm(N,10,1)
> set.seed(15)
> Y<-.60*X+.50*rnorm(N,10,1)
> Data<-data.frame(ID,Y,X)
>
> ## Outlier
> ID<-101
> X<-10
> set.seed(15)
> Y<-17
> O<-data.frame(ID,Y,X)
> D_O<-rbind(Data,O)
>
> M<-lm(Y~X, data = D_O)
> summary(M)

Call:
lm(formula = Y ~ X, data = D_O)

Residuals:
   Min     1Q Median     3Q    Max
-1.329 -0.369 -0.098  0.340  5.890

Coefficients:
            Estimate Std. Error t value  Pr(>|t|)
(Intercept)   5.1243     0.9011    5.69 0.0000001313 ***
X             0.5986     0.0904    6.62 0.0000000018 ***
---
Signif. codes:  0 '***' 0.001 '**' 0.01 '*' 0.05 '.' 0.1 ' ' 1
```

```
Residual standard error: 0.775 on 99 degrees of freedom
Multiple R-squared:  0.307, Adjusted R-squared:   0.3
F-statistic: 43.9 on 1 and 99 DF,  p-value: 0.00000000183

> lm.beta(M)

Call:
lm(formula = Y ~ X, data = D_O)

Standardized Coefficients::
(Intercept)           X
     0.0000       0.5542
```

### Example 4: Outliers on Y but not X (Y = 50; X = 10 [mean of X])

This process was replicated again, except the outlier was specified to have a value of 50 for *Y* (*X* remained 10). After simulating the data, our linear OLS regression model was reestimated. The results, as you can now expect, illustrated that the inclusion of the outlier in our analytical sample slightly biased the slope coefficient, upwardly biased the standard error, downwardly biased the standardized slope coefficient (*b* = .6271; SE = .4577; *β* = .1364) and no effect on the *p*-value.

```
> ## Data
> N<-100
> ID<-c(1:100)
> set.seed(32)
> X<-rnorm(N,10,1)
> set.seed(15)
> Y<-.60*X+.50*rnorm(N,10,1)
> Data<-data.frame(ID,Y,X)
>
> ## Outlier
> ID<-101
> X<-10
> set.seed(15)
> Y<-50
> O<-data.frame(ID,Y,X)
> D_O<-rbind(Data,O)
>
> M<-lm(Y~X, data = D_O)
> print(summary(M), digits = 4)

Call:
lm(formula = Y ~ X, data = D_O)

Residuals:
   Min     1Q Median     3Q    Max
-1.657 -0.717 -0.433 -0.008 38.562

Coefficients:
            Estimate Std. Error t value Pr(>|t|)
(Intercept)   5.1674     4.5642   1.132    0.260
X             0.6271     0.4577   1.370    0.174

Residual standard error: 3.927 on 99 degrees of freedom
Multiple R-squared:  0.01861,   Adjusted R-squared:  0.0087
F-statistic: 1.878 on 1 and 99 DF,  p-value: 0.1737

> lm.beta(M)

Call:
lm(formula = Y ~ X, data = D_O)

Standardized Coefficients::
(Intercept)           X
     0.0000       0.1364
```

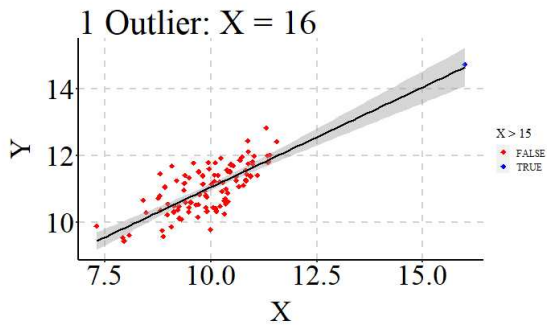
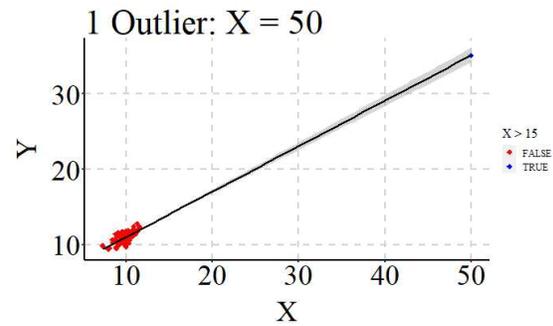
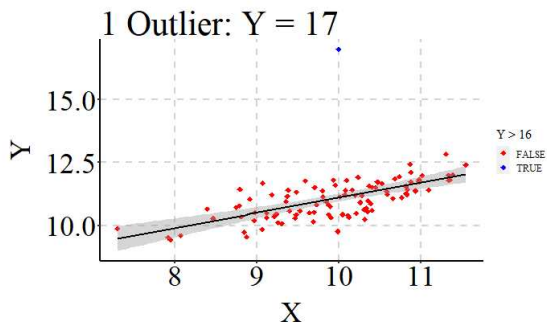
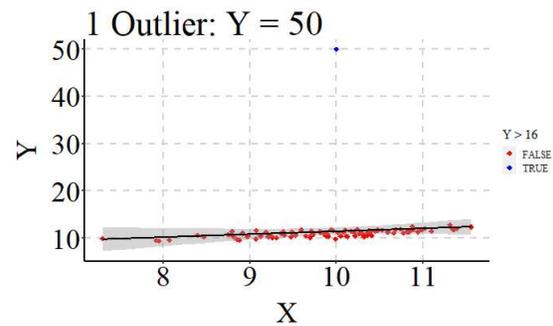

**Outlier on *X* and *Y* at $Y_{bar}$ (N = 101)**

You might have been asking yourself – as we went through the above examples – why did we not just set the *Y* value to the mean when exploring outliers on *X* but not *Y* (i.e., examples 1 and 2)? While that would be logical, specifying an *X* outlier at the mean of *Y* ($Y_{bar}$) will commonly downwardly bias the slope coefficient of a regression line. This downward bias occurs because while *Y* is not an outlier, we still violate the normality assumption (i.e., the residual distribution of *Y* values is not normally distributed). Let's explore this using two examples. While these examples employ a positive association, I encourage readers to use the available *R*-code to observe how *X* outlier at the mean of *Y* can downwardly bias negative associations (make the association closer to zero than reality).

*Example 5: X = 16; Y = $Y_{bar}$*

Consistent with the previous examples, we replicated the no outlier code except added a single outlier (ID: 101) to the data frame that was specified to have a value of 16 on *X* and 11.01 ($Y_{bar}$) on *Y*. After creating the sample, we replicated the linear OLS regression model. Distinct from the previous examples, however, the slope coefficient is downwardly biased (*b* = .3970) compared to the true slope coefficient (.5934). The downward bias in the slope coefficient becomes more substantial, however, when the *X* value is increased.

```
> ## Data
> N<-100
> ID<-c(1:100)
> set.seed(32)
> X<-rnorm(N,10,1)
> set.seed(15)
> Y<-.60*X+.50*rnorm(N,10,1)
```

```
> Data<-data.frame(ID,Y,X)
>
> ## Outlier
> ID<-101
> X<-16
> set.seed(15)
> Y<-mean(Y)
> O<-data.frame(ID,Y,X)
> D_O<-rbind(Data,O)
>
> M<-lm(Y~X, data = D_O)
> summary(M)

Call:
lm(formula = Y ~ X, data = D_O)

Residuals:
    Min      1Q  Median      3Q     Max
-2.3838 -0.3676 -0.0247  0.4197  1.3037

Coefficients:
            Estimate Std. Error t value             Pr(>|t|)
(Intercept)   7.0440     0.5534   12.73 < 0.0000000000000002 ***
X             0.3970     0.0551    7.21         0.00000000011 ***
---
Signif. codes:  0 '***' 0.001 '**' 0.01 '*' 0.05 '.' 0.1 ' ' 1

Residual standard error: 0.578 on 99 degrees of freedom
Multiple R-squared:  0.344,  Adjusted R-squared:  0.338
F-statistic:    52 on 1 and 99 DF,  p-value: 0.000000000113

> lm.beta(M)

Call:
lm(formula = Y ~ X, data = D_O)

Standardized Coefficients::
(Intercept)           X
     0.0000      0.5868
```

*Example 6: X = 50; Y = $Y_{bar}$*

When the outlying case is specified to have a value of 50 on *X* and 11.01 ($Y_{bar}$) on *Y*, the slope coefficient estimated using the simulated data is .0263 and is no longer statistically significant. Moreover, the standardized slope coefficient from the model was equal to .1509, which is a meaningful departure from the true standardized slope coefficient of .7174.

```
> ## Data
> N<-100
> ID<-c(1:100)
> set.seed(32)
> X<-rnorm(N,10,1)
> set.seed(15)
> Y<-.60*X+.50*rnorm(N,10,1)
> Data<-data.frame(ID,Y,X)
>
> ## Outlier
> ID<-101
> X<-50
> set.seed(15)
> Y<-mean(Y)
> O<-data.frame(ID,Y,X)
> D_O<-rbind(Data,O)
>
> M<-lm(Y~X, data = D_O)
> summary(M)
```

```
Call:
lm(formula = Y ~ X, data = D_O)

Residuals:
   Min     1Q Median     3Q    Max
-1.520 -0.528  0.080  0.516  1.800

Coefficients:
            Estimate Std. Error t value             Pr(>|t|)
(Intercept)  10.7411     0.1919   55.96 <0.0000000000000002 ***
X             0.0263     0.0173    1.52                0.13
---
Signif. codes:  0 '***' 0.001 '**' 0.01 '*' 0.05 '.' 0.1 ' ' 1

Residual standard error: 0.705 on 99 degrees of freedom
Multiple R-squared:  0.0228,    Adjusted R-squared:  0.0129
F-statistic: 2.31 on 1 and 99 DF,  p-value: 0.132

> lm.beta(M)

Call:
lm(formula = Y ~ X, data = D_O)

Standardized Coefficients::
(Intercept)           X
     0.0000      0.1509
```

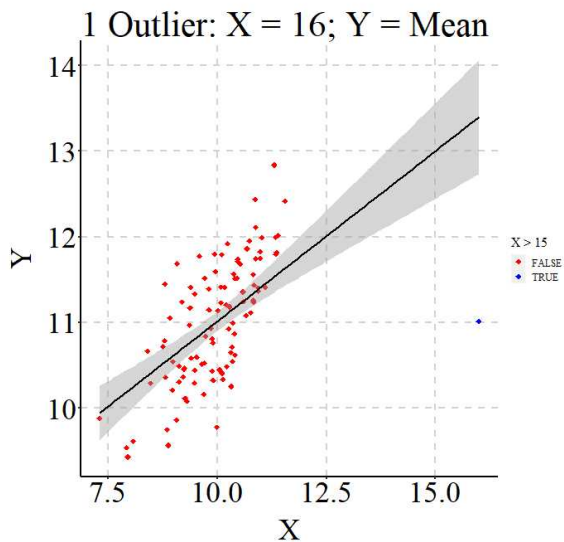
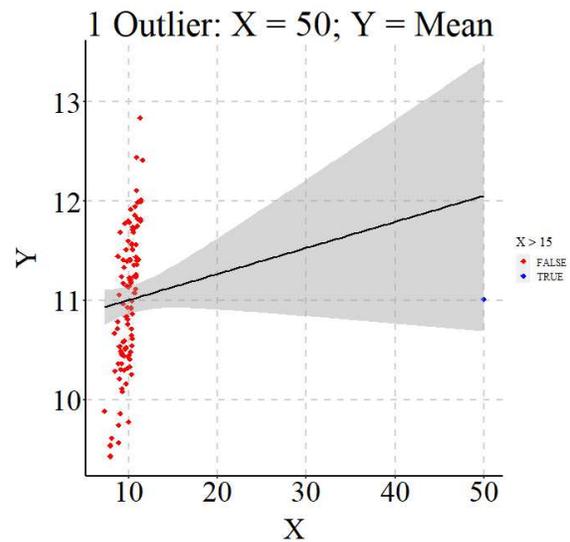

Consistent with the two examples above, the slope coefficient will commonly be downwardly biased – closer to zero – if a case is an outlier on $X$ at $Y_{bar}$. As such, outliers on $X$ at $Y_{bar}$ can only serve to downwardly bias the unstandardized and standardized slope coefficients.

**Outlier on $X$ and $Y$ (N = 101)**
Finally, let's explore the effects of an outlier on both $X$ and $Y$ on the estimates derived from a linear OLS regression model. In the first example we specified that the outlier would have a value of 16 on $X$ and 20 on $Y$, in the second example we specified that the outlier would have a value of 50 on $X$ and 100 on $Y$, in the third example we specified that the outlier would have a value of 16 on $X$ and -25 on *Y*. When the outlier was specified to exist at 16 on *X* and 25 on *Y* the slope of the association estimated using an OLS regression model was upwardly biased (*b* = 1.16), as well as the standardized slope estimate (*β* = .7791). When the outlier possessed a value of 50 on *X* and 100 on *Y*, the degree of upward bias was substantially increased. Specifically, the estimated slope coefficient was 2.149 and the standardized slope coefficient was .9866. Finally, when the outlier was specified to exist at 16 on *X* and -25 on *Y* the slope of the association estimated using an OLS regression model was upwardly biased (further from zero) but the sign was flipped (*b* = -1.586), as well as the standardized slope estimate (*β* = -.4503). Overall, a common outlier on *X* and *Y* will frequently upwardly bias the slope coefficient.

## *Example 7: X = 16; Y = 25*

```
> ## Data
> N<-100
> ID<-c(1:100)
> set.seed(32)
> X<-rnorm(N,10,1)
> set.seed(15)
> Y<-.60*X+.50*rnorm(N,10,1)
> Data<-data.frame(ID,Y,X)
> 
> ## Outlier
> ID<-101
> X<-16
> set.seed(15)
> Y<-25
> O<-data.frame(ID,Y,X)
> D_O<-rbind(Data,O)
> 
> M<-lm(Y~X, data = D_O)
> summary(M)

Call:
lm(formula = Y ~ X, data = D_O)

Residuals:
   Min     1Q Median     3Q    Max
-1.369 -0.600 -0.092  0.463  6.882

Coefficients:
            Estimate Std. Error t value            Pr(>|t|)
(Intercept)  -0.4478     0.9432   -0.47                0.64
X             1.1604     0.0939   12.36 <0.0000000000000002 ***
---
Signif. codes:  0 '***' 0.001 '**' 0.01 '*' 0.05 '.' 0.1 ' ' 1

Residual standard error: 0.984 on 99 degrees of freedom
Multiple R-squared:  0.607, Adjusted R-squared:  0.603
F-statistic:   153 on 1 and 99 DF,  p-value: <0.0000000000000002

> lm.beta(M)

Call:
lm(formula = Y ~ X, data = D_O)

Standardized Coefficients::
(Intercept)           X
     0.0000      0.7791
```

## *Example 8: X = 50; Y = 100*

```
> ## Data
```

```
> N<-100
> ID<-c(1:100)
> set.seed(32)
> X<-rnorm(N,10,1)
> set.seed(15)
> Y<-.60*X+.50*rnorm(N,10,1)
> Data<-data.frame(ID,Y,X)
>
> ## Outlier
> ID<-101
> X<-50
> set.seed(15)
> Y<-100
> O<-data.frame(ID,Y,X)
> D_O<-rbind(Data,O)
>
> M<-lm(Y~X, data = D_O)
> summary(M)

Call:
lm(formula = Y ~ X, data = D_O)

Residuals:
    Min      1Q  Median      3Q     Max
-2.285  -1.164  -0.132   0.919   4.500

Coefficients:
             Estimate Std. Error t value             Pr(>|t|)
(Intercept) -10.3113     0.3973     -26 <0.0000000000000002 ***
X             2.1491     0.0358      60 <0.0000000000000002 ***
---
Signif. codes:  0 '***' 0.001 '**' 0.01 '*' 0.05 '.' 0.1 ' ' 1

Residual standard error: 1.46 on 99 degrees of freedom
Multiple R-squared:  0.973,  Adjusted R-squared:  0.973
F-statistic: 3.61e+03 on 1 and 99 DF,  p-value: <0.0000000000000002

> lm.beta(M)

Call:
lm(formula = Y ~ X, data = D_O)

Standardized Coefficients::
(Intercept)           X
     0.0000      0.9866
```

## *Example 9: X = 16; Y = -25*

```
> ## Data
> N<-100
> ID<-c(1:100)
> set.seed(32)
> X<-rnorm(N,10,1)
> set.seed(15)
> Y<-.60*X+.50*rnorm(N,10,1)
> Data<-data.frame(ID,Y,X)
>
>
>
> ## Outlier
> ID<-101
> X<-16
> set.seed(15)
> Y<--25
> O<-data.frame(ID,Y,X)
>
> D_O<-rbind(Data,O)
>
```

```
> M<-lm(Y~X, data = D_O)
> summary(M)

Call:
lm(formula = Y ~ X, data = D_O)

Residuals:
    Min      1Q  Median      3Q     Max
-26.239  -1.019   0.407   1.644   4.243

Coefficients:
            Estimate Std. Error t value       Pr(>|t|)
(Intercept)   26.333      3.141    8.38 0.00000000000036 ***
X             -1.568      0.313   -5.02 0.00000229958141 ***
---
Signif. codes:  0 '***' 0.001 '**' 0.01 '*' 0.05 '.' 0.1 ' ' 1

Residual standard error: 3.28 on 99 degrees of freedom
Multiple R-squared:  0.203, Adjusted R-squared:  0.195
F-statistic: 25.2 on 1 and 99 DF,  p-value: 0.0000023

> lm.beta(M)

Call:
lm(formula = Y ~ X, data = D_O)

Standardized Coefficients::
(Intercept)           X
     0.0000      -0.4503
```

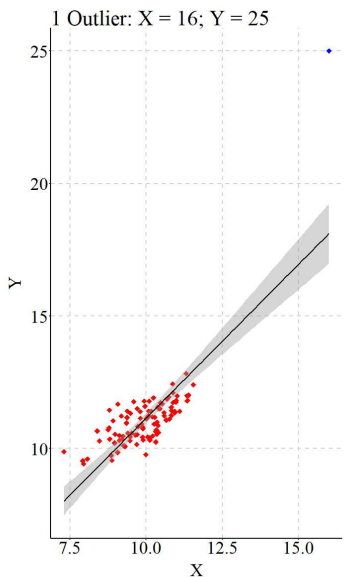
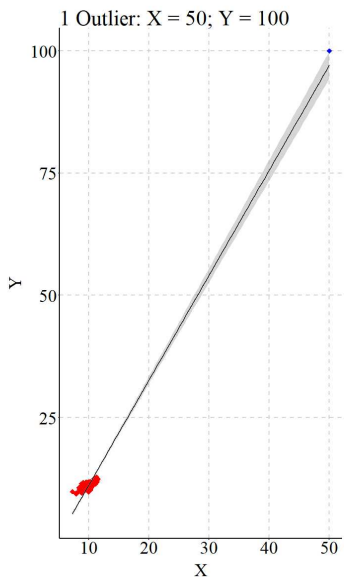
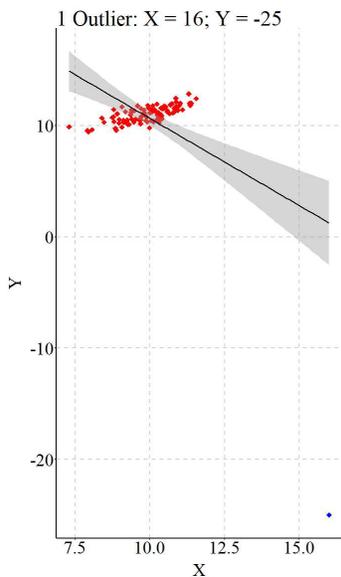

### Conclusions

This entry explored the effects of outliers at different locations on the slope coefficients, standard errors, standardized slope coefficients, and *p*-values from linear OLS regression models. As demonstrated: (1) outliers on *X* but not *Y* will only result in slight biases to the estimates from a regression model, (2) outliers on *Y* but not *X* will only result in slight biases to the estimates from a regression model , (3) outliers on *X* but within the distribution of *Y* will commonly downwardly bias the unstandardized and standardized slope coefficients, (4) and outliers on both *X* and *Y* will commonly upwardly bias the unstandardized and standardized slope coefficients.[iii] While discussing the effects of potential outliers is important when reviewing the results of a regression model, only an outlier on both *X* and *Y* can result in meaningful upward bias in the unstandardized and standardized slope coefficients and, in turn, increase the likelihood of rejecting the null hypothesis when the null hypothesis should have been retained. Moreover, the effects of an outlier on both *X* and *Y* depends upon the sample size (Illustrated in the available R-Code). As such, sensitivity analyses should be estimated to explore if an outlier on both *X* and *Y* exists in the sample and the effects of the outlier on the estimated coefficients. If the effects are statistically significant when the outlier is included and not statistically significant when the outlier is excluded, justification should be provided for the differences between the models.

I, however, believe it is important to remind readers about the caution that has to exist when handling outliers in a dataset. Primarily, a datapoint is recognizably different from the remainder of the distribution for three reasons: (1) measurement error, (2) random error, and (3) these datapoints exist in the broader population. All three present difficulties for defining and handling outliers. The existence of known measurement error represents the situation where outliers are the easiest to handle. Specifically, if scores are unreasonable, unlikely, or not realistic we can justifiably remove those cases from the data due to the inability to properly score particular cases. Nevertheless, if there is no apparent measurement error, the regression model estimated without an outlier included could be more biased – further from the population estimates – than the regression model estimated with the outlier included. Moreover, the removal of multiple datapoints perceived as outliers could meaningfully alter the results of the model. Considering this, extreme caution should be taken when evaluating if it is or is not appropriate to remove perceived outliers from a dataset prior to estimating a regression model. Only in undoubtedly justifiable circumstances should you non-randomly remove perceived outliers from a dataset.[iv]

---

[i] The inclusion of random error ensures that values on *X* do not perfectly predict values on *Y*.

[ii] All of the examples below are replicated in the *R*-code with 1,001 and 10,001 cases.

[iii] In addition to the examples reviewed above, the available *R*-code also replicates the examples using samples comprised of 1,001 and 10,001 cases.

[iv] The effects of non-randomly removing cases from a dataset will be reviewed when discussing additional assumptions of regression models.

# Statistical Biases When Examining Causal Associations

The next section of the Sources of Statistical Biases Series will provide simulated explorations of key assumptions of causal analysis. It is important, however, to understand that **two** frameworks are primarily used to evaluate for causal associations: *Causal Inference Through Experimental Design* and *Causal Inference Through Directed Acyclic Graphs (DAG).* Distinct from popular belief, these frameworks are not mutually exclusive and can be complimentary if integrated properly. Nevertheless, that discussion is for another series. Importantly, both frameworks define causality the exact same: logical, temporally ordered, and non-spurious. Considering the focus of the current section is on violating assumptions when examining causal associations, we must talk about assumptions within both causal frameworks. These include statistical assumptions of (1) randomized controlled trials (Rubin Causal Models) and (2) structural causal models/DAGs. Before going into the deep end of the causal pool, let's briefly review the causal frameworks.

In the vast majority of sciences, we teach that causal inference can only be achieved through experimental design. This framework was initially proposed by Ronald Fisher (1936) and popularized by Paul Rosenbaum and Donald Rubin. In the simplest terms, a randomized experiment – randomized controlled trial – allows one to estimate the causal association between two variables by randomly assigning scores on the independent variable. This random assignment, if conducted properly, makes the independent variable unpredictable, in turn allowing you to achieve the three requirements of causality. Although related, quasi-experimental designs will be discussed in a subsequent section.

In addition to experimental designs, directed acyclic graphs (DAGs) can provide the ability to make inferences about causal associations. DAGs are graphical displays with nodes and edges, where the nodes represent variables and the edges represent the relationship between the nodes. As argued by Judea Pearl, the estimation process of structural models as well as the theoretically guided selection of edges directly and indirectly connecting nodes permit DAGs to be used to develop causal inferences. Nevertheless, it is important to note that causal inference can only be achieved when theory, logic, and a transparent decision-making process is used to guide the development of a DAG. Moreover, unlike randomized controlled trials, for DAGs to permit the ability to make inferences about causal associations potential confounders must be included in the model or justifiably ruled out.

Now that we introduced the two causal frameworks, let's begin the current section talking about randomized controlled trials. I hope you enjoy!

# Entry 5: Unknown Interactions (Randomized Controlled Trials)

**Introduction**
Conducting a randomized controlled trial is the foremost strategy for estimating a causal association between a treatment variable and an outcome. In the context of a randomized controlled trial, a treatment variable is oftentimes a dichotomous, categorical, or semi-continuous construct that is differentially provided to a sample of participants.[i] For instance, something as common as aspirin can be defined as a treatment variable and provided to participants at different dosages (e.g., 0 pills, 1 pill, 2 pills, etc.). I would note that aspirin is not fully continuous because dosages beyond 10 pills would inevitably result in stomach ulcers. An outcome, or the dependent variable, in a randomized controlled trial is defined as any concept we expect to vary after the introduction of the treatment. To continue with our aspirin example, we would expect to observe differences in back pain after the introduction of aspirin, and the level of pain to vary by the dosage of aspirin.

The most common way of conducting a randomized controlled trial is to randomly select half of the participants to be exposed to a treatment (e.g., the aspirin) and half of the participants to not be exposed to the treatment, otherwise known as the control group (e.g., no aspirin). By randomly assigning participants to the treatment and control groups, we can create a condition where nothing predicts exposure to the treatment. This essentially removes any possibility of an unknown variable confounding the association between exposure to the treatment and an outcome of interest. After randomly assigning participants, we can calculate the difference in the outcome between the treatment cases (i.e., individuals exposed to the treatment) and control cases (i.e., individuals not exposed to the treatment) and evaluate the influence of the treatment on the outcome. The process of conducting a randomized controlled trial, or true experiment, satisfies the three requirements of causality (logical, temporally ordered, and non-spurious) and permits scholars to infer about the causal association between the treatment and the outcome of interest. Moreover, it allows us to assume that the estimates derived from any statistical analysis (e.g., *t*-test, Anova, OLS regression models) approach causality better than observational studies. Nevertheless, for the estimates derived from a single, or multiple, randomized controlled trials to permit inferences of causal associations in the population two key assumptions must be satisfied: (1) the assumption of known treatment interactions and (2) the assumption of balance between the treatment and control cases.[ii]

*A Research Note: The Importance of Replication to Causal Inference*
Causal inference through experimental design can not be achieved without replication. A fundamental assumption of experimental research is that the experiment will be conducted multiple times and the findings will be aggregated to create a distribution of findings illuminating the causal effects of the treatment on the outcome. Although the findings of a single randomized controlled trial can be used to make inferences about causal associations, the true causal effects in the population will remain unknown until replications are conducted. As such, replications are key to all research, but are especially important for experimental research. Let this brief note serve as a reminder that more replications need to be conducted. Reminders about the importance of replications will appear throughout the Sources of Statistical Biases Series. Now back to our regularly scheduled program.

**The Assumption of Known Interactions**

An interaction refers to a condition where the effects of an independent variable on a dependent variable change across levels of a third variable. For example, the magnitude of the effect of GPA on placement in a law firm will vary by the law school an individual attended. In theory, higher GPA would predict better placement, but the magnitude of the effect will change if an individual attended Harvard law or the University of Cincinnati law school (Go Bearcats!). To use statistical terms, the third variable (e.g., the law school attended) moderates the association between the independent variable and the dependent variable. In complex systems, like topics studied in the social sciences, numerous interactions between constructs exist.

When conducting analyses on data collected during randomized controlled trials, we oftentimes either assume that the effects of the treatment on the outcome is not moderated by a third construct or that we know the moderating mechanisms. This is the assumption of known interactions. The assumption of known interactions, however, is difficult to satisfy because it requires the identification of the possible moderators before the randomized controlled trial is conducted. As such, we can never know if we have identified all of the possible constructs that influence the association between the treatment and outcome of interest.

Violating the known interaction assumption does not hinder the ability of randomized controlled trials to estimate causal associations. Violations, however, could substantially hinder our ability to generate causal inferences using data from a randomized controlled trial. More specifically, the estimates from randomized controlled trials that violate the assumption of known interactions could limit the generalizability of the findings, as well as diminish our understanding of the causal association between the treatment and the outcome of interest. Nevertheless, the detrimental effects of unknown interactions only become important when the initial sample is pulled from the broader population using non-random sampling techniques. As illustrated below, only a random sample from the population will produce a sampling distribution that encompasses the causal association in the broader population. Additionally, data from randomized controlled trials employing non-random sampling techniques could produce sampling distributions contrasting with the true effects of the treatment on the outcome.

**Simulating the Population**
For the purpose of illustrating the effects of unknown interactions, we will conduct a simulated – I know surprising, right? – randomized controlled trial evaluating the effects of a hypothetical employment program on hypothetical success in employment scale. For this example, an interaction between the employment program and parent education will exist in the population. In this simulation, we will specify that the employment program is less effective for individuals with higher parent education and more effective for individuals with lower parent education.

Now that we grasp the example, let me begin the simulation by stating a simple truth; *It will take some mental gymnastics to work through this example*. Two things in particular are important to remember or more specifically forget. First, even though we build an interaction between the employment program and parent education, we are going to imagine that we don't know that interaction exists. So, after we build the population forget about that interaction! Second, it is difficult to simulate a randomized controlled trial because scores on the treatment variable are not influenced by any other construct, as well as the need to specify exposure to the treatment variable (i.e., the education program) before we create the outcome (i.e., success in employment). Although not important in the population, when samples are pulled from the population, we are unable to reassign exposure to the treatment variable. As such, we need to imagine that the randomized controlled trial is being conducted in each sample we pull.

For the population, we will randomly assign 500,000 cases to be exposed to the employment program or not exposed to the employment program (250,000 each). After randomly designating cases as treatment or control, the cases were assigned scores on parental education – conceptualized as highest grade completed. The scores were simulated to be normally distributed, with a mean of 12 (representing 12$^{th}$ grade), and a standard deviation of 2.5 grades. Although scores on parental education initially ranged between 0 and 23, we truncated the distribution to range between 4$^{th}$ grade (all scores below 4 received a score of 4) and 19$^{th}$ grade (all scores above 19 received a score of 19). This decision was made to keep the distribution of scores realistic and serves no purpose for the actual illustration.

```
> ## Building Population (N = 500,000) ####
> ### (X) Example Variable: Employment Program
> set.seed(7)
> EP<-rep(c(0,1),times=250000)
> table(EP)
EP
     0      1
250000 250000

> ### (M[moderator]) Example Variable: Parental Educational Attainment
> set.seed(1121)
> PEA<-trunc(rnorm(500000,12,2.5))
> PEA[PEA<=4]<-4 # Making Lowest Grade Completed 4th grade (For example purposes)
> PEA[PEA>=19]<-19 # Making Highest Grade Completed 19th grade (For example purposes)
> table(PEA)
PEA
    4     5     6     7     8     9    10    11    12    13    14    15    16    17    18    19
 1295  2904  7122 15874 30260 48594 66516 77422 78113 66457 48261 29883 16004  7200  2759  1336
```

After simulating exposure to the employment program and parental education, we can create the success in employment construct. The modifying values in the initial specification are uninformative because of the inclusion of an interaction term [*(-.50\*(PEA\*EP))*] and the inclusion of some normally distributed variation [*rnorm(500000,5,.25)*]. As such, we estimate a baseline model to identify the effect of the employment program on success in employment for the population (N = 500,000). The estimated coefficient, *b* = 1.25, represents the difference on success in employment between the treatment and control cases across the entire population, where the treatment cases will score 1.25 points higher than control cases on success in employment.

```
> ### (Y) Example Variable: Success In Employment
> ### Higher scores means more success, Lower scores means less success
> set.seed(39817)
> SIEM<-7.00*EP+0.0*PEA+(-.50*(PEA*EP))+rnorm(500000,5,.25)
>
> A<-data.frame(EP,PEA,SIEM)
>
> lm(SIEM~EP, data = A)

Call:
lm(formula = SIEM ~ EP, data = A)

Coefficients:
(Intercept)                 EP
```

```
     5.00          1.25
```

For illustrative purposes, let's take a moment to look at the distribution of scores on success in employment for the treatment (the red distribution) and control groups (the light blue distribution). As you probably noticed, the distribution of success in employment for the treatment group is wider (more platykurtic), while the distribution of success in employment for the control cases is narrower (more leptokurtic). The distributional differences between the two groups is a direct result of: (1) the effects of the employment program, (2) the interaction between the employment program and parental education.

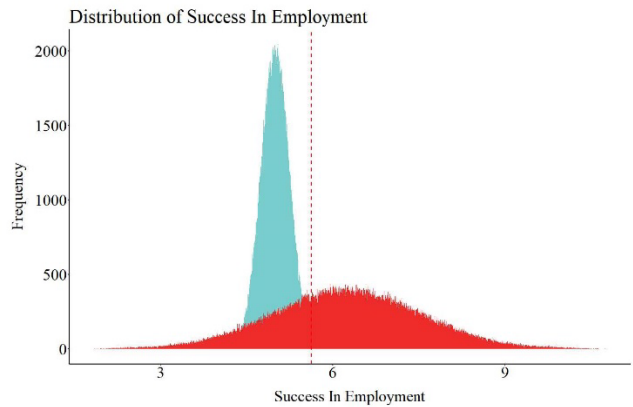

*Notes:* Red represents the treatment group (individuals exposed to the employment program) and light blue represents the control group (individuals not exposed to the employment program).

**Sampling Distribution of Association: Random Sample of Population**
For the first illustration, we will randomly sample 1,000 cases from the population and estimate the effects of the employment program on success in employment. However, while one sample can be informative, repeating this process various times could help us develop a sampling distribution of the effects of the employment program on success in employment. As such, using a loop in R, a random sample of 1,000 cases was drawn from the population 10,000 times. Each time a random sample was drawn, our code informed *R* to estimate and record the slope coefficient and standard error of the association between the employment program and success in employment. This looped simulation simply mirrors how we would conduct randomized controlled trials if money and time were of no concern: (1) randomly sample from the population, (2) randomly expose cases to a treatment or control condition, (3) evaluate differences in the outcome, and (4) replicate the process.

```
>N<-10000
>RSDF1 <- new.env()
>### Loop
>for(i in 1:N)
>{
>RS1<-sample_n(A, 1000, replace = F)
>
>LM_RS<-lm(SIEM~EP, data = RS1)
>summary(LM_RS)
>
>RSDF1$Slope<-c(RSDF1$Slope, summary(LM_RS)$coefficients[2])
>RSDF1$SE<-c(RSDF1$SE, summary(LM_RS)$coefficients[4])
>}
```

The distribution of estimated slope coefficients (*b*) – the sampling distribution – derived from the 10,000 random samples of 1,000 respondents is provided in the figure below. The red dashed line represents the mean slope coefficient, while the blue, yellow, and red areas represent 1, 2, and 3 (respectively) standard deviations from the mean. As illustrated by the figure, the mean of the sampling distribution for the slope coefficient (*b* = 1.25) – the difference on success in employment between the treatment and control cases – is almost identical to the difference observed for the entire population (*b* = 1.25). Moreover, only a limited number of random samples drawn from the population were observed to have differences on success in employment between the treatment and control cases (i.e., slope coefficients) smaller than 1.07 and larger than 1.44. Overall, this finding suggests that participation in the employment program will cause an increase of 1.25 on success in employment independent of who receives the program.

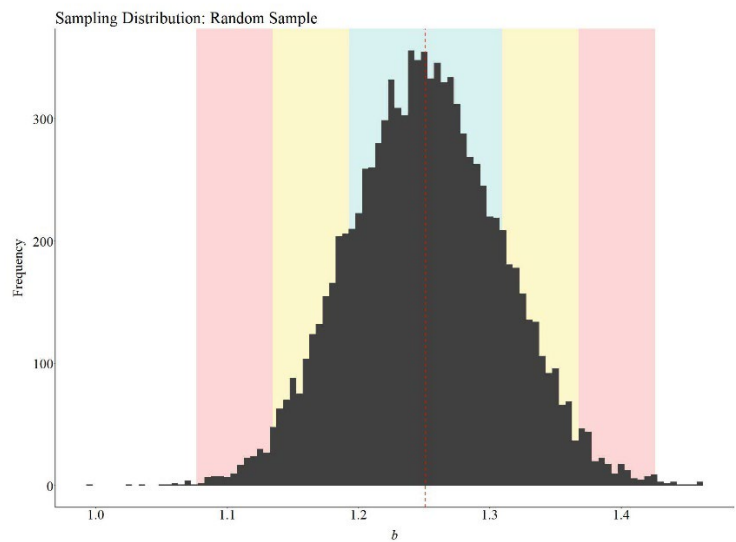

*Notes*: The red dashed line represents the mean slope coefficient, while the blue, yellow, and red areas represent 1, 2, and 3 (respectively) standard deviations from the mean.

**Sampling Distribution of Association: Non-Random Samples of the Population**

*High Levels Parental Education (PEA>=15)*
Now let's repeat the process with a non-random convenience sample.[iii] How about we only sample from individuals with high levels of parental education? When the slope coefficients from the 10,000 non-random samples of 1,000 participants with high parental education are plotted the findings suggest that participation in the employment program causes a -.88 reduction in success in employment. Moreover, the range of the sampling distribution suggests that the employment program will approximately cause between a 1 to .75 reduction in success in employment.

```
>N<-10000
>RSDF1 <- new.env()
>### Loop
>for(i in 1:N)
>{
>   NRS1<-sample_n(A[which(A$PEA >= 15),] ,1000, replace = F)
>
```

```
>   LM_NRS<-lm(SIEM~EP, data = NRS1)
>   summary(LM_NRS)
>
>   NRS1DF1$Slope<-c(NRS1DF1$Slope, summary(LM_NRS)$coefficients[2])
>   NRS1DF1$SE<-c(NRS1DF1$SE, summary(LM_NRS)$coefficients[4])
>
>}
```

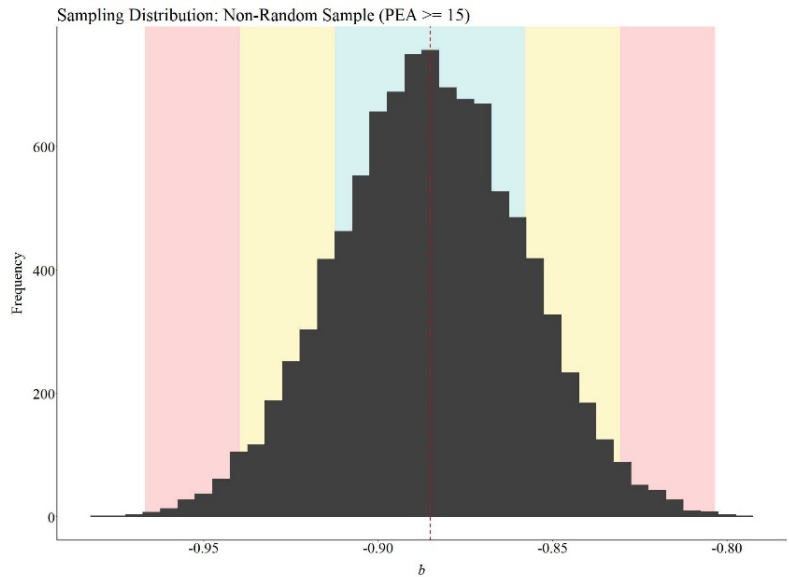

*Notes*: The red dashed line represents the mean slope coefficient, while the blue, yellow, and red areas represent 1, 2, and 3 (respectively) standard deviations from the mean.

The difference between the random sampling distribution and the first non-random sampling distribution occurred because we restricted our sampling process to only a proportion of the treatment and control distribution. The non-random sampling process, as a product of that unknown interaction, impacts the location of cases drawn from the treatment distribution but not the control distribution (pay attention to the location on the X-axis). This is illustrated in the figure below.

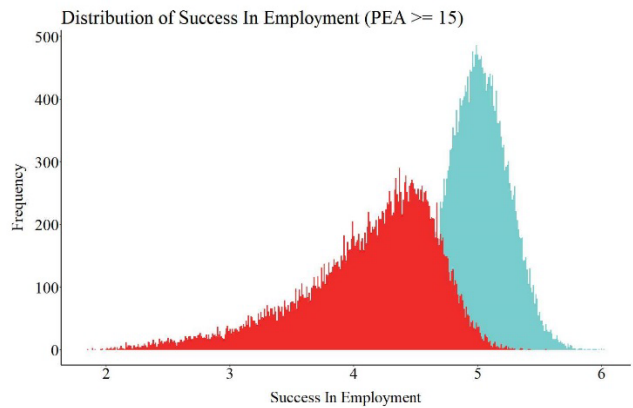

*Notes:* Red represents the treatment group (individuals exposed to the employment program) and light blue represents the control group (individuals not exposed to the employment program).

### *Above Average Parental Education (PEA >= 11.5 and PEA <= 15)*

This time let's sample a group of individuals with above average parental education. When the slope coefficients from the 10,000 samples of 1,000 participants with above average parental education are plotted the findings suggest that participation in the employment program causes a .40 increase in success in employment. Additionally, the sampling distribution of the difference between the treatment and control cases on success in employment (i.e., *b*) ranges between .30 and .55.

```
>N<-10000
>RSDF1 <- new.env()
>### Loop
>for(i in 1:N)
>{
>   NRS3<-sample_n(A[which(A$PEA >= 11.5 & A$PEA <= 15),] ,1000, replace = F)
>
>   LM_NRS<-lm(SIEM~EP, data = NRS3)
>   summary(LM_NRS)
>
>   NRS3DF1$Slope<-c(NRS3DF1$Slope, summary(LM_NRS)$coefficients[2])
>   NRS3DF1$SE<-c(NRS3DF1$SE, summary(LM_NRS)$coefficients[4])
>
>}
```

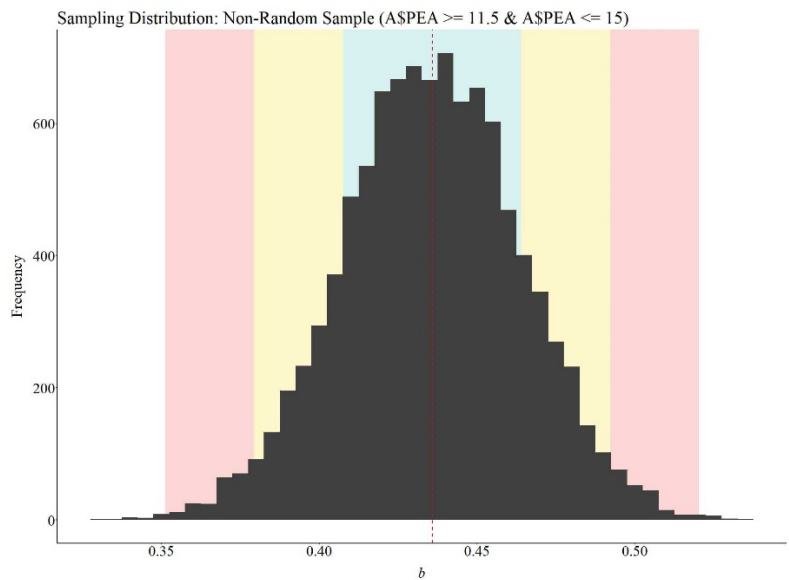

*Notes*: The red dashed line represents the mean slope coefficient, while the blue, yellow, and red areas represent 1, 2, and 3 (respectively) standard deviations from the mean.

The figure below was created to illustrate the location of the above average parental education treatment and control cases. Again, pay attention to the location of the treatment distribution on the X-axis.

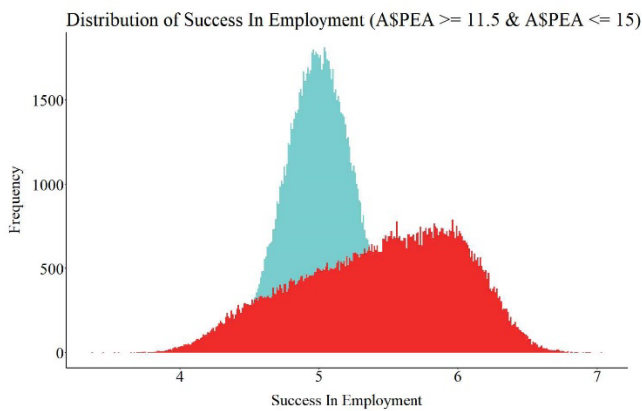

*Notes:* Red represents the treatment group (individuals exposed to the employment program) and light blue represents the control group (individuals not exposed to the employment program).

### *Below Average Parental Education (PEA >= 8 and PEA <= 11.4)*

What about cases with below average parental education? Let's draw 10,000 samples of 1,000 participants from this proportion of the distribution. When the slope coefficients from the 10,000 samples of 1,000 participants with below average parental education are plotted the findings suggest that participation in the employment program causes a 2.07 increase in success in employment. Additionally, the sampling distribution of the difference between the treatment and control cases on success in employment (i.e., *b*) ranges between 1.90 and 2.20.

```
>N<-10000
>RSDF1 <- new.env()
>### Loop
>for(i in 1:N)
>{
>   NRS4<-sample_n(A[which(A$PEA >= 8 & A$PEA <= 11.4),] ,1000, replace = F)
>
>   LM_NRS<-lm(SIEM~EP, data = NRS4)
>   summary(LM_NRS)
>
>   NRS4DF1$Slope<-c(NRS4DF1$Slope, summary(LM_NRS)$coefficients[2])
>   NRS4DF1$SE<-c(NRS4DF1$SE, summary(LM_NRS)$coefficients[4])
>
>}
```

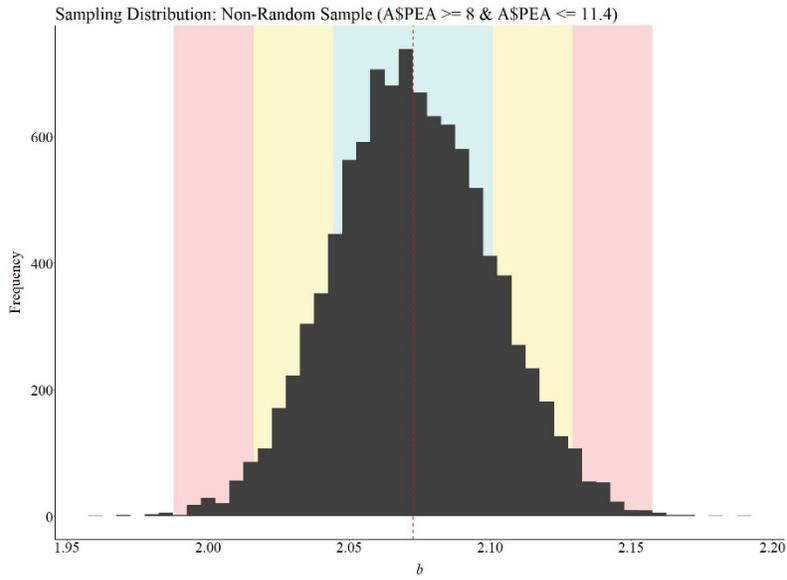

*Notes*: The red dashed line represents the mean slope coefficient, while the blue, yellow, and red areas represent 1, 2, and 3 (respectively) standard deviations from the mean.

The figure below was created to illustrate the location of the below average parental education treatment and control cases.

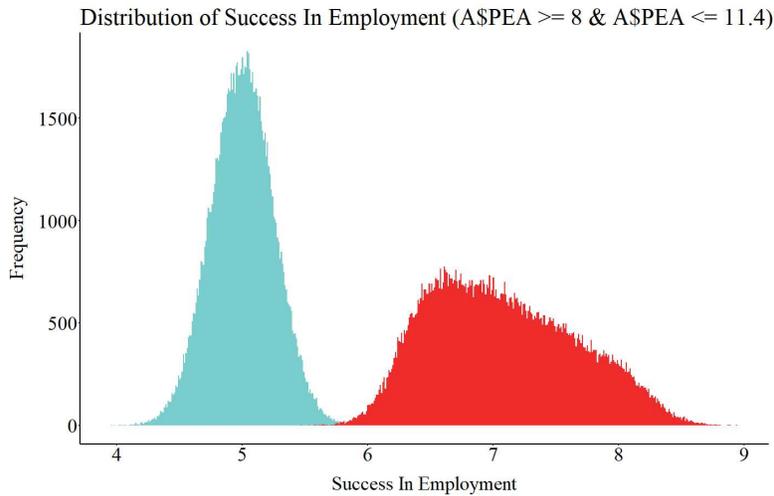

*Notes:* Red represents the treatment group (individuals exposed to the employment program) and light blue represents the control group (individuals not exposed to the employment program).

### *Low Parental Education (PEA <= 8)*

Finally, let's take a sample of individuals with low parental education. When the slope coefficients from the 10,000 samples of 1,000 participants with low parental education are plotted the findings suggest that participation in the employment program causes a 3.37 increase in success in employment. Additionally, the sampling distribution of the difference between the treatment and control cases on success in employment (i.e., *b*) ranges between 3.29 and 3.50.

```
>N<-10000
>RSDF1 <- new.env()
>### Loop
>for(i in 1:N)
>{
>   NRS2<-sample_n(A[which(A$PEA <= 8),] ,1000, replace = F)
>
>   LM_NRS<-lm(SIEM~EP, data = NRS2)
>   summary(LM_NRS)
>
>   NRS2DF1$Slope<-c(NRS2DF1$Slope, summary(LM_NRS)$coefficients[2])
>   NRS2DF1$SE<-c(NRS2DF1$SE, summary(LM_NRS)$coefficients[4])
>
>}
```

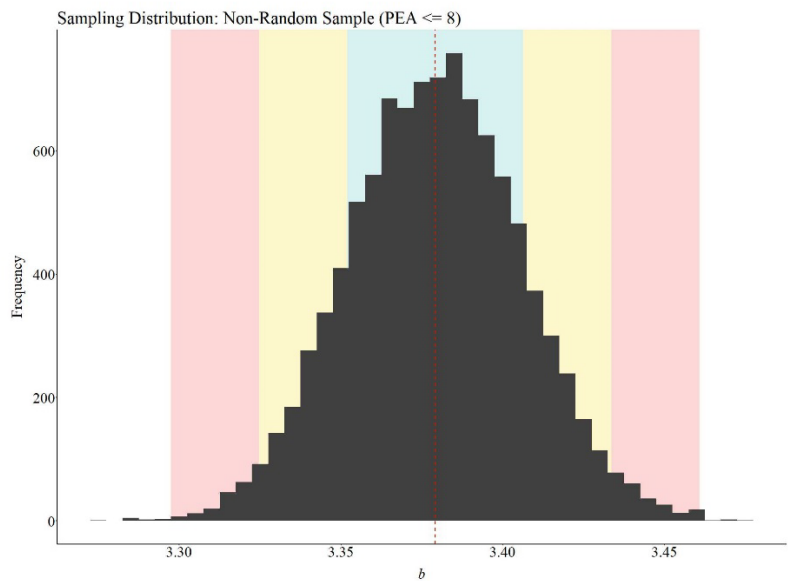

*Notes*: The red dashed line represents the mean slope coefficient, while the blue, yellow, and red areas represent 1, 2, and 3 (respectively) standard deviations from the mean.

The figure below was created to illustrate the location of the low parental education treatment and control cases.

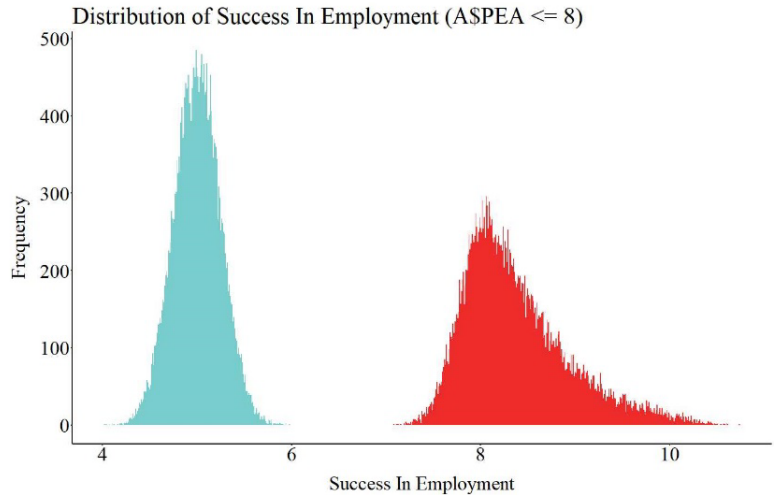

*Notes:* Red represents the treatment group (individuals exposed to the employment program) and light blue represents the control group (individuals not exposed to the employment program).

**Conclusion**

Taken together, these findings illustrate that the effectiveness of the employment program varies by parental education. Nevertheless, when conducting a randomized controlled trial, we likely only have the time and funds to draw one sample from the population. Moreover, the cost of conducting a randomized controlled trial oftentimes encourages scholars to sacrifice the generalizability of the findings. Sacrificing the generalizability of a study is not a problem when we satisfy the assumption of known interactions. However, when an interaction between the treatment program and a third variable is unknown, drawing a convenience sample could drastically alter the perceived effectiveness of a program.

Going back to our non-random sampling distributions, if we mistakenly drew our sample from the high parental education distribution, we would perceive that the employment program is detrimental or ineffective at best. Specifically, we would conclude that individuals exposed to the employment program had lower levels of success in employment than individuals that did not participated in the employment program. In opposition, if we mistakenly drew our sample from the low parental education distribution, we would conclude that the employment program is extremely effective at increasing success in employment. While all of the non-random sampling distributions represent causal estimates (as a product of random assignment), none of the distributions captured the difference on success in employment across the entire population. Explicitly, the causal effects observed in each non-random sampling distribution would only generalize to individuals contained within that group. This highlights the potential problem with conducting experimental research on non-generalizable samples. As such, I would caution scholars to consider the effects of violating the known interaction assumption the next time they conduct a randomized controlled trial.

---

[i] Continuous treatment variables don't inherently permit themselves to be examined through randomized controlled trials.

[ii] As a reminder, the focus of the series is on violating statistical assumptions. However, these assumptions also talk to methodological limitations that could exist in a randomized controlled trial.

[iii] We assume that the interaction between parental education and the employment program is unknown to us. Even though this loop is coded as a random sample, it emulates a convivence sample by only drawing cases from a proportion of the distribution on parental education.

**Entry 6: Covariate Imbalance (Randomized Controlled Trials)**

**Introduction**
As introduced in Entry 5, the random assignment of respondents to a treatment or a control is fundamental to the process of conducting a true experiment. We assume that random assignment will remove the possibility of an unknown variable confounding the association between exposure to the treatment and an outcome of interest. Under this assumption, random assignment permits the ability to make *causal inferences about the effects of the treatment*. This assumption, however, does not provide us with the ability to *direct observe the causal effects of a treatment* (the variation in the outcome directly caused by the treatment). Specifically, covariate balance across the treatment and control cases is needed to observe the causal effects of a treatment. In the current context, covariates refer to any observable or unobservable construct that can influence the outcome of interest. Covariate balance oftentimes is not achieved within a single randomized controlled trial. Only when enough replications of the randomized controlled trial are conducted can we assume – through reliance on central limit theorem – that covariate balance between the treatment and control cases is established. As illustrated throughout the current entry, when covariate balance is not achieved – violations of the assumption of covariate balance – the treatment effects estimated from a single true experiment can deviate substantively from the true causal effects of the treatment in the population.

**The Counterfactual**
Although it can be described in a variety of manners, a counterfactual can be considered the "what if?" statement to an experiment. Imagine we are conducting a basic experiment in a lab. For simplicity, we are interested in observing what happens when rubbing alcohol is exposed to a flame. We distribute rubbing alcohol from the same bottle evenly into two cups, light a match, and place the match close to the rubbing alcohol in one of the cups. The rubbing alcohol lit on fire. Nevertheless, we can not determine if the match caused the rubbing alcohol to light on fire without observing what happened to the rubbing alcohol in the other cup. As such, we turn to the other cup and observe what happened when the rubbing alcohol is not exposed to the flame. The rubbing alcohol did not appear to light on fire without being exposed to the flame. Considering everything else is equal, we can determine that exposing rubbing alcohol to a flame caused it to light on fire.

As alluded to throughout the example, a true counterfactual condition can only be established when all of the characteristics of the experiment are replicated except for the treatment being varied (e.g., flame vs. no flame). Due to this requirement, a true counterfactual can not be established when conducting an experiment with humans because the participants exposed to the treatment and the participants exposed to the control will differ. These differences could arise for a host of reasons, including sociological differences, biological differences, different life-course histories, etc. Moreover, differences on these key characteristics are likely to influence our outcome of interest, generating a need to establish covariate balance.

*The Assumption of Covariate Balance*
Our best opportunity to approximate a true counterfactual condition when conducting an experiment with humans is through the establishment of covariate balance. Covariate balance in the context of a counterfactual refers to the condition where cases assigned to the treatment group vary identically to cases assigned to the control group across all key characteristics. To provide an example, our treatment and control groups will be balanced on age if the mean, standard deviation, skewness, and kurtosis of the distribution is identical across the two groups. Any differences, however, will be indicative of covariate imbalance (i.e., the characteristics of the participants are not equal across the treatment and control groups).

When conducting a true experiment, we randomly assign participants to the treatment and control groups under the assumption that (1) variation in the treatment will not be predicted by a confounding variable and (2) covariate balance will be established. Satisfying the first assumption permits causal inferences to be made, while satisfying the second assumption permits the identification of the true effect size of the treatment in the population (i.e., the causal effects of a treatment in the population).[i] Random assignment for a single experiment, however, commonly does not satisfy the second assumption. Specifically, simply randomly assigning participants increases the likelihood of covariate imbalance across the treatment and control groups in a single experiment. Nevertheless, while the assumption of covariate balance is commonly violated in a single experiment, multiple replications using random assignment will ensure that the second assumption is satisfied when a meta-analysis is conducted.

**Violations of The Assumption of Covariate Balance**
Let's begin the illustration by simulating the treatment population and the control population separately. By simulating the treatment and control populations separately, we can ensure that the specified covariates are identical – balanced – across the two populations. First, let us begin with the treatment population. The specification below simulates 50,000 cases with a value of 1 on the treatment ("Tr" in the *R*-code) variable. In addition to the treatment variable, four additional variables are simulated for the treatment cases. These include a dichotomous variable ("DI1" in the *R*-code), a categorical variable ("DV1" in the *R*-code), a semi-continuous variable ("SCV1" in the *R*-code), and a continuous variable ("CV1" in the *R*-code). After which, we merge all of the measures into a single dataframe. Importantly, the seed for the simulation was set to 1992 to ensure the process can be replicated.

```
> set.seed(1992)
> N<-50000
> Tr<-rep(1,N)
> DI1<-rep(c(0,1),length.out = N) # Sex
> DV1<-trunc(rnorm(N,2.5,.75)) # Categorical variable
> SCV1<-trunc(rnorm(N,35,5)) # Age
> CV1<-trunc(rnorm(N,35000,5000)) # Income
>
> Treat.DF<-data.frame(Tr,DI1,DV1,SCV1,CV1)
```

After simulating the treatment population, the simulation is replicated for the control population except for scores on the treatment variable being equal to "0" instead of "1" ("Tr" in the *R*-code).

```
> set.seed(1992)
> N<-50000
> Tr<-rep(0,N)
> DI1<-rep(c(0,1),length.out = N) # Dichotomous Variable
> DV1<-trunc(rnorm(N,2.5,.75)) # Categorical Variable
> SCV1<-trunc(rnorm(N,35,5)) # Semi-continuous Variable
> CV1<-trunc(rnorm(N,35000,5000)) # Continuous Variable
>
> Control.DF<-data.frame(Tr,DI1,DV1,SCV1,CV1)
```

As a brief check, let's make sure that the mean, standard deviation, skew, and kurtosis for the four additional variables (i.e., DI1; DV1; SCV1; CV1) are identical between the treatment and control populations. This check is conducted by subtracting the mean for the control group from the mean of the treatment group, the standard deviation for the control group from the standard deviation of the treatment group, and so on. The zeros across the board indicate that the treatment and control populations are identical.

```
> mean(Treat.DF$DI1)-mean(Control.DF$DI1)
[1] 0
> sd(Treat.DF$DI1)-sd(Control.DF$DI1)
[1] 0
> skew(Treat.DF$DI1)-skew(Control.DF$DI1)
[1] 0
> kurtosi(Treat.DF$DI1)-kurtosi(Control.DF$DI1)
[1] 0
>
> mean(Treat.DF$DV1)-mean(Control.DF$DV1)
[1] 0
> sd(Treat.DF$DV1)-sd(Control.DF$DV1)
[1] 0
> skew(Treat.DF$DV1)-skew(Control.DF$DV1)
[1] 0
> kurtosi(Treat.DF$DV1)-kurtosi(Control.DF$DV1)
[1] 0
>
> mean(Treat.DF$SCV1)-mean(Control.DF$SCV1)
[1] 0
> sd(Treat.DF$SCV1)-sd(Control.DF$SCV1)
[1] 0
> skew(Treat.DF$SCV1)-skew(Control.DF$SCV1)
[1] 0
> kurtosi(Treat.DF$SCV1)-kurtosi(Control.DF$SCV1)
[1] 0
>
> mean(Treat.DF$CV1)-mean(Control.DF$CV1)
[1] 0
> sd(Treat.DF$CV1)-sd(Control.DF$CV1)
[1] 0
> skew(Treat.DF$CV1)-skew(Control.DF$CV1)
[1] 0
> kurtosi(Treat.DF$CV1)-kurtosi(Control.DF$CV1)
[1] 0
```

Now that we have identical treatment and control populations, we can merge the dataframes – creating the overall population – and specify that scores on the dependent variable ("*Y*") were equal to 10 times the treatment, 1.25 times DI1, .25 times DV1, .0075 times SCV1, and .0075 times CV1.[ii]

```
> Pop<-rbind(Treat.DF,Control.DF)
> Pop$Y<-(10.00*Pop$Tr)+(1.25)*Pop$DI1+(.25*Pop$DV1)+(.0075*Pop$SCV1)+(.0075*Pop$CV1)
```

Now that we have created our simulated dataset, let's run some randomized controlled trials. As a reminder, it takes some mental gymnastics to work through simulated randomized controlled trials. Specifically, it is difficult to simulate a randomized controlled trial because scores on the treatment variable are not influenced by any other construct. As such, we are unable to reassign exposure to the treatment variable when samples are pulled from the population. For each sample we pull, just imagine that a randomized controlled trial is being conducted.

*Randomized Controlled Trial 1: Sample of 100*

To conduct our first randomized controlled trial, we sampled 50 treatment cases and 50 control cases from the population. The seed for this sampling procedure was specified as 633487.[iii] After randomly sampling treatment and control cases from the population, we need to evaluate if the covariates are balanced across the treatment and control cases. As a reminder, for us to define the covariates as balanced the characteristics of the participants must be equal across the two groups. While definitions of equal vary, this is generally defined as non-significant differences between the two groups. Personally, I don't like to define distributional differences based on *p*-values, so for this case we will define balance as the evidence *generally* – I use this term loosely – suggests that there are no distributional differences on the four covariates between the treatment and control cases. Specifically, the differences between the mean, standard deviation, skew, and kurtosis would have to be negligible after considering the range of scores on the covariate.

Considering the results of all of the distributional comparisons between our treatment and control cases, our random selection of cases from the population did not achieve covariate balance, violating the assumption of covariate balance. Specifically, the distributional differences appeared minimal for the dichotomous and categorical variable, but more substantive for the semi-continuous and continuous variables. By violating the assumption of covariate balance, we have unfortunately biased the estimates derived from our randomized controlled trial. In this case, the estimated effect of the treatment on the dependent variable (*Y*) was 18.93, which is almost double the true effect size of the treatment in the population (10.00). These results illustrate the importance of satisfying the covariate balance assumption when conducting a true experiment.

```
> set.seed(633487)
> TMC<-sample_n(Pop[which(Pop$Tr == 1),] ,50, replace = F)
> CMC<-sample_n(Pop[which(Pop$Tr == 0),] ,50, replace = F)
>
> # Dichotomous Variable
> mean(TMC$DI1)-mean(CMC$DI1)
[1] -0.06
> sd(TMC$DI1)-sd(CMC$DI1)
[1] 0.001215
> skew(TMC$DI1)-skew(CMC$DI1)
[1] 0.2334
> kurtosi(TMC$DI1)-kurtosi(CMC$DI1)
[1] -0.01859
>
> # Categorical Variable
> mean(TMC$DV1)-mean(CMC$DV1)
[1] 0.24
> sd(TMC$DV1)-sd(CMC$DV1)
[1] -0.0829
> skew(TMC$DV1)-skew(CMC$DV1)
[1] -0.4097
> kurtosi(TMC$DV1)-kurtosi(CMC$DV1)
[1] 0.8232
>
> # Semi-continuous Variable
> mean(TMC$SCV1)-mean(CMC$SCV1)
[1] -1.14
> sd(TMC$SCV1)-sd(CMC$SCV1)
[1] 0.8317
> skew(TMC$SCV1)-skew(CMC$SCV1)
[1] -0.2257
> kurtosi(TMC$SCV1)-kurtosi(CMC$SCV1)
[1] 0.1349
>
> # Continuous Variable
> mean(TMC$CV1)-mean(CMC$CV1)
[1] 1193
> sd(TMC$CV1)-sd(CMC$CV1)
[1] -701.5
> skew(TMC$CV1)-skew(CMC$CV1)
[1] -0.869
> kurtosi(TMC$CV1)-kurtosi(CMC$CV1)
[1] -1.204
>
> DF<-rbind(TMC,CMC)
>
> LM_NRS<-lm(Y~Tr, data = DF)
> summary(LM_NRS)

Call:
```

```
lm(formula = Y ~ Tr, data = DF)

Residuals:
    Min      1Q  Median      3Q     Max
 -86.36  -23.48   -6.25   24.51  129.68

Coefficients:
            Estimate Std. Error t value             Pr(>|t|)
(Intercept)   254.96       5.24   48.69 <0.0000000000000002 ***
Tr             18.93       7.41    2.56                0.012 *
---
Signif. codes:  0 '***' 0.001 '**' 0.01 '*' 0.05 '.' 0.1 ' ' 1

Residual standard error: 37 on 98 degrees of freedom
Multiple R-squared:  0.0625,    Adjusted R-squared:  0.0529
F-statistic: 6.53 on 1 and 98 DF,  p-value: 0.0121

>
```

One question that remains: does the sample size of the treatment and the control groups in a randomized controlled trial matter for satisfying the assumption of covariate balance? As demonstrated by the three replications of the first simulation with varying sample sizes (N = 200; N = 500; N = 1000), it does.

### *Randomized Controlled Trial 2: Sample of 200*

```
> set.seed(267714)
> TMC<-sample_n(Pop[which(Pop$Tr == 1),] ,100, replace = F)
> CMC<-sample_n(Pop[which(Pop$Tr == 0),] ,100, replace = F)
>
>
> # Dichotomous Variable
> mean(TMC$DI1)-mean(CMC$DI1)
[1] -0.03
> sd(TMC$DI1)-sd(CMC$DI1)
[1] -0.0003017
> skew(TMC$DI1)-skew(CMC$DI1)
[1] 0.1183
> kurtosi(TMC$DI1)-kurtosi(CMC$DI1)
[1] 0.004714
>
> # Categorical Variable
> mean(TMC$DV1)-mean(CMC$DV1)
[1] -0.01
> sd(TMC$DV1)-sd(CMC$DV1)
[1] 0.07611
> skew(TMC$DV1)-skew(CMC$DV1)
[1] -0.1198
> kurtosi(TMC$DV1)-kurtosi(CMC$DV1)
[1] 0.6161
>
> # Semi-continuous Variable
> mean(TMC$SCV1)-mean(CMC$SCV1)
[1] -0.15
> sd(TMC$SCV1)-sd(CMC$SCV1)
[1] -0.3808
> skew(TMC$SCV1)-skew(CMC$SCV1)
[1] -0.1472
> kurtosi(TMC$SCV1)-kurtosi(CMC$SCV1)
[1] 0.4915
>
> # Continuous Variable
> mean(TMC$CV1)-mean(CMC$CV1)
[1] 1212
> sd(TMC$CV1)-sd(CMC$CV1)
[1] 128.8
> skew(TMC$CV1)-skew(CMC$CV1)
[1] 0.2267
> kurtosi(TMC$CV1)-kurtosi(CMC$CV1)
[1] -0.1588
>
> DF<-rbind(TMC,CMC)
>
> LM_NRS<-lm(Y~Tr, data = DF)
> summary(LM_NRS)
```

```
Call:
lm(formula = Y ~ Tr, data = DF)

Residuals:
    Min     1Q Median     3Q    Max
-84.28 -21.64   0.95  21.63 100.18

Coefficients:
            Estimate Std. Error t value             Pr(>|t|)
(Intercept)   258.48       3.52   73.49 < 0.0000000000000002 ***
Tr             19.05       4.97    3.83              0.00017 ***
---
Signif. codes:  0 '***' 0.001 '**' 0.01 '*' 0.05 '.' 0.1 ' ' 1

Residual standard error: 35.2 on 198 degrees of freedom
Multiple R-squared:  0.069,     Adjusted R-squared:  0.0643
F-statistic: 14.7 on 1 and 198 DF,  p-value: 0.000172

> 
```

## *Randomized Controlled Trial 3: Sample of 500*

```
> set.seed(92021)
> 
> TMC<-sample_n(Pop[which(Pop$Tr == 1),] ,250, replace = F)
> CMC<-sample_n(Pop[which(Pop$Tr == 0),] ,250, replace = F)
> 
> 
> # Dichotomous Variable
> mean(TMC$DI1)-mean(CMC$DI1)
[1] 0.028
> sd(TMC$DI1)-sd(CMC$DI1)
[1] 0.0005615
> skew(TMC$DI1)-skew(CMC$DI1)
[1] -0.1114
> kurtosi(TMC$DI1)-kurtosi(CMC$DI1)
[1] -0.00891
> 
> # Categorical Variable
> mean(TMC$DV1)-mean(CMC$DV1)
[1] -0.064
> sd(TMC$DV1)-sd(CMC$DV1)
[1] 0.09333
> skew(TMC$DV1)-skew(CMC$DV1)
[1] 0.08768
> kurtosi(TMC$DV1)-kurtosi(CMC$DV1)
[1] -0.5585
> 
> # Semi-continuous Variable
> mean(TMC$SCV1)-mean(CMC$SCV1)
[1] 0.032
> sd(TMC$SCV1)-sd(CMC$SCV1)
[1] -0.1224
> skew(TMC$SCV1)-skew(CMC$SCV1)
[1] -0.07692
> kurtosi(TMC$SCV1)-kurtosi(CMC$SCV1)
[1] -0.6315
> 
> # Continuous Variable
> mean(TMC$CV1)-mean(CMC$CV1)
[1] 879.8
> sd(TMC$CV1)-sd(CMC$CV1)
[1] -459.2
> skew(TMC$CV1)-skew(CMC$CV1)
[1] 0.2811
> kurtosi(TMC$CV1)-kurtosi(CMC$CV1)
[1] 0.2042
> 
> DF<-rbind(TMC,CMC)
> 
> LM_NRS<-lm(Y~Tr, data = DF)
> summary(LM_NRS)

Call:
lm(formula = Y ~ Tr, data = DF)

Residuals:
    Min      1Q  Median      3Q     Max
```

```
-130.17  -22.38    0.95   23.50  149.28

Coefficients:
            Estimate Std. Error t value             Pr(>|t|)
(Intercept)   257.98       2.26   114.2 < 0.0000000000000002 ***
Tr             16.62       3.19     5.2           0.00000029 ***
---
Signif. codes:  0 '***' 0.001 '**' 0.01 '*' 0.05 '.' 0.1 ' ' 1

Residual standard error: 35.7 on 498 degrees of freedom
Multiple R-squared:  0.0516,	Adjusted R-squared:  0.0496
F-statistic: 27.1 on 1 and 498 DF,  p-value: 0.000000287

>
```

## *Randomized Controlled Trial 4: Sample of 1000*

```
> set.seed(21873)
>
> TMC<-sample_n(Pop[which(Pop$Tr == 1),] ,500, replace = F)
> CMC<-sample_n(Pop[which(Pop$Tr == 0),] ,500, replace = F)
>
>
> # Dichotomous Variable
> mean(TMC$DI1)-mean(CMC$DI1)
[1] -0.056
> sd(TMC$DI1)-sd(CMC$DI1)
[1] 0
> skew(TMC$DI1)-skew(CMC$DI1)
[1] 0.2237
> kurtosi(TMC$DI1)-kurtosi(CMC$DI1)
[1] 0
>
> # Categorical Variable
> mean(TMC$DV1)-mean(CMC$DV1)
[1] 0.046
> sd(TMC$DV1)-sd(CMC$DV1)
[1] -0.02973
> skew(TMC$DV1)-skew(CMC$DV1)
[1] -0.1608
> kurtosi(TMC$DV1)-kurtosi(CMC$DV1)
[1] 0.149
>
> # Semi-continuous Variable
> mean(TMC$SCV1)-mean(CMC$SCV1)
[1] 0.276
> sd(TMC$SCV1)-sd(CMC$SCV1)
[1] 0.07067
> skew(TMC$SCV1)-skew(CMC$SCV1)
[1] 0.06591
> kurtosi(TMC$SCV1)-kurtosi(CMC$SCV1)
[1] -0.159
>
> # Continuous Variable
> mean(TMC$CV1)-mean(CMC$CV1)
[1] -345.6
> sd(TMC$CV1)-sd(CMC$CV1)
[1] -144
> skew(TMC$CV1)-skew(CMC$CV1)
[1] 0.2711
> kurtosi(TMC$CV1)-kurtosi(CMC$CV1)
[1] 0.3549
>
> DF<-rbind(TMC,CMC)
>
> LM_NRS<-lm(Y~Tr, data = DF)
> summary(LM_NRS)

Call:
lm(formula = Y ~ Tr, data = DF)

Residuals:
    Min      1Q  Median      3Q     Max
-119.25  -25.64   -2.59   25.84  145.00

Coefficients:
            Estimate Std. Error t value            Pr(>|t|)
(Intercept)   265.45       1.66  159.43 <0.0000000000000002 ***
Tr              7.35       2.35    3.12              0.0018 **
```

```
---
Signif. codes:  0 '***' 0.001 '**' 0.01 '*' 0.05 '.' 0.1 ' ' 1

Residual standard error: 37.2 on 998 degrees of freedom
Multiple R-squared:  0.00967,   Adjusted R-squared:  0.00868
F-statistic: 9.75 on 1 and 998 DF,  p-value: 0.00185

>
```

**Discussion**

As illustrated by the four simulations, randomization does not always satisfy the assumption of covariate balance. In most situations, randomly assigning cases to the treatment and control groups actually generates a condition where we have imbalance – the mean, standard deviation, skewness, and kurtosis of the distribution is different – between our treatment and control cases on known and unknown covariates. Violations of the assumption of covariate balance will bias the observed effectiveness of a treatment, limiting our ability to identify the true effect size of the treatment in the population. Increasing the sample size of the randomized controlled trial is one potential method for satisfying the assumption of covariate balance. Nevertheless, conducting a large randomized controlled trial is quite expensive. There, however, are alternative methods for achieving covariate balance when conducting randomized controlled trials.

One method to increase covariate balance during a randomized controlled trial is blocked randomization (Efird, 2011; Lachin, Matts, and Wei, 1988). Briefly, blocked randomization is the process of stratifying a sample into strata with similar scores on key covariates and randomly assigning participants from each stratum to the treatment and control groups. For example, a sample of 12 individuals could be stratified into two groups based upon biological sex (6 males in strata 1 and 6 females in strata 2). Then 3 individuals from stratum 1 and 3 individuals from stratum 2 can be randomly assigned to the treatment group, while 3 individuals from stratum 1 and 3 individuals from stratum 2 can be randomly assigned to the control group. Blocked randomization requires the observation of key covariates to identify appropriate strata. While blocked randomization can increase covariate balance, replications are fundamental to satisfying the assumption of covariate balance.

*Achieving Covariate Balance Through Replications*

To provide an illustration of how replications can satisfy the assumption of covariate balance, three additional simulations were conducted. For these simulations, we (1) randomly sampled 100 cases (50 treatment cases & 50 control cases), (2) estimated the mean difference between the treatment and control case on the key covariates, (3) estimated the effects of the treatment on the outcome of interest, and (4) repeated the process 10, 100, and 1000 times (respectively).[iv] The results illustrate that the assumption of covariate balance is partially satisfied when a randomized controlled trial is replicated 10 times, but fully satisfied when a randomized controlled trial is replicated 100 or 1000 times. These findings directly illustrate the principles of the central limit theorem and indicate that random assignment can achieve covariate balance if enough replications are conducted. This – in addition to various others – is just another reason why replications are extremely important when conducting experimental research.

*10 replications*

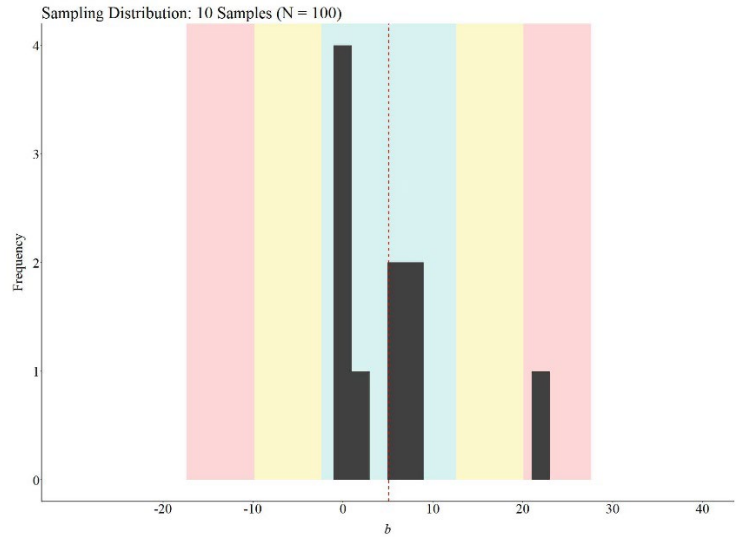

*Notes:* "*b*" represents the estimated effects of the treatment and dependent variable. The red dashed line represents the mean effect size of the treatment on the dependent variable across the 10 replications of the randomized controlled trial (N = 100; true causal estimate in the population is 10.00). The blue, yellow, and red areas represent 1, 2, and 3 (respectively) standard deviations from the mean.

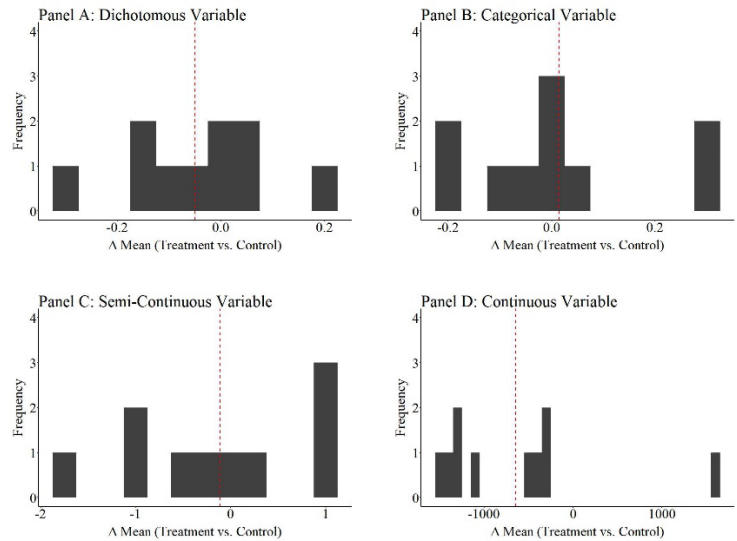

*Notes:* "Δ mean" is the mean difference between the treatment and control cases on the specified covariate. The red dashed line represents the average of the mean differences across the 10 replications of the randomized controlled trial (N = 100) on the specified covariate.

*100 replications*

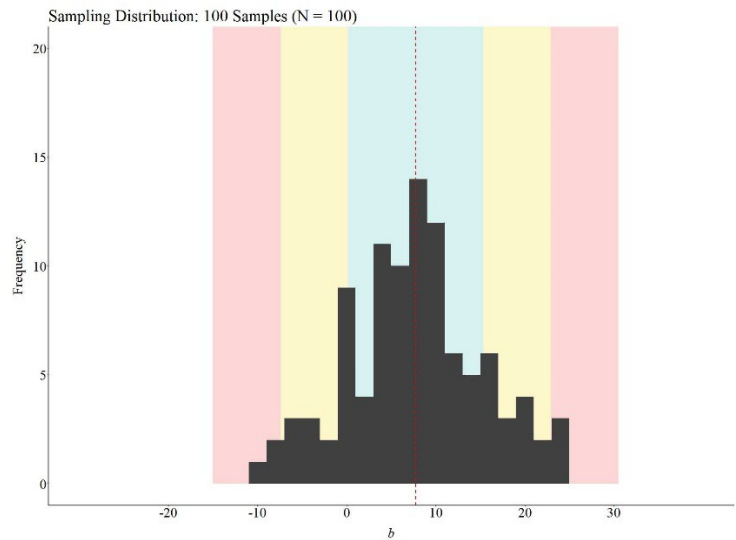

*Notes:* "*b*" represents the estimated effects of the treatment and dependent variable. The red dashed line represents the mean effect size of the treatment on the dependent variable across the 100 replications of the randomized controlled trial (N = 100; true causal estimate in the population is 10.00). The blue, yellow, and red areas represent 1, 2, and 3 (respectively) standard deviations from the mean.

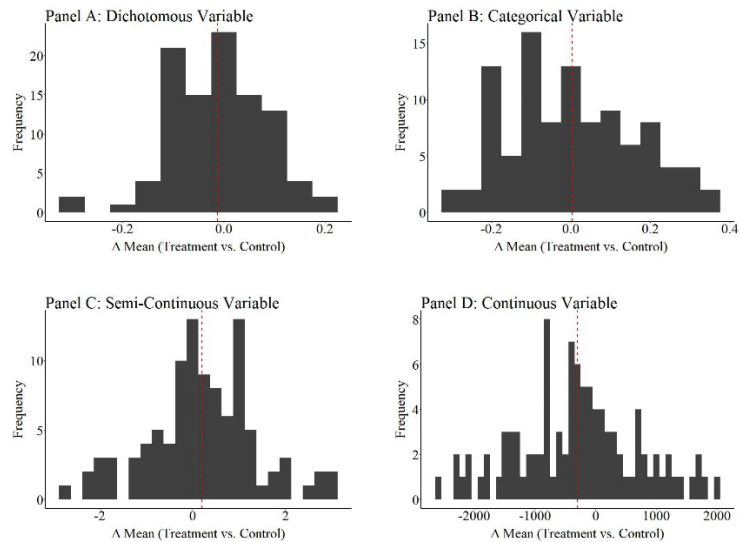

*Notes:* "Δ mean" is the mean difference between the treatment and control cases on the specified covariate. The red dashed line represents the average of the mean differences across the 100 replications of the randomized controlled trial (N = 100) on the specified covariate.

*1000 replications*

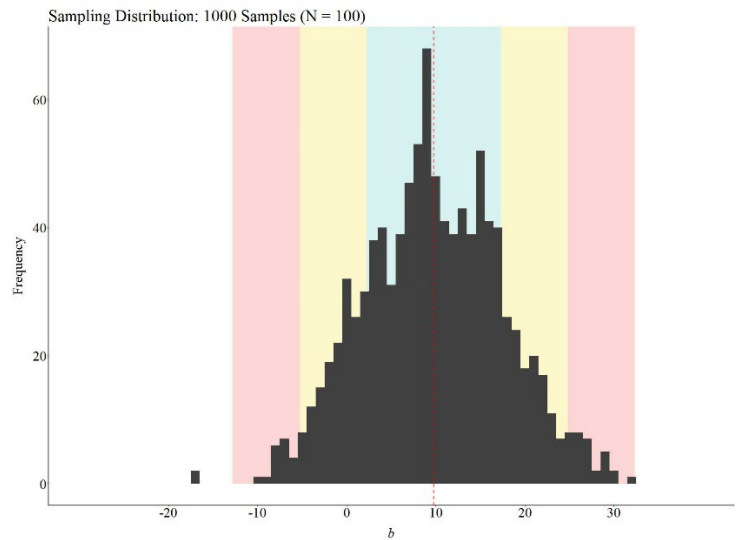

*Notes:* "b" represents the estimated effects of the treatment and dependent variable. The red dashed line represents the mean effect size of the treatment on the dependent variable across the 1000 replications of the randomized controlled trial (N = 100; true causal estimate in the population is 10.00). The blue, yellow, and red areas represent 1, 2, and 3 (respectively) standard deviations from the mean.

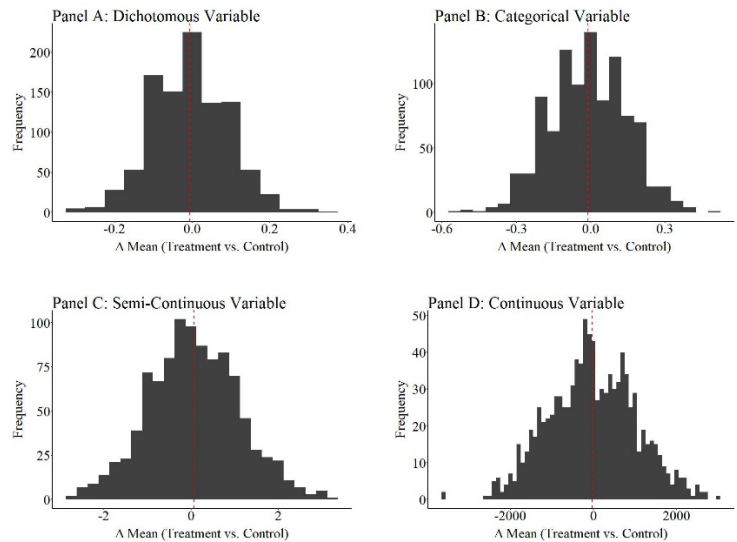

*Notes:* "Δ mean" is the mean difference between the treatment and control cases on the specified covariate. The red dashed line represents the average of the mean differences across the 1000 replications of the randomized controlled trial (N = 100) on the specified covariate.

**Conclusion**

As illustrated throughout the current entry, conducting a randomized controlled trial might not satisfy the assumption of covariate balance. Specifically, covariate imbalance between the treatment and control cases is likely to exist when participants are simply randomly assigned to the treatment or control conditions. When covariate imbalance exists – a violation of the assumption of covariate balance – the estimated effects of the treatment will likely be biased. While the direction and magnitude of the bias could vary substantively, violations of the assumption of covariate balance hinder our ability to identify the true effect size of the treatment in the population (i.e., the causal effects). Considering the importance of satisfying the assumption of covariate balance, we should (1) proactively employ techniques that increase our ability to satisfy the assumption of covariate balance (e.g., blocked randomization) and (2) replicate randomized controlled trials as frequently as possible.

---

[i] Importantly, the assumption of covariate balance only matters when variation in the outcome of interest is predicted by known or unknown variation in the population. Considering that almost all outcomes of interest are predicted by known or unknown variation in the population, this assumption must be satisfied to observed the causal effects of a treatment in the population when conducting a true experiment.

[ii] These values were selected based upon the level of measurement for the specified variable, as well as illustrative purposes.

[iii] The seed for these simulations are extremely important, as the seed influences if the randomization procedure generates balance or imbalanced on the covariates across the treatment and control cases. I suggest altering the seed to explore covariate balance when other seed values are used to conduct the random sampling.

[iv] The *R*-code for this procedure is provided in the *Sources of Statistical Bias Series* on ianasilver.com.

# Entry 7: The Exclusion of Confounders (Confounder Bias)

**Introduction**

Unlike analyses employing experimental data, conducting research with observational data requires us to consider the potential mechanisms confounding the association between two variables of interest. Briefly, a confounder is a variable that causes variation in two or more distinct constructs. For example, height can cause variation in partners' height, while simultaneously causing variation in the likelihood of playing professional sports. Considering that we can not randomly assign partners' height (unless you somehow create a love potion), if we were interested in the association between playing professional sports and partners' height we would have to adjust our model – or condition our analysis – upon the height of the participant. This is because height in the current scenario represents a confounder and can bias the association between partners' height and playing professional sports.

Numerous observed and unobserved confounders can exist when studying any association with observational data. When these confounders are not adjusted for, they can upwardly bias the slope coefficient – generate a slope coefficient further from zero than reality – and make us believe an association exists when in reality the association is spurious. Spurious in the current context is used to describe a statistical association that only exists due to an unadjusted for confounder. Nevertheless, when an association does exist (e.g., playing professional sports does influence partner's height), not adjusting for a confounder can upwardly or downwardly (generate a slope coefficient closer to zero than reality) bias the slope coefficient . Importantly, the slope coefficient being upwardly or downwardly biased, as well as the direction – positive or negative – and magnitude of the bias, is conditional upon the true slope coefficient of the association of interest and the direction and magnitude of the effects of the confounder on the constructs being examined.

**Visualizing The Directionality of Slope Coefficient Confounder Bias**

Although defined above, the best method for identifying and determining the bias generated by an unadjusted for confounder is through visualization, or more formally a Directed Acyclic Graph (DAG). Panel A of Figure 1 provides an illustration of a confounder or C causally influencing (indicated by the solid single headed arrow) variation in X and variation in Y. When the influence of the confounder is left unadjusted for, a covariance – illustrated using the double headed dashed arrow – between X and Y exists, when in reality X and Y are unrelated. Briefly, this covariance exists because X and Y share variation, but the shared variation is only attributable to the common cause (i.e., C) rather than X causing variation in Y or Y causing variation in X. In this scenario, a linear regression model will produce upwardly biased slope coefficients for the association between X and Y when the model does not adjust for the influence of the confounder on Y. That is, the estimates will be further from zero than reality, in turn increasing the likelihood of committing a type 1 error and rejecting the null hypothesis when in reality we should have retained the null hypothesis.

Nevertheless, let's consider that the slope coefficient of the true association between X and Y is not zero, but rather 1. In this scenario, not adjusting for the effects of a confounder in a statistical model could downwardly or upwardly bias the slope coefficients. Similar to the previous scenario, upward bias is generated by a confounder increasing the covariation between X and Y, where a portion of the covariation is attributable to the causal pathway between X and Y and a portion of the covariation is attributable to the common cause (i.e., C). Focusing on Panel B of Figure 1, an unadjusted for confounder can upwardly bias slope coefficients – in this scenario make the slope coefficient larger than 1 – when increased scores on the confounder cause scores on both X and Y to increase or decrease simultaneously. The type of confounder described above acts as an amplifier, where the confounded association is stronger than the unconfounded association.

Confounders, however, can upwardly or downwardly bias the observed slope coefficients by differentially influencing scores on X and Y. Specifically, when a confounder (C) negatively influences scores on X but positively influences scores on Y (or the alternative), the confounded association will possess a slope coefficient smaller than 1. Under certain conditions, the slope coefficient can be downwardly bias (closer to zero than the absolute value of 1) and range between -.99 and .99, or upwardly biased in the opposite direction (i.e., $b > -1.00$; further from zero than the absolute value of 1). The downward bias is generated by the confounder reducing the covariation between X and Y, while the upward bias in the opposite direction is generated by the confounder increasing the covariation between X and Y in the opposite direction. The observation of a downwardly or upwardly biased slope coefficient is conditional upon the magnitude of the differential effects the confounder has on X and Y.

This, contrary to popular belief, means that not adjusting for a confounder in a statistical model can increase the likelihood of committing a type 2 error and/or produce an estimate resulting in a distinct interpretation from the true association between X and Y. Considering these effects, it is extremely important to understand how confounders could bias the results of a statistical model. As such, let's conduct a detailed exploration of the bias generated by not adjusting for confounders in a linear regression model.

Panel A: Confounder Bias        Panel B: Direction of Confounder Bias

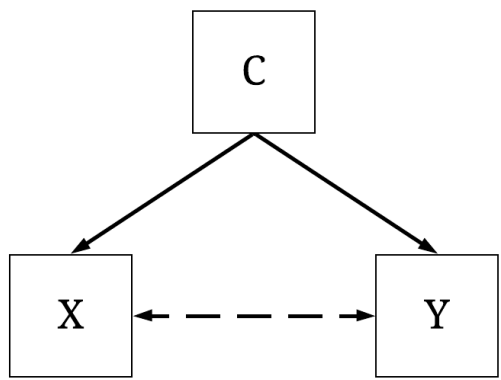
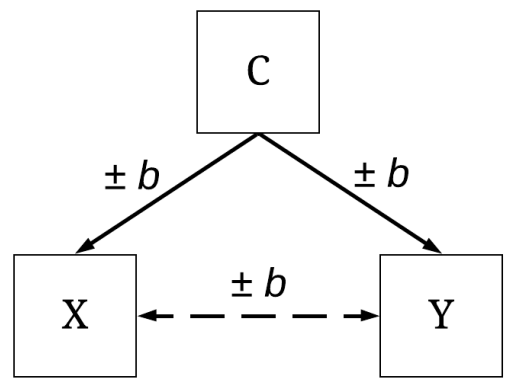

**True Association Between X and Y = 0**
The examples below are specified in a manner where X has no causal influence on Y. As such, the true slope coefficient between X and Y is equal to zero. Nevertheless, we specified that the association between X and Y will be confounded by C, where the confounder will cause variation in both X and Y. Importantly, the direction of influence (e.g., positive or negative specification) of the confounder on X and/or Y will vary across our simulations. Although not adjusting for the confounder will inevitably generate an upward bias in the slope coefficient (slope coefficients further from zero than reality [the true $b = 0$]), the direction of the bias will vary depending upon the direction of the effects of the confounder on X and Y.

*Confounder (+,+)*

Let's start with the confounder having a positive influence on both X and Y, where increased (or decreased) scores on C cause increased (or decreased) scores on X and Y simultaneously. In this scenario, as specified in the code below, we simulated the confounder or $c$ to be a normally distributed variable with a mean of 0 and a SD of 2.5. Our sample size for this simulation was 500 cases. After simulating the confounder, X and Y were specified in an identical manner (excluding the different set.seed), where a 1 point increase (or decrease) in C corresponded to a 2 point increase (or decrease) in X and a 2 point increase (or decrease) in Y. The confounder was specified to explain ~ 50 percent of the variation in both X and Y. The remaining – or residual – variation was specified to be normally distributed with a mean of 0 and a SD of 2.5. After simulating the three constructs, a dataframe was created and titled Data (creative, I know).

```
> ## Simulating a Confounder (+,+) ####
>
> n<-500 # Sample size
> set.seed(1001) # Seed
> c<-rnorm(n,0,2.5)
> set.seed(42125) # Seed
> x<-2*c+2*rnorm(n,0,2.5) # Specification of the independent variable
> set.seed(31) # Seed
> y<-2*c+2*rnorm(n,0,2.5)
>
> Data<-data.frame(c,x,y)
```

Using the data, we then estimated the unconfounded association (Panel A; i.e., the model adjusted for the influence of the confounder) and confounded association (Panel B; i.e., the model not adjusted for the influence of the confounder). As demonstrated, the unconfounded model produced a slope coefficient of .017 (Panel A) and suggested that X has no influence on Y ($p = .701$). The slope coefficient not being perfectly zero is due to random error in the seeds selected for the simulation. Unsurprisingly, the confounded model produced a slope coefficient of .463 (Panel B) and suggested that X has a strong and statistically significant positive influence on Y ($p < .001$).

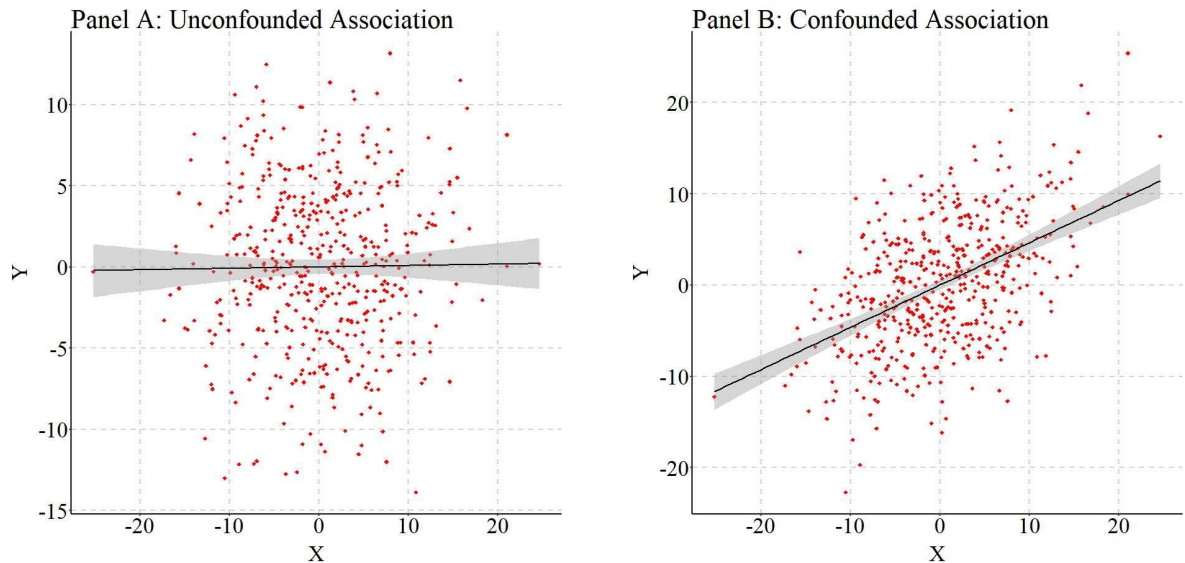

These findings, however, only represent the selected scenario. As such, using the code below, we replicated the simulation 10,000 times. In this looped simulation, we allow the number of cases to randomly vary (on a uniform distribution) between 100 and 1000. Additionally, we permit the strength of the association between the confounder and X, and the confounder and Y, to be randomly specified as any value between 1 and 100. Moreover, the influence of the residual variation in X and Y was randomly specified as any value between 1 and 100. The residual variation was specified to be normally distributed with a mean randomly selected from any value between -5 and 5 and a standard deviation randomly selected from any value between 1 and 5. All of the random specifications were conducted by drawing a single value from a uniform distribution – where all values have an equal likelihood of being selected – using the runif command in R. Importantly, the specified association between X and Y remained zero across all of the simulations.

The loop was specified using the foreach package, which permits parallel processing and the aggregation of data from each loop into a single dataframe. Briefly, the foreach package is extremely useful for running a simulation analysis like the one specified below. For the current loops, we recorded the run number (represented by i) and the estimated slope coefficient of the confounded association. Conducting a simulation in the manner below permits us to develop a comprehensive understanding of the bias generated by not adjusting for confounders that have a positive influence on both X and Y across randomly specified scenarios.

*Brief Note*: I just want to take a second to briefly note that this analysis might appear complex, but we are simply replacing the values in the code above (i.e., 2) with values generated by the computer between thresholds we set. Moreover, the loop just replicates the process as many times as we decide.

```
n<-10000

DATA1 = foreach (i=1:n, .packages='lm.beta', .combine=rbind) %dopar%
  {
    N<-sample(100:1000, 1)
    c<-rnorm(n,runif(1,-5,5),runif(1,1,5))
    x<-runif(1,1,100)*c+runif(1,1,100)*rnorm(n,runif(1,-5,5),runif(1,1,5))
    y<-runif(1,1,100)*c+runif(1,1,100)*rnorm(n,runif(1,-5,5),runif(1,1,5))
```

```
        Data<-data.frame(c,x,y)
        M<-lm(y~x, data = Data)
        bXY<-M$coefficients[2]
        Distance<-0-bXY
        data.frame(i,bXY)
}
```

The findings of the simulation loop are provided in the figure below and overwhelmingly indicated that the confounded slope coefficient will be a positive value. This suggests that confounders that have a positive influence on both the independent and dependent variable will commonly be upwardly bias and generate a positive slope coefficient when the influence of the confounder is not adjusted for in the model. Moreover, the magnitude of the bias that exists in the estimate depends upon the magnitude of the effects the confounder has on the variation in both X and Y.

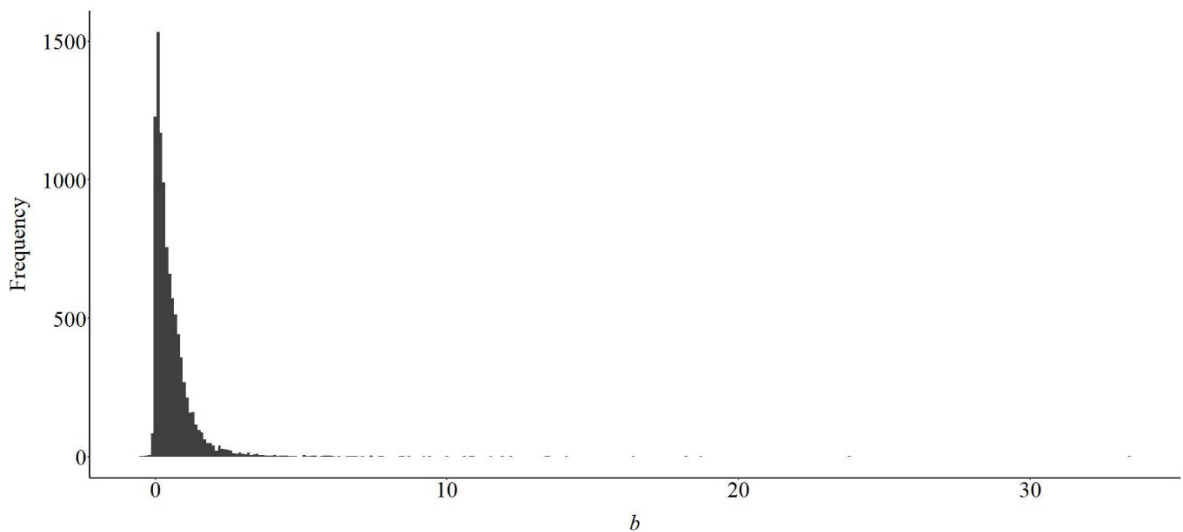

Now let's replicate the analysis a couple of times, but alter the direction of the influence of the confounder on X and Y.

*Confounder (-,+)*
For this replication, everything remained the same excluding the influence of the confounder (C) on X. Specifically, now every one-point increase (or decrease) in the confounder was associated with a two-point decrease (or increase) in X. The code for this replication (and the subsequent replications) is provided at www.ianasilver.com. Although the unconfounded model produced a slope coefficient identical to the previous example ($b = .017$; $p = .701$), the confounded model produced a negative slope coefficient ($b = -.446$; $p < .001$). Considering that the current example is identical to the previous example except for one difference, the findings suggest that the direction of the influence of the confounder on X influences the direction of the upward bias observed in the slope coefficient of the confounded association.

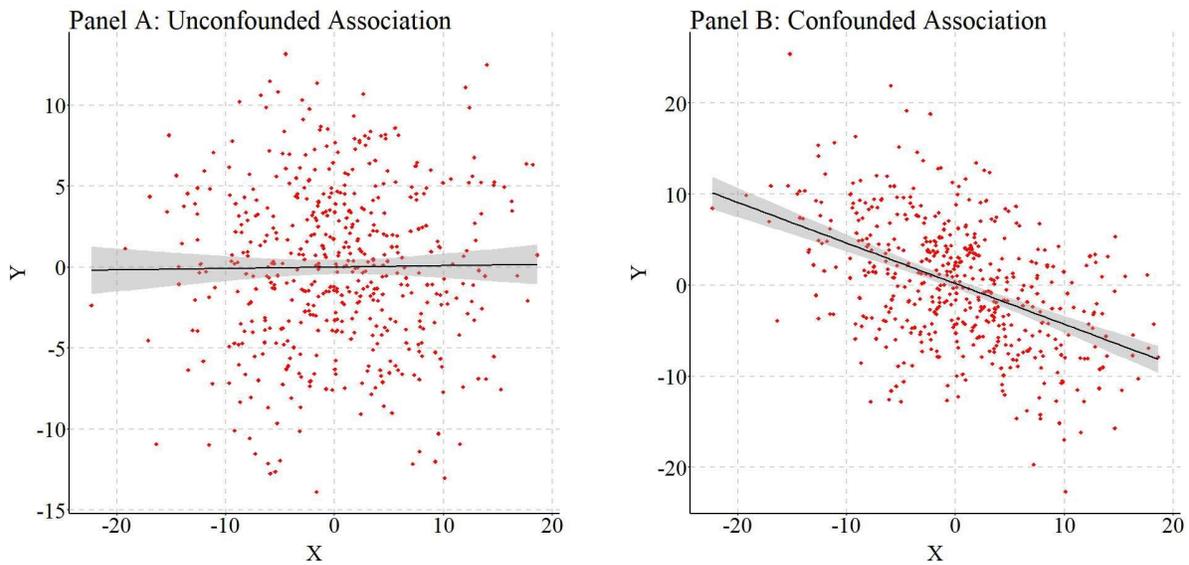

That interpretation was reconfirmed by the looped analysis. Specifically, distinct from the previous example, the findings of the simulation loop – provided in the figure below – overwhelmingly indicated that the confounded slope coefficient will be a negative value. This suggests that confounders that have a negative influence on the independent variable, but a positive influence on the dependent variable will commonly be upwardly bias and generate a negative slope coefficient when the influence of the confounder is not adjusted for in the model.

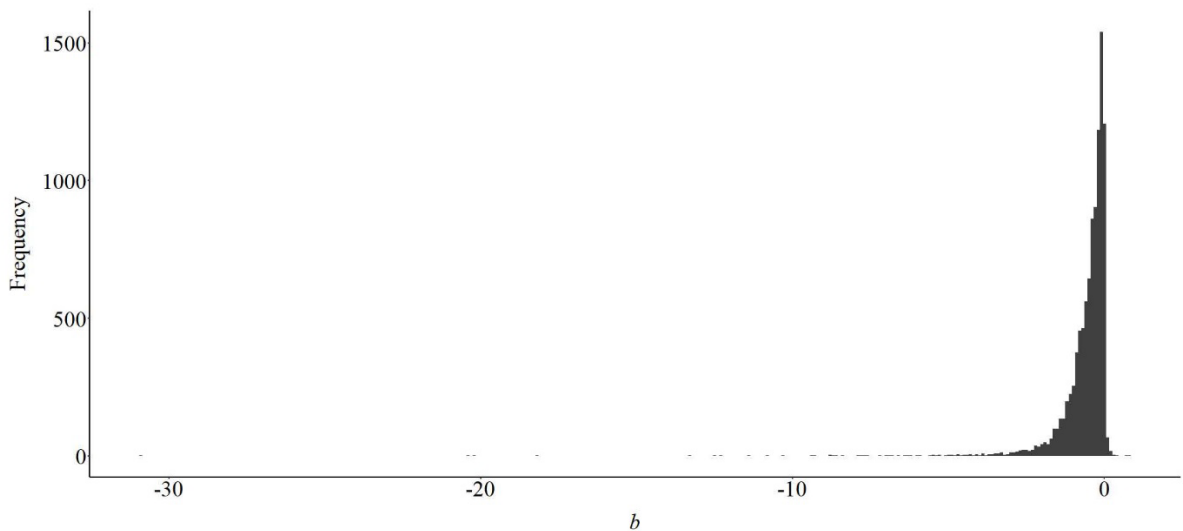

*Confounder (+,-)*
To observe the effects of the opposite specification, the current replication was specified where a one-point increase (or decrease) in the confounder (C) was associated with a two-point decrease (or increase) in Y. Similar to the previous example, the negative association between the confounder and Y generated a negative confounded association between X and Y (Panel B).

Specifically, the findings suggested that a one-point increase (or decrease) in X was associated with a .497 decrease (or increase) in Y ($p < .001$). The unconfounded association (Panel A) produced a slope coefficient identical to the previous examples ($b = .017$; $p = .701$).

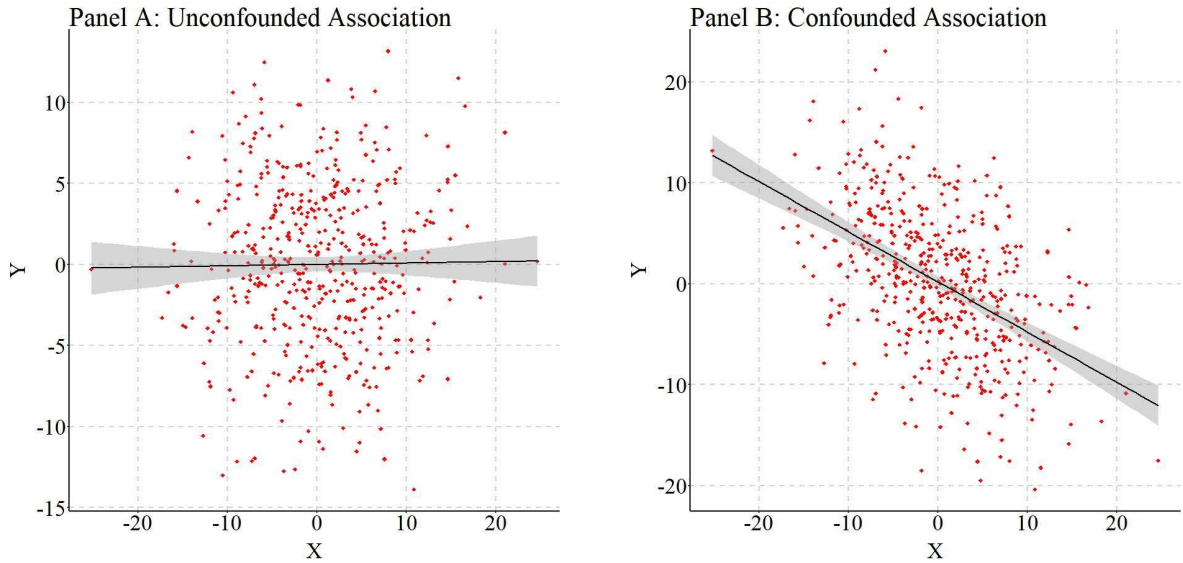

Switching to the looped simulation analysis, the findings indicated that the confounded slope coefficient will be a negative value. This suggests that confounders that have a negative influence on the dependent variable, but a positive influence on the independent variable will commonly be upwardly bias and generate a negative slope coefficient when the influence of the confounder is not adjusted for in the model.

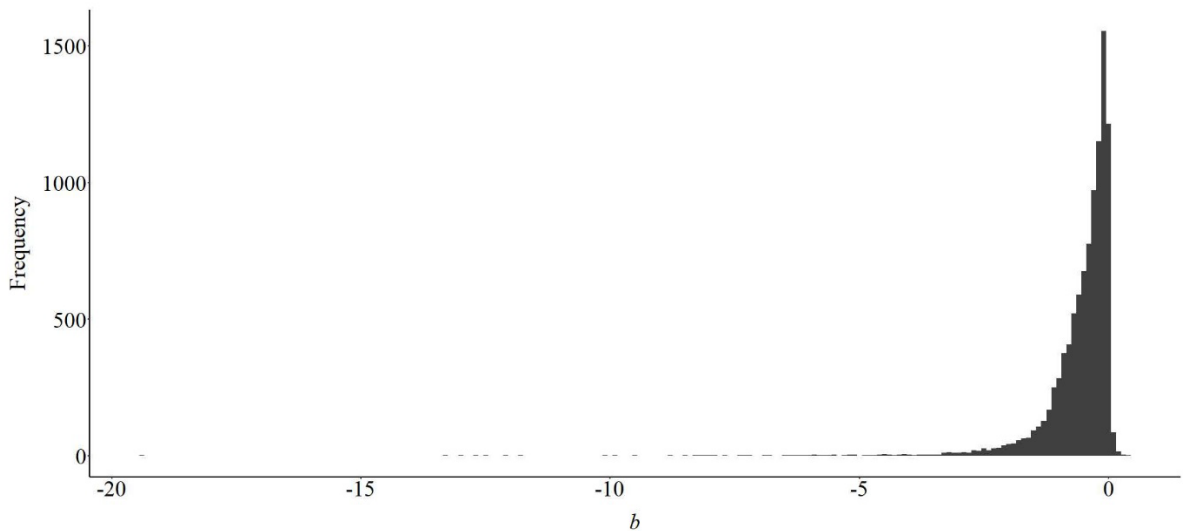

***Confounder (-,-)***

Focusing on a confounder with a negative causal influence on both X (the independent variable) and Y (the dependent variable), the results of the single simulation and the looped simulation demonstrated that the confounded slope coefficient will commonly be upwardly biased – further from zero – and a positive value.

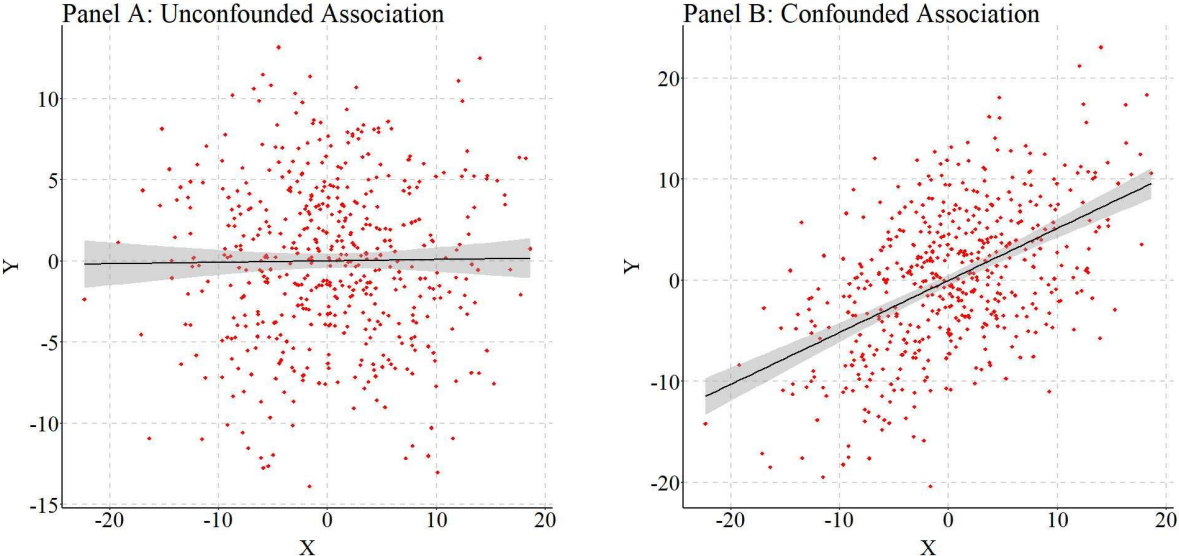

This suggests that confounders that have a negative influence on both the independent and dependent variable will commonly be upwardly bias – further from zero – and generate a positive slope coefficient when the influence of the confounder is not adjusted for in the model.

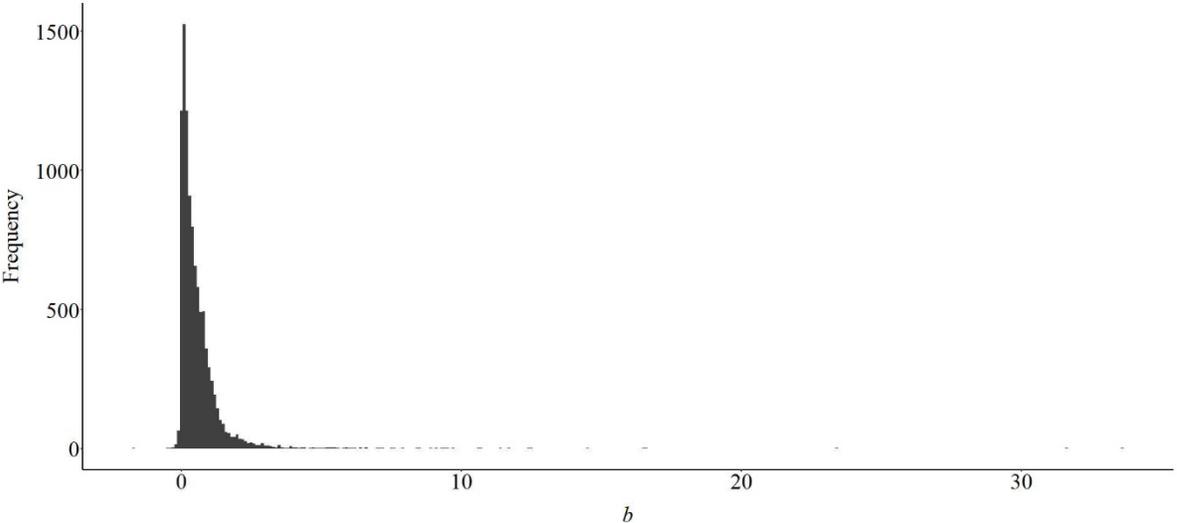

**Intermission Discussion**

So … you might be asking yourself why is the directionality of the bias important when the association between X and Y is confounded (the model does not adjust for the influence of the confounder or confounders). When X has no influence on Y, not adjusting for the influence of a confounder will upwardly bias the association of interest because – logically – you can not have an estimate closer to zero than the true association ($b = 0$). Nevertheless, when X does influence Y, not adjusting for the influence of a confounder can upwardly or downwardly bias the slope coefficient. This means that under certain conditions we can actually commit a type 2 error – retaining the null hypothesis, when in reality we should have rejected the null hypothesis – by not adjusting for the influence of a confounder. Moreover, we can also commit a Type S error and generate interpretations distinct from the true association when a confounder is not adjusted for in our statistical models. Let's demonstrate using looped simulations.

**True Association Between X and Y = 1**
Following the simulations above, we have conducted four looped simulations. Importantly, X was specified to be positively associated with Y, with a slope coefficient of 1.00. This slope coefficient is statistically significant for continuous constructs with a mean of 5 (or -5) and a SD = 5 at a sample size of 100, which are the extreme values of our simulation thresholds. Moreover, the only difference between the simulations is the direction of the influence of the confounder on X and Y across the 10,000 randomly specified scenarios. Previous analyses demonstrated that the se for the confounded association between X and Y was on average .05 (this is conservative at N = 100, but generous at N = 1,000). As such, in each figure we identified the *Region of Nullification*, otherwise known as the scenarios where we could potentially commit a type 2 error. As demonstrated below, the direction of influence the confounder has on X and Y dictates if the confounded association between X and Y crosses over the region of nullification.

*Confounder (+,+)*
For the current example, the confounder (C) was specified to have a positive influence on variation in both Y and X, while the slope coefficient between X and Y was specified to be 1. As demonstrated by the findings, when C has a positive influence on the variation in both Y and X, the confounded slope coefficient will generally be upwardly biased and a positive value. On some occasions the slope coefficient can be closer to zero – increasing the likelihood of a Type 2 error – or a negative value. But generally, the interpretation of the estimated association – confounded or unconfounded – will be that X has a positive influence on Y, but the magnitude of the association will vary conditional upon the magnitude of the effects of the confounder on X and Y.

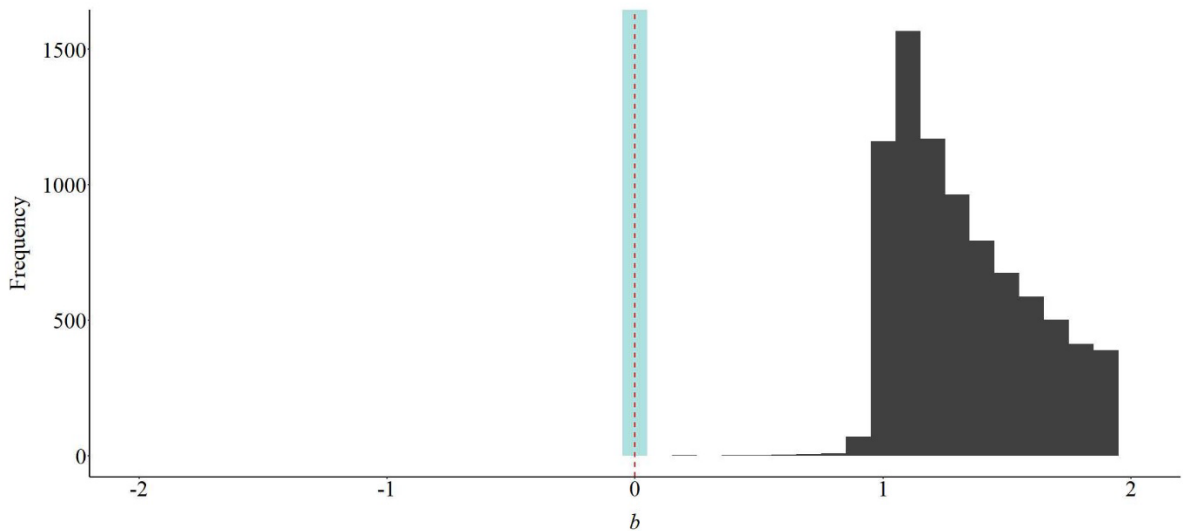

*Confounder (-,+)*
Altering the specification, the confounder (C) was now designated to have a positive influence on the variation in Y, but a negative influence on the variation in X. The slope coefficient between X and Y remained 1. As demonstrated by the findings, when the confounder has a positive influence on the variation in Y but a negative influence on the variation in X, the confounded slope coefficient could be upwardly or downwardly biased, as well as a positive or negative value. Moreover, a subset of the random scenarios produced slope coefficients that fell within the *Region of Nullification*. Considering this evidence, the interpretation of the confounded association could vary depending upon the magnitude of the negative effects of the confounder on X and the magnitude of the positive effects of the confounder on Y. Three distinct interpretations can be drawn from the confounded association: X has a statistically significant negative influence on Y, X and Y are unrelated, and X has a statistically significant positive influence on Y. Only the last interpretation captures the true association between X and Y.

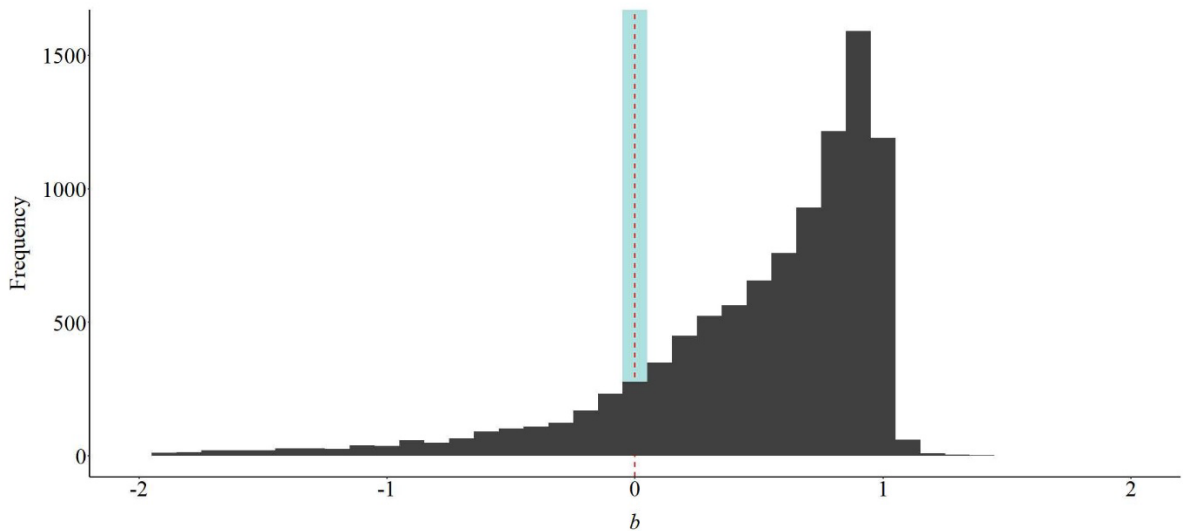

*Confounder (+,-)*

Opposite of the previous specification, the confounder (C) was now designated to have a positive influence on the variation in X, but a negative influence on the variation in Y. The slope coefficient between X and Y remained 1. The findings of the randomly specified 10,000 simulations were almost identical to the previous specification. That is, when the confounder had a positive influence on the variation in X but a negative influence on the variation in Y, the confounded slope coefficient could be upwardly or downwardly biased, as well as a positive or negative value. Moreover, a subset of the random scenarios produced slope coefficients that fell within the *Region of Nullification*. Similar to the previous example, three distinct interpretations can be drawn from the confounded association: X has a statistically significant negative influence on Y, X and Y are unrelated, and X has a statistically significant positive influence on Y. Again, only the last interpretation captures the true association between X and Y.

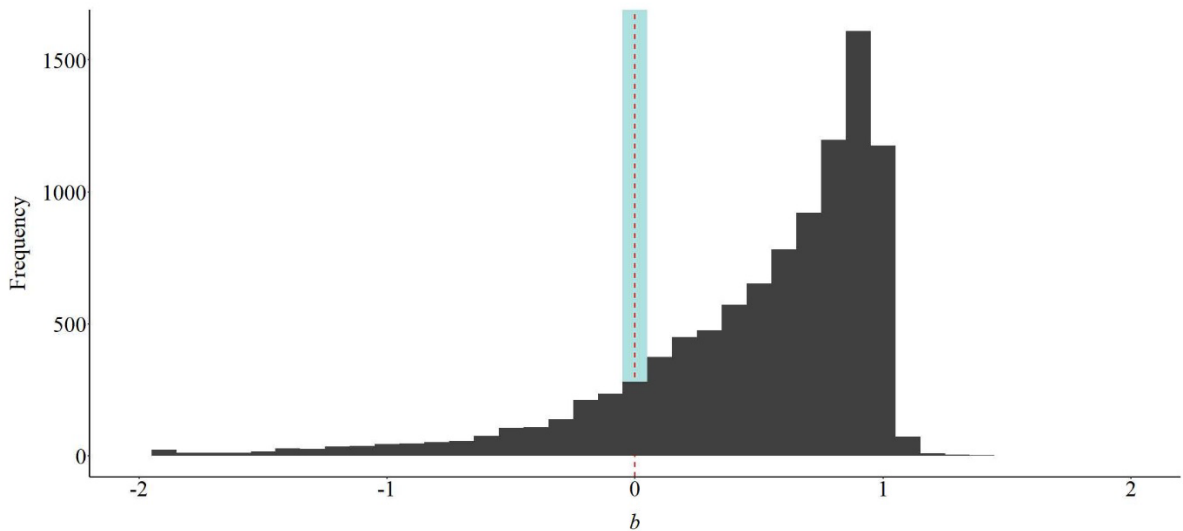

*Confounder (-,-)*
Finally, when the confounder was specified to have a negative influence on the variation in both X and Y, the results were identical to the simulated loops where the confounder was specified to have a positive influence on the variation in both X and Y. Specifically, while on some occasions the slope coefficient can be closer to zero – increasing the likelihood of a Type 2 error – or a negative value, the interpretation of the association – confounded or unconfounded – will generally be that X has a positive influence on Y, but the magnitude of the association will vary conditional upon the magnitude of the effects of the confounder on X and Y.

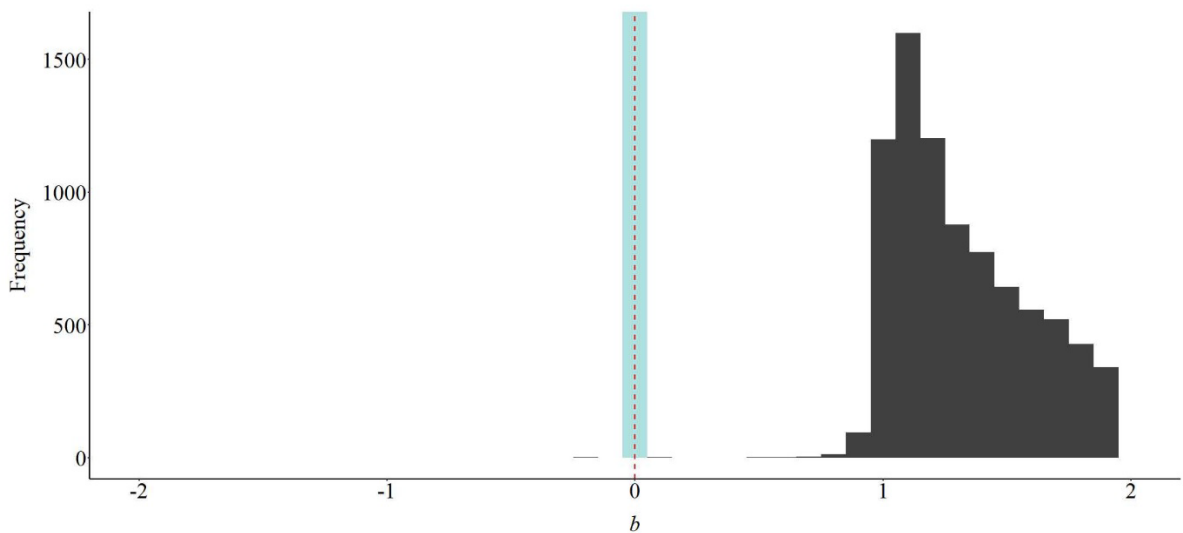

**Conclusion**

Confounder bias is taught in statistics courses across a wide variety of fields due to the increased likelihood of committing a Type 1 error. While this is correct – not adjusting for the influence of a confounders can only upwardly bias the slope coefficient of null associations –, it is important to remember that confounders can upwardly or downwardly bias the slope coefficient of variables that are causally associated. Moreover, under some circumstances we can commit a Type 2 error (retain the null hypothesis, when in reality we should have rejected the null hypothesis) or observe a slope coefficient that is negative (or positive) when the true association is positive (or negative; a Type S error). These effects are conditional upon the direction and the magnitude of the influence of a confounder on both the independent and the dependent variables of interest. While the conclusion remains the same as we have been previously taught, always make efforts to adjust for confounders in our methodological or statistical designs, the intricacies of confounder bias should be taught more broadly. If there is nothing else taken away from this entry: not adjusting for confounders can generate statistically significant slope coefficients when in reality no association exists between the variables of interest, ***but it can also generate null slope coefficients or slope coefficients in the opposite direction when in reality an association exists between the variables of interest***!

# Entry 8: The Inclusion of Colliders (Collider Bias)

**Introduction**
You might be saying that Entry 7 provides good evidence for why we should include numerous variables in a statistical model and hope they adjust for confounder bias. I mean… in some sense… climate change can confound the association between unemployment rates and crime rates. While it does suggest that, it would be negligent for me to recommend adjusting our estimates for any variable that is empirically related to the independent and dependent variable of interest. This is because any of the variables we introduce into our statistical model could function as a collider. And, opposite of a confounder, when a collider variable is *introduced* into a multivariable model, it can bias the estimated association between the independent and dependent variable of interest.

**Collider Variables**
It is important to develop an understanding of what a collider variable is before illustrating the effects of collider variables in multivariable models. A collider variable is the opposite of a confounder, where variation in the Col is caused by both variation in X and variation in Y (illustrated below). For example, it can be suspected that both professional appearance during a job interview and educational attainment (e.g., degree earned) can causally influence the likelihood of an individual being employed (i.e., a collider). Assuming that professional appearance during a job interview has no effect on educational attainment, estimating the bivariate association will produce estimates consistent with the null association. Nevertheless, if we estimated a multivariable model adjusting for the influence of being employed – the theorized collider –, the slope coefficient of the association between professional appearance during a job interview and educational attainment will become biased. That is, adjusting for the influence of a collider in a multivariable model will bias the observed association between the variables of interest. Similar to confounder bias, the direction and the magnitude of the collision dictate the direction and magnitude of the bias in the slope coefficient. As such, conditional upon the true association, adjusting a multivariable model for the influence of a collider can generate Type 1, Type 2, Type M (magnitude biases), and Type S errors (slope coefficients with signs opposite of the true association).

Panel A: Collider Bias         Panel B: Direction of Collider Bias

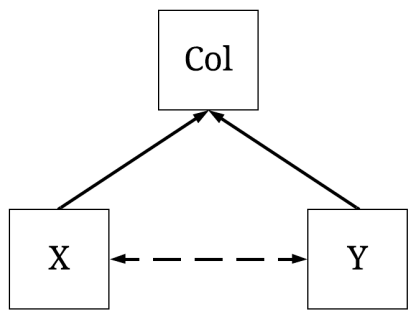
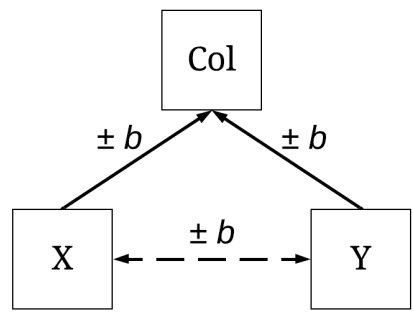

*Important Note: Identifying a Collider – Good Luck!*
So… in the previous entry we did not discuss the variety of statistical techniques used to identify a confounder between an independent and dependent variable. This is because scholars commonly rely on bivariate statistics to identify potential confounders. This is not an appropriate technique for a variety of reasons – including variables can become confounders only within the context of a multivariable model (this will be discussed in future entries) –, but the primary reason is that colliders and confounders look extremely similar in correlation matrices and other bivariate statistics. Look at the figure below and (1) identify the relationship confounded and which variable is confounding the association?, as well as (2) identify the collider and the two variables causing variation in the collider? Any ideas? I bet you thought V3 is the collider between V1 and V2… hmm, though V5 is related to every variable? If you can correctly identify both the confounder and the collider from a correlation matrix you have a super power and should monetize that immediately! For the rest of us, identifying the confounders and colliders requires a sound theoretical model and a directed acyclic graph (DAG). While the stepwise estimation of regression models including and excluding every variable (one at a time) might appear to be a valid technique for identifying confounders and colliders, be careful as you could end up *p*-hacking a model.

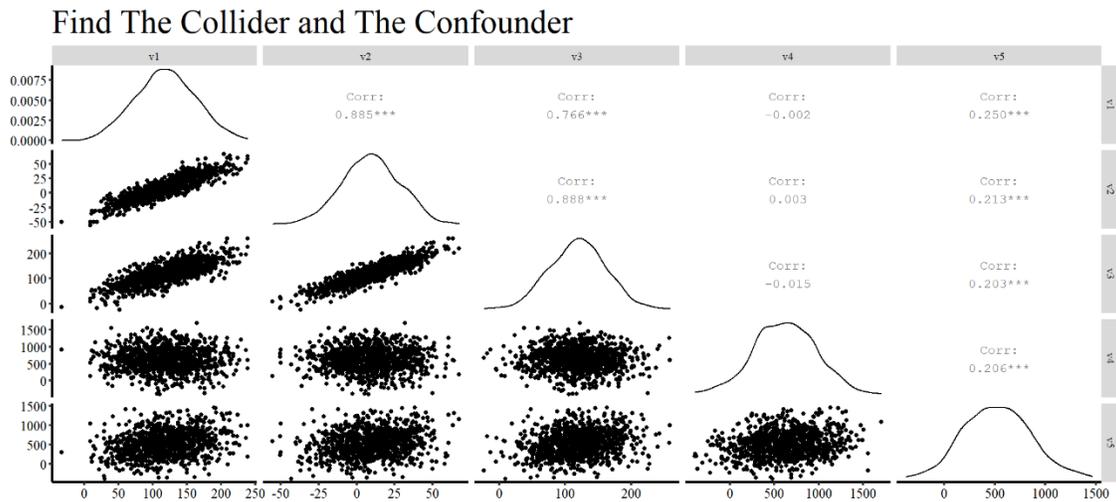

**True Association Between X and Y = 0**
Consistent with the entire series, we will conduct simulation analyses to evaluate the direction (upwardly or downwardly) and magnitude of the bias in the slope coefficient between X and Y when the association is conditioned – or adjusted for – the influence of a collider. For the first set of simulations, we will specify that the independent variable (X) and dependent variable (Y) are unrelated. That is, variation in X does not cause variation in Y and the slope of the association between the variables is 0 (i.e., $b = 0$). Notably, while these simulations will be very similar to entry 7, X and Y will now causally influence variation in the collider.

*Collider (+,+)*
Let's start the simulations by having variation in the independent variable (X) and variation in the dependent variable (Y) causally predict variation in the collider. To begin the specification, we simulate two normally distributed variables – using separate seeds – with a mean of 0 and a standard deviation of 2.5. The first variable we will label as X, while the second variable will be labeled as Y. Additionally, each variable will be specified to have 500 cases. After simulating X and Y, we can specify that C – the collider – is equal to 2*X plus 2*Y with a normally distributed residual variation specified to have a mean of 0 and a standard deviation of 2.5. This specification means that a 1-point increase in X and Y is associated with a 2-point increase in C. Importantly, as specified the true association between X and Y is not statistically significant and assumed to be 0.

```
## Simulating a Collider (+,+) ####

n<-500 # Sample size

set.seed(42125) # Seed
x<-rnorm(n,0,2.5) # Specification of the independent variable
set.seed(31) # Seed
y<-rnorm(n,0,2.5) # Specification of the dependent variable
set.seed(1001) # Seed
c<-2*x+2*y+rnorm(n,0,2.5) # Specification of the Collider

Data<-data.frame(c,x,y)
```

Now let's estimate the bivariate association between X and Y, as well as the association between X and Y after adjusting for C (i.e., the collider). Just to make sure you believe me, I want to provide you with the results as both syntax and figures. As demonstrated in the first model – Y regressed on X – slope coefficient of $b = .016$ ($p = .706$) provides evidence suggesting that Y and X are unrelated (illustrated in Panel A of the figure below). This result is consistent with the specification of the data were variation in X was not specified to cause variation in Y (or vice versa). Nevertheless, when Y is regressed on X and C – the second model – the estimated slope coefficient suggests that X does cause variation in Y. Specifically, a 1-point increase (or decrease) in X appears to be associated with a .816 decrease (or increase) in Y. The observed negative association is statistically significant at the $p < .001$ level and the visual illustration (Panel B) suggests that a strong association between X and Y exists.

```
> summary(lm(y~x))

Call:
lm(formula = y ~ x)

Residuals:
        Min          1Q      Median          3Q         Max
-7.139868316 -1.667583254  0.038898907  1.769745904  6.482678410

Coefficients:
              Estimate    Std. Error t value Pr(>|t|)
(Intercept) 0.0472380429 0.1122933055 0.42067  0.67418
x           0.0164779548 0.0436787351 0.37725  0.70615

Residual standard error: 2.50941291 on 498 degrees of freedom
Multiple R-squared:  0.000285701863,    Adjusted R-squared:  -0.00172175657
F-statistic: 0.142320189 on 1 and 498 DF,  p-value: 0.70614595

> summary(lm(y~x+c))

Call:
lm(formula = y ~ x + c)

Residuals:
       Min          1Q     Median          3Q         Max
-3.529044556 -0.769445860  0.021351340  0.739786582  3.216551591

Coefficients:
               Estimate    Std. Error   t value         Pr(>|t|)
(Intercept)  0.0101428998  0.0507314166   0.19993             0.84161
x           -0.8164554579  0.0273172709 -29.88789 < 0.0000000000000002 ***
c            0.4088164989  0.0092730205  44.08666 < 0.0000000000000002 ***
```

```
---
Signif. codes:  0 '***' 0.001 '**' 0.01 '*' 0.05 '.' 0.1 ' ' 1

Residual standard error: 1.1335365 on 497 degrees of freedom
Multiple R-squared:  0.79642253,	Adjusted R-squared:  0.795603305
F-statistic: 972.165527 on 2 and 497 DF,  p-value: < 0.00000000000000002220446

>
```

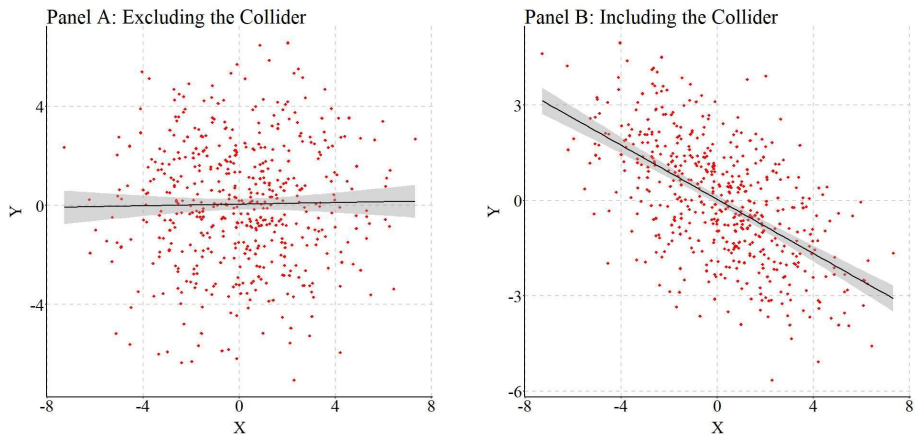

To further demonstrate the bias generated from the inclusion of a collider in a multivariable regression model, we can specify a looped simulation analysis. For the looped simulation analysis, we told the computer to conduct 10,000 simulations (n), where the slope coefficient of the association between X and C, and Y and C, randomly varied between 1 and 100 on a uniform distribution. Moreover, the number of cases in the simulation were randomly selected between 100 and 1000 (on a uniform distribution), and the distribution of X, Y, and C were randomly specified to have a mean between -5 and 5 on a uniform distribution and a standard deviation between 1 and 5 on a uniform distribution. After simulating the data, a regression model was estimated where the slope coefficient of the association between X and Y was adjusted for the influence of C. The estimated slope coefficient was than recorded and used to generate the plot below.

```
n<-10000

DATA1 = foreach (i=1:n, .packages='lm.beta', .combine=rbind) %dopar%
  {
    N<-sample(100:1000, 1)
    x<-runif(1,1,100)*rnorm(N,runif(1,-5,5),runif(1,1,5))
    y<-runif(1,1,100)*rnorm(N,runif(1,-5,5),runif(1,1,5))
    c<-runif(1,1,100)*x+runif(1,1,100)*y+runif(1,1,100)*rnorm(N,runif(1,-5,5),runif(1,1,5))
    Data<-data.frame(c,x,y)

    M<-lm(y~x+c, data = Data)
    bXY<-M$coefficients[2]

    data.frame(i,bXY)

  }
```

As demonstrated in the plot below, adjusting for a collider (C) between X and Y when estimating the bivariate effects of X on Y will commonly be upwardly (further from zero than reality) biased and regularly generate a negative slope coefficient. The magnitude of the bias, as demonstrated by the illustration, depends upon the magnitude of the collision between X and Y on C. This finding suggests that adjusting for a collider in a multivariable model increases our likelihood of committing a Type 1 error (i.e., reject the null hypothesis when in reality the null hypothesis should have been retained).

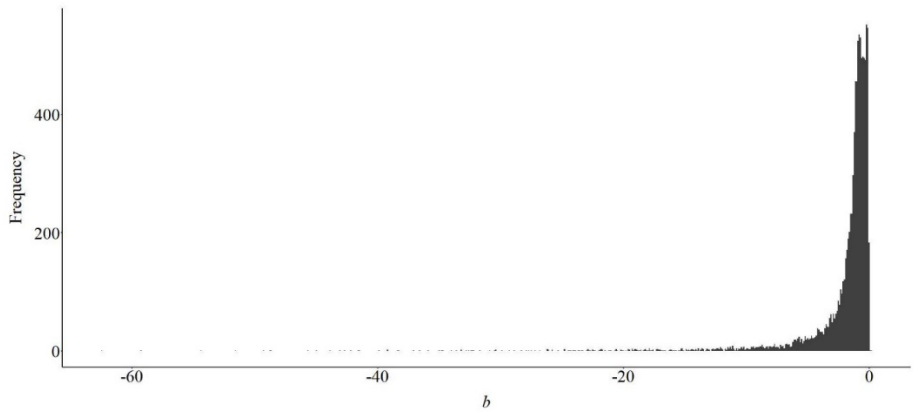

*Collider (-,+)*
As stated before, however, the direction (i.e., positive or negative association) of the causal influence of X on C, and Y on C, dictates the direction of the bias generated when we condition the slope coefficient of the association between X and Y on C. To illustrate this, we replicated the above simulation, except specified that X was negatively associated with C, where a 1-point increase (or decrease) in X caused a 2-point decrease (or increase) in C. The findings, provided in the figure below, demonstrated that when X and Y have opposite causal effects on C, the slope of the association between X and Y after adjusting for C becomes upwardly biased and positive. Specifically, a 1-point increase in X was associated with a .819 increase in Y ($p < .001$).

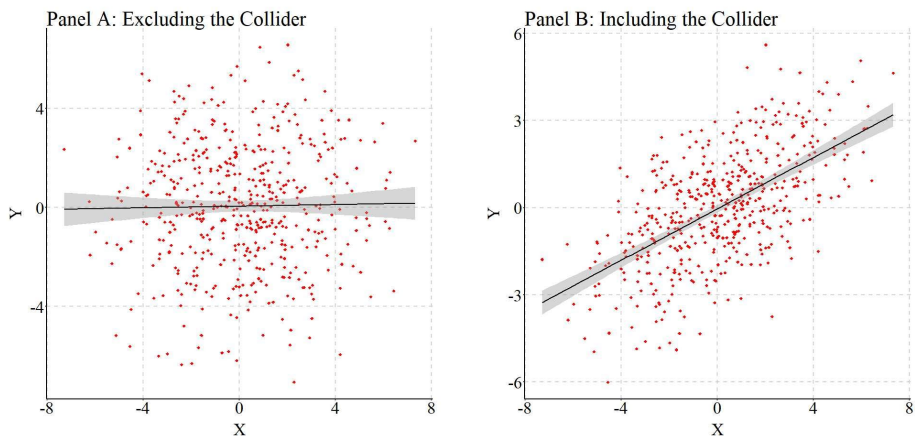

To replicate the looped simulation, we permitted the computer to randomly select any value between -100 and -1 on a uniform distribution to serve as the slope coefficient for the causal influence of X on C. The remainder of the looped simulation was identical to the looped simulation syntax provided above. Consistent with the findings presented by the non-looped simulation analysis, the looped simulation suggested that the slope coefficient of the association between X and Y will commonly become upwardly biased (further from zero than reality) and positive when a collider (C) is adjusted for in a multivariable model. Moreover, similar to the previous specification, we possess an increased likelihood of committing a Type 1 error when a model estimating the effects of an independent variable on a dependent variable is conditioned upon a collider.

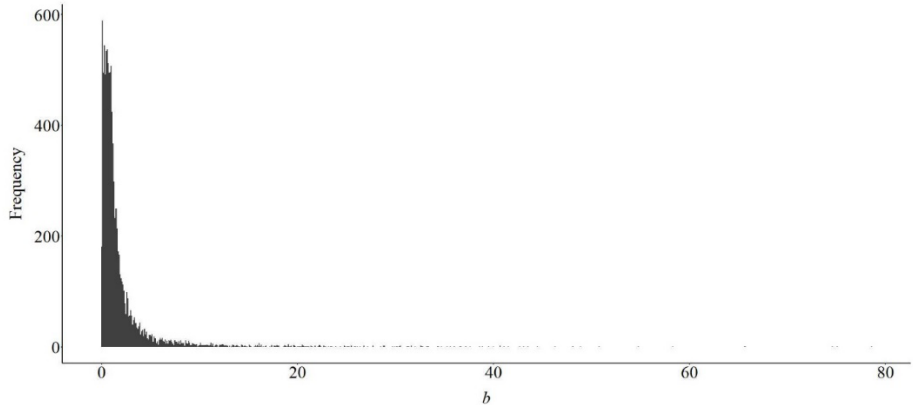

*Collider (+,-)*
To observe if the opposite specification produced similar findings, we replicated the simulation again. This time, we specified that the association between Y and C was negative, where a 1-point increase (or decrease) in Y was associated with a 2-point decrease (or increase) in C. The remainder of the simulation was identical to the previous examples. The findings further supported the previous demonstration, suggesting that when X and Y have opposite causal effects on C, the slope of the association between X and Y conditioned upon the collider becomes upwardly biased and positive. In this case, the slope coefficient was close to the previous example and suggested that a 1-point increase in X was associated with a .797 increase in Y ($p < .001$).

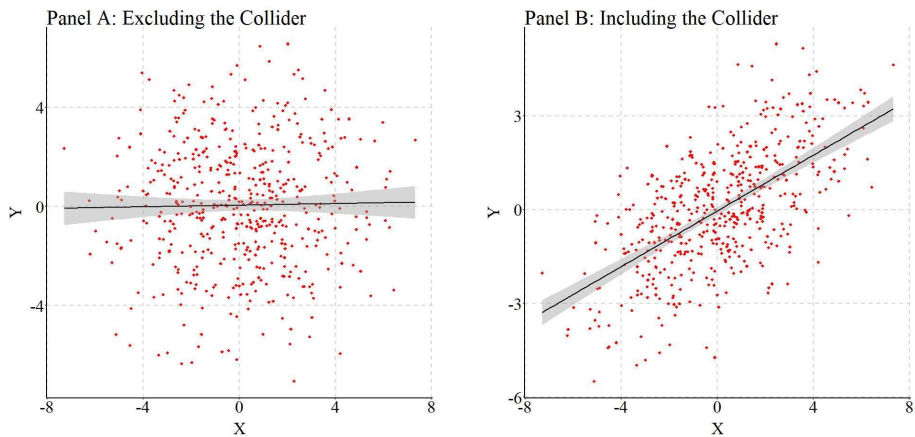

Moreover, the looped simulations again demonstrated that the slope coefficient of the association between X and Y will commonly become upwardly biased (further from zero than reality) and positive when the association between X and Y is estimated after adjusting for the effects of the collider (C) in a multivariable model. The specification for the looped simulations permitted the computer to randomly select any value between -100 and -1 on a uniform distribution to serve as the slope coefficient for the causal influence of Y on C, while the slope coefficient of the causal influence of X on C was randomly selected by the computer on a uniform distribution between 1 and 100.

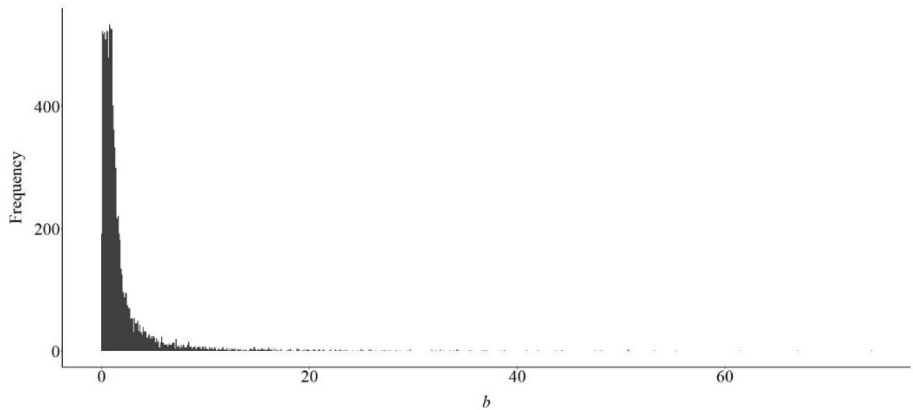

*Collider (-,-)*
For the final simulation – where the true slope coefficient of the association between X and Y equaled 0 – we replicated the first simulation, but specified that both X and Y had a negative causal influence on the collider (C). Specifically, a 1-point increase (or decrease) in X or Y was specified to cause a 2-point decrease (or increase) in the collider. After simulating the data using code similar to the code above, we estimated the slope coefficient of the association between X and Y with and without adjusting for the influence of the collider (C). Similar to the findings of the first simulation, the slope coefficient of the association between X and Y adjusted for the influence of the collider (C) was upwardly biased (further from zero than reality) and a negative value. Specifically, the results suggested that a 1-point increase (or decrease) in X predicted a .787 decrease (or increase) in Y.

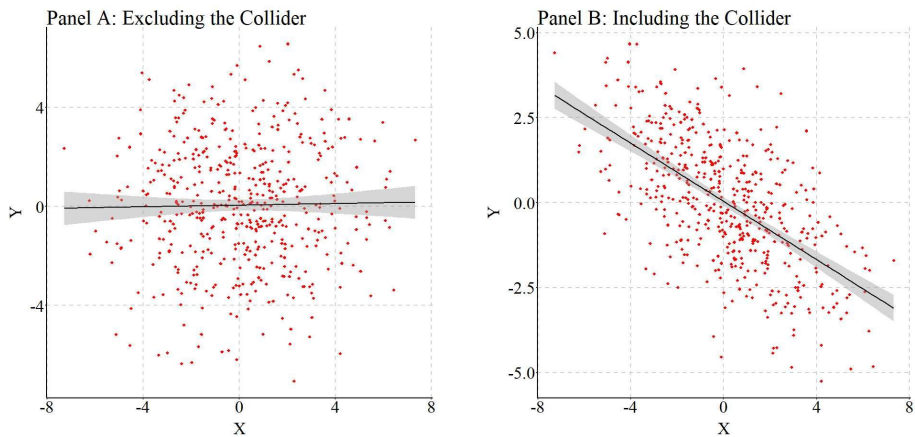

The looped simulations supported these findings, indicating that when X and Y have a negative collision on C, the slope of the association between X and Y adjusted for C commonly becomes negative and further from zero.

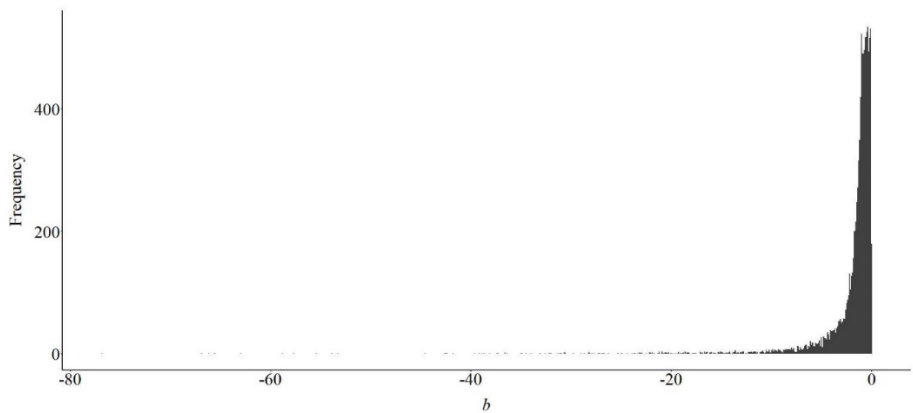

**Intermission Discussion**

As demonstrated above, the inclusion of a collider when estimating an association between an independent and dependent variable will commonly upwardly bias (further from zero than reality) the slope coefficient if the true slope coefficient is zero.[i] The direction – positive or negative values – of that bias, however, varies depending upon the direction of the effects of the independent and dependent variables on the collider. If the effects are in the same direction (e.g., the independent and dependent variables both positively [or negatively] influence variation in the collider) the upward bias will generally be in the negative direction. If the effects are in different directions (e.g., the independent variable positively [or negatively] influences variation in the collider, while the dependent variable negatively [or positively] influences variation in the collider) the upward bias will generally be in the positive direction.

To briefly discuss, the inclusion of a collider biases the slope coefficient – and other estimates – of the association between the independent and dependent variable because the remaining variation in the model after stratification is just the variation in the collider facing the opposite direction. Explicitly, multivariable regression models remove all of the shared variation in the collider – the variation caused by the independent and dependent variables. Nevertheless, the variation that remains after adjusting for the collider is just the *shared variation in the opposite direction*. This is because the unadjusted for variation in the collider is still common variation between the independent and dependent variables. The shared variation being in the opposite direction is the reason that the slope coefficient between the independent variable and dependent variable is the opposite direction of the collision (e.g., X and Y are positively associated with the collider, but negatively associated with each other when a multivariable model adjusts for said collider). I know this was not a relaxing intermission, but let's continue with our regularly scheduled program.

**True Association Between X and Y = 1**

Similar to confounder bias, collider bias can generate Type 2 errors and interpretations in the opposite direction when the true slope coefficient of the association between the independent variable and dependent variable is not equal to zero (Type S error). To evaluate the likelihood of committing a Type 2 error or an interpretation in the incorrect direction, the four looped simulations discussed above were replicated except now the association between X and Y was specified to be equal to 1.00.

*Collider (+,+)*
As demonstrated in the figure below, when the association between X and Y is conditioned on a collider, where both the independent variable and the dependent variable positively influence variation in the collider, the estimated slope coefficient will generally be lower than the true slope coefficient ($b = 1$). Under these circumstances, we have a higher likelihood of committing a Type 2 error – retaining the null hypothesis when it should have been rejected – or perceiving that the independent variable (X) has a negative influence on the dependent variable (Y). Overall, this finding indicates that collider bias can both downwardly (closer to zero than the true association) or upwardly bias (further from zero than the true association) the estimated slope coefficient in the wrong direction when the association between X and Y is conditioned upon the variation in a collider.

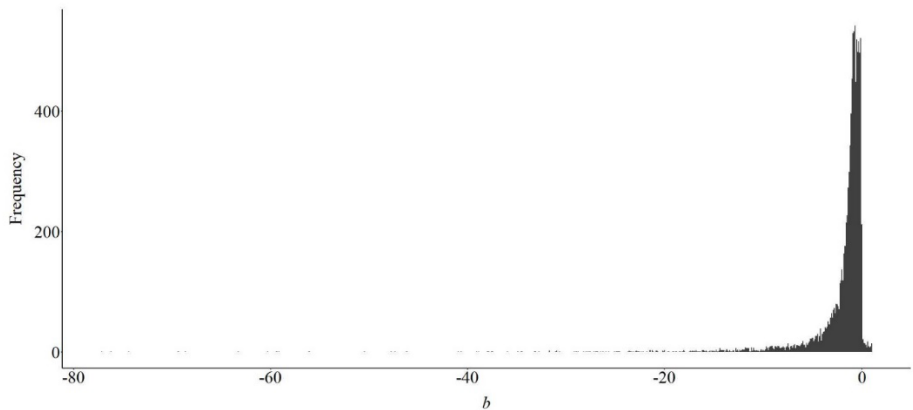

*Collider (-,+)*
Similar to the previous replication, the current simulation specified that the collider was differentially influenced by the independent and dependent variables. Specifically, increased scores on X were specified to be associated with decreased scores on the collider, while increased scores on Y were specified to be associated with increased scores on the collider. Evident by the findings, it appears that the model conditioned upon the collider will generally produce a slope coefficient that is larger than the true slope coefficient ($b = 1.00$). However, on some occasions the slope coefficient can be closer to zero than the true slope coefficient ($b = 1.00$).

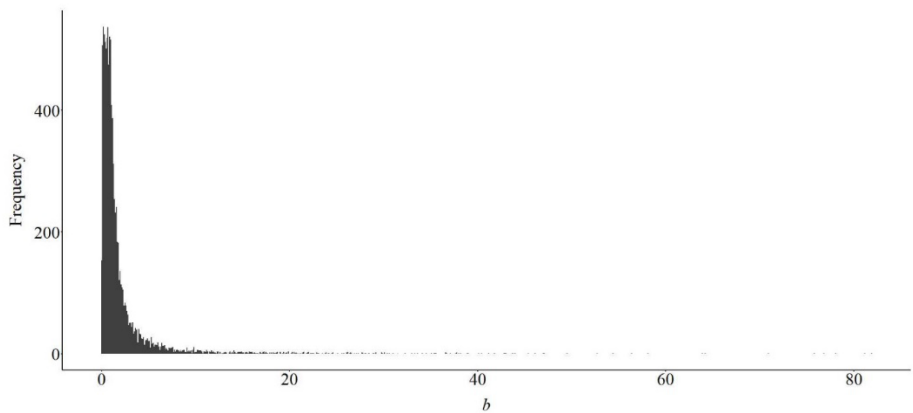

*Collider (+,-)*
Conditioning upon a collider where the independent variable is positively associated with the collider, but the dependent variable is negatively associated with the collider appeared to commonly upwardly bias the estimated slope coefficient. That is, the estimated slope coefficient was generally further from zero than the true slope coefficient ($b = 1.00$). Nevertheless, on some occasions the estimated association between X and Y was closer to zero than the true slope coefficient ($b = 1.00$). The estimated slope coefficient always appeared to be a positive value, but conditioning upon a differentially influenced collider – where X and Y cause variation in opposite directions – can generate both Type 1 and Type 2 errors.

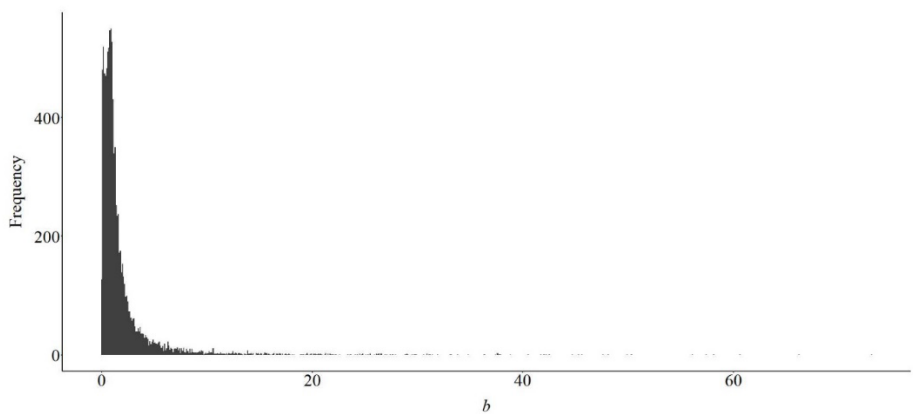

*Collider (-,-)*
Finally, when both the independent variable and dependent variable negatively influence variation in the collider, conditioning a model upon the collider will generally produce a slope coefficient smaller than the true slope coefficient. Similar to the first simulation, this generates an increased likelihood of committing a Type 2 error – retaining the null hypothesis when it should have been rejected – or perceiving that the independent variable (X) has a negative influence on the dependent variable (Y).

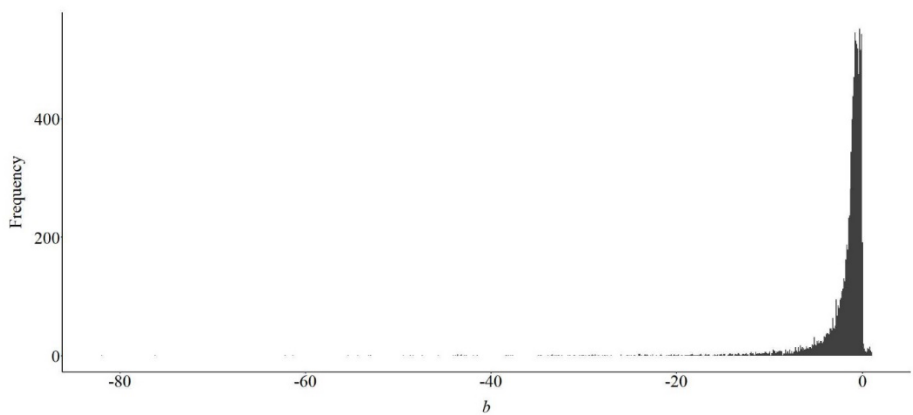

**Conclusion**
As demonstrated throughout Entry 8, accidentally conditioning an association of interest upon a collider – a variable causally influenced by both the independent and dependent variable – can introduce bias into the statistical estimates generated from our models. The estimated slope coefficients can be upwardly biased (further from zero than the true slope coefficient) or downwardly biased (closer to zero than the true slope coefficient) and in the same direction (e.g., positive true *b* and positive estimated *b*) or opposite direction (e.g., positive true *b*, but negative estimated *b*) conditional upon the interrelationship between the variables. Considering this, scholars should always express caution when selecting variables to include in multivariable statistical models. The likelihood of accidentally including a collider between an independent variable and the dependent variable in a multivariable regression model is substantively higher than we perceive.

---

[i] I have now implemented caution in my statements, because there are some anomalies that appear to be bugs in the simulations. Specifically, some data specifications in the looped simulations appear to produce slope coefficients on zero. These bugs could exist for a variety of reasons and, unfortunately, I have not had the opportunity to explore why or how these bugs exist. In theory though, collider bias – similar to confounder bias – can only serve to upwardly bias estimates if the true causal association is 0.

**Entry 9: The Inclusion and Exclusion of Mediating and Moderating Constructs**

**Introduction**
When I state "**X** is a cause of **Y**", I am inherently implying that variation in **X** caused variation in **Y**. This statement, however, does not provide any indication of *how* **X** caused variation in **Y**. The *how* is just as important, if not more important, than knowing the causal association. Take for example the statement "genetics cause cancer". Well, genetics are causally related to cancer, but that causal relationship is extremely complicated. Variation in your genetic code can **directly cause** variation in cancer by creating an increased predisposition to cellular mutations, variation in your genetic code can **indirectly cause** variation in cancer by creating variation in environmental factors, or variation in your genetic code can **directly or indirectly only cause** variation in cancer in the presence of certain environmental factors. The total causal pathway between any **X** and **Y** can be broken down into the: (1) direct causal pathway, (2) indirect causal pathway, (3) direct causal pathway under certain conditions, and (4) indirect causal pathway under certain conditions.

The causal pathway of interest, inherent in our hypotheses and research questions, generates assumptions about what the coefficients in a regression model represent. For example, if I state "*it is hypothesized that gene XX causes an increased risk of cancer*" I am intrinsically stating my interest in the total causal effects of gene XX on cancer. Here, assuming the association is not confounded, I can simply estimate the bivariate association. However, if I state that "*it is hypothesized that gene XX causes an increased risk of cancer through dietary habits*" I am implicitly interested in the indirect causal effects of gene XX on cancer. Estimating a regression model of the bivariate association between gene XX and cancer does not produce coefficients representative of the indirect causal effects. Similarly, if I state that "*it is hypothesized that gene XX directly causes an increased risk of cancer*", my regression model would have to adjust for the variation in any indirect pathway to ensure the coefficients only represent the direct causal pathway. Evident by the examples, *the manner in which we state our hypotheses and research questions generates an assumption about what causal pathway the produced coefficients represent.* If the model is accidentally misspecified, the observed coefficients will not represent the causal pathway of interest and, in turn, violate the assumptions within our hypotheses. But before illustrating this principal, let's discuss the distinction between the causal pathways.

**Causal Pathways**
As introduced above, four unique causal pathways can theoretically exist between two constructs, the direction and magnitude of which contributes to the direction and magnitude of the total causal pathway. Let's start with the easiest, the ***direct causal pathway***. For a direct causal pathway to exist no other mechanism can mediate the relationship between the input – or independent variable – and the outcome – or the dependent variable. For example, the golf head speed and location of a swing for Brooks Koepka – team Koepka all the way! – is a direct cause of the distance a golf ball traveled. Importantly, while this example is immediate in time, a direct causal pathway does not have any time requirements, only the requirement that no mechanism mediates the association. Direct causal pathways not immediate in time, however, are uncommon within the social sciences.

Direct Causal Pathway

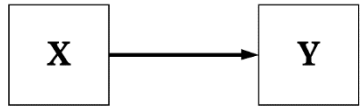

An ***indirect causal pathway*** is the combination of multiple direct causal pathways. For example, an indirect causal pathway with one mediator exists when variation in a distant input causes variation in an intermediate mechanism (the first direct causal pathway) and variation in the intermediate mechanism causes variation in an outcome (the second direct causal pathway). The intermediate mechanism serves as a mediator, where the indirect causal effects of the input on the outcome exists as a product of the mediator. In theory, an infinite number of mediators can exist within an indirect causal pathway. To provide an example, genetics has an indirect causal pathway to human behavior through a host of mechanisms including, but not limited to, brain development, hormone production, environmental selection, and psychological functioning. Moreover, to revert back to a sports example, Aaron Rodgers training has an indirect causal effect on how he performs in a game, through his muscle memory, knowledge of the defense, muscle fatigue, and etc. Again, similar to direct causal pathways, indirect causal pathways do not have any time requirements and can occur in extremely short periods of time. For instance, electric signals in one neuron indirectly cause electric signals in another neuron through the release and uptake of chemicals within the synapse, all of which happens in the span of milliseconds.

Indirect Causal Pathway

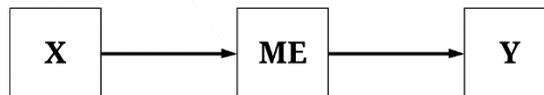

A ***moderated direct causal pathway*** exists when the magnitude, direction, or existence of a direct causal pathway differs in the presence of one or more mechanisms. To revert back to the golf example, while the golf head speed and location of a swing directly causes the distance a golf ball travels, the effects vary by the air quality, humidity, temperature, windspeed, and etc. Similarly, a ***moderated indirect causal pathway*** exists when the magnitude, direction, or existence of an indirect causal pathway differs in the presence of one or more mechanisms. Explicitly, this means that an indirect causal pathway can become a moderated indirect causal pathway when the magnitude, direction, or existence of one or more direct pathways vary by the presence of an external mechanism. For instance, Aaron Rodgers training has an indirect causal effect on how he performs in a game, but that effect of fatigue on how he performs differs by air quality, humidity, temperature, windspeed, and how well everyone around him is playing.

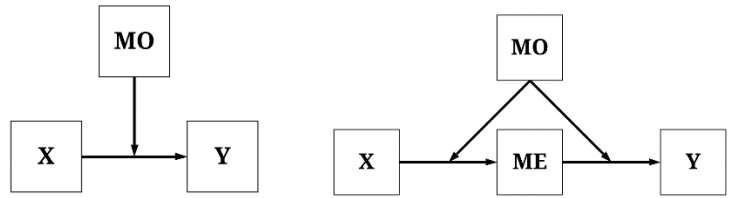

Moderated Direct Causal Pathway    Moderated Indirect Causal Pathway

Each of these causal pathways can exist simultaneously, generating the observation of unique causal effects between **X** and **Y** depending upon how a regression model is estimated. As such, it is important to theoretically consider *how* **X** causes variation in **Y** preceding the estimation of any regression model.

**Estimating the Causal Effects of Each Pathway**
For the purposes of the current demonstration, the data simulation will specify that the total causal effects of **X** on **Y** is a combination of all of the causal pathways that can exist. Explicitly, following the diagram below, a direct causal effect and an indirect causal effect – through **ME** – were specified to exist from **X** to **Y**, the effects of which are moderated by scores on **MO**. The employment of a single structural causal network permits us to generate estimates for all of the causal effects of **X** on **Y**, as well as illustrate the differences between the causal estimates.

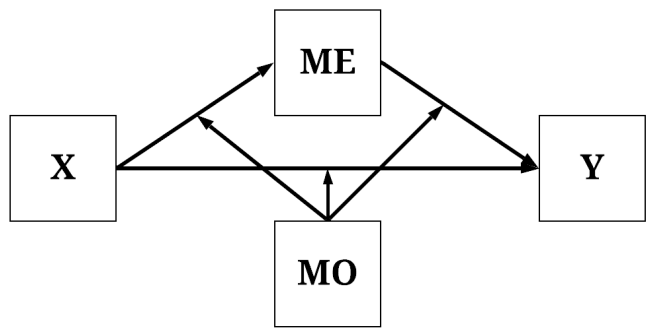

Following our reliance on directed equation simulations, the data was simulated using the syntax below. Briefly, **X** and **MO** were specified as normally distributed variables with a mean of 0 and a standard deviation of 10. Variation in **ME** was specified to be directly causally influenced by variation in **X** and random error with a mean of 0 and a standard deviation of 10. The magnitude of the direct causal effects of **X** on **ME** was moderated by scores on **MO**. Similarly, variation in **Y** was specified to be directly causally influenced by variation in **X,** variation in **ME**, and random error with a mean of 0 and a standard deviation of 10. The magnitude of the direct causal effects of **X** and **ME** on **Y**, however, was specified to be moderated by scores on **MO**.

```
set.seed(1992)
n<-10000 # Sample Size
X<-rnorm(n,0,10) # Independent Variable
MO<-rnorm(n,0,10)
ME<-1.00*X+1.00*(X*MO)+2.00*rnorm(n,0,10) # Mediator Variable
Y<-1.00*ME+1.00*(ME*MO)+1.00*X+1.00*(X*MO)+2.00*rnorm(n,0,10) # Dependent Variable

DF<-data.frame(X,MO,ME,Y)
```

*Total Causal Effects (Unconditional)*

Now let's estimate some effects, starting with the unconditional total causal effects. The total unconditional causal effects represent the combination of all of the effects of **X** on **Y** and is calculated using the equations below.[i] Explicitly, the unconditional total causal effect of **X** on **Y** is equal to the direct causal effect of **X** on **Y** plus the indirect causal effect of **X** on **Y**. The direct causal effect of **X** on **Y** is the value of $b_{YX}$ across the entire distribution of **MO**, while the indirect causal effects of **X** on **Y** is $b_{MEX}$ multiplied by $B_{YME}$ across the entire distribution of **MO**.

[Equation 1]

$$YX_{total} = YX_{direct} + YX_{indirect}$$

$$YX_{direct} = b_{YX} | (MO)$$

$$YX_{indirect} = (b_{MEX} * b_{YME} | (MO))$$

To observe the unconditional total causal effects, we can simply estimate a bivariate OLS regression model of **Y** on **X**. Evident by the findings, a 1 point increase in **X** causes a 104.378 increase in **Y** ($\beta$ = .557).

```
> summary(lm(Y~X, data = DF))

Call:
lm(formula = Y ~ X, data = DF)

Residuals:
        Min          1Q      Median          3Q         Max
-36956.51320   -492.65349    10.91097    506.05138  27639.21191

Coefficients:
              Estimate   Std. Error  t value            Pr(>|t|)
(Intercept) -17.58947245  15.57680124 -1.12921            0.25884
X           104.37890336   1.55709982 67.03418 < 0.0000000000000002 ***
---
Signif. codes:  0 '***' 0.001 '**' 0.01 '*' 0.05 '.' 0.1 ' ' 1

Residual standard error: 1557.54009 on 9998 degrees of freedom
Multiple R-squared:  0.310082167, Adjusted R-squared:  0.310013161
F-statistic: 4493.58076 on 1 and 9998 DF,  p-value: < 0.00000000000000002220446

> lm.beta(lm(Y~X, data = DF))

Call:
lm(formula = Y ~ X, data = DF)

Standardized Coefficients::
    (Intercept)                X
0.000000000000 0.556850219132

>
```

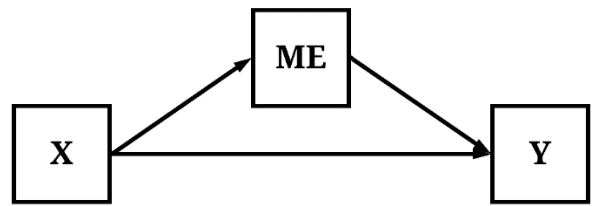

Total Causal Effects of X on Y = 104.38*

***Total Causal Effects (Moderated)***

The simplest way to calculate the conditional total causal effects is to regress **Y** on **X** at specific values of **MO**. For example, using the following equation, we can calculate the total causal effects of **X** on **Y** when **MO** is greater than 1 SD away from the mean.

[Equation 2]

$$YX_{total} = YX_{direct} + YX_{indirect}$$

$$YX_{direct} = b_{YX} | (MO > 10)$$

$$YX_{indirect} = (b_{MEX} * b_{YME} | (MO > 10))$$

By estimating the OLS regression model of **Y** on **X** with only cases that had a value on **MO** > 10, a unique estimate for the total causal effects of **X** on **Y** is produced. Specifically, for cases with values on **MO** > 10, a 1 point increase in **X** causes a 296.721 increase in **Y** ($\beta$ = .843). Given that **MO** is a continuous construct, we can calculate conditional total causal effects of **X** on **Y** across the entire distribution of **MO** – we, however, won't do that because it requires the estimation of numerous regression models.

```
> summary(lm(Y~X, data = DF[which(DF$MO>10),]))

Call:
lm(formula = Y ~ X, data = DF[which(DF$MO > 10), ])

Residuals:
        Min          1Q      Median          3Q         Max
-17755.378780  -665.205154  -22.661785   664.391620  22361.476927

Coefficients:
               Estimate   Std. Error  t value           Pr(>|t|)
(Intercept)    2.71454150  48.41920601  0.05606           0.9553
X            296.72099681   4.78082742 62.06478 <0.0000000000000002 ***
---
Signif. codes:  0 '***' 0.001 '**' 0.01 '*' 0.05 '.' 0.1 ' ' 1

Residual standard error: 1919.41272 on 1571 degrees of freedom
Multiple R-squared:  0.710309913,  Adjusted R-squared:  0.710125514
F-statistic: 3852.03679 on 1 and 1571 DF,  p-value: < 0.00000000000000002220446

> lm.beta(lm(Y~X, data = DF[which(DF$MO>10),]))

Call:
lm(formula = Y ~ X, data = DF[which(DF$MO > 10), ])

Standardized Coefficients::
   (Intercept)                X
0.000000000000 0.842798856595

>
```

Instead, we can calculate the conditional total causal effects of **X** on **Y** by regressing **Y** on **X**, **MO**, and an interaction between **MO** and **X**. Ignoring the direct estimates – they provide limited value in an interaction model –, the interaction between **MO** and **X** indicates that as scores on **X** increase and scores on **MO** increase the magnitude of the total causal effects of **X** on **Y** becomes stronger to a degree of 2.099 ($b$ = 2.099, $\beta$ = .152).

```
> summary(lm(Y~X+MO+(X*MO), data = DF))

Call:
lm(formula = Y ~ X + MO + (X * MO), data = DF)

Residuals:
        Min         1Q      Median          3Q        Max
-39002.57746  -478.66599   26.09335   499.73783  28115.66175

Coefficients:
               Estimate    Std. Error  t value           Pr(>|t|)
(Intercept)  -20.934662684  15.423450011 -1.35733           0.17471
```

```
X            104.370148587   1.541756629 67.69561 < 0.000000000000000222 ***
MO             6.171780872   1.541903304  4.00270         0.000063075 ***
X:MO           2.099475702   0.152219126 13.79246 < 0.000000000000000222 ***
---
Signif. codes:  0 '***' 0.001 '**' 0.01 '*' 0.05 '.' 0.1 ' ' 1

Residual standard error: 1541.92432 on 9996 degrees of freedom
Multiple R-squared:  0.323982194, Adjusted R-squared:  0.323779307
F-statistic: 1596.86425 on 3 and 9996 DF,  p-value: < 0.00000000000000002220446

> lm.beta(lm(Y~X+MO+(X*MO), data = DF))

Call:
lm(formula = Y ~ X + MO + (X * MO), data = DF)

Standardized Coefficients::
    (Intercept)                X               MO             X:MO
0.0000000000000 0.5568035133416 0.0329230065591 0.1134284720746

>
```

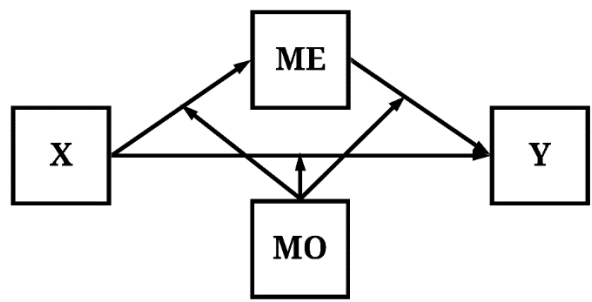

Moderated Total Causal Effects of X
on Y = 2.099*

### *Direct Causal Pathway (Unconditional)*

We can calculate the unconditional direct causal effects of **X** on **Y** by removing the unconditional indirect causal effect between **X** and **Y**. This can be achieved by adjusting the regression model of **Y** on **X** for the variation in **ME**. To briefly explain why this works, let's make some logic statements about the data:

    The variation in **ME** can be broken down into two parts:
        (1) Variation directly caused by **X**
        (2) Variation caused by random error
    The variation in **Y** can be broken down into three parts:
        (1) Variation directly caused by **X**
        (2) Variation directly caused by **ME**
        (3) Variation caused by random error

Using these statements, it is logical that some of the variation in **ME** is caused by variation in **X** and causes variation in **Y**. This shared variation between the three constructs is the unconditional indirect causal pathway. By simply introducing **ME** as a statistical control, the variation in **Y** directly caused by **ME** is removed – including the variation shared between the three constructs – from the estimated association between **X** and **Y**, permitting the slope coefficient to only represent the unconditional direct causal effects. But let's see our estimates. As it can be observed, the unconditional direct causal effects of **X** on **Y** is 102.434 ($\beta$ = .546).

```
> summary(lm(Y~X+ME, data = DF))

Call:
lm(formula = Y ~ X + ME, data = DF)

Residuals:
       Min         1Q     Median         3Q        Max
-39229.61629  -476.95047   20.82690   498.69863  27816.06452

Coefficients:
              Estimate    Std. Error  t value         Pr(>|t|)
(Intercept) -20.288729973 15.429626706 -1.31492          0.18857
X           102.434803473  1.548561553 66.14836 < 0.0000000000000002 ***
ME            2.084127472  0.149536892 13.93721 < 0.0000000000000002 ***
---
Signif. codes:  0 '***' 0.001 '**' 0.01 '*' 0.05 '.' 0.1 ' ' 1

Residual standard error: 1542.70242 on 9997 degrees of freedom
Multiple R-squared:  0.323232051, Adjusted R-squared:  0.323096657
F-statistic: 2387.34032 on 2 and 9997 DF,  p-value: < 0.00000000000000002220446

> lm.beta(lm(Y~X+ME, data = DF))

Call:
lm(formula = Y ~ X + ME, data = DF)

Standardized Coefficients::
    (Intercept)              X             ME
0.000000000000 0.546478655382 0.115141018013

>
```

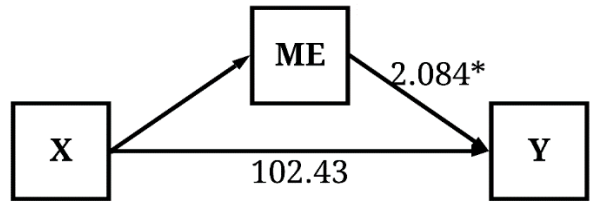

Total Causal Effects of X on Y = 104.38*

### *Direct Causal Pathway (Moderated)*

Similar to the preceding example, to estimate the conditional direct causal effects of **X** on **Y** we have to condition the entire equation upon **MO**. That is, we need to include an interaction term between **X** and **MO,** and an interaction term between **ME** and **MO** to observe an estimate representative of the conditional direct causal effects of **X** on **Y**. Take a second and guess what the estimate will be? If you guessed it will be ~ 1.00, you must've payed attention to the process used to simulate the data. Explicitly, the estimation below employs the formula used to create variation in **Y**, which approximately reproduces the estimates used to simulate the data.

```
> summary(lm(Y~X+ME+MO+(X*MO)+(MO*ME), data = DF))

Call:
lm(formula = Y ~ X + ME + MO + (X * MO) + (MO * ME), data = DF)

Residuals:
       Min         1Q     Median         3Q        Max
-71.52270176 -13.74040107  -0.04610956  13.56569193  69.35878533

Coefficients:
              Estimate    Std. Error  t value         Pr(>|t|)
(Intercept) 0.087740584106 0.199775785840  0.43920          0.66053
X           0.989644628777 0.025885238207 38.23201 < 0.0000000000000002 ***
```

```
ME            1.001267146613   0.010095571104    99.17885 < 0.0000000000000002 ***
MO           -0.029651948625   0.019980866923    -1.48402              0.13784
X:MO          1.001169783134   0.010281344249    97.37732 < 0.0000000000000002 ***
ME:MO         1.000064148382   0.000129552884  7719.35071 < 0.0000000000000002 ***
---
Signif. codes:  0 '***' 0.001 '**' 0.01 '*' 0.05 '.' 0.1 ' ' 1

Residual standard error: 19.9638423 on 9994 degrees of freedom
Multiple R-squared:  0.999886699, Adjusted R-squared:  0.999886642
F-statistic: 17639519.8 on 5 and 9994 DF,  p-value: < 0.00000000000000002220446

> lm.beta(lm(Y~X+ME+MO+(X*MO)+(MO*ME), data = DF))

Call:
lm(formula = Y ~ X + ME + MO + (X * MO) + (MO * ME), data = DF)

Standardized Coefficients::
    (Intercept)                   X                   ME                   MO                  X:MO
 0.000000000000000   0.005279647616876   0.055316634951478  -0.000158176597529   0.054090246754991
          ME:MO
 0.987820735912706

>
```

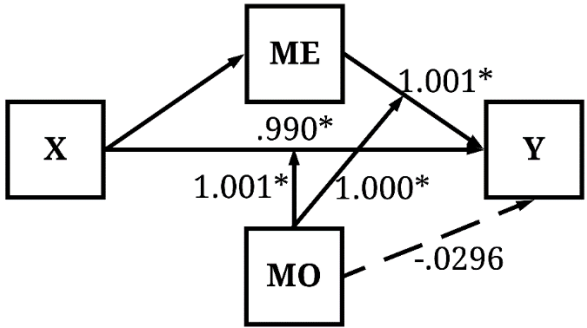

### *Indirect Causal Pathway (Unconditional)*

Although separate regression models can be estimated to calculate indirect causal pathways, the easiest process is through the implementation of path analysis. Path analysis – to quickly introduce – is a method in which regression equations are stacked on top of one another (or plugged into one another).[ii] Although not directly observed, the logic and process of specifying a path model is the same logic and process we use to conduct our simulations (directed equations). In this circumstance and anytime we estimate a path model in this series, we will use Lavaan. Since we are working with non-clustered continuous constructs, the Maximum Likelihood estimator in Lavaan works best.

To estimate the indirect effects in Lavaan, we will specify two equations within F1 – our formula. Within F1, we regress (Eq.1) **ME** on **X** and (Eq.2) **Y** on **ME** and **X**. The letters a, b, and c inform Lavaan to save the unstandardized slope coefficient as the corresponding letter, which we then use to calculate the indirect effects and total effects. To calculate the indirect effects, the slope coefficient of the association between **ME** and **X** (saved as b) is multiplied by the association between **Y** and **ME**. The total effects, consistent with our discussion above, is calculated by adding the slope of the direct association between **Y** and **X** to the indirect calculation.

```
>F1<-'
>ME~b*X
>Y~c*ME+a*X
>
>bc:=b*c
>abc:=(b*c)+a
```

Once we specify our formula, we can estimate the path model in Lavaan using the *sem* command with the ML estimator. To simplify the output, only the regression estimates and the defined parameters are presented here. You can see the full Lavaan output if you estimate the association using the provided *R*-code. Focusing on key estimates, it can be observed that the direct effects of **X** on **Y** (labeled with a) is almost identical to the estimate produced above. The indirect slope coefficient of **X** on **Y** through **ME** (bc within the *defined parameters* section) is 1.944, which has a standardized effect of β = .010. The total slope coefficient of **X** on **Y** (abc within the *defined parameters* section) – which is identical to regressing **Y** on **X** – is 104.379, which has a standardized effect of β = .557. As a brief note, the significance of the defined parameters in Lavaan is calculated using the delta method.

```
>M1<-sem(F1, data=DF , estimator = "ML")
>summary(M1, standardized = TRUE, ci = TRUE, rsquare = T)

Regressions:
                   Estimate   Std.Err   z-value  P(>|z|)  ci.lower   ci.upper    Std.lv    Std.all
  ME ~
    X        (b)      0.933     0.103     9.044    0.000     0.731      1.135     0.933      0.090
  Y ~
    ME       (c)      2.084     0.150    13.939    0.000     1.791      2.377     2.084      0.115
    X        (a)    102.435     1.548    66.158    0.000    99.400    105.469   102.435      0.546

Defined Parameters:
                   Estimate   Std.Err   z-value  P(>|z|)  ci.lower   ci.upper    Std.lv    Std.all
    bc                1.944     0.256     7.587    0.000     1.442      2.446     1.944      0.010
    abc             104.379     1.557    67.041    0.000   101.327    107.430   104.379      0.557
```

Total Causal Effects of X on Y = 104.379*

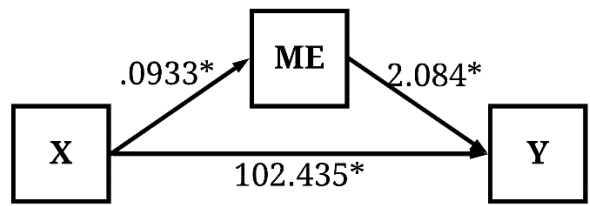

Indirect Causal Effects of X on Y = 1.944*

*Indirect Causal Pathway (Moderated)*
To calculate the moderated indirect causal pathways, we first have to create **XMO** (specified as **X*MO**) and **MEMO** (**ME*MO**) because Lavaan does not permit the specification of interaction terms within the formula (e.g., the way the interactions were specified when using the lm command). Focusing on the moderated indirect effects, it can be observed that the effects of **X** on **Y** through **ME** conditional on **MO** is equal to 1.000. Take a second and identify the rest of the causal effects of **X** on **Y** observed below.

```
DF$XMO<-DF$X*DF$MO
DF$MEMO<-DF$ME*DF$MO

F1<-'
ME~X+MO+b*(XMO)
Y~ME+MO+c*(MEMO)+X+a*(XMO)

bc:=b*c
abc:=(b*c)+a

'

Regressions:
                   Estimate  Std.Err  z-value  P(>|z|) ci.lower  ci.upper   Std.lv    Std.all
  ME ~
    X                 0.982    0.020   49.663    0.000    0.943     1.021    0.982      0.095
    MO               -0.019    0.020   -0.980    0.327   -0.058     0.019   -0.019     -0.002
    XMO       (b)     1.000    0.002  511.980    0.000    0.996     1.003    1.000      0.977
  Y ~
    ME                1.001    0.010   99.214    0.000    0.981     1.021    1.001      0.055
    MO               -0.030    0.020   -1.484    0.138   -0.069     0.009   -0.030     -0.000
    MEMO      (c)     1.000    0.000 7722.066    0.000    1.000     1.000    1.000      0.988
    X                 0.990    0.026   38.168    0.000    0.939     1.040    0.990      0.005
    XMO       (a)     1.001    0.010   97.410    0.000    0.981     1.021    1.001      0.054

Defined Parameters:
                   Estimate  Std.Err  z-value  P(>|z|) ci.lower  ci.upper   Std.lv    Std.all
    bc                1.000    0.002  510.858    0.000    0.996     1.003    1.000      0.966
    abc               2.001    0.010  191.235    0.000    1.980     2.021    2.001      1.020
```

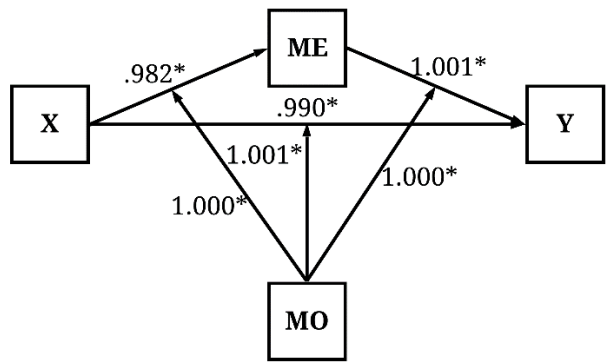

Moderated Total Causal Effects of X
on Y = 2.001*

Moderated Indirect Causal Effects of
X on Y = 1.000*

### *Discussion & Conclusion*

So you might be asking, what does this have to do with assumption violations? This is a great question, which I ask you to think about the research process. When we estimate a regression model, we generally follow the guidance of a hypothesis or research question. For instance, *it is hypothesized that a felony conviction decreases the ability to obtain a job*. Although not explicit, the manner in which this hypothesis is stated generates an assumption that we are interested in the total causal effects of a felony conviction on employment. Similarly, if we stated *it is hypothesized that a felony conviction decreases the ability to obtain a job through the inability to find transportation* it is assumed that we are interested in the indirect causal effects of a felony conviction on employment. The assumptions inherent in our stated hypotheses possess important implications for the specification of a regression model, where alternative specifications could produce estimates not representative of the causal effects of interest.

To provide an example, let's focus on a simulation without a moderating effect and no-direct effect of **X** on **Y**. Within this specification, variation in **X** only indirectly causes variation in **Y**. This means that **X** does cause variation in **Y** (i.e., the total causal effects), but only through **ME**.

```
> set.seed(1992)
> n<-10000 # Sample Size
> X<-rnorm(n,0,10) # Independent Variable
> ME<-1.00*X+2.00*rnorm(n,0,10) # Mediator Variable
> Y<-1.00*ME+0*X+2.00*rnorm(n,0,10) # Dependent Variable
>
> DF<-data.frame(X,ME,Y)
>
```

Now imagine we are interested in the total causal effects of **X** on **Y**. As presented in the results below, the total causal effects of **X** on **Y** is statistically significant ($p < .001$), suggesting that a 1 point increase in **X** corresponds to a 1.018 increase in **Y** (abc within the *defined parameters* section). The observed coefficients, as demonstrated above, is identical to regressing **Y** on **X** without any statistical controls. These results would be interpreted as variation in **X** causally influences variation in **Y**.

Now, imagine we regressed **Y** on **X** but accidentally included **ME** as a statistical control. The inclusion of **ME** as a statistical control adjusts the regression estimates, where the observed coefficients of the effects of **X** on **Y** only represent the direct causal effects. In this example, consistent with the results presented below, variation in **X** does not directly cause variation in **Y**. As such, the accidental inclusion of **ME** in the regression analysis would generate the interpretation – if we are only interested in the total causal effects – that variation in **X** does not causally influence variation in **Y**. This interpretation, however, only exists because the estimated coefficients represent the direct causal effects – not the total causal effects –, which occurred because **ME** was included as a covariate in the regression model. As such, the statistical coefficient violates the assumption stated within our hypothesis or research question.

```
F1<-'
ME~b*X
Y~c*ME+a*X

bc:=b*c
abc:=(b*c)+a

'

M1<-sem(F1, data=DF , estimator = "ML")
summary(M1, standardized = TRUE, ci = TRUE, rsquare = T)

Regressions:
                   Estimate  Std.Err  z-value  P(>|z|) ci.lower ci.upper   Std.lv  Std.all
  ME ~
    X          (b)    1.036    0.020   51.817    0.000    0.997    1.075    1.036    0.460
  Y ~
    ME         (c)    0.990    0.010  100.158    0.000    0.971    1.010    0.990    0.749
    X          (a)   -0.008    0.022   -0.372    0.710   -0.052    0.035   -0.008   -0.003

Defined Parameters:
                   Estimate  Std.Err  z-value  P(>|z|) ci.lower ci.upper   Std.lv  Std.all
    bc                1.026    0.022   46.023    0.000    0.982    1.070    1.026    0.345
    abc               1.018    0.028   36.373    0.000    0.963    1.073    1.018    0.342
```

Total Causal Effects of X on Y =
1.018*

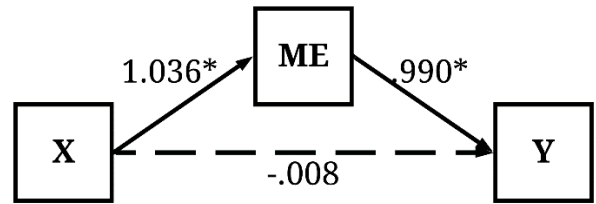

Indirect Causal Effects of X on Y =
1.026*

The assumption violation does not come from the estimated association, but rather is a product of our initial research question and the process used to interpret the estimate. To avoid this violation, it is always important to: (1) define the causal effects of interest (total, direct, or indirect), (2) theorize about and identify potential mediators and moderators using the principles of structural causal modeling (Pearl, 2009), (3) properly specify the regression equation to accurately estimate the causal effects of interest. Through this process you will produce statistical coefficients representative of the causal effects of interest to you!

---

[i] I know I generally avoid formulas in this series, but it is really important to know how each of these effects is calculated.

[ii] A more detailed discussion of path analysis will be provided in later entries.

# Entry 10: The Inclusion and Exclusion of Descendants

**Introduction**
You might have the thought of: *I know the constructs that can confound, mediate, or moderate the association I am interested in, and I am surely not going to include a collider as a covariate in my regression model*! You also, like a great student of causal inference, have embraced the process of creating a directed acyclic graph (DAG) to guide the selection of your covariates. I, however, have to ask, did you consider if any of those covariates are descendants of confounders, mediators, moderators, or colliders? If not, let's spend some time talking about the importance of descendants and, if so, you can probably skip over this entry … just kidding. Descendants, just like their ancestors (terms describe below), under some conditions can be used to reduce bias, while under other conditions, the introduction of a descendant can increase the bias in our statistical estimates. In this entry we will review how the inclusion of descendants in our statistical models can increase or decrease – depending upon the ancestor – the bias in the estimates corresponding with the association of interest.

**Descendants**
Within most scientific disciplines, we use the terms independent variable and dependent variable – or exogenous, lagged endogenous, and endogenous variables – to describe the construct we believe is causing variation and the construct that has variation being caused. These terms, although really designed to ease model specification, present problems when describing causal pathways in structural systems. That is, it is confusing to use terms like lagged endogenous variables or dependent variables to describe all of the causes and mechanisms being caused when discussing complicated structural systems. As such, the kinship terms of *parent*, *child*, *ancestor*, and *descendent* have been adopted to ease the description of structural systems. Briefly, a *parent* is a construct that causes variation in a *child* (a direct *descendant*). If the *parent* is also a *child* (i.e., variation caused by a $1^{st}$ generation), the indirect cause of the variation to the $2^{nd}$ generation is an *ancestor*, while the child in the $2^{nd}$ generation is a *descendant* of both the *parent* and the *ancestor*. I know it can be confusing, so let's apply these kinship terms to the causal pathway illustrated below.

If we assume the pathways below represent causal relationships, the figure demonstrates that variation in W causes variation in X, which causes variation in Y, which in turn causes variation in Z. Using the kinship terms, W is a parent to X and an ancestor to Y and Z because X is the parent of Y and Y is the parent of Z. Alternatively, Y and Z are descendants of W because X is a child to W, Y is a child to X, and Z is a child to Y. While it might take some time to get used to the nomenclature, out of my own preference I typically only use the terms *ancestors* and *descendants* because in practice we can not easily identify a child or a parent. As such, X, Y, and Z are descendants of W because variation in these constructs is causally influenced by variation in W.

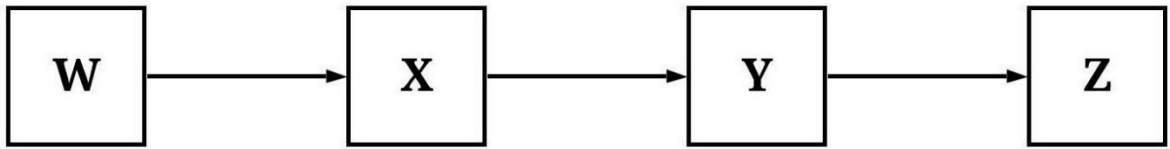

**The Importance of Descendants**

The best way to understand the importance of descendants when specifying a statistical model is through a series of simulated illustrations. Nevertheless, before we work through our illustrations, let's briefly define the variation in a descendant. The variation in a descendant is comprised of two components: (1) shared variation with the ancestors and (2) variation unique to the descendant. The further a descendant is from an ancestor (in the causal pathway), the less variation a descendant shares with that ancestor. The amount of variation shared is proportional to the variation shared by ancestor-descendant pairs and the number of mediating constructs along the causal pathway. Given the shared variation, adjusting for a descendant in a statistical model can indirectly adjust for the variation associated with the ancestors. By adjusting for the descendants of key constructs within a statistical model, and indirectly adjusting for the variation in the ancestor, we can reduce or increase the bias in the estimates corresponding to the association of interest. Let's begin our illustrations by focusing on the descendants of confounders.

*Descendants of Confounders*

When referencing a descendant of a confounder, we are discussing a mechanism causally influenced by the variation in the confounder, but not causally associated with the association of interest. This is demonstrated in the figure below, where Con (short for confounder) causally influences both X and Y (i.e., X and Y are descendants of Con) but the variation in X does not causally influence the variation in Y. Under these conditions, a statistical model will suggest that X causes variation in Y unless the model is conditioned upon Con. Importantly, Des is also causally influenced by Con, creating identical conditions where the lack of causal influence between X and Des and Y and Des will not be revealed unless the statistical model is conditioned upon Con. Although these identical conditions exist, the focus of the example below will be the causal association between X and Y.

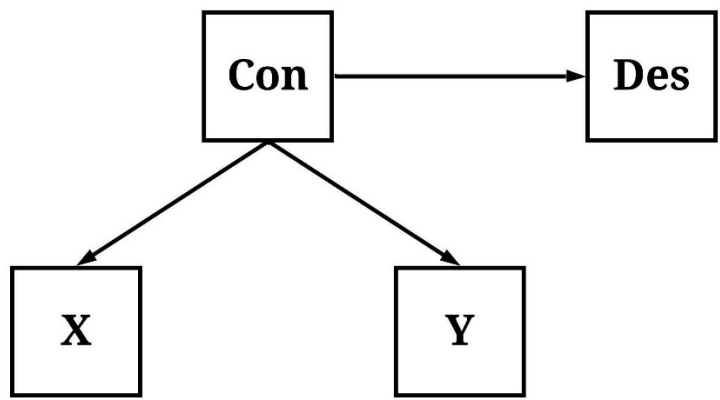

To simulate our structural system – the causal pathways within the figure –, we will simulate Con first and then simulate the descendants (X, Y, and Des). Evident in the syntax below, we simulated a dataset of 1000 cases, where all of the measures were specified to be normally distributed with a mean of zero. Moreover, variation in X, Y, and Des were specified to be causally influenced by variation in Con, but the amount of variation in DES causally influenced by Con was substantially larger than the amount of variation in X and Y causally influenced by Con. This specification makes Des an extremely good approximation of Con. Moreover, using the syntax below we can ensure that the variation in X has no causal influence on the variation in Y.

```
> ## Example 1 ####
> n<-1000
>
> set.seed(1992)
> Con<-rnorm(n,0,10)
> X<--2*Con+.50*rnorm(n,0,10)
> Y<--2*Con+.00*X+.50*rnorm(n,0,10)
> Des<-20*Con+1*rnorm(n,0,10)
```

Now that we have our data, let's estimate three models: (1) the bivariate association between X and Y, (2) the association between X and Y adjusting for Con, and (3) the association between X and Y adjusting for Des. Focusing on the first model, when Y is regressed on X without adjusting for Con the statistical estimates suggest that variation in X causally influences variation in Y. The slope coefficient of .930 suggests that a 1 point change in X is associated with a .930 change in Y.

```
> summary(lm(Y~X))

Call:
lm(formula = Y ~ X)

Residuals:
         Min             1Q        Median             3Q           Max
-23.626283367  -4.734664574  -0.164041357    4.520330368   25.280209778

Coefficients:
              Estimate    Std. Error   t value           Pr(>|t|)
(Intercept) -0.3675314299  0.2235938024 -1.64375            0.10054
X            0.9295740269  0.0111141701 83.63864  < 0.0000000000000002 ***
---
Signif. codes:  0 '***' 0.001 '**' 0.01 '*' 0.05 '.' 0.1 ' ' 1

Residual standard error: 7.0705672 on 998 degrees of freedom
Multiple R-squared:  0.87514733,   Adjusted R-squared:  0.875022227
F-statistic: 6995.42136 on 1 and 998 DF,  p-value: < 0.00000000000000002220446

> lm.beta(lm(Y~X))

Call:
lm(formula = Y ~ X)

Standardized Coefficients::
   (Intercept)                X
0.000000000000 0.935493094548

>
```

After adjusting for the variation in Con, however, the findings suggest that variation in X does not causally influence variation in Y. The slope coefficient of -.043 should not come as a surprise, given our previous discussions about confounder bias ([Entry 7](#)). The estimates presented below represent the unconfounded association and is a substantial departure from the estimates representing the confounded association (the preceding model).

```
> summary(lm(Y~X+Con))

Call:
```

```
lm(formula = Y ~ X + Con)

Residuals:
        Min              1Q         Median              3Q            Max
-19.455680213  -3.487189970  -0.301747267   3.282097622  16.863658189

Coefficients:
              Estimate     Std. Error  t value            Pr(>|t|)
(Intercept) -0.0980793972  0.1598106304 -0.61372              0.53954
X           -0.0432708159  0.0323468275 -1.33771              0.18129
Con          2.0710151227  0.0667582688 31.02260 < 0.0000000000000002 ***
---
Signif. codes:  0 '***' 0.001 '**' 0.01 '*' 0.05 '.' 0.1 ' ' 1

Residual standard error: 5.04612237 on 997 degrees of freedom
Multiple R-squared:  0.936471369, Adjusted R-squared:  0.936343929
F-statistic: 7348.35566 on 2 and 997 DF,  p-value: < 0.00000000000000002220446

> lm.beta(lm(Y~X+Con))

Call:
lm(formula = Y ~ X + Con)

Standardized Coefficients::
     (Intercept)                 X                 Con
 0.0000000000000 -0.0435463430992  1.0098723974868
```

Just for a second, though, let's imagine that we can not observe or capture variation in Con. As such, we are left searching for an approximation and stumble upon Des, which we then plug into our model. Oh my, the slope coefficient and standardized slope coefficient for the association between X and Y are almost identical to the estimates produced when we adjusted for Con in our statistical model. Although differences do exist – because the variation in Des is not identical to the variation in Con –, the slope coefficient produced when adjusting for the variation in Des is substantially closer to the unconfounded model than the confounded model. Importantly, the findings suggest that variation in Des causally influences variation in Y – which we know is false –, requiring the development of a Directed Acyclic Graph (DAG) to accurately interpret the estimated associations.

```
> summary(lm(Y~X+Des))

Call:
lm(formula = Y ~ X + Des)

Residuals:
         Min            1Q        Median            3Q           Max
-18.47467968  -3.63240155  -0.36160646   3.37157472  17.52413891

Coefficients:
              Estimate     Std. Error  t value            Pr(>|t|)
(Intercept) -0.14713163065  0.16408631510 -0.89667              0.37011
X            0.00625869336  0.03252152764  0.19245              0.84743
Des          0.09837471729  0.00335450446 29.32615 < 0.0000000000000002 ***
---
Signif. codes:  0 '***' 0.001 '**' 0.01 '*' 0.05 '.' 0.1 ' ' 1

Residual standard error: 5.18335365 on 997 degrees of freedom
Multiple R-squared:  0.932969011, Adjusted R-squared:  0.932834546
F-statistic: 6938.35882 on 2 and 997 DF,  p-value: < 0.00000000000000002220446

> lm.beta(lm(Y~X+Des))

Call:
lm(formula = Y ~ X + Des)

Standardized Coefficients::
     (Intercept)                 X                 Des
 0.00000000000000  0.00629854562082  0.95980424619948

>
```

The differences between the confounded model and the model adjusting for Des are further illustrated in the figure below.

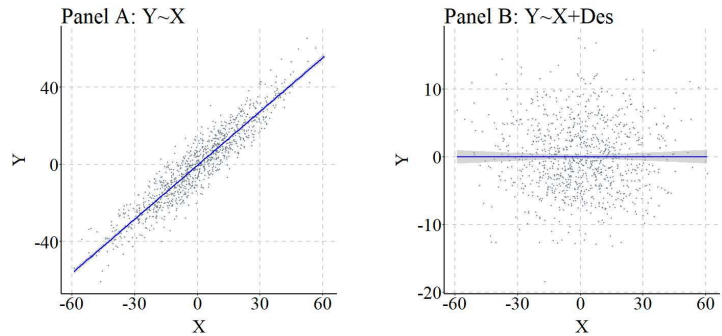

As a replication, we can simulate the data again but weakened the causal influence of Con on Des, reducing the ability of Des to serve as an approximation of Con. Evident by the findings of the replication, the amount of variation shared between Con and Des directly influences the ability of Des to adjust for confounder bias when estimating the causal effects of X on Y. In particular, while the estimated association between X and Y is attenuated, under these circumstances adjusting for the variation in Des can not produce estimates representative of the specified causal association.

```
> ## Example 2 ####
> n<-1000
>
> set.seed(1992)
> Con<-rnorm(n,0,10)
> X<--2*Con+.50*rnorm(n,0,10)
> Y<--2*Con+.00*X+.50*rnorm(n,0,10)
> Des<--2*Con+1*rnorm(n,0,10)
>
> summary(lm(Y~X))

Call:
lm(formula = Y ~ X)

Residuals:
        Min           1Q       Median           3Q          Max
-23.626283367  -4.734664574  -0.164041357   4.520330368  25.280209778

Coefficients:
              Estimate    Std. Error   t value             Pr(>|t|)
(Intercept) -0.3675314299  0.2235938024 -1.64375             0.10054
X            0.9295740269  0.0111141701 83.63864 < 0.0000000000000002 ***
---
Signif. codes:  0 '***' 0.001 '**' 0.01 '*' 0.05 '.' 0.1 ' ' 1

Residual standard error: 7.0705672 on 998 degrees of freedom
Multiple R-squared:  0.87514733,   Adjusted R-squared:  0.875022227
F-statistic: 6995.42136 on 1 and 998 DF,  p-value: < 0.00000000000000002220446

> lm.beta(lm(Y~X))

Call:
lm(formula = Y ~ X)

Standardized Coefficients::
   (Intercept)              X
0.000000000000 0.935493094548

> summary(lm(Y~X+Des))

Call:
lm(formula = Y ~ X + Des)

Residuals:
        Min           1Q       Median           3Q          Max
```

```
-21.281840236  -4.740428993  -0.087244974   4.542584785  24.360961565

Coefficients:
              Estimate    Std. Error  t value          Pr(>|t|)
(Intercept) -0.3845427252  0.2153325869 -1.78581          0.074434 .
X            0.7736553380  0.0205372816 37.67078 < 0.0000000000000002 ***
Des          0.1675961926  0.0188405306  8.89551 < 0.0000000000000002 ***
---
Signif. codes:  0 '***' 0.001 '**' 0.01 '*' 0.05 '.' 0.1 ' ' 1

Residual standard error: 6.80905941 on 997 degrees of freedom
Multiple R-squared:  0.884328015, Adjusted R-squared:  0.884095975
F-statistic:  3811.1001 on 2 and 997 DF,  p-value: < 0.00000000000000002220446

> lm.beta(lm(Y~X+Des))

Call:
lm(formula = Y ~ X + Des)

Standardized Coefficients::
   (Intercept)              X              Des
0.000000000000 0.778581592531 0.183852941585
```

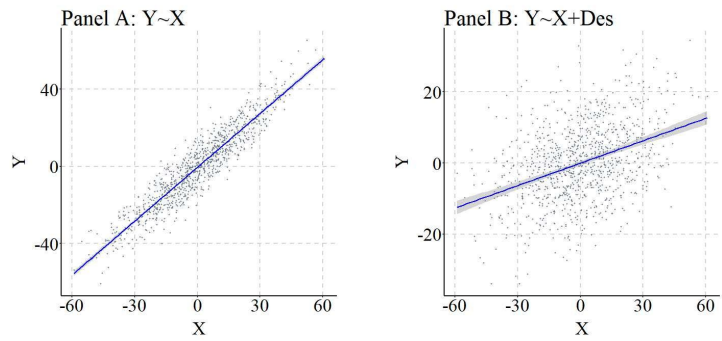

## *Descendants of Colliders*

Similar to a descendant of a confounder, a descendant of a collider refers to a construct causally influenced by the variation in a collider unrelated to the association of interest. As demonstrated below, variation in Col (short for collider) causally influences variation in Des, but variation in Col is now causally influenced by variation in both X and Y.

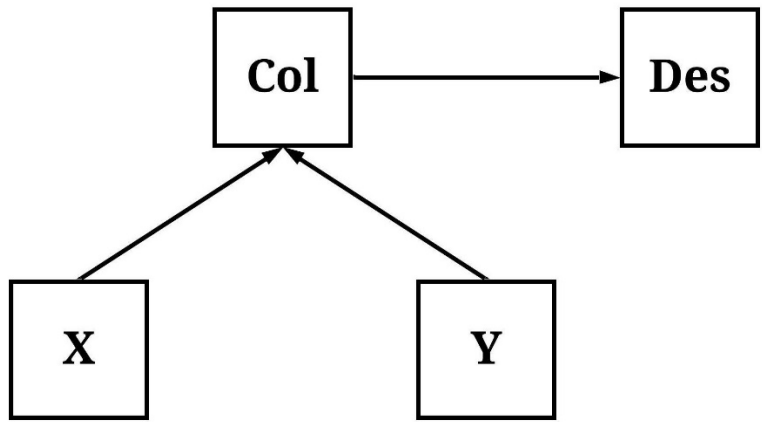

To simulate data following the structural system presented above, we reverse the order of the simulation and specify the distribution of scores on X and Y first. After simulating X and Y, we can specify that Col is causally influenced by the variation in X and Y, and causally influences the variation in Des. The specification of the association between Col and Des below generates a condition where Des is almost a perfect approximation of Col.

```
## Example 1 ####
n<-1000

set.seed(1992)
X<-rnorm(n,0,10)
Y<-rnorm(n,0,10)
Col<-2*X+2*Y+rnorm(n,0,10)
Des<-20*Col+1*rnorm(n,0,10)
```

Following the process detailed above, three linear regression models were estimated where: (1) Y was regressed on X, (2) Y was regressed on X and Col, and (3) Y was regressed on X and Des. As presented below, the estimates produced in the bivariate regression model present evidence suggesting that X does not causally influence Y, while the regression model adjusting for Col suggests that X does causally influence Y. The bias in the estimates, as expected given Entry 8 in the series, corresponds with the introduction of the collider into a regression model. Importantly, Model 3 produced estimates almost identical to the model adjusting for the variation in Col, suggesting that the variation in Des can generate collider bias if Des is a descendant of a collider. The bias generated by adjusting for Des – a descendant of a collider – is further demonstrated in a figure below.

*Bivariate Model*
```
> summary(lm(Y~X))

Call:
lm(formula = Y ~ X)

Residuals:
        Min         1Q      Median         3Q         Max
-33.39823634  -6.39223268  -0.12454071   6.77441730  32.32615833

Coefficients:
              Estimate    Std. Error  t value  Pr(>|t|)
(Intercept) 0.5328217473 0.3123247274  1.70599  0.088322 .
X           0.0016311375 0.0320395916  0.05091  0.959407
---
Signif. codes:  0 '***' 0.001 '**' 0.01 '*' 0.05 '.' 0.1 ' ' 1

Residual standard error: 9.87622125 on 998 degrees of freedom
Multiple R-squared:  2.59702143e-06,    Adjusted R-squared:  -0.000999404384
F-statistic: 0.00259183411 on 1 and 998 DF,  p-value: 0.959407377

> lm.beta(lm(Y~X))

Call:
lm(formula = Y ~ X)

Standardized Coefficients::
     (Intercept)                  X
0.00000000000000  0.00161152766817
```

*Model Adjusting for the Collider*
```
> summary(lm(Y~X+Col))

Call:
lm(formula = Y ~ X + Col)

Residuals:
         Min          1Q      Median         3Q         Max
-13.543830672  -2.946857580  0.158945553  2.951317200  16.910867430

Coefficients:
              Estimate    Std. Error  t value           Pr(>|t|)
(Intercept)  0.19287028087 0.14470345773  1.33287            0.18288
X           -0.79043929094 0.01978553859 -39.95036 < 0.0000000000000002 ***
Col          0.40163018863 0.00663937684  60.49215 < 0.0000000000000002 ***
```

```
---
Signif. codes:  0 '***' 0.001 '**' 0.01 '*' 0.05 '.' 0.1 ' ' 1

Residual standard error: 4.57230915 on 997 degrees of freedom
Multiple R-squared:  0.785882067, Adjusted R-squared:  0.785452543
F-statistic: 1829.65624 on 2 and 997 DF,  p-value: < 0.00000000000000002220446

> lm.beta(lm(Y~X+Col))

Call:
lm(formula = Y ~ X + Col)

Standardized Coefficients::
    (Intercept)              X            Col
 0.000000000000 -0.780936486139  1.182480809249

>
```

*Model Adjusting for the Descendant of the Collider*

```
> summary(lm(Y~X+Des))

Call:
lm(formula = Y ~ X + Des)

Residuals:
         Min            1Q        Median            3Q           Max
-13.509839731  -2.973642429   0.149793824   2.944323483  17.154745121

Coefficients:
                 Estimate     Std. Error   t value               Pr(>|t|)
(Intercept)  0.185652395314  0.144691697215   1.28309               0.19976
X           -0.790051816042  0.019778785457 -39.94440 < 0.00000000000000002 ***
Des          0.020081846598  0.000331926437  60.50090 < 0.00000000000000002 ***
---
Signif. codes:  0 '***' 0.001 '**' 0.01 '*' 0.05 '.' 0.1 ' ' 1

Residual standard error: 4.57178981 on 997 degrees of freedom
Multiple R-squared:  0.785930705, Adjusted R-squared:  0.785501278
F-statistic: 1830.18521 on 2 and 997 DF,  p-value: < 0.00000000000000002220446

> lm.beta(lm(Y~X+Des))

Call:
lm(formula = Y ~ X + Des)

Standardized Coefficients::
    (Intercept)              X            Des
 0.000000000000 -0.780553669535  1.182248072119

>
```

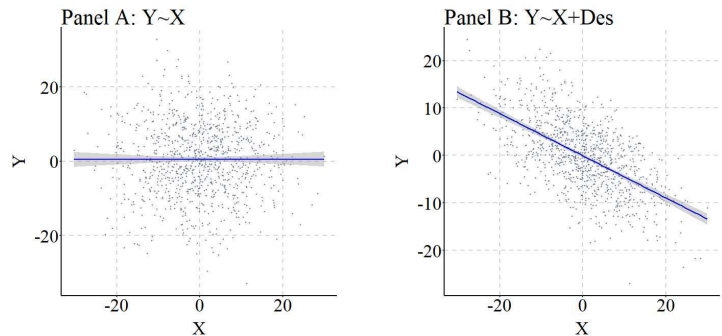

Nevertheless, similar to the previous example this is an extreme case where almost all of the variation in Des is caused by Col. As such, we can replicate the previous simulation but weaken the causal influence of Col on Des. The findings and figures are presented below and further demonstrate that adjusting for the descendant of a collider can introduce collider bias into the estimates corresponding with the association of interest.

```
> ## Example 2 ####
> n<-1000
>
> set.seed(1992)
> X<-rnorm(n,0,10)
> Y<-rnorm(n,0,10)
> Col<--2*X+2*Y+rnorm(n,0,10)
> Des<-.25*Col+1*rnorm(n,0,10)
>
>
> summary(lm(Y~X))

Call:
lm(formula = Y ~ X)

Residuals:
         Min          1Q      Median          3Q         Max
-33.39823634  -6.39223268  -0.12454071   6.77441730  32.32615833

Coefficients:
                 Estimate   Std. Error t value Pr(>|t|)
(Intercept) 0.5328217473 0.3123247274 1.70599 0.088322 .
X           0.0016311375 0.0320395916 0.05091 0.959407
---
Signif. codes:  0 '***' 0.001 '**' 0.01 '*' 0.05 '.' 0.1 ' ' 1

Residual standard error: 9.87622125 on 998 degrees of freedom
Multiple R-squared:  2.59702143e-06,    Adjusted R-squared:  -0.000999404384
F-statistic: 0.00259183411 on 1 and 998 DF,  p-value: 0.959407377

> lm.beta(lm(Y~X))

Call:
lm(formula = Y ~ X)

Standardized Coefficients::
      (Intercept)                X
0.000000000000000 0.00161152766817

>
> summary(lm(Y~X+Des))

Call:
lm(formula = Y ~ X + Des)

Residuals:
          Min           1Q       Median          3Q         Max
-28.76071481  -5.82984012   0.04863643   6.06431850  27.29574583

Coefficients:
                 Estimate    Std. Error   t value                  Pr(>|t|)
(Intercept)  0.3333244798  0.2847473164   1.17060                   0.24204
X           -0.1637272299  0.0313627762  -5.22043                0.0000002172 ***
Des          0.3495374707  0.0243220826  14.37120 < 0.0000000000000000222 ***
---
Signif. codes:  0 '***' 0.001 '**' 0.01 '*' 0.05 '.' 0.1 ' ' 1

Residual standard error: 8.99347184 on 997 degrees of freedom
Multiple R-squared:  0.171606606, Adjusted R-squared:  0.169944834
F-statistic: 103.267233 on 2 and 997 DF,  p-value: < 0.00000000000000002220446

> lm.beta(lm(Y~X+Des))

Call:
lm(formula = Y ~ X + Des)

Standardized Coefficients::
     (Intercept)                 X                Des
 0.000000000000 -0.161758871352    0.445302028509

>
```
122

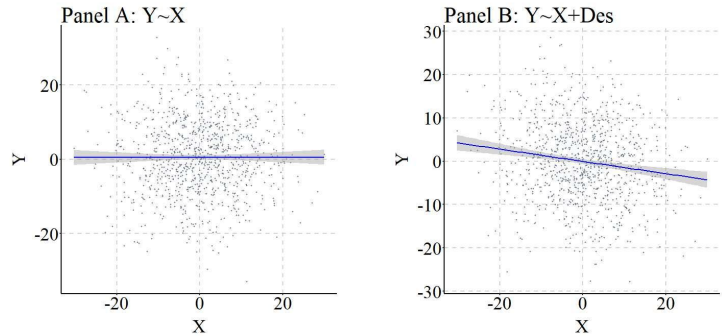

*Descendants of Mediators*

Descendants of mediators are constructs causally influenced by the mechanism mediating the association of interest. Take for example the figure below, where the variation in X causally influences the variation in Me which, in turn, causally influences the variation in Des and the variation in Y. As the trend as begun to appear, adjusting for Des – under these conditions – will function similarly to adjusting for Me, reducing our ability to observe the total causal effects of X on Y and only permitting the observation of the direct causal effects.

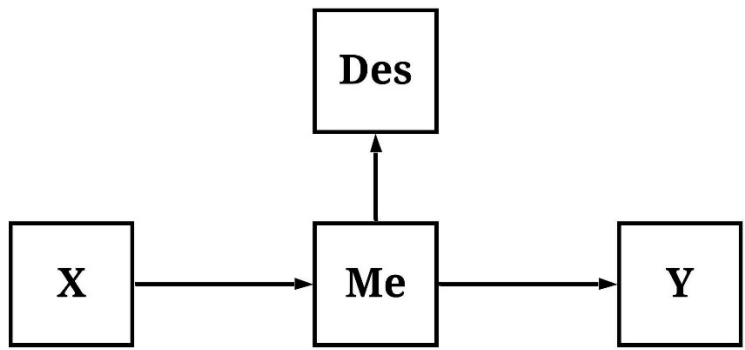

To simulate data following the structural specification outlined in the figure, we must first simulate variation in X. After which, we can specify that X causally influences variation in M, which can be specified to causally influence variation in Y and variation in Des. Following a consistent pattern, the specification below is designed to make the variation in Des almost identical to the variation in M.

```
## Example 1 ####
n<-1000

set.seed(1992)
X<-rnorm(n,0,10)
M<-2*X+rnorm(n,0,10)
Y<-2*M+rnorm(n,0,10)
Des<-20*M+1*rnorm(n,0,10)
```

After simulating the data, we can again estimate our three regression models: (1) Y regressed on X, (2) Y regressed on X and M, and (3) Y regressed on X and Des. As a reminder, the first model will provide estimates representative of the total causal effects of X on Y, while the second and third models will produce estimates representative of only the direct causal effects of X on Y. If we are interested in the total causal effects of X on Y, adjusting for the variation in M or the variation in Des will provide estimates not representative of the desired causal effects.

The effects of adjusting for the mediator and the descendant of a mediator is demonstrated below, where the bivariate association highlights that variation in X does causally influence variation in Y, but the models adjusting for M or Des suggest that X does not causally influence variation in Y (because X does not *directly* cause variation in Y). Importantly, the association between M and Y is substantially larger than the association between Des and Y, suggesting that the variation shared between Des and M might not perfectly emulate the variation shared between M and Y. The effects of adjusting for Des, under these conditions, is further demonstrated in the figure below.

*Bivariate Model*
```
> summary(lm(Y~X))

Call:
lm(formula = Y ~ X)

Residuals:
        Min          1Q      Median          3Q         Max
-69.81609609 -14.16806065  0.07581825 15.30999351 63.58971146

Coefficients:
               Estimate   Std. Error  t value            Pr(>|t|)
(Intercept) 0.8464290685 0.6893798040  1.22781             0.21981
X           3.9721386760 0.0707194962 56.16752 < 0.0000000000000002 ***
---
Signif. codes:  0 '***' 0.001 '**' 0.01 '*' 0.05 '.' 0.1 ' ' 1

Residual standard error: 21.7993225 on 998 degrees of freedom
Multiple R-squared:  0.759679651, Adjusted R-squared:  0.759438849
F-statistic: 3154.79024 on 1 and 998 DF,  p-value: < 0.00000000000000002220446

> lm.beta(lm(Y~X))

Call:
lm(formula = Y ~ X)

Standardized Coefficients::
   (Intercept)                X
0.000000000000 0.871596036565

> 0.00000000000000 0.00161152766817
```

*Model Adjusting for the Mediator*
```
> summary(lm(Y~X+M))

Call:
lm(formula = Y ~ X + M)

Residuals:
        Min          1Q      Median          3Q         Max
-38.91136043  -6.97437994 -0.60349453  6.56419524 33.72731638

Coefficients:
                Estimate   Std. Error  t value            Pr(>|t|)
(Intercept) -0.1961587943 0.3196212609 -0.61372             0.53954
X            0.0554886135 0.0725536497  0.76479             0.44458
M            1.9567291841 0.0323468275 60.49215 < 0.0000000000000002 ***
---
Signif. codes:  0 '***' 0.001 '**' 0.01 '*' 0.05 '.' 0.1 ' ' 1

Residual standard error: 10.0922447 on 997 degrees of freedom
Multiple R-squared:  0.94854297,  Adjusted R-squared:  0.948439746
F-statistic: 9189.19477 on 2 and 997 DF,  p-value: < 0.00000000000000002220446

> lm.beta(lm(Y~X+M))

Call:
lm(formula = Y ~ X + M)

Standardized Coefficients::
    (Intercept)                X                M
0.0000000000000 0.0121757218428 0.9630506718006
```

## Model Adjusting for the Descendant of the Mediator

```
> summary(lm(Y~X+Des))

Call:
lm(formula = Y ~ X + Des)

Residuals:
        Min           1Q       Median           3Q          Max
-37.68591523  -6.95585465  -0.62557652   6.60579653  33.72367847

Coefficients:
                  Estimate       Std. Error  t value              Pr(>|t|)
(Intercept)   -0.22738900718  0.32216328928 -0.70582               0.48047
X              0.07163169784  0.07300878543  0.98114               0.32676
Des            0.09748180748  0.00162764264 59.89141 < 0.0000000000000002 ***
---
Signif. codes:  0 '***' 0.001 '**' 0.01 '*' 0.05 '.' 0.1 ' ' 1

Residual standard error: 10.171544 on 997 degrees of freedom
Multiple R-squared:  0.947731151,  Adjusted R-squared:  0.947626299
F-statistic:  9038.7294 on 2 and 997 DF,  p-value: < 0.00000000000000002220446

> lm.beta(lm(Y~X+Des))

Call:
lm(formula = Y ~ X + Des)

Standardized Coefficients::
    (Intercept)              X             Des
0.000000000000  0.015717956753  0.959467971278

>
```

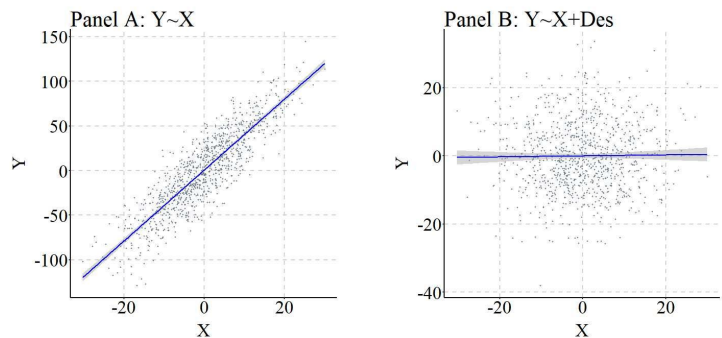

Again, we can replicate this simulation while reducing the causal effects of M on Des. The findings of the replication are presented below.

```
> ## Example 2 ####
> n<-1000
>
>
> set.seed(1992)
> X<-rnorm(n,0,10)
> M<-2*X+rnorm(n,0,10)
> Y<-2*M+rnorm(n,0,10)
> Des<-1*M+1*rnorm(n,0,10)
>
>
> summary(lm(Y~X))

Call:
lm(formula = Y ~ X)

Residuals:
        Min           1Q       Median           3Q          Max
```

```
-69.81609609 -14.16806065    0.07581825   15.30999351   63.58971146

Coefficients:
              Estimate   Std. Error   t value            Pr(>|t|)
(Intercept) 0.8464290685 0.6893798040  1.22781            0.21981
X           3.9721386760 0.0707194962 56.16752 < 0.00000000000000002 ***
---
Signif. codes:  0 '***' 0.001 '**' 0.01 '*' 0.05 '.' 0.1 ' ' 1

Residual standard error: 21.7993225 on 998 degrees of freedom
Multiple R-squared:  0.759679651, Adjusted R-squared:  0.759438849
F-statistic: 3154.79024 on 1 and 998 DF,  p-value: < 0.00000000000000002220446

> lm.beta(lm(Y~X))

Call:
lm(formula = Y ~ X)

Standardized Coefficients::
   (Intercept)              X
0.000000000000 0.871596036565

>
> summary(lm(Y~X+Des))

Call:
lm(formula = Y ~ X + Des)

Residuals:
       Min          1Q      Median          3Q         Max
-55.60249098 -11.80836898  -0.33076703  12.56735897  54.59397296

Coefficients:
              Estimate   Std. Error   t value            Pr(>|t|)
(Intercept) 0.0330722183 0.5522890161  0.05988            0.95226
X           2.1650996171 0.0948040714 22.83762 < 0.00000000000000002 ***
Des         0.9118749831 0.0383986518 23.74758 < 0.00000000000000002 ***
---
Signif. codes:  0 '***' 0.001 '**' 0.01 '*' 0.05 '.' 0.1 ' ' 1

Residual standard error: 17.4306717 on 997 degrees of freedom
Multiple R-squared:  0.846503875, Adjusted R-squared:  0.846195959
F-statistic: 2749.13898 on 2 and 997 DF,  p-value: < 0.00000000000000002220446

> lm.beta(lm(Y~X+Des))

Call:
lm(formula = Y ~ X + Des)

Standardized Coefficients::
   (Intercept)              X            Des
0.000000000000 0.475082165797 0.494011612561

>
```

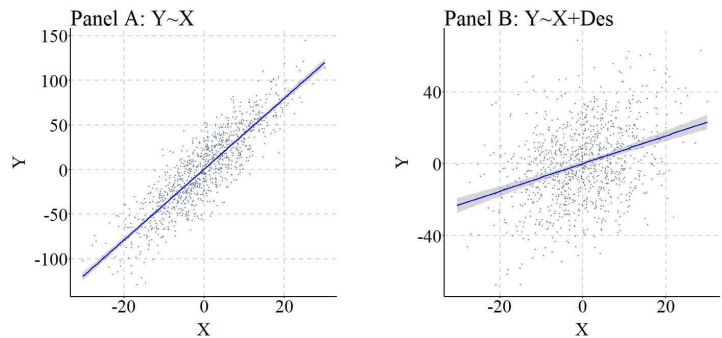

Panel A: Y~X    Panel B: Y~X+Des

### *Descendants of Moderators*
Consistent with the previous discussions, in the current context a descendant of a moderator refers to a construct causally influenced by the variation in a mechanism moderating the association of interest. This, as illustrated below, means that variation in X does causally influence variation in Y, but the magnitude of the association changes at different levels of Mo. Moreover, the variation in Mo causally influences the variation in Des.

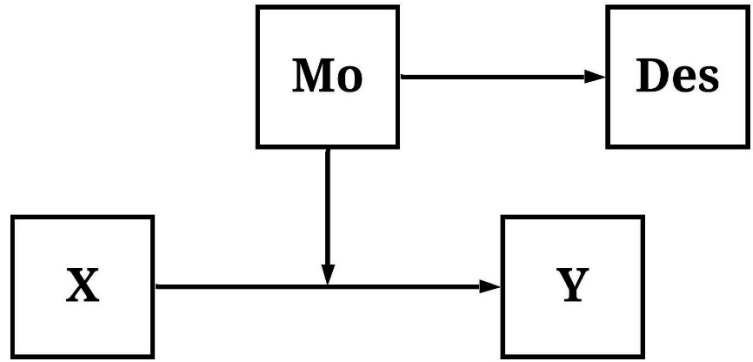

To simulate the data, X and M (sorry the syntax uses M for Mo) are specified as continuous normally distributed variables, while Y is specified to be normally distributed and causally influenced by only the interaction between X and M. This is evident by the specification of X*M. Des was specified to be causally influenced by M. Three regression models were specified to examine the effects of including an interaction between X and Des in the model: (1) Y regressed on X, (2) Y regressed on X, M, and an interaction between X and M, and (3) Y regressed on X, Des, and an interaction between X and Des. Evident by the results, Des – when variation in Des is causally influenced by variation in M – functions similarly to M in that including an interaction between Des and X demonstrates that X does causally influence Y. The magnitude of the causal effects of X on Y does vary by M. It is important, however, to note that the magnitude of the interaction effects is substantially diminished when comparing the interaction estimates corresponding to X*M to the interaction estimates corresponding to X*Des. This attenuation is likely the product of the residual variation in Des (i.e., the variation unique to Des).

*Bivariate Model*

```
> summary(lm(Y~X))

Call:
lm(formula = Y ~ X)

Residuals:
        Min           1Q       Median           3Q          Max
-263.41802778  -20.08017285   0.26256427   21.08051513  242.35063198

Coefficients:
              Estimate   Std. Error  t value Pr(>|t|)
(Intercept) -0.147056842  1.561068597 -0.09420  0.92497
X            0.170546456  0.160141020  1.06498  0.28714

Residual standard error: 49.363555 on 998 degrees of freedom
Multiple R-squared:  0.00113515824,    Adjusted R-squared:  0.000134291665
F-statistic: 1.13417539 on 1 and 998 DF,  p-value: 0.287144056

> lm.beta(lm(Y~X))

Call:
lm(formula = Y ~ X)

Standardized Coefficients::
    (Intercept)              X
0.0000000000000 0.0336921094564
```

*Model Adjusting for the Moderator*

```
> summary(lm(Y~X+M+X*M))

Call:
lm(formula = Y ~ X + M + X * M)

Residuals:
        Min          1Q      Median          3Q         Max
-38.91513976  -6.97454924  -0.60176279   6.56371310  33.72800861

Coefficients:
              Estimate      Std. Error    t value        Pr(>|t|)
(Intercept) -0.19611564507  0.31978882353  -0.61327        0.53984
X           -0.03102597597  0.03278451867  -0.94636        0.34419
M           -0.04333370793  0.03251416397  -1.33276        0.18291
X:M          0.49993320824  0.00332520586 150.34654 < 0.0000000000000002 ***
---
Signif. codes:  0 '***' 0.001 '**' 0.01 '*' 0.05 '.' 0.1 ' ' 1

Residual standard error: 10.0973078 on 996 degrees of freedom
Multiple R-squared:  0.958290608, Adjusted R-squared:  0.958164977
F-statistic: 7627.83793 on 3 and 996 DF,  p-value: < 0.00000000000000002220446

> lm.beta(lm(Y~X+M+X*M))

Call:
lm(formula = Y ~ X + M + X * M)

Standardized Coefficients::
      (Intercept)                X                M              X:M
 0.00000000000000 -0.00612930108789 -0.00866491210276 0.97828216503063
```

*Model Adjusting for the Descendant of the Moderator*

```
> summary(lm(Y~X+Des+X*Des))

Call:
lm(formula = Y ~ X + Des + X * Des)

Residuals:
        Min          1Q      Median          3Q         Max
-42.04562007  -7.07697013  -0.54281243   6.92872633  38.91376855

Coefficients:
               Estimate       Std. Error   t value        Pr(>|t|)
(Intercept) -0.149368925888  0.329261126425  -0.45365        0.65018
X           -0.034387581711  0.033753411897  -1.01879        0.30855
Des         -0.001992587263  0.001671596363  -1.19203        0.23353
X:Des        0.024945398603  0.000171069152 145.82055 < 0.0000000000000002 ***
---
Signif. codes:  0 '***' 0.001 '**' 0.01 '*' 0.05 '.' 0.1 ' ' 1

Residual standard error: 10.3954536 on 996 degrees of freedom
Multiple R-squared:  0.955791116, Adjusted R-squared:  0.955657957
F-statistic: 7177.80278 on 3 and 996 DF,  p-value: < 0.00000000000000002220446

> lm.beta(lm(Y~X+Des+X*Des))

Call:
lm(formula = Y ~ X + Des + X * Des)

Standardized Coefficients::
      (Intercept)                X              Des            X:Des
 0.00000000000000 -0.00679339925381 -0.00798044157255 0.97708618766751
```

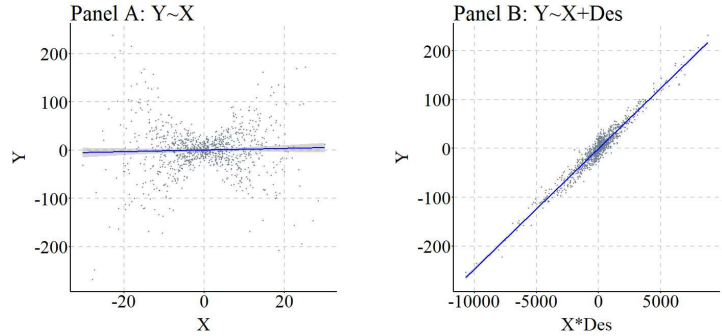

Again, we can replicate this simulation analysis but weaken the causal influence of M on Des.

```
> ## Example 2 ####
> n<-1000
>
> set.seed(1992)
> X<-rnorm(n,0,10)
> M<-rnorm(n,0,10)
> Y<-0*X+0*M+.5*(X*M)+rnorm(n,0,10)
> Des<-3*M+1*rnorm(n,0,10)
>
> summary(lm(Y~X))

Call:
lm(formula = Y ~ X)

Residuals:
         Min           1Q       Median          3Q         Max
-263.41802778  -20.08017285   0.26256427   21.08051513  242.35063198

Coefficients:
              Estimate   Std. Error   t value  Pr(>|t|)
(Intercept) -0.147056842 1.561068597 -0.09420   0.92497
X            0.170546456 0.160141020  1.06498   0.28714

Residual standard error: 49.363555 on 998 degrees of freedom
Multiple R-squared:  0.00113515824,    Adjusted R-squared:  0.000134291665
F-statistic: 1.13417539 on 1 and 998 DF,  p-value: 0.287144056

> lm.beta(lm(Y~X))

Call:
lm(formula = Y ~ X)

Standardized Coefficients::
    (Intercept)               X
0.0000000000000 0.0336921094564

>
> summary(lm(Y~X+Des+X*Des))

Call:
lm(formula = Y ~ X + Des + X * Des)

Residuals:
       Min          1Q       Median          3Q         Max
-87.12503840 -10.88789202  -0.70743204   9.75046825  96.39895051

Coefficients:
               Estimate    Std. Error   t value           Pr(>|t|)
(Intercept)  0.11494833452 0.61376507350  0.18728             0.85148
X           -0.03038750205 0.06289904139 -0.48312             0.62912
Des         -0.02223004781 0.01961135313 -1.13353             0.25726
X:Des        0.14754988719 0.00200565836 73.56681 < 0.0000000000000002 ***
---
Signif. codes:  0 '***' 0.001 '**' 0.01 '*' 0.05 '.' 0.1 ' ' 1

Residual standard error: 19.3706177 on 996 degrees of freedom
Multiple R-squared:  0.846499566, Adjusted R-squared:  0.846037215
F-statistic: 1830.86033 on 3 and 996 DF,  p-value: < 0.00000000000000002220446
```

```
> lm.beta(lm(Y~X+Des+X*Des))

Call:
lm(formula = Y ~ X + Des + X * Des)

Standardized Coefficients::
      (Intercept)                X               Des             X:Des
 0.00000000000000 -0.00600316810493 -0.01414348400394  0.91876640659984

>
```

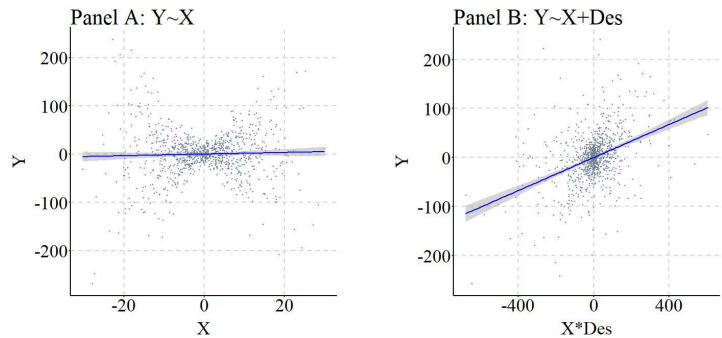

## Discussion & Conclusions

Descendants, conditional upon their ancestors, can reduce or introduce biases into the estimated association between two variables. Under certain circumstances, adjusting for the variation in a descendant can be advantageous. For example, when we can not directly observe the variation associated with a confounder, introducing a descendant of the confounder into a regression model can reduce confounder bias in the estimates corresponding to the association of interest. Additionally, when we can not directly observe variation in a moderator, a descendant of the moderator can permit us to observe how the causal effects of the independent variable on the dependent variable change across scores on the descendant/moderator. Nevertheless, adjusting for the variation in a descendant can also be disadvantageous. For instance, introducing the descendant of collider into a regression model can increase the bias in the estimates corresponding to the association of interest. Alternatively, introducing the descendant of a mediator into a regression model could generate estimates only representative of the direct causal effects. Before concluding, it is important to note with a brief example that adjusting for both the descendant and the ancestor removes the influence of the descendant on the bias within the model (i.e., the ancestor is now the mechanism generating or reducing the bias). This is demonstrated in the syntax below. Given the potential for generating statistical biases and the potential for reducing statistical biases, it is important to identify descendants of key constructs. After identifying the descendants, one can use the principles of Structural Causal Modeling to guide the specification of the regression model and reduce the bias in the estimates produced.

```
> n<-1000
>
> set.seed(1992)
> Con<-rnorm(n,0,10)
> X<-2*Con+.50*rnorm(n,0,10)
> Y<-2*Con+.00*X+.50*rnorm(n,0,10)
> Des<-20*Con+1*rnorm(n,0,10)
>
>
> summary(lm(Y~X+Con+Des))

Call:
```

```
lm(formula = Y ~ X + Con + Des)

Residuals:
         Min             1Q         Median             3Q            Max
-19.618816079  -3.479404832  -0.284815154   3.283236361  16.866716153

Coefficients:
                Estimate    Std. Error  t value              Pr(>|t|)
(Intercept) -0.0932681968  0.1599235600 -0.58320              0.55989
X           -0.0432234038  0.0323506426 -1.33609              0.18182
Con          2.3392851900  0.3133850695  7.46457 0.00000000000018202 ***
Des         -0.0134316493  0.0153302256 -0.87615              0.38116
---
Signif. codes:  0 '***' 0.001 '**' 0.01 '*' 0.05 '.' 0.1 ' ' 1

Residual standard error: 5.04671047 on 996 degrees of freedom
Multiple R-squared:  0.936520294, Adjusted R-squared:  0.93632909
F-statistic: 4898.01795 on 3 and 996 DF,  p-value: < 0.00000000000000002220446

> lm.beta(lm(Y~X+Con+Des))

Call:
lm(formula = Y ~ X + Con + Des)

Standardized Coefficients::
    (Intercept)               X               Con              Des
 0.0000000000000 -0.0434986291032  1.1406867662775 -0.1310474314437

>
```

# Entry 11: Instrumental Variables

**Introduction**
Generating causal inferences is a difficult process. We can run a true experiment (i.e., a randomized controlled trial), however that can be ethically concerning when randomizing certain *treatments*. Alternatively, we can develop a Directed Acyclic Graph (DAG) and reduce the influence of confounders and colliders on our association of interest. Or we can conduct complex statistical analyses, such as a regression discontinuity model, a difference-in-difference model, or a lagged panel model. Nevertheless, these techniques are subject to various assumptions – discussions planned in the future – and might not be applicable for the association of interest. As such, we might have to rely on an alternative approach for generating causal inferences: *instrumental variable analysis*.

*Instrumental Variables*
Let's take a moment to define instrumental variables and identify constructs that can be categorized as instruments. A construct can be employed as an instrument when: (1) the variation in the construct directly causes variation in the independent variable, (2) the variation in a construct is uncorrelated with any confounders, (3) the variation in the construct does not directly cause variation in the dependent variable, and (4) the variation in the construct does not have any other indirect causal influence on the dependent variable (Figure 1). As illustrated below, **IN** satisfies these requirements, making it an instrument for the association between **X** and **Y**, which permits the estimation of an instrumental variable analysis.

[Figure 1]

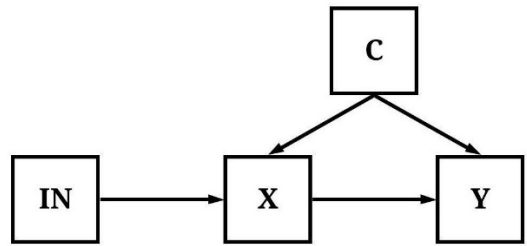

In figures 2-5, **IN** would not qualify as an instrumental variable. Explicitly, in Figure 2, **IN** has an indirect effect on **X** and **Y** through **C**.

[Figure 2]

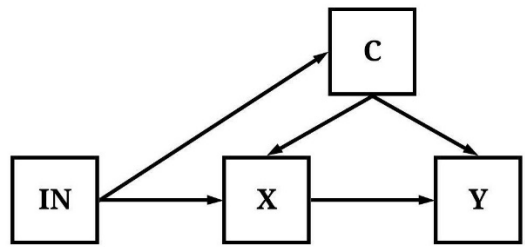

In Figure 3, the variation in **IN** is correlated with the variation in **C**.

[Figure 3]

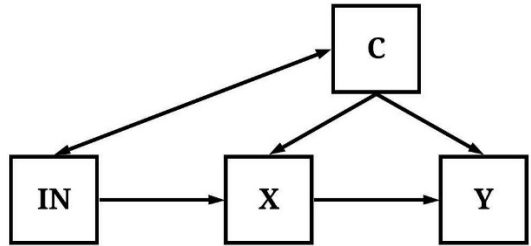

In Figure 4, variation in **IN** has a direct causal effect on both the variation in **X** and the variation in **Y**.

[Figure 4]

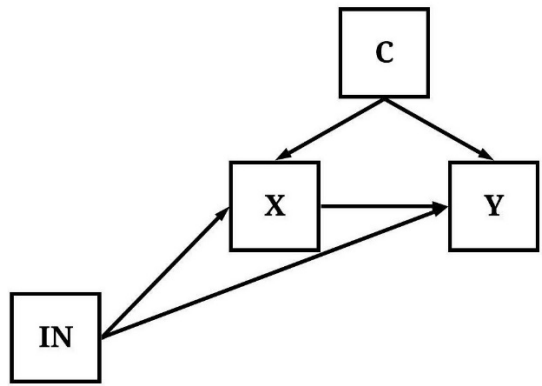

In Figure 5, variation in **IN** causally influences variation in **Y** indirectly through both **X** and **M**.

[Figure 5]

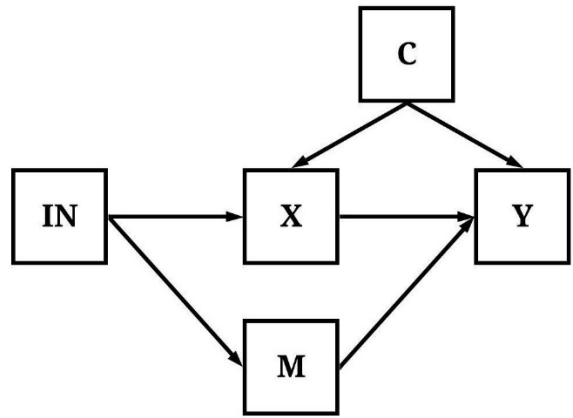

As one can recognized, it is quite difficult to identify a construct that satisfies the requirements of an instrumental variable. These requirements, however, serve a valuable purpose when trying to generate causal inferences about the effects of **X** on **Y**. In particular, by finding the perfect instrument, one can estimate the causal effects of **X** on **Y** ($b_{YX}$) without observing the known or unknown mechanisms confounding the association. To demonstrate how instrumental variable analysis works, let's briefly review how to calculate the *total*, *direct*, and *indirect effects* of one construct on another construct. As presented in Equation 1, the total effect of **X** on **Y** is equal to the direct effect ($b_{YX}$) of **X** on **Y** plus the indirect effect of **X** on **Y** ($b_{MEX} * b_{YME}$). The indirect effect is calculated as the effects of **X** on the mediator (**ME**) multiplied by the effects of the mediator (**ME**) on **Y**.

[Equation 1]

$$YX_{total} = YX_{direct} + YX_{indirect}$$
$$YX_{direct} = b_{YX}$$
$$YX_{indirect} = (b_{MEX} * b_{YME})$$

Translating this to an instrumental variable analysis, we can replace **X** with **IN**.

[Equation 2]

$$YIN_{total} = YIN_{direct} + YIN_{indirect}$$
$$YIN_{direct} = b_{YIN}$$
$$YIN_{indirect} = (b_{MEIN} * b_{YME})$$

Now that we have translated the equation for **Y** and **IN**, we can observe how the requirements of an instrument can permit the observation of the causal effects of **X** on **Y**. First, as identified above, the instrument can not have a direct influence on **Y**. By satisfying this requirement, we can remove the direct effects portion of the equation because $b_{YIN}$ is assumed to be equal to 0.

[Equation 3]

$$YIN_{total} = YIN_{indirect}$$
$$YIN_{indirect} = (b_{MEIN} * b_{YME})$$

Second, the instrumental variable can not have any other indirect causal influence on the dependent variable. By satisfying this requirement, we can set the indirect equation to only represent the indirect pathway through the independent construct of interest (**X**).

[Equation 3]

$$YIN_{total} = YIN_{indirect}$$
$$YIN_{indirect} = (b_{XIN} * b_{YX})$$

Third, the variation in the instrumental variable is required to be uncorrelated with any mechanisms that confound the association of interest. By satisfying this requirement, the total effects of **IN** on **Y** – $b_{YIN}$ – will remain unbiased. Equation 4 represents the calculation of the total effects of **IN** on **Y** when a construct satisfies the requirements for being designated as an instrumental variable.

[Equation 4]

$$YIN_{total} = b_{XIN} * b_{YX}$$

Through reverse estimation, we can solve Equation 4 to identify the causal effects of **X** on **Y** ($b_{YX}$). The reverse estimation can be done by: (1) regressing **Y** on **IN** to observe the slope coefficient representative of the total effects of **IN** on **Y** and (2) regressing **X** on **IN** to observe the slope coefficient representative of the direct effects of **IN** on **X**. After estimating these regression models, we can solve for $b_{YX}$ by setting $b_{YX}$ equal to $b_{YIN}$ divided by $b_{XIN}$.

[Equation 5]

$$\frac{b_{YIN}}{b_{XIN}} = b_{YX}$$

To provide an example, if the slope coefficient of the total effects of **Y** on **IN** is equal to 1 and the slope coefficient of the **IN** on **X** is equal to 2, $b_{YX}$ is equal to .50 even if the association between **X** and **Y** is confounded by **C**. But let's work through some simulations!

**An Appropriate Instrument**

Using the diagram presented in Figure 1, we can simulate a structural system with an association between **X** and **Y**, where variation in **X** is causally influenced by variation in the instrumental variable (**IN**) and a confounder (**C**) and the variation in **Y** is causally influenced by the variation in **X** and the confounder (**C**). For simplicity, the distribution of scores on the confounder (**C**) and the instrumental variable (**IN**) were specified to have a mean of 0 and a standard deviation of 10, while the residual variation for **X** and **Y** was specified to have a mean of 0 and a standard deviation of 10. The slope coefficient of **C** and **IN** on **X** was set equal to 1, while the slope coefficient of **C** and **X** on **Y** was set equal to 1.

```
> ## Example 1 ####
> n<-1000
>
> set.seed(1992)
>
> C<-rnorm(n,0,10)
> IN<-rnorm(n,0,10)
> X<-1*C+1*IN+1*rnorm(n,0,10)
> Y<-1*C+1*X+1*rnorm(n,0,10)
```

After simulating the data, we can estimate the confounded and unconfounded association between **X** and **Y**. Evident by the estimates, the confounded slope coefficient was $b_{YX}$=1.313 and the slope coefficient produced by the model adjusting for the influence of C was $b_{YX}$=.986. This demonstrates that not adjusting for C in the regression equation produces confounder bias when estimating the effects of **X** on **Y**.

```
# Confounded Model ###
> summary(lm(Y~X))

Call:
lm(formula = Y ~ X)

Residuals:
        Min          1Q      Median          3Q         Max
-42.48178035  -8.95502545   0.05464166   9.34051095  42.09044496

Coefficients:
               Estimate    Std. Error    t value         Pr(>|t|)
(Intercept) 0.2052258593 0.4154580586    0.49397          0.62143
X           1.3127961362 0.0248194464   52.89385 < 0.0000000000000002 ***
---
Signif. codes:  0 '***' 0.001 '**' 0.01 '*' 0.05 '.' 0.1 ' ' 1

Residual standard error: 13.1366588 on 998 degrees of freedom
Multiple R-squared:  0.737075026, Adjusted R-squared:  0.736811574
F-statistic: 2797.75962 on 1 and 998 DF,  p-value: < 0.00000000000000002220446

> lm.beta(lm(Y~X))

Call:
lm(formula = Y ~ X)

Standardized Coefficients::
   (Intercept)                X
0.000000000000 0.858530736884
```

```
# Unconfounded Model ###
>
> summary(lm(Y~X+C))

Call:
lm(formula = Y ~ X + C)

Residuals:
         Min          1Q      Median          3Q         Max
-32.87338972  -6.48966902  0.36362255  6.78372717  39.13992447

Coefficients:
               Estimate    Std. Error    t value              Pr(>|t|)
(Intercept) 0.3636575998 0.3297320015    1.10289               0.27034
X           0.9855920916 0.0238734227   41.28407 < 0.00000000000000002 ***
C           0.9940259708 0.0409924314   24.24901 < 0.00000000000000002 ***
---
Signif. codes:  0 '***' 0.001 '**' 0.01 '*' 0.05 '.' 0.1 ' ' 1

Residual standard error: 10.4239797 on 997 degrees of freedom
Multiple R-squared:  0.834615908, Adjusted R-squared:  0.834284145
F-statistic: 2515.69559 on 2 and 997 DF,  p-value: < 0.00000000000000002220446

> lm.beta(lm(Y~X+C))

Call:
lm(formula = Y ~ X + C)

Standardized Coefficients::
   (Intercept)                X                C
0.000000000000 0.644548747046 0.378588396674

>
```

Now let's turn our attention to the instrumental variable analysis by regressing **Y** on **IN** and **X** on **IN**. The slope coefficient produced by the model regressing **Y** on the instrumental variable was $b_{YIN}$ =.962 and the slope coefficient produced by the model regressing **X** on the instrumental variable was $b_{XIN}$ =.958. Using the calculation above, we can estimate that the causal effects of **X** on **Y** is equal to approximately $b_{YX}$ = 1.004, which is close to the specification of the simulated data ($b_{YX}$ = 1.000) and approximately .018 larger than the slope coefficient produced by the unconfounded model.

```
> # Y on Instrumental Variable ###
> summary(lm(Y~IN))

Call:
lm(formula = Y ~ IN)

Residuals:
         Min          1Q      Median          3Q         Max
-70.73574803 -17.40492518  1.02465114  16.11752345  75.55079914

Coefficients:
                 Estimate      Std. Error    t value              Pr(>|t|)
(Intercept) -0.000430986651 0.753532943015   -0.00057               0.99954
IN           0.961597210437 0.076262940367   12.60897 < 0.00000000000000002 ***
---
Signif. codes:  0 '***' 0.001 '**' 0.01 '*' 0.05 '.' 0.1 ' ' 1

Residual standard error: 23.7941498 on 998 degrees of freedom
Multiple R-squared:  0.137414049, Adjusted R-squared:  0.136549735
F-statistic: 158.986152 on 1 and 998 DF,  p-value: < 0.00000000000000002220446

> lm.beta(lm(Y~IN))

Call:
lm(formula = Y ~ IN)

Standardized Coefficients::
   (Intercept)             IN
0.000000000000 0.370694010171

>
> # X on Instrumental Variable ###
```

```
> summary(lm(X~IN))

Call:
lm(formula = X ~ IN)

Residuals:
        Min          1Q      Median          3Q         Max
-48.69473518  -9.55124948  -0.46506143   9.14557423  46.89763854

Coefficients:
                Estimate    Std. Error  t value            Pr(>|t|)
(Intercept) -0.2769319073 0.4378308474 -0.63251              0.5272
IN           0.9582718965 0.0443116232 21.62575 <0.0000000000000002 ***
---
Signif. codes:  0 '***' 0.001 '**' 0.01 '*' 0.05 '.' 0.1 ' ' 1

Residual standard error: 13.8252917 on 998 degrees of freedom
Multiple R-squared:  0.319084078, Adjusted R-squared:  0.318401798
F-statistic:  467.67288 on 1 and 998 DF,  p-value: < 0.00000000000000002220446

> lm.beta(lm(X~IN))

Call:
lm(formula = X ~ IN)

Standardized Coefficients::
   (Intercept)               IN
0.000000000000 0.564875276591

>
```

You might be asking why are their slight differences for the **b**yx between the instrumental variable analysis, the unconfounded model, and the simulation specification. These slight differences exist because we can not generate random variables that are perfectly uncorrelated. For example, a slight non-causal correlation exists between the confounder and the instrumental variable due to the random selection of scores from a uniform distribution. However, as demonstrated above, an instrumental variable analysis provides the ability to deduce the causal effects of one construct on another construct when (1) the association of interest is confounded by unobserved mechanisms and (2) when the instrument satisfies the requirements of an instrumental variable analysis.

```
> # Test ###
> cor(IN,C)
[1] 0.00161152766817
```

However, let's replicate this analysis 50,000 with random estimates to demonstrate the average distance between the slope coefficient produced by the instrumental variable analysis and the slope coefficient produced by the unconfounded model. The syntax below was used to conduct the randomly specified simulation analysis.

```
## Replicated Example 1 ####
n<-50000

set.seed(1992)
Example1_DATA = foreach (i=1:n, .packages=c('lm.beta'), .combine=rbind) %dorng%
  {
    ### Value Specifications ####
    ## Example 1 ####
    N<-sample(150:10000, 1)
    
    set.seed(1992)
    
    Con<-(rnorm(N,runif(1,-5,5),runif(1,1,30)))
    IN<-(rnorm(N,runif(1,-5,5),runif(1,1,30)))
    X<-(runif(1,.1,10))*Con+(runif(1,.1,10))*IN+(runif(1,.1,10))*(rnorm(N,runif(1,-5,5),runif(1,1,30)))
    Y<-(runif(1,.1,10))*Con+(runif(1,.1,10))*X+(runif(1,.1,10))*(rnorm(N,runif(1,-5,5),runif(1,1,30)))
    
    ### Models
    M1<-summary(lm(Y~X+Con))
```

```
    M2<-summary(lm(Y~IN))
    M3<-summary(lm(X~IN))

    M1_byx<-coef(M1)[2, 1]
    M2_byin<-coef(M2)[2, 1]
    M3_bxin<-coef(M3)[2, 1]

    IN_byx<-M2_byin/M3_bxin

    # Data Frame ####

    data.frame(i,N,M1_byx,M2_byin,M3_bxin,IN_byx)

  }
```

As demonstrated in Figure 6, the slope coefficient for the association between **X** and **Y** produced by the unconfounded model has almost a perfect positive correlation with the slope coefficient produced by the instrumental variable analysis across the 50,000 randomly specified simulations. To be precise, the correlation is .93 after removing 291 simulations due to the production of simulation anomalies. Moreover, the median difference between the slope coefficient produced by the unconfounded model and the slope coefficient produced by the instrumental variable analysis was 0.0000969833, while the mean difference was 0.0248374101. Altogether, the randomly specified simulation analysis provides evidence supporting the validity of the instrumental variable approach and demonstrates that instrumental variables can be used to estimate the causal effects of **X** on **Y**.

```
> corr.test(DF$IN_byx,DF$M1_byx)
Call:corr.test(x = DF$IN_byx, y = DF$M1_byx)
Correlation matrix
[1] 0.93
Sample Size
[1] 49799
These are the unadjusted probability values.
  The probability values  adjusted for multiple tests are in the p.adj object.
[1] 0

 To see confidence intervals of the correlations, print with the short=FALSE option
>
> summary(DF$IN_byx-DF$M1_byx)
         Min.         1st Qu.          Median            Mean         3rd Qu.            Max.
 -6.7471599646   -0.0137575791    0.0000969833    0.0248374101    0.0151523267  121.1289942258
>
```

[Figure 6]

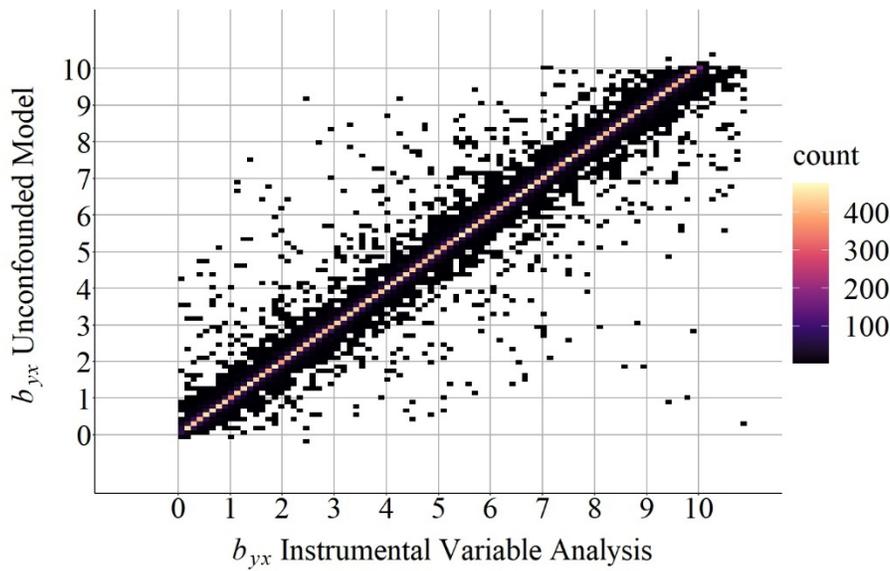

But let's be honest... you did not read this entry because you were worried about the validity of an instrumental variable analysis. You came here because you want to see what happens when we mis-identify an instrumental variable. That is, you want to see the bias that exists when we include one of the instruments in figures 2-5 in our analysis! Don't worry I do too! Although not exhaustive, we will continue our discussion by conducting four randomly specified simulation analyses.

**Misidentified Instrument: IN causes variation in C**
Using the diagram presented in Figure 2, an updated simulation analysis was conducted where variation in the *instrumental variable* (**IN**) caused variation in the confounder (**C**). Identical to the previous simulation analysis, 50,000 randomly specified simulations were estimated and the slope coefficients for the causal effects of **X** on **Y** ($b_{YX}$) produced by the unconfounded model and the instrumental variable models were compared.

```
> n<-50000
> 
> set.seed(1992)
> Example2_DATA = foreach (i=1:n, .packages=c('lm.beta'), .combine=rbind) %dorng%
+   {
+     ### Value Specifications ####
+     ## Example 2 ####
+     N<-sample(150:10000, 1)
+     
+     IN<-(rnorm(N,runif(1,-5,5),runif(1,1,30)))
+     Con<-runif(1,.1,10)*IN+runif(1,.1,10)*(rnorm(N,runif(1,-5,5),runif(1,1,30)))
+     X<-runif(1,.1,10)*Con+runif(1,.1,10)*IN+runif(1,.1,10)*(rnorm(N,runif(1,-5,5),runif(1,1,30)))
+     Y<-runif(1,.1,10)*Con+runif(1,.1,10)*X+runif(1,.1,10)*(rnorm(N,runif(1,-5,5),runif(1,1,30)))
+     
+     
+     ### Models
+     M1<-summary(lm(Y~X+Con))
+     M2<-summary(lm(Y~IN))
+     M3<-summary(lm(X~IN))
+     
+     M1_byx<-coef(M1)[2, 1]
+     M2_byin<-coef(M2)[2, 1]
+     M3_bxin<-coef(M3)[2, 1]
+     
+     IN_byx<-M2_byin/M3_bxin
```

```
+
+       # Data Frame ####
+
+       data.frame(i,N,M1_byx,M2_byin,M3_bxin,IN_byx)
+
+   }
>
```

As demonstrated below, the correlation coefficient of the slope coefficient produced by the unconfounded model and the slope coefficient produced by the instrumental variable analysis was substantially weaker than the previous simulation analysis. More importantly, the median difference between the slope coefficient produced by the unconfounded model and the slope coefficient produced by the instrumental variable analysis was 0.744839750, while the mean difference was 1.185461148. These values suggest that additional bias existed when estimating the causal effects (i.e., slope coefficient) of **X** on **Y** using the instrument in Figure 2, which can be further evaluated by comparing the results in Figure 7 to the results in Figure 6.

```
> DF<-Example2_DATA[which(Example2_DATA$IN_byx>=0),]
>
> corr.test(DF$IN_byx,DF$M1_byx)
Call:corr.test(x = DF$IN_byx, y = DF$M1_byx)
Correlation matrix
[1] 0.71
Sample Size
[1] 49985
These are the unadjusted probability values.
  The probability values  adjusted for multiple tests are in the p.adj object.
[1] 0

 To see confidence intervals of the correlations, print with the short=FALSE option
>
> summary(DF$IN_byx-DF$M1_byx)
        Min.      1st Qu.       Median         Mean      3rd Qu.         Max.
  -6.661808399   0.355579409   0.744839750   1.185461148   1.316089008  482.423168807
>
```

[Figure 7]

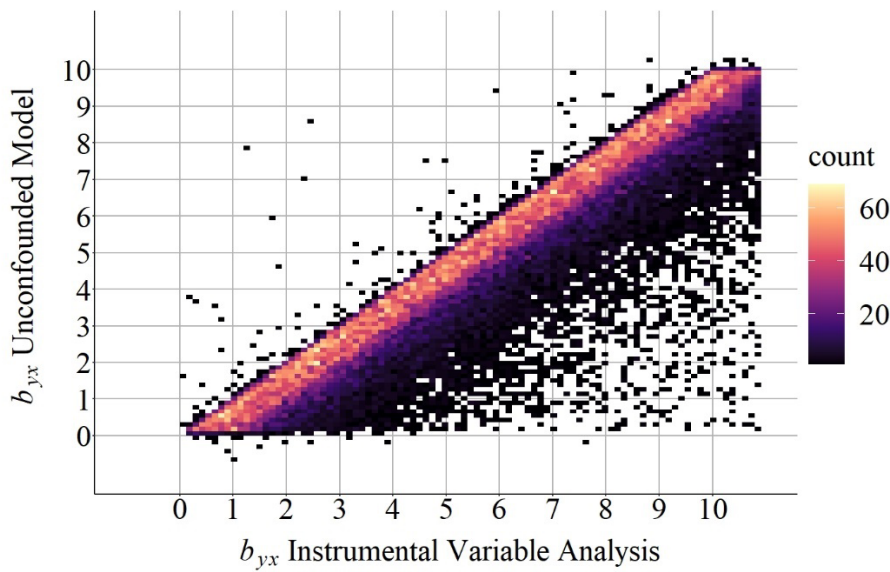

*Important note: Figure 7 through Figure 10 were set on the same scale as Figure 6 to permit an observation of the pattern of findings. The linear pattern extends beyond the 10,10 coordinate in figures 7-10 and can be observed by editing the available code.*

**Misidentified Instrument: Variation in IN correlated with C**

Using the diagram presented in Figure 3, an updated simulation analysis was conducted where variation in the instrumental variable (**IN**) was correlated with variation in the confounder (**C**).

```
> n<-50000
>
> set.seed(1992)
> Example3_DATA = foreach (i=1:n, .packages=c('lm.beta'), .combine=rbind) %dorng%
+   {
+     ### Value Specifications ####
+     ## Example 3 ####
+     N<-sample(150:10000, 1)
+     COR<-(rnorm(N,runif(1,-5,5),runif(1,1,30)))
+
+     IN<-runif(1,.1,10)*(rnorm(N,runif(1,-5,5),runif(1,1,30)))+runif(1,.1,10)*COR
+     Con<-runif(1,.1,10)*(rnorm(N,runif(1,-5,5),runif(1,1,30)))+runif(1,.1,10)*COR
+     X<-runif(1,.1,10)*Con+runif(1,.1,10)*IN+runif(1,.1,10)*(rnorm(N,runif(1,-5,5),runif(1,1,30)))
+     Y<-runif(1,.1,10)*Con+runif(1,.1,10)*X+runif(1,.1,10)*(rnorm(N,runif(1,-5,5),runif(1,1,30)))
+
+
+     ### Models
+     M1<-summary(lm(Y~X+Con))
+     M2<-summary(lm(Y~IN))
+     M3<-summary(lm(X~IN))
+
+     M1_byx<-coef(M1)[2, 1]
+     M2_byin<-coef(M2)[2, 1]
+     M3_bxin<-coef(M3)[2, 1]
+
+     IN_byx<-M2_byin/M3_bxin
+
+     # Data Frame ####
+
+     data.frame(i,N,M1_byx,M2_byin,M3_bxin,IN_byx)
+
+   }
```

Focusing on the difference calculations, the median difference between the slope coefficient produced by the unconfounded model and the slope coefficient produced by the instrumental variable analysis was 0.1989677123 , while the mean difference was 0.3920520418. These values, again, suggest that additional bias existed when estimating the causal effects of **X** on **Y** using the instrument in Figure 3, which can be further evaluated by comparing the results in Figure 8 to the results in Figure 6. Nevertheless, the bias generated when the instrumental variable (**IN**) is correlated with the confounder (**C**) appears to be less than the bias generated when the instrumental variable (**IN**) causes variation in the confounder (**C**).

```
> DF<-Example3_DATA[which(Example3_DATA$IN_byx>=0),]
>
> corr.test(DF$IN_byx,DF$M1_byx)
Call:corr.test(x = DF$IN_byx, y = DF$M1_byx)
Correlation matrix
[1] 0.97
Sample Size
[1] 49993
These are the unadjusted probability values.
  The probability values  adjusted for multiple tests are in the p.adj object.
[1] 0

 To see confidence intervals of the correlations, print with the short=FALSE option
>
> summary(DF$IN_byx-DF$M1_byx)
        Min.      1st Qu.       Median         Mean      3rd Qu.         Max.
-5.2422226730  0.0532766452  0.1989677123  0.3920520418  0.4953741926  26.0382490036
```

[Figure 8]

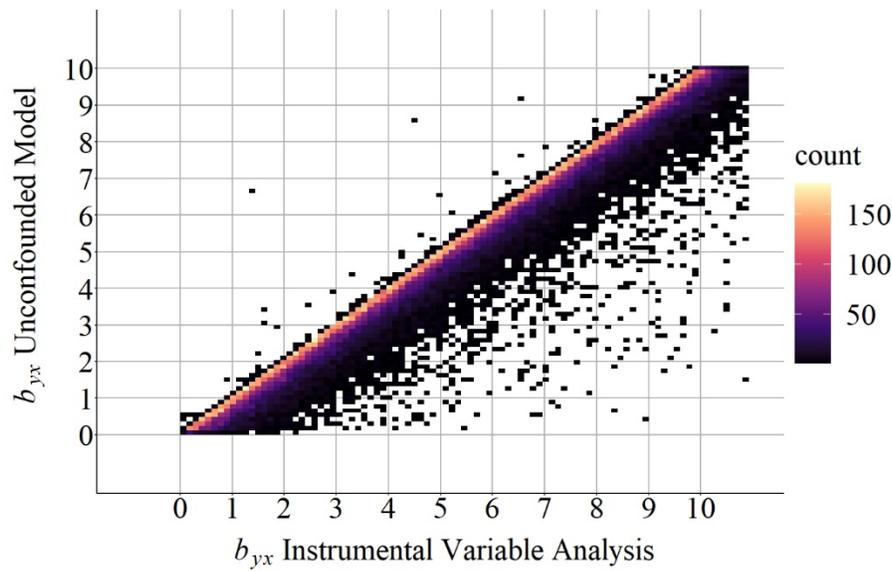

**Misidentified Instrument: IN causes variation in Y**

Using the diagram presented in Figure 4, an updated simulation analysis was conducted where variation in the *instrumental variable* (**IN**) directly caused variation in the dependent variable (**Y**).

```
> n<-50000
>
> set.seed(1992)
> Example4_DATA = foreach (i=1:n, .packages=c('lm.beta'), .combine=rbind) %dorng%
+   {
+     ### Value Specifications ####
+     ## Example 4 ####
+     N<-sample(150:10000, 1)
+
+     IN<-(rnorm(N,runif(1,-5,5),runif(1,1,30)))
+     Con<-(rnorm(N,runif(1,-5,5),runif(1,1,30)))
+     X<-runif(1,.1,10)*Con+runif(1,.1,10)*IN+runif(1,.1,10)*(rnorm(N,runif(1,-5,5),runif(1,1,30)))
+     Y<-runif(1,.1,10)*Con+runif(1,.1,10)*X+runif(1,.1,10)*IN+runif(1,.1,10)*(rnorm(N,runif(1,-5,5),runif(1,1,30))))
+
+
+     ### Models
+     M1<-summary(lm(Y~X+Con))
+     M2<-summary(lm(Y~IN))
+     M3<-summary(lm(X~IN))
+
+     M1_byx<-coef(M1)[2, 1]
+     M2_byin<-coef(M2)[2, 1]
+     M3_bxin<-coef(M3)[2, 1]
+
+     IN_byx<-M2_byin/M3_bxin
+
+     # Data Frame ####
+
+     data.frame(i,N,M1_byx,M2_byin,M3_bxin,IN_byx)
+
+   }
```

As demonstrated below, the median difference between the slope coefficient produced by the unconfounded model and the slope coefficient produced by the instrumental variable analysis was 0.32667350, while the mean difference was 2.18199589. These values suggest that additional bias existed when estimating the causal effects of **X** on **Y** using the instrument in Figure 4, which can be further evaluated by comparing the results in Figure 9 to the results in Figure 6.

```
> DF<-Example4_DATA[which(Example4_DATA$IN_byx>=0),]
> corr.test(DF$IN_byx,DF$M1_byx)
Call:corr.test(x = DF$IN_byx, y = DF$M1_byx)
Correlation matrix
[1] 0.14
Sample Size
[1] 49713
These are the unadjusted probability values.
  The probability values  adjusted for multiple tests are in the p.adj object.
[1] 0

 To see confidence intervals of the correlations, print with the short=FALSE option
> summary(DF$IN_byx-DF$M1_byx)
       Min.     1st Qu.      Median        Mean     3rd Qu.        Max.
-12.16465547  0.05619586  0.32667350  2.18199589  1.17082275 3179.94559972
>
```

[Figure 9]

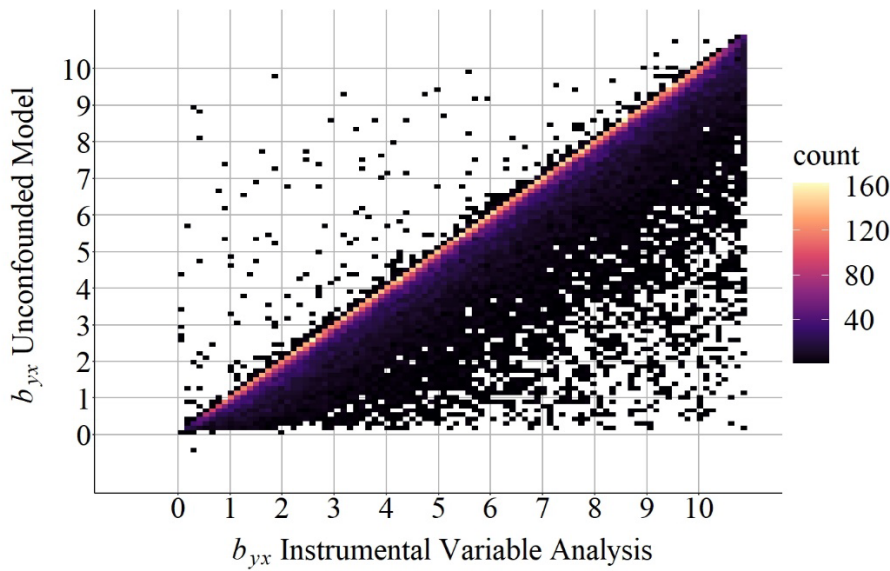

### Misidentified Instrument: Alternate Indirect Pathway Exists

Using the diagram presented in Figure 5, an updated simulation analysis was conducted where variation in the *instrumental variable* (**IN**) indirectly caused variation in the dependent variable (**Y**) through another mediator.

```
> n<-50000
>
> set.seed(1992)
> Example5_DATA = foreach (i=1:n, .packages=c('lm.beta'), .combine=rbind) %dorng%
+   {
+     ### Value Specifications ####
+     ## Example 5 ####
+     N<-sample(150:10000, 1)
+
+     IN<-(rnorm(N,runif(1,-5,5),runif(1,1,30)))
+     Con<-(rnorm(N,runif(1,-5,5),runif(1,1,30)))
+     M<-runif(1,.1,10)*IN+runif(1,.1,10)*(rnorm(N,runif(1,-5,5),runif(1,1,30)))
+     X<-runif(1,.1,10)*Con+runif(1,.1,10)*IN+runif(1,.1,10)*(rnorm(N,runif(1,-5,5),runif(1,1,30)))
+     Y<-runif(1,.1,10)*Con+runif(1,.1,10)*X+runif(1,.1,10)*M+runif(1,.1,10)*(rnorm(N,runif(1,-
5,5),runif(1,1,30)))
+
+
+     ### Models
+     M1<-summary(lm(Y~X+Con))
```

```
+     M2<-summary(lm(Y~IN))
+     M3<-summary(lm(X~IN))
+
+     M1_byx<-coef(M1)[2, 1]
+     M2_byin<-coef(M2)[2, 1]
+     M3_bxin<-coef(M3)[2, 1]
+
+     IN_byx<-M2_byin/M3_bxin
+
+     # Data Frame ####
+
+     data.frame(i,N,M1_byx,M2_byin,M3_bxin,IN_byx)
+
+   }
```

As demonstrated by the results, the median difference between the slope coefficient produced by the unconfounded model and the slope coefficient produced by the instrumental variable analysis was 1.2321460, while the mean difference was 11.5971274. These values suggest that additional bias existed when estimating the causal effects of **X** on **Y** using the instrument in Figure 5, which can be further evaluated by comparing the results in Figure 10 to the results in Figure 6.

```
> DF<-Example5_DATA[which(Example5_DATA$IN_byx>=0),]
> corr.test(DF$IN_byx,DF$M1_byx)
Call:corr.test(x = DF$IN_byx, y = DF$M1_byx)
Correlation matrix
[1] 0.06
Sample Size
[1] 49688
These are the unadjusted probability values.
  The probability values  adjusted for multiple tests are in the p.adj object.
[1] 0

 To see confidence intervals of the correlations, print with the short=FALSE option
> summary(DF$IN_byx-DF$M1_byx)
       Min.       1st Qu.        Median          Mean       3rd Qu.          Max.
 -69.4275307     0.1881808     1.2321460    11.5971274     5.1498372  37904.7566381
>
```

[Figure 10]

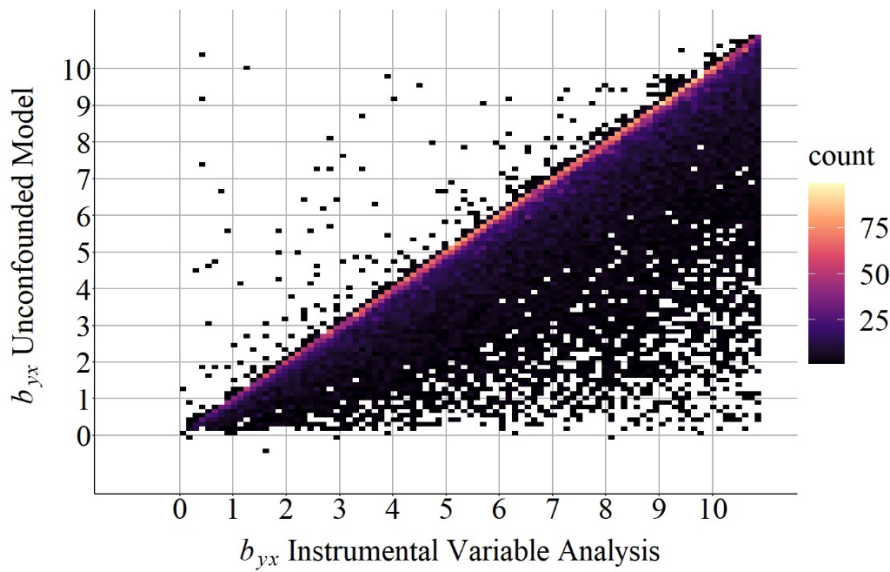

**Summary of Results, Discussion, and Conclusion**
Due to concerns about the length of this entry – and the complexity of the analyses – I could not provide a comprehensive review of the results that is warranted. However, there are two important patterns that need to be highlighted to guide the conclusion. First, evident in Figure 6, a normal distribution appeared in the plot, where the highest density of $b_{YX}$ were the same value for both the unconfounded model and the instrumental variable analysis. This pattern is distinct from the results in figures 7-10, where it appears that the slope coefficient produced by the instrumental variable analysis was primarily upwardly biased (i.e., further from zero than the slope coefficient produced by the unconfounded model). Second, although not visually displayed, the range of differences between the $b_{YX}$ produced by the unconfounded model and the $b_{YX}$ produced by the instrumental variable analysis was substantially wider for the simulations with instrumental variables that did not satisfy the requirements of an instrument. These results, overall, demonstrate that the causal effects of **X** on **Y** can be estimated using an instrumental variable analysis when employing a construct that satisfies the requirements of an instrument. Nevertheless, when the construct does not satisfy the requirements of an instrument, the estimated effects of **X** on **Y** produced by an instrumental variable analysis is commonly upwardly biased – further from zero than the true causal effects.

Typically, I do not introduce philosophical debates into this series, but I do believe that it is required when discussing sources of bias in an instrumental variable analysis. If you have ever participated in a research methods or statistics course, one of the first things you are taught is that you can not rule out unobserved confounders in non-experimental research. This directly translates to, *you can not rule out unobserved causes and causal pathways*. Now, take a glance at figures 1-5 again and think of a scenario where you can *definitively* identify an instrument. That is, it can not cause or be correlated with variation in the confounder, it can not directly cause variation in the dependent variable, and it can not indirectly cause variation in the dependent variable through an alternative path. Can you truly think of or find any? So I ask, knowing that violating the requirements of an instrument can reduce your ability to observe the causal effects between constructs, are the benefits of an instrumental variable analysis diminished by the inability to intrinsically define any construct as a true instrument? Do the benefits of the technique outweigh the shortcomings?

*p.s. I have no bearing on your implementation of instrumental variable analysis – as long as you think it is the best approach for the research at hand please do not let me dissuade you. It, however, is always good practice to note the assumptions and limitations of the technique.*

## Entry 12: The Inclusion of Non-Causally Associated Constructs and Reverse Causal Specifications

**Introduction**
The previous entries have focused on the biases that can exist when generating causal inferences through methodological and statistical approaches. I know it is considerably difficult for researchers trained in experimental methods – *such as myself in criminology* – to discern that causal inferences can be generated without randomly assigning participants or creating a counterfactual. Specifically, the appropriate selection of variables in a linear regression model estimated using non-experimental data can provide the ability to generate inferences about causal processes in the population (See Pearl 2009). Nevertheless, this is easier said than done. As such, an extensive number of statistical techniques have been created to generate causal inferences from non-experimental data. The foundational knowledge provided throughout these entries, however, does not cover those techniques and, in turn, the mechanisms generating biases with-in those models. That said, let me be frank for a moment, after much thought we will have to revisit causal inference in the future.[i] There is just too much to talk about, too many sources of biases, and the methods are too complicated for discussions at this point of the series. Additionally, we need to talk about other foundational sources of statistical bias – such as, measurement, missing data, and imputation – before discussing the advanced statistical approaches commonly used to generate causal inferences: propensity score weighting and matching, difference-in-difference modeling, regression discontinuity, synthetic controls, marginal structural models, and lagged panel models. But for now, let's finish up our first discussion of causal inference by talking about: (1) the inclusion of non-causally associated constructs in our models and (2) reverse causal specifications.

*Brief Note: In addition to the need to have other conversations before discussing complicated analytical techniques, I decided to end our first visit to causal inference here in an effort to keep the series interesting by switching topics regularly. I hope you understand and I promise we will finish up causal inference when we revisit it! In the meantime, I have been writing a lot about causal inference in my recent research, which you should check out when it is published soon, hopefully!*

**Non-Causally Associated Constructs: Constructs with No Association or a Non-Causal Association**
In the current context, non-causally associated constructs refers to the condition where the variation in one construct does not cause variation in another construct. These constructs can be completely independent (i.e., no covariation between the constructs) or covary due to either a common cause or measurement error.[ii] In this sense, the vast majority of constructs that exist in the generation of data across the universe do not cause variation in one another. It is extremely important to remember this, because prior knowledge is used guide the research we conduct, which means that most of the time we only focus our efforts on things we believe are causally associated. Take for example my own research, I am not going to study how incarceration influences computer programming skills because most incarcerated individuals do not have access to a computer during or after incarceration.[iii] I, however, have published research on how incarceration influences employment because discussions with formerly incarcerated individuals highlight the difficulties that exist when seeking employment after incarceration. That said, it is important to consider the effects of including a construct in a multivariable regression model unrelated to the constructs of interests, as well as including a construct in a multivariable regression model that covaries with either the independent or dependent variable of interest. Our discussion will focus on *either or* because Entry 10 demonstrates the effects of including and excluding a construct that covaries with the independent and dependent variable on the slope coefficient for the association of interest. Keeping up with the theme of the series, let's get right into our simulations.

*Non-Causally Associated Constructs: No Covariation*
So, if you have followed the series, you should know that simulating a dataset with a causal association between two constructs, and no association – causal or non-causal – with a third construct is relatively simple. Below, is all the syntax we need to specify that **X** and **Z** are normally distributed constructs with a mean of 0 and a standard deviation of 10, and **Y** is a normally distributed construct causally influenced by **X** with a residual distribution that possesses a mean of 0 and a standard deviation of 10. The variation in **X** is linearly associated with the variation in **Y**, where a 1 point increase in **X** corresponds to a 1 point increase in **Y**. Nevertheless, using this simulation specification, the variation in **X** and **Y** will not be statistically associated with the variation in **Z**. Before we estimate the models, I want you to take a second and think about what would happen if **Z** is introduced as a covariate when **Y** is regressed on **X**.

```
> ## Example ####
> n<-1000
> 
> set.seed(1992)
> 
> Z<-rnorm(n,0,10)
> X<-1*rnorm(n,0,10)
> Y<-1*X+1*rnorm(n,0,10)
> 
> Data<-data.frame(Z,X,Y)
```

If you guessed *"the estimated effects will slightly vary"* you would be correct. You can see below that *b* = .95668 when **Y** was regressed on **X**, while *b* = .95673 when **Y** was regressed on **X** and **Z**. The slight differences in the estimates can be attributed to the existence of shared error due to the random variation used to specify the distribution of each construct. Alternatively, the estimated effects are essentially the same because only the variation that is shared between **X**, **Y**, and **Z** is removed from the calculation of the slope coefficient (*b*) for **Y** on **X**. That is, only a covariate that shares variation with both the independent and dependent variable will reduce the bias (*when the covariate confounds the association of interest*), misrepresent the effects (*when the covariate mediates the association of interest*), or increase the bias (*when the covariate is a collider for the association of interest*) observed in the estimated effects of the independent variable on the dependent variable.

```
> summary(lm(Y~X, data = Data))

Call:
lm(formula = Y ~ X, data = Data)

Residuals:
    Min      1Q  Median      3Q     Max
-38.598  -7.038  -0.491   6.679  33.637

Coefficients:
             Estimate Std. Error t value Pr(>|t|)
(Intercept)  -0.19357    0.31959  -0.606    0.545
X             0.95668    0.03235  29.577   <2e-16 ***
---
Signif. codes:  0 '***' 0.001 '**' 0.01 '*' 0.05 '.' 0.1 ' ' 1

Residual standard error: 10.09 on 998 degrees of freedom
```

```
Multiple R-squared:  0.4671,	Adjusted R-squared:  0.4666 
F-statistic: 874.8 on 1 and 998 DF,  p-value: < 2.2e-16

> summary(lm(Y~X+Z, data = Data))

Call:
lm(formula = Y ~ X + Z, data = Data)

Residuals:
    Min      1Q  Median      3Q     Max 
-38.911  -6.974  -0.603   6.564  33.727 

Coefficients:
             Estimate Std. Error t value Pr(>|t|)    
(Intercept) -0.19616    0.31962  -0.614    0.540    
X            0.95673    0.03235  29.577   <2e-16 ***
Z           -0.03105    0.03274  -0.948    0.343    
---
Signif. codes:  0 '***' 0.001 '**' 0.01 '*' 0.05 '.' 0.1 ' ' 1

Residual standard error: 10.09 on 997 degrees of freedom
Multiple R-squared:  0.4676,	Adjusted R-squared:  0.4665 
F-statistic: 437.8 on 2 and 997 DF,  p-value: < 2.2e-16
```

The similarities between the two models are illustrated in Figure 1, where the regression lines are almost identical.

[Figure 1]

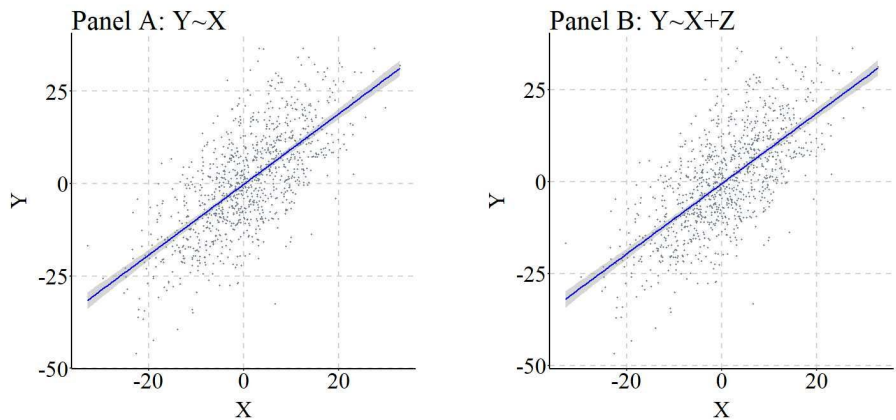

Furthermore, to provide a comprehensive demonstration of this principle, we can replicate the simulation 10,000 times, randomly specify all of the key values – *b*, mean, and standard deviation –, and calculate the difference between the slope coefficient estimated when **Y** was regressed on **X** (i.e., M1_byx in the syntax) and the slope coefficient estimated when **Y** was regressed on **X** and **Z** (i.e., M2_byx in the syntax). As it can be observed, the difference between M1_byx and M2_byx – or D_byx – is extremely small at the $1^{st}$ quartile, median, mean, and $3^{rd}$ quartile, with only the minimum and maximum values demonstrating the existence of some deviation. These deviations, as discussed below, are likely the product of simulation anomalies.

```
> n<-10000
> 
> set.seed(1992)
> Example1_DATA = foreach (i=1:n, .packages=c('lm.beta'), .combine=rbind) %dorng%
+   {
+     ### Value Specifications ####
+     N<-sample(150:10000, 1)
+     Z<-rnorm(N,runif(1,-5,5),runif(1,1,30))
+     X<-(runif(1,-10,10)*rnorm(N,runif(1,-5,5),runif(1,1,30)))
+     Y<-(runif(1,-10,10)*X)+(runif(1,-10,10)*rnorm(N,runif(1,-5,5),runif(1,1,30)))
```

```
+
+       Data<-data.frame(Z,X,Y)
+
+       ### Models ####
+       M1<-summary(lm(Y~X, data = Data))
+       M2<-summary(lm(Y~X+Z, data = Data))
+
+       M1_byx<-coef(M1)[2, 1]
+       M2_byx<-coef(M2)[2, 1]
+
+       D_byx<-M1_byx-M2_byx
+
+       # Data Frame ####
+
+       data.frame(i,N,M1_byx,M2_byx,D_byx)
+
+   }
>
> summary(Example1_DATA$D_byx)
          Min.         1st Qu.          Median            Mean         3rd Qu.            Max.
-1.759304446361 -0.000080997271 -0.000000297547 -0.000229557172  0.000073507961  0.782685990625
>
```

### *Non-Causally Associated Constructs: Covarying with Dependent Variable*

Now, we have already reviewed that only the variation that is shared between **X**, **Y**, and **Z** is removed from the calculation of the slope coefficient (*b*), so if **Z** only shares variation with **Y** we can expect the same results as the previous simulation. We can simulate a non-causal correlation between two constructs by either (1) creating a construct that causes variation in both of the correlated constructs (i.e., a confounder) or by specifying shared residual variation. Personally, I prefer the former technique, but both techniques are extremely similar. In this simulation, we create a non-causal correlation between **Z** and **Y** by specifying that **Cor** – a normally distributed construct with a mean of 0 and standard deviation of 10 – causes variation in both **Z** and **Y**. When modeling the associations, let's just forget **Cor** exists.

```
> ## Example ####
> n<-1000
>
> set.seed(1992)
>
> Cor<-rnorm(n,0,10)
> Z<-1*rnorm(n,0,10)+1*Cor
> X<-1*rnorm(n,0,10)
> Y<-1*X+1*rnorm(n,0,10)+1*Cor
>
> Data<-data.frame(Z,X,Y)
```

After simulating the data, we can estimate our models. Similar to the first simulation, the slope coefficient of the association between **X** and **Y** was .94287 when **Z** was not included in the model, and .97698 when **Z** was included as a covariate in the model. Again, the slight difference is the product of shared variation between **X**, **Y**, and **Z**.

```
> summary(lm(Y~X, data = Data))

Call:
lm(formula = Y ~ X, data = Data)

Residuals:
        Min          1Q      Median          3Q         Max
-47.55754362 -9.40598359  0.26206352  9.76044280 46.06588793

Coefficients:
             Estimate  Std. Error  t value              Pr(>|t|)
(Intercept) 0.265895531 0.446985198  0.59486              0.55207
X           0.942871210 0.044286559 21.29023 < 0.0000000000000002 ***
---
Signif. codes:  0 '***' 0.001 '**' 0.01 '*' 0.05 '.' 0.1 ' ' 1

Residual standard error: 14.1316564 on 998 degrees of freedom
```

```
Multiple R-squared:  0.312328374, Adjusted R-squared:  0.311639324
F-statistic:   453.27407 on 1 and 998 DF,  p-value: < 0.00000000000000002220446

> 
> summary(lm(Y~X+Z, data = Data))

Call:
lm(formula = Y ~ X + Z, data = Data)

Residuals:
        Min          1Q      Median          3Q         Max 
-40.23321543  -8.52102295   0.58367414   8.82508959  40.15749172 

Coefficients:
              Estimate    Std. Error  t value             Pr(>|t|)    
(Intercept) 0.0553391656 0.3936365735  0.14058              0.88823    
X           0.9769833340 0.0390329312 25.02972 < 0.00000000000000002 ***
Z           0.4841391536 0.0283753707 17.06195 < 0.00000000000000002 ***
---
Signif. codes:  0 '***' 0.001 '**' 0.01 '*' 0.05 '.' 0.1 ' ' 1

Residual standard error: 12.4388961 on 997 degrees of freedom
Multiple R-squared:  0.46774069, Adjusted R-squared:  0.466672968 
F-statistic: 438.073566 on 2 and 997 DF,  p-value: < 0.00000000000000002220446
```

Figure 2, similar to Figure 1, further illustrates the similarities between the results of the regression models.

[Figure 2]

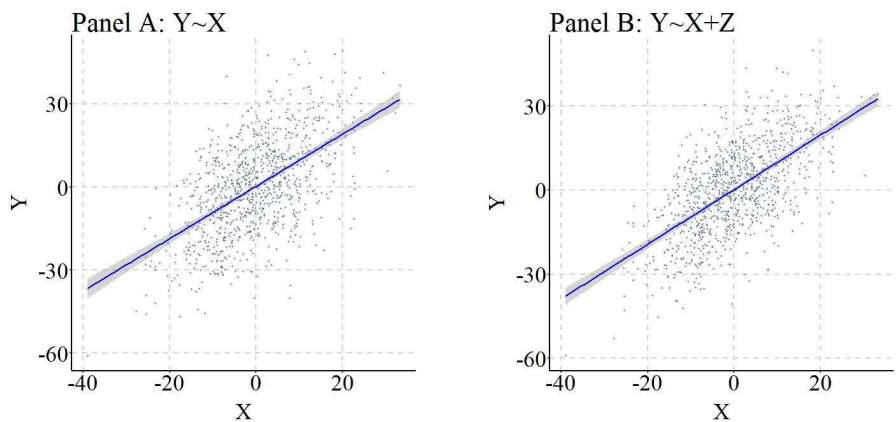

Once again, replicating the simulation 10,000 times generally resulted in extremely small differences between the slope coefficient estimated when **Y** was regressed on **X** (i.e., M1_byx in the syntax) and the slope coefficient estimated when **Y** was regressed on **X** and **Z** (i.e., M2_byx in the syntax). Nevertheless, it is important to point out that the minimum difference between the slope coefficients was -209 and the maximum difference between the slope coefficients was 320. You might ask, why would differences that large occur? Well, to let you know the truth it is likely a simulation anomaly. That is, as the simulation is replicated and the values for the *b*, mean, and standard deviation are randomly sampled from uniform distributions, a perfect condition is sampled from the simulation space to create a statistical anomaly. Generally, these are rare, but occur at higher rates when the values for the *b* and standard deviation are selected from uniform distributions that have the ability to sample numbers close to 0. It is always important to consider the existence of anomalies when conducting simulations.

```
> n<-10000
> 
> set.seed(1992)
```

```
> Example2_DATA = foreach (i=1:n, .packages=c('lm.beta'), .combine=rbind) %dorng%
+   {
+     ### Value Specifications ####
+     N<-sample(150:10000, 1)
+     Cor<-rnorm(N,runif(1,-5,5),runif(1,1,30))
+     Z<-(runif(1,-10,10)*rnorm(N,runif(1,-5,5),runif(1,1,30)))+(runif(1,-10,10)*Cor)
+     X<-(runif(1,-10,10)*rnorm(N,runif(1,-5,5),runif(1,1,30)))
+     Y<-(runif(1,-10,10)*X)+(runif(1,-10,10)*rnorm(N,runif(1,-5,5),runif(1,1,30)))+(runif(1,-10,10)*Cor)
+
+     Data<-data.frame(Z,X,Y)
+
+     ### Models ####
+     M1<-summary(lm(Y~X, data = Data))
+     M2<-summary(lm(Y~X+Z, data = Data))
+
+     M1_byx<-coef(M1)[2, 1]
+     M2_byx<-coef(M2)[2, 1]
+
+     D_byx<-M1_byx-M2_byx
+
+     # Data Frame ####
+
+     data.frame(i,N,M1_byx,M2_byx,D_byx)
+
+   }
>
> summary(Example2_DATA$D_byx)
          Min.      1st Qu.        Median          Mean       3rd Qu.          Max.
-209.741442187  -0.005051558  -0.000004274  -0.013072247   0.004472406  320.331011725
```

### Non-Causally Associated Constructs: Covarying with Independent Variable

Let's finish up this section by exploring the effects of a non-causal correlation between **Z** and **X**. Following the previous simulation, we create a non-causal correlation between **Z** and **X** by specifying that **Cor** – a normally distributed construct with a mean of 0 and standard deviation of 10 – causes variation in both **Z** and **X**.

```
> n<-1000
>
> set.seed(1992)
>
> Cor<-rnorm(n,0,10)
> Z<-1*rnorm(n,0,10)+1*Cor
> X<-1*rnorm(n,0,10)+1*Cor
> Y<-1*X+1*rnorm(n,0,10)
>
> Data<-data.frame(Z,X,Y)
>
```

Evident by the results of the models, the slope coefficient of the association between **X** and **Y** was .97509 when **Z** was not included in the model, and .97368 when **Z** was included as a covariate in the model. Again, the slight difference is the product of shared variation between **X**, **Y**, and **Z**.

```
> summary(lm(Y~X, data = Data))

Call:
lm(formula = Y ~ X, data = Data)

Residuals:
        Min          1Q      Median          3Q         Max
-32.64604854  -6.56097850  0.28267451  6.70606963  39.12979940

Coefficients:
             Estimate  Std. Error  t value          Pr(>|t|)
(Intercept) 0.353333856 0.329484762  1.07238           0.28381
X           0.975090101 0.023839719 40.90191 < 0.0000000000000002 ***
---
Signif. codes:  0 '***' 0.001 '**' 0.01 '*' 0.05 '.' 0.1 ' ' 1

Residual standard error: 10.4167819 on 998 degrees of freedom
Multiple R-squared:  0.626352477, Adjusted R-squared:  0.62597808
F-statistic: 1672.96645 on 1 and 998 DF,  p-value: < 0.00000000000000002220446
```

```
> 
> summary(lm(Y~X+Z, data = Data))

Call:
lm(formula = Y ~ X + Z, data = Data)

Residuals:
        Min          1Q      Median          3Q         Max
-32.60805564 -6.55249618  0.26957028  6.73053562 39.13236765

Coefficients:
               Estimate    Std. Error  t value              Pr(>|t|)
(Intercept) 0.35153822414 0.33002211971  1.06520              0.28705
X           0.97368272558 0.02684369244 36.27231 < 0.0000000000000002 ***
Z           0.00305349263 0.02672169109  0.11427              0.90905
---
Signif. codes:  0 '***' 0.001 '**' 0.01 '*' 0.05 '.' 0.1 ' ' 1

Residual standard error: 10.4219364 on 997 degrees of freedom
Multiple R-squared:  0.62635737, Adjusted R-squared:  0.625607836
F-statistic:  835.66254 on 2 and 997 DF,  p-value: < 0.00000000000000002220446
```

To illustrate the similarities between the results of the regression models, Figure 3 was produced.

[Figure 3]

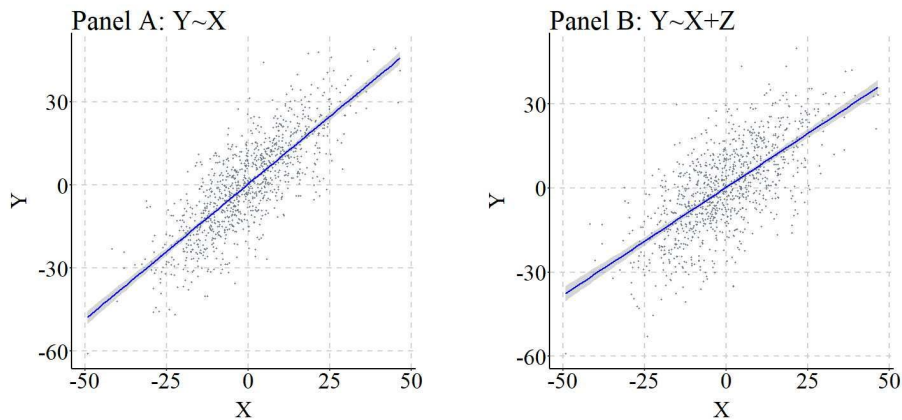

We again can replicate the simulation 10,000 times, randomly specify all of the key values – $b$, mean, and standard deviation –, and calculate the difference between the slope coefficient estimated when **Y** was regressed on **X** (i.e., M1_byx in the syntax) and the slope coefficient estimated when **Y** was regressed on **X** and **Z** (i.e., M2_byx in the syntax). As it can be observed, the difference between M1_byx and M2_byx – or D_byx – is extremely small at the 1st quartile, median, mean, and 3rd quartile, with only the minimum and maximum values demonstrating the existence of some deviation. These deviations, again, are likely the product of simulation anomalies.

```
> n<-10000
> 
> set.seed(1992)
> Example3_DATA = foreach (i=1:n, .packages=c('lm.beta'), .combine=rbind) %dorng%
+   {
+     ### Value Specifications ####
+     N<-sample(150:10000, 1)
+     Cor<-rnorm(N,runif(1,-5,5),runif(1,1,30))
+     Z<-(runif(1,-10,10)*rnorm(N,runif(1,-5,5),runif(1,1,30)))+(runif(1,-10,10)*Cor)
+     X<-(runif(1,-10,10)*rnorm(N,runif(1,-5,5),runif(1,1,30)))+(runif(1,-10,10)*Cor)
+     Y<-(runif(1,-10,10)*X)+(runif(1,-10,10)*rnorm(N,runif(1,-5,5),runif(1,1,30)))
```

```
+
+       Data<-data.frame(Z,X,Y)
+
+       ### Models ####
+       M1<-summary(lm(Y~X, data = Data))
+       M2<-summary(lm(Y~X+Z, data = Data))
+
+       M1_byx<-coef(M1)[2, 1]
+       M2_byx<-coef(M2)[2, 1]
+
+       D_byx<-M1_byx-M2_byx
+
+       # Data Frame ####
+
+       data.frame(i,N,M1_byx,M2_byx,D_byx)
+
+   }
>
> summary(Example3_DATA$D_byx)
          Min.         1st Qu.          Median            Mean         3rd Qu.            Max.
-2.389818583275 -0.001469675403 -0.000000033681 -0.000511453956  0.001380548570  1.467722891503
```

*Discussion*

As demonstrated by the results, adjusting for the variation in a construct correlated with neither or one of the constructs of interest will have limited impact on the estimated slope coefficient. That is, if **Z** does not share variation with both **X** and **Y**, the slope coefficient of the association between **X** and **Y** will not become biased by excluding **Z** in the estimated regression model. Although the simulations reviewed above do not demonstrate a source of statistical bias, it does demonstrate that adjusting for a variable that does not share variation with both **X** and **Y** has limited impact on the estimated effects of **X** on **Y**. As such, for the sake of parsimony, it is best we exclude constructs from our regression models that do not covary with both the independent and dependent variables of interest.

Now let's take a brief intermission and then move onto the second topic of this entry: *misspecified causal structure*.

**Intermission: Theoretical Discussion of Bivariate and Multivariable Regression Models**
Okay, so this is not much of an intermission, but I really wanted to have a theoretical discussion about multivariable regression models at some point in the *Statistical Biases when Examining Causal Associations* section. So here we are…

Let me start off by stating this, when we estimate a bivariate regression model we are intrinsically stating that we believe that variation in **X** (or the identified independent variable) causes variation in **Y** (or the identified dependent variable). Here, I am using cause to mean either a direct or indirect cause. This intrinsic statement exists for a variety of reasons, the primary being that your bivariate regression model is a directed equation from **X** to **Y**. That is, Equation 1 is equal to Figure 4.

[Equation 1]

$$Y = \beta X + \varepsilon$$

[Figure 4]

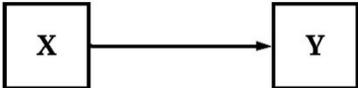

It is easy to consider this an overstatement, but when have we ever estimated a regression equation with an independent variable that was believed to not cause variation in the dependent variable. I for one have never been shocked when the slope coefficient produced by a model suggests a meaningful effect of the independent variable on the dependent variable. I, however, have experienced that feeling when the resulting model coefficients suggest that an association does not exist between the independent variable and the dependent variable. These feelings occur because our prior knowledge and experiences informs the hypotheses we develop about how the universe works and, in turn, guide the specification of our regression models. If you still disagree, consider the research published or grant applications. Often they are relatively straight forward research questions, like incarceration influences employment. Or an empirical evaluation of a theoretical model of how the universe works. Or a grant to help an agency implement evidence based training in hopes that it causes positive change in the clienteles' lives.

If Equation 1 intrinsically expresses our belief that variation in **X** causes variation in **Y** – i.e., a causal statement of how the world works –, what does Equation 2 express? Following the same logic, the inclusion of a covariate inherently expresses the believe that variation in $Z_k$ (i.e., covariates) causes variation in both **X** and **Y**. Specifically, the inclusion of any $Z_k$ expresses that we believe the association between **X** and **Y** can not be properly estimated without adjusting for the variation associated with $Z_k$. Here, Equation 2 is equivalent to Figure 5.

[Equation 2]

$$Y = \beta X + \beta Z_k + \varepsilon$$

[Figure 5]

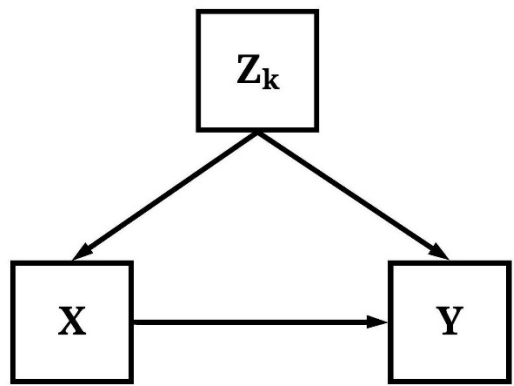

Now, you might be asking, why is it important for us to acknowledge that a regression equation is inherently an expression of causal beliefs? At the forefront, acknowledging the underlying causal expression of a regression equation increases our care when estimating a multivariable model. That is, by acknowledging that Equation 2 represents the belief that $Z_k$ (i.e., covariates) cause variation in both **X** and **Y**, we can be more careful about the covariates that are included in our equations. Only constructs that are believed to cause variation in both **X** and **Y**, or as covered in Entry 10 is a descendant of a construct that causes variation in both **X** and **Y**, should be included in a multivariable regression model. Everything else need not be included as additional covariates because they can: (1) reduce the degrees of freedom, (2) introduce bias or misrepresent the causal association between constructs, (3) have limited influence on estimating the causal effects of **X** on **Y**, and (4) increase the likelihood of overfitting a model. In the end, recognizing our causal claims within our statistics can only serve to benefit the advancement of our knowledge.

*I apologize for getting on my high horse but I do believe this is extremely important. Being careful with statistics is important as we are making causal claims about how the world works within our equations.*

Now, let's talk about misspecifying the causal structure in a regression model!

**Misspecified Causal Structure: Reverse Causal Specification**
Considering our intermission discussion, if a regression equation is inherently a statement of causal processes, the specification of our regression model will either (1) capture the correct causal pathway from one variable to another variable or (2) misspecify the causal pathway from one variable to another variable. The latter is much more likely, hence the discussions dedicated to Colliders, Confounders, Descendants, and Mediators/ Moderators. But there is one more misspecification that we might include in a regression model, the reverse causal specification. As implied, the reverse causal specification refers to the condition where we flip the two constructs in the equation and estimate the effects of the true dependent construct (i.e., the construct where variation is being caused) on the true independent construct (i.e., the construct causing variation). To provide an easy example, we can estimate the effects of college major on career employment or career employment on college major. The first regression model will produce estimates capturing the causal effects, while the second regression model will produce estimates capturing the reverse effects. In this example, we are assuming that the measurement of college major only captures the first attempt at college.

While it is easy to recognize that the second regression model in the example above implies the reverse causal specification, it is extremely difficult in practice to identify reverse causal specifications in cross-sectional data. For example, did marijuana smoking cause an individual to express symptoms of depression or did symptoms of depression cause an individual to smoke marijuana? If both constructs are measured at the same time period, it is difficult to discern and estimate the causal effects of one on the other. Furthermore, we condition the statement with "*cross-sectional data*" because causal effects can not work backwards in time and longitudinal methods permit the estimation of simultaneous regression models to examine reciprocal associations. Given this, the likelihood of estimating a regression model with the reverse causal specification is substantially higher when employing cross-sectional data. Independent of the data structure, estimating a regression model with the reverse causal specification can produce findings supportive of the misspecified causal structure. To demonstrate, let's again conduct a data simulation.

For this data simulation, we are going to simply use a bivariate directed equation specification to simulate the causal effects of **X** on **Y**. **X**, following the simulations above, is a normally distributed construct with a mean of 0 and a standard deviation of 10, where a 1 point increase in **X** causes a 1 point increase in **Y**. The residual variation in **Y** is normally distributed with a mean of 0 and a standard deviation of 10.

```
> ## Example ####
> n<-1000
>
```

```
> set.seed(1992)
>
> X<-1*rnorm(n,0,10)
> Y<-1*X+1*rnorm(n,0,10)
>
> Data<-data.frame(X,Y)
```

After simulating the data, we can regress **Y** on **X** and **X** on **Y** and observe the slope coefficients produced by the OLS models. Evident by the findings, regressing **Y** on **X** suggests that a 1 point increase in **X** causes a 1.002 point increase in **Y**, which is consistent with the specification of the data simulation. Alternatively, regressing **X** on **Y** suggests that a 1 point increase in **Y** causes a .4939 increase in **X**, suggesting the existence of a reciprocal causal pathway. This reciprocal causal pathway, however, does not exist as our simulation specification dictated that the variation in **Y** was the product of the variation in **X**, and not vice versa.

```
> summary(lm(Y~X, data = Data))

Call:
lm(formula = Y ~ X, data = Data)

Residuals:
        Min          1Q      Median          3Q         Max
-33.39823634  -6.39223268  -0.12454071   6.77441730  32.32615833

Coefficients:
              Estimate    Std. Error   t value           Pr(>|t|)
(Intercept) 0.5328217473 0.3123247274   1.70599           0.088322 .
X           1.0016311375 0.0320395916  31.26229 < 0.0000000000000002 ***
---
Signif. codes:  0 '***' 0.001 '**' 0.01 '*' 0.05 '.' 0.1 ' ' 1

Residual standard error: 9.87622125 on 998 degrees of freedom
Multiple R-squared:  0.494768228, Adjusted R-squared:  0.494261984
F-statistic: 977.331038 on 1 and 998 DF,  p-value: < 0.00000000000000002220446

>
> summary(lm(X~Y, data = Data))

Call:
lm(formula = X ~ Y, data = Data)

Residuals:
         Min           1Q       Median           3Q          Max
-22.946438389  -4.195328463  -0.078404385   4.594062381  22.235582859

Coefficients:
               Estimate   Std. Error   t value           Pr(>|t|)
(Intercept) -0.3048824723 0.2194381704  -1.38938           0.16503
Y            0.4939625074 0.0158005841  31.26229 < 0.0000000000000002 ***
---
Signif. codes:  0 '***' 0.001 '**' 0.01 '*' 0.05 '.' 0.1 ' ' 1

Residual standard error: 6.93559772 on 998 degrees of freedom
Multiple R-squared:  0.494768228, Adjusted R-squared:  0.494261984
F-statistic: 977.331038 on 1 and 998 DF,  p-value: < 0.00000000000000002220446

>
```

Although the lines appear to be identical, it is important to note that the x and y-axes switch from Panel A to Panel B. Notably, independent of the exact slope coefficients, the results in Figure 6 further solidify the suggestion that variation in **X** causes variation in **Y** and variation in **Y** causes variation in **X**. This interpretation, however, is incorrect as it does not align with the simulation specification.

[Figure 6]

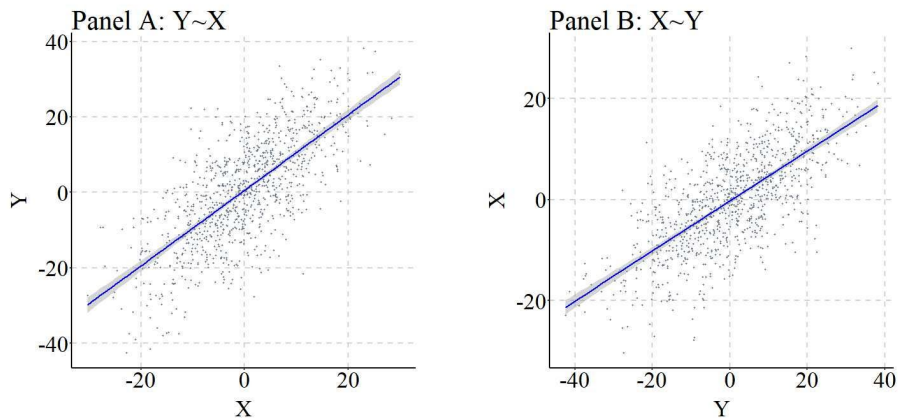

You might be asking, how often can the reverse specification produce results supporting the reciprocal causal pathway? Well it is all conditional upon the true causal effects of **X** on **Y**. Specifically, the likelihood of observing results supporting the reciprocal causal pathway is diminished if the variation in **X** has a weak (*relative to the scale of the constructs*) causal influence on the variation in **Y**, while the likelihood of observing results supporting the reciprocal causal pathway is increased if the variation in **X** has a strong (*relative to the scale of the constructs*) causal influence on the variation in **Y**. Given that the reverse causal pathway is conditional upon the specified causal effects of **X** on **Y**, it is extremely difficult to discern a rate at which this can happen. Coinciding with this difficulty, we will not conduct replications of the simulation because the resulting information could misguide or be misinterpreted as evidence of the rate at which the results from a misspecified regression model supports the reciprocal causal pathway.

**Conclusion**
Overall, I believe this entry represents a perfect end to our first simulated exploration of the statistical biases that could exist when examining causal associations. As reviewed, a construct correlated with neither or either the independent or dependent variable produces limited biases when estimating the association of interest. Moreover, as discussed in the second half of this entry, the estimation of a regression model with the reverse causal specification using cross-sectional data could provide evidence supporting the existence of a reciprocal causal pathway. Although relatively straightforward, it is important to consider this when identifying and estimating the causal pathway from one construct to another construct. Now, let's move on to examining the statistical biases that could be generated by factors related to measurement!

---

[i] This is a theme that will continue until the end of the Sources of Statistical Biases Series! There will always be issues and topics to revisit!

[ii] For a great depiction of the latter situation, simply review Tyler Vigen's seminal work on spurious correlations (http://www.tylervigen.com/spurious-correlations). I think my favorite is the number of people who drowned in a pool and the number of films Nicholas Cage appeared in each year, which has a correlation of (r = .66). The observed correlation is likely the product of measurement error – I can't think of anything that could be a common cause –, but you never know…

[iii] I do like the idea of a computer programming employment opportunity for incarcerated individuals, but a lot of kinks would have to be worked out.

# Statistical Biases and Measurement

Welcome to what I believe will be one of the most important sections of the ***Sources of Statistical Biases Series***: Measurement. Besides the existence of confounders, I strongly believe that the measurement of a construct represents one of the largest sources of statistical bias in all scientific disciplines. This belief stems from the numerous researcher degrees of freedom – decision points – that exist when conceptualizing and operationalizing a construct.

The conceptualization of a construct is discipline specific and, often times, is based on the data collection techniques or the available data. Numerous conceptualizations can exist for a single construct, the differences between which contribute to the observation of distinct statistical estimates when conducting a regression analysis. Meaning that how we define our constructs is important when examining the influence of one construct on another construct.

To provide a criminal justice example, our conceptualization of *recidivism* can drastically impact the outcome of an evaluation. In particular, three important components exist when defining *recidivism*. First, a conceptualization of recidivism requires one to define a start date for what counts as recidivism. Here are just a few examples of what can be considered a start date for a recidivism event:

1. Date arrested for a crime
2. Date convicted of a crime
3. Date sentenced for a crime
4. Date placed on probation
5. Date placed in prison
6. Date place on parole
7. Date probation was successfully completed
8. Date probation was successfully or unsuccessfully completed
9. Date released from prison
10. Date parole was successfully completed
11. Date parole was successfully or unsuccessfully completed

The distinctions between the possible start dates will influence the period in which the recidivism could have occurred and, in turn, increase or decrease the number of individuals that could have recidivated. After defining the start date, we have to define what counts as recidivism. Here are some options used throughout the correctional literature:

1. Rearrest for non-criminal/ non-probation or parole related violation
2. Rearrest for a technical violation or new crime
3. Rearrest for a new crime (misdemeanor or felony)
4. Rearrest for a new crime (felony)
5. Reconviction for a new crime (misdemeanor or felony)
6. Reconviction for a new crime (felony)
7. Reincarceration for a technical violation or new crime
8. Reincarceration for a new crime (misdemeanor or felony)

9. Reincarceration for a new crime (felony)

As demonstrated throughout the literature, rearrest and measures including technical violations tend to increase the rate of recidivism within a sample. This is because individuals previously involved in the criminal justice system tend to have an increased likelihood of being arrested for a non-criminal behavior and/or experience a probation violation resulting from non-criminal activities (disobeying officer orders, missing probation/parole meetings, failing to maintain employment, etc.). Finally, we have to define the timeframe that we are examining:

1. 1-year from start date
2. 2-years from start date
3. 3-years from start date
4. Etc.

The longer the timeframe, the higher likelihood of an individual recidivating, which is extremely important when studying specialized populations. For instance, sex-offenders tend to sexually-recidivate at an extremely low rate for approximately three years when released back into the community. All of these decision points when conceptualizing *recidivism* can result in meaningful variation in the magnitude of the association for the causes and effects of recidivism. Recidivism, however, is not unique as many constructs across scientific disciplines can have multiple conceptualizations. Nevertheless, while understanding the distinctions between conceptualizations is important to evaluating and interpreting the resulting coefficients, the specificity this discussion requires is beyond the scope of this series. In particular, simulation analyses would be useless, as we would have to focus in on key constructs across scientific disciplines and discuss how definitional differences bias statistical estimates. As such, the Measurement section of the series will primarily focus on how we operationalize constructs.

The operationalization of a construct is extremely important when generating inferences about causal associations within a population. The importance begins no sooner than with the level of measurement and the creation of a construct. The level of measurement can be (1) dichotomous, (2) ordinal, or (3) continuous,[i] while a construct can be created various different ways, either being derived from a single item or a representation of multiple items (e.g., aggregation or average). The first five entries of the ***Measurement*** section will focus on these topics:

1. Level of Measurement: Dependent Variable
2. Level of Measurement: Independent Variable
3. Measurement Creation: Dichotomies, Ordinal Measures, Aggregate Events, & Variety Scores
4. Measurement Creation: Aggregate Scale, Average Scale, Standardized Scale, Weighted Scale & CFA
5. Measurement Error

After discussing issues related to the measurement of a construct, our discussion will switch gears and focus on how the distribution of a dependent, lagged endogenous, or endogenous variable can influence the estimates derived from various statistical models.

6. Distributional Assumptions for the Dependent Variable (Linear Regression Models)
7. Distributional Assumptions for the Dependent Variable (Mixed-Effects Linear Models)
8. Distributional Assumptions for the Dependent Variable (Non-Parametric Regression Models)
9. Distributional Assumptions for the Lagged-Endogenous Variables (Structural Equation Modeling)
10. Distributional Assumptions for the Endogenous Variable (Structural Equation Modeling)

Finally, we will end the Measurement section by discussing the clustering of data across space and time. This discussion will focus on how clustering influences statistical estimates, as well as the interaction between clustering and measurement when estimating regression models.

11. Longitudinal Clustering
12. Spatial Clustering

Overall, the discussions and data simulations should provide insight into how measurement can bias the statistical estimates produced by a regression analysis. Now, let's get into our discussions by focusing on biases related to the level of measurement for the dependent and independent variables.

---

[i] Two brief points, when estimating a statistical model the level of measurement can not traditionally be nominal – except for when estimating multinomial regression model – and often times requires the dichotomization of each option. Interval and ratio level of measurement both describe continuous constructs, but traditionally have distinct distributional assumptions.

## Entry 13: Level of Measurement of The Dependent Variable

*Just a warning, this is a long entry! Overall message, operationalize variables on a continuous scale when possible and consider the ramifications of transforming continuous distributions.*

**1. Introduction**

I can see no place more fitting to start a discussion of Statistical Biases and Measurement than focusing on the measurement of our dependent variables (i.e., endogenous variables). In this entry we will discuss altering the level of measurement and data transformations. However, before we get started I want to take a brief moment to justify the structure of these data simulations, as well as identify all of the estimates we will be discussing.

For this entry, we want our simulations to isolate the bias in statistical estimates generated by altering the operationalization of our dependent variable (Y) across levels of measurement. In particular, the continuous variation in Y will be caused by variation in X, with the focus being on observing if the estimated effects of X on Y differ across various operationalizations of Y. That is, does the magnitude of the effects of X on Y differ if Y is coded as a continuous construct, ordered construct, or dichotomous construct, as well as if the scores on Y are altered using continuous data transformations common within the literature.

This is easier said than done. To begin every simulation, we will estimate a *Spearman* – not Pearson – correlation. Spearman's rank correlation is appropriate for both continuous and ordinal operationalizations of constructs, and is only slightly biased when estimating the correlation using a dichotomous operationalization of a construct.[i] Given the utility of Spearman's rank correlation, we can begin to develop a standardized process of the differences between operationalizations of Y. Nevertheless, if we do not account for the distributional assumptions of the dependent variable when estimating the effects of X on Y using a linear regression model, we will not be able to tell if the bias is due to the operationalization of Y or violating a key assumption of the specified regression model. As such, we will estimate three types of linear regression models to ensure we do not violate the distributional assumptions of the specified models. These are:

1. Gaussian generalized linear regression model for the continuous operationalization of Y
2. Ordered logistic regression model for the ordered operationalization of Y
3. Binary logistic regression model for the dichotomous operationalization of Y

It is expected that the following estimates will vary simply due to distinctions between the three models.

1. Slope coefficient (*b*)
2. Standard error (SE)
3. Confidence intervals (CI)
4. Exponential value of the slope coefficient (exp[*b*])
5. Standardized slope coefficient (*β*)

Nevertheless, if altering the operationalization of Y does not bias the estimated association between X and Y, the test statistics (*t*-statistic, *z*-statistic, and $x^2$-statistic) should remain relatively stable across the statistical models. As a reminder – because I needed one also – the *t*-value, *z*-value, $x^2$-value are interpreted as the departure of the estimated slope coefficient from the hypothesized slope coefficient (i.e., $b = 0$) conditional on the standard error of X. Briefly, *t*-statistic, *z*-statistic are calculated by dividing the estimated slope coefficient by the estimated standard error of X, while the $x^2$-statistic is calculated by squaring the difference between the observed value and the expected value and dividing by the expected value. The characteristics of a *t*-statistic and *z*-statistic permit comparisons across models with different assumptions as the ratio of the slope coefficient to the standard error for X should remain the same if the magnitude of the association between X and Y remains the same. Moreover, the $x^2$-statistic compares estimated slope coefficient to the hypothesized slope coefficient (i.e., $b = 0$) creating an identical comparison across all analyses.

In theory, if the amount of variation in Y explained by the variation in X is equal across the operationalizations of Y, the *t*-value, *z*-value, and $x^2$-value should be relatively stable because

    1) The standard error of X should not change
    2) The difference between the magnitude of the slope coefficient between X and Y and a slope coefficient of 0 should not change

However, if altering the operationalization of Y reduces the amount of variation in Y explained by the variation in X, the *t*-statistic, *z*-statistic, and $x^2$-statistic should become attenuated because the 1) ratio of the estimated slope coefficient to the standard error will become smaller and 2) the estimated slope coefficient will become closer to the hypothesized slope coefficient (i.e., $b = 0$). Importantly, the estimates for the *t*-test and *z*-test can be directly compared as the simulated sample size (N = 10,000) and, in turn, the degrees of freedom will make the test distributions almost identical.

## 2. Altering the Measurement of Y[ii]
### 2.1. Definitions
To provide my informal definition, a dependent variable is a construct where the variation across the population is caused by the variation within another construct and random error. If conducting a statistical analysis, the variation that exists within a dependent variable should be measured – or operationalized – one of three general ways: as a continuous construct, as an ordered construct, or as a dichotomous construct. Operationalizing a dependent variable as a *continuous construct* means that the variation within the construct will cross over more than a designated number of expected values (definitions are subject to vary by scholar). Generally, I consider constructs to approach continuous variation when the scale contains more than 10 expected values within the population (e.g., income). Unique to my work, I label constructs that contain between 10 and 25 expected values within the population (e.g., educational attainment) as semi-continuous constructs. Generally, I do not treat them any differently than a continuous construct, but employ the term to provide clarity to the reader about how I would designate the scale. Importantly, the term continuous, however, does not provide any information about the distribution of a construct.

Operationalizing a dependent variable as an *ordered construct* means that the variation within the construct will cross over a limited number of expected values. For example, your class grade might only be categorized within one of five expected values (F = 0; D = 1; C = 2; B = 3; A = 4) if the professor is a jerk and does not provide plus or minus grades. The employment of a Likert scale is the most common process for measuring a construct with ordered variation in the social sciences.

Operationalizing a dependent variable as a *dichotomous construct* means that the variation within the construct will cross over only two expected values. Often times the expected values are limited to 0 and 1, although this need not be the expected values used in the construction of the measure. While defining the differences between the terms is important for this entry, it is also important to talk about the information captured at each level of measurement.

***Brief Note:*** *I define a nominal construct that requires the introduction of multiple dichotomies into a statistical model as a dichotomous operationalization. For example, Race is inherently a nominal construct and requires the introduction of multiple dichotomous variables to capture the variation associated with more than two racial groups. As such, to introduce measures that capture variation in race into a statistical model, we might include two dichotomies operationalizing white and black participants to permit comparisons to other racial groups (e.g., Asian). While this is my personal definition, I understand that you might disagree.*

## *2.2. Information Captured and Measurement*
"The devil is in the details" is the perfect idiom for describing how important the level of measurement is when capturing information about the variation in a construct. Explicitly, the more expected values contained within the operationalization of a construct, the more likely we are to capture the variation of the construct across the population. Which, in turn, translates to a more finite understanding of the variation within the population, as well as an increased ability to make finite predictions about the causes and effects of variation in a construct.

Figure 1 provides an example of the reduction in information captured in a construct as the operationalization moves from continuous to dichotomous. Briefly, Y was created as a uniform distribution ranging between 0 and 1 with up to 6 digits. This continuous operationalization of Y means that the number of expected values in the distribution is $(5^6)^{10}$ or 8.673617e+41 value units.[iii] Panel A below was created with .01 scaled bins, meaning that we are not visually evaluating the distribution beyond 2 decimal places. That, however, does not matter as the general form of the continuous distribution is evident below. As demonstrated, it appears that only a limited number of cases fall within the 2 decimal place bins, with some bins possessing less than 10 cases and some bins possessing more than 30 cases. Moreover, we can see a substantial amount of variation in the frequency of scores on Y across the expected values of Y.

Now, let's focus on Panel B, where scores on Y were rounded to the nearest whole number. If we treat the information about Y contained within Panel B separate from the information in Panel A, we still observe the frequency of scores on Y across the expected values of Y. Nevertheless, when compared to Panel A, the variation in the frequency of scores on Y across the expected values of Y is substantially limited,. In particular, approximately 2000 cases score 1, 2, 3, or 4 on Y (respectively) and approximately 1000 cases score 0 or 5 on Y (respectively). The reduction in the variation in the frequency of scores on Y across the expected values of Y – a loss of variation – corresponds to the reduction in information that occurs when moving from a continuous operationalization to an ordered operationalization. The reduction in the variation in the frequency of scores on Y across the expected values of Y is even further reduced in Panel C – a dichotomous operationalization of Y –, were an equal number of cases scored 0 and 1.

[Figure 1]

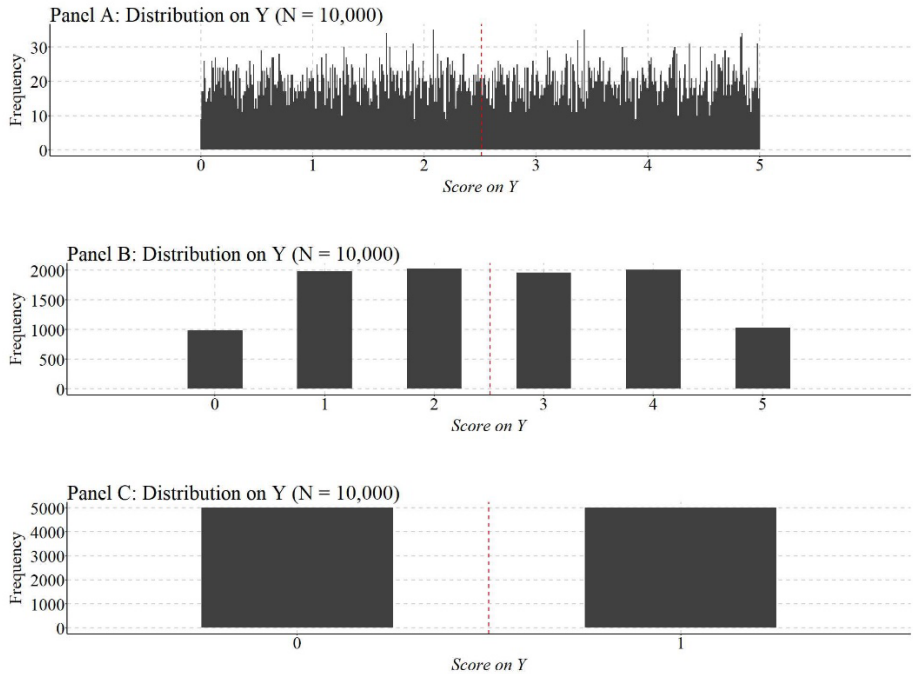

Now, that we understand this loss of information let us conduct various simulations to explore how altering the level of measurement and/or conducting data transformations on the dependent variable can influence the estimated effects of **X** on **Y** in our statistical models.

***Brief Note***: *If a construct is initially operationalized at a lower level of measurement (dichotomous or ordered), you can not regain the lost variation. That is, you can not transform scores on a construct from a dichotomy to ordered to continuous or from ordered to continuous. You can only transform scores on constructs from higher levels of measurement to lower levels of measurement. Considering this, it is best to measure constructs as continuous when collecting data and then create ordered or dichotomous operationalizations if theoretically and/or empirically relevant.*

### 2.3. Continuous Operationalization of Y
Whew, I never thought we would get here! So, let's make this simulation as simple as possible. We are going to conduct a simple bivariate simulation, where a 1 point change in X (a normally distributed variable with a mean of 0 and standard deviation of 10) causes a 1 point change in Y (a normally distributed variable with a mean of 0 and standard deviation of 30). We place both variables in a dataframe labeled as DF. Oh, by the way, we are working with 10,000 cases.

```
# Continuous Construct (Linear Association) ####
n<-10000

set.seed(1992)

X<-1*rnorm(n,0,10)
Y<-1*X+1*rnorm(n,0,30)
```

```
DF<-data.frame(X,Y)
```

Briefly, let's use stat.desc(DF) to ensure both constructs are continuous with the specified means and standard deviations.

```
> stat.desc(DF)
                           X                    Y
nbr.val       10000.00000000000000 10000.0000000000000
nbr.null          0.00000000000000     0.0000000000000
nbr.na            0.00000000000000     0.0000000000000
min             -39.07330748828923  -145.9427130332870
max              39.36370511012955   122.8730438722921
range            78.43701259841879   268.8157569055791
sum            1341.34631652977782  -724.4569705432502
median            0.00749561623815     0.0266760098151
mean              0.13413463165298    -0.0724456970543
SE.mean           0.10003327704085     0.3180169884958
CI.mean.0.95      0.19608535605966     0.6233773027025
var             100.06656515531016  1011.3480497191558
std.dev          10.00332770408478    31.8016988495765
coef.var         74.57677097115814  -438.9729154752876
```

As estimated below, the Spearman correlation coefficient between X and Y is equal to .31.

```
> corr.test(DF$X,DF$Y, method = "spearman")
Call:corr.test(x = DF$X, y = DF$Y, method = "spearman")
Correlation matrix
[1] 0.31
Sample Size
[1] 10000
These are the unadjusted probability values.
  The probability values  adjusted for multiple tests are in the p.adj object.
[1] 0

 To see confidence intervals of the correlations, print with the short=FALSE option
```

Now, let's estimate our regression model and pull the key statistics. First, focusing on the specification of the regression, it can be observed that we estimated a generalized linear model (glm) while specifying that the distribution of the dependent variable is gaussian. Looking at the model results, the slope coefficient ($b$) for the association between X and Y was 1.054 (SE = .030), suggesting a 1 point change in X caused a 1.054 change in Y.

Briefly, the linearHypothesis function directly tests the likelihood of the slope coefficient being different from a specified value – in this case 0 – using the $x^2$-statistics. This is extremely useful, as the $x^2$-statistic can be used to test the similarity of coefficients across like models (e.g., a single type of model) or provide a standardized comparison of the estimated slope coefficient to a slope coefficient of 0.

The $t$- statistic of the estimated effects of X on Y was 35.143, while the $x^2$- statistic of the estimated effects of X on Y was 1235.034. In this case, both tests provide evidence that the slope coefficient of the association between X and Y was statistically different from zero.

```
> M1<-glm(Y~X, data = DF, family = gaussian(link = "identity"))
>
> # Model Results
> summary(M1)

Call:
glm(formula = Y ~ X, family = gaussian(link = "identity"), data = DF)

Deviance Residuals:
```

```
              Min                 1Q              Median                  3Q                 Max
 -129.75743633    -20.01790880           0.07065325         20.26479953         115.13226023

Coefficients:
               Estimate   Std. Error   t value           Pr(>|t|)
(Intercept) -0.213841897  0.300067599  -0.71265              0.47608
X            1.054136418  0.029995581  35.14306 < 0.0000000000000002 ***
---
Signif. codes:  0 '***' 0.001 '**' 0.01 '*' 0.05 '.' 0.1 ' ' 1

(Dispersion parameter for gaussian family taken to be 900.243756876)

    Null deviance: 10112469.149  on 9999  degrees of freedom
Residual deviance:  9000637.081  on 9998  degrees of freedom
AIC: 96409.42614

Number of Fisher Scoring iterations: 2

>
> # Chi-Square Test
> linearHypothesis(M1,c("X = 0"))
Linear hypothesis test

Hypothesis:
X = 0

Model 1: restricted model
Model 2: Y ~ X

  Res.Df Df      Chisq            Pr(>Chisq)
1   9999
2   9998  1 1235.03447 < 0.0000000000000000222 ***
---
Signif. codes:  0 '***' 0.001 '**' 0.01 '*' 0.05 '.' 0.1 ' ' 1
>
```

Now that we have the estimates corresponding to the effects of X on the continuous operationalization of Y, let's estimate the effects of X on dichotomous and ordered operationalizations of Y.

### 2.4. Dichotomous Y: Version 1
For the first dichotomous operationalization of Y, we can split the distribution at the median providing a value of "1" to cases that scored above the median of Y and a value of "0" to cases that scored below or equal to the median of Y. Given that we split the distribution of Y at the median, an equal number of cases received a "0" or "1."

```
> ## Dichotomous Recode 1 ####
> summary(DF$Y)
         Min.         1st Qu.           Median              Mean         3rd Qu.                Max.
-145.9427130333  -21.3951226685       0.0266760098    -0.0724456971   21.3101884312   122.8730438723
> DF$Y_DI1<-NA
> DF$Y_DI1[DF$Y<=median(DF$Y)]<-0
> DF$Y_DI1[DF$Y>median(DF$Y)]<-1
> summary(DF$Y_DI1)
   Min. 1st Qu.  Median    Mean 3rd Qu.    Max.
    0.0     0.0     0.5     0.5     1.0     1.0
> table(DF$Y_DI1)

   0    1
5000 5000
>
```

Now, we can estimate the Spearman correlation coefficient between X and Y_DI1 (the dichotomous operationalization of Y). The Spearman correlation coefficient is equal to .26, which is approximately .05 smaller than the Spearman correlation coefficient between X and Y.

```
> corr.test(DF$X,DF$Y_DI1, method = "spearman")
Call:corr.test(x = DF$X, y = DF$Y_DI1, method = "spearman")
Correlation matrix
[1] 0.26
Sample Size
[1] 10000
These are the unadjusted probability values.
  The probability values  adjusted for multiple tests are in the p.adj object.
[1] 0

 To see confidence intervals of the correlations, print with the short=FALSE option
```

Now, let's estimate a binary logistic regression model and pull the key statistics. For the binary operationalization of Y, we estimated a generalized linear model (glm) while specifying that the dependent variable has a binomial distribution. Looking at the model results, the slope coefficient (*b*) for the association between X and Y_DI1 was 0.057 (SE = .002), suggesting a 1 point change in X caused a .057 change in Y_DI1. These estimates, however, can not be directly compared to the results of Y regressed on X because the slope coefficient will vary due to the estimator of the model.

That said, we can look at the *z*-statistic and $x^2$-statistic to understand if the magnitude of the effects of X on Y_DI1 varies between the two models. The *z*-statistic of the estimated effects of X on Y_DI1 was 25.730, while the $x^2$-statistic was 662.046. While the results of these tests still provide evidence that the slope coefficient of the association between X and Y_DI1 was statistically different from zero, the magnitude of the slope coefficient of X on Y_DI1 being different than zero is attenuated when compared to the magnitude of the slope coefficient of X on Y being different than zero. This suggests that recoding Y into Y_DI1 resulted in a loss of variation.

```
> ### Appropriate Model
> M2<-glm(Y_DI1~X,data = DF, family = binomial(link="logit"))
>
> # Model Results
> summary(M2)

Call:
glm(formula = Y_DI1 ~ X, family = binomial(link = "logit"), data = DF)

Deviance Residuals:
         Min           1Q        Median            3Q           Max
-1.9309245663 -1.1113985191 -0.0001803579    1.1130285916    1.9772226014

Coefficients:
                 Estimate    Std. Error   z value              Pr(>|z|)
(Intercept) -0.00747164118  0.02075348924 -0.36002               0.71883
X            0.05728121156  0.00222621940 25.73026 < 0.0000000000000002 ***
---
Signif. codes:  0 '***' 0.001 '**' 0.01 '*' 0.05 '.' 0.1 ' ' 1

(Dispersion parameter for binomial family taken to be 1)

    Null deviance: 13862.94361  on 9999  degrees of freedom
Residual deviance: 13127.49903  on 9998  degrees of freedom
AIC: 13131.49903

Number of Fisher Scoring iterations: 4

>
> # Chi-Square Test
> linearHypothesis(M2,c("X = 0"))
Linear hypothesis test

Hypothesis:
X = 0

Model 1: restricted model
Model 2: Y_DI1 ~ X

  Res.Df Df    Chisq         Pr(>Chisq)
1   9999
```

```
2    9998 1 662.04646 < 0.000000000000000222 ***
---
Signif. codes:  0 '***' 0.001 '**' 0.01 '*' 0.05 '.' 0.1 ' ' 1
```

## 2.5. Dichotomous Y: Version 2

For the second dichotomous operationalization of Y, we split the distribution at the 1st quartile of the distribution on Y. Specifically, we provided a value of "1" to cases that scored above the 1st quartile on Y and a value of "0" to cases that scored below or equal to the 1st quartile on Y. Using this recoding process, three-quarters of the cases received a value of "1" while 1-quarter of cases received a value of "0."

```
> ## Dichotomous Recode 2 ####
> summary(DF$Y)
          Min.         1st Qu.          Median            Mean         3rd Qu.           Max.
-145.9427130333  -21.3951226685    0.0266760098   -0.0724456971   21.3101884312  122.8730438723
> DF$Y_DI2<-NA
> DF$Y_DI2[DF$Y<=quantile(DF$Y,.25)]<-0
> DF$Y_DI2[DF$Y>quantile(DF$Y,.25)]<-1
> summary(DF$Y_DI2)
   Min. 1st Qu.  Median    Mean 3rd Qu.    Max.
   0.00    0.75    1.00    0.75    1.00    1.00
> table(DF$Y_DI2)

   0    1
2500 7500
>
```

The Spearman correlation coefficient between X and Y_DI2 was equal to .23, which is approximately .08 smaller than the Spearman correlation coefficient between X and Y.

```
> corr.test(DF$X,DF$Y_DI2, method = "spearman")
Call:corr.test(x = DF$X, y = DF$Y_DI2, method = "spearman")
Correlation matrix
[1] 0.23
Sample Size
[1] 10000
These are the unadjusted probability values.
  The probability values  adjusted for multiple tests are in the p.adj object.
[1] 0

 To see confidence intervals of the correlations, print with the short=FALSE option
>
```

Turning to our binary logistic regression model, the slope coefficient ($b$) for the association between X and Y_DI2 was 0.058 (SE = .003), suggesting a 1 point change in X caused a .058 change in Y_DI2. The $z$-statistic of the estimated effects of X on Y_DI2 was 22.925, while the $x^2$-statistic was 525.560. Similar to the first recode, these results suggest that recoding Y into Y_DI2 resulted in a loss of variation.

```
> ### Appropriate Model
> M3<-glm(Y_DI2~X,data = DF, family = binomial(link="logit"))
>
> # Model Results
> summary(M3)

Call:
glm(formula = Y_DI2 ~ X, family = binomial(link = "logit"), data = DF)

Deviance Residuals:
          Min              1Q          Median              3Q             Max
-2.4283659845  -0.0026699768    0.6343058479    0.7897773859    1.5895968726

Coefficients:
                  Estimate      Std. Error  z value                Pr(>|z|)
(Intercept) 1.17040134187  0.02459766919  47.5818 < 0.000000000000000222 ***
X           0.05798629529  0.00252938007  22.9251 < 0.000000000000000222 ***
---
```

```
Signif. codes:  0 '***' 0.001 '**' 0.01 '*' 0.05 '.' 0.1 ' ' 1

(Dispersion parameter for binomial family taken to be 1)

    Null deviance: 11246.70289  on 9999  degrees of freedom
Residual deviance: 10671.48756  on 9998  degrees of freedom
AIC: 10675.48756

Number of Fisher Scoring iterations: 4

> # Chi-Square Test
> linearHypothesis(M3,c("X = 0"))
Linear hypothesis test

Hypothesis:
X = 0

Model 1: restricted model
Model 2: Y_DI2 ~ X

  Res.Df Df    Chisq            Pr(>Chisq)
1   9999
2   9998  1 525.56029 < 0.0000000000000000222 ***
---
Signif. codes:  0 '***' 0.001 '**' 0.01 '*' 0.05 '.' 0.1 ' ' 1
>
```

## 2.6. Dichotomous Y: Version 3

For the third dichotomous operationalization of Y, we split the distribution at the score of 90. Specifically, we provided a value of "1" to cases that scored above 90 on Y and a value of "0" to cases that scored below or equal to 90 on Y. Using this recoding process, 24 cases received a value of "1" while 9,976 cases received a value of "0."

```
> ## Dichotomous Recode 3 ####
> summary(DF$Y)
          Min.         1st Qu.          Median            Mean         3rd Qu.            Max.
-145.9427130333  -21.3951226685    0.0266760098   -0.0724456971   21.3101884312  122.8730438723
> DF$Y_DI3<-NA
> DF$Y_DI3[DF$Y<=90]<-0
> DF$Y_DI3[DF$Y>90]<-1
> summary(DF$Y_DI3)
   Min. 1st Qu.  Median    Mean 3rd Qu.    Max.
 0.0000  0.0000  0.0000  0.0024  0.0000  1.0000
> table(DF$Y_DI3)

   0    1
9976   24
>
```

We can now estimate the Spearman correlation coefficient between X and Y_DI3. The Spearman correlation coefficient is equal to .03, which is approximately .28 smaller than the Spearman correlation coefficient between X and Y.

```
> corr.test(DF$X,DF$Y_DI3, method = "spearman")
Call:corr.test(x = DF$X, y = DF$Y_DI3, method = "spearman")
Correlation matrix
[1] 0.03
Sample Size
[1] 10000
These are the unadjusted probability values.
  The probability values  adjusted for multiple tests are in the p.adj object.
[1] 0

 To see confidence intervals of the correlations, print with the short=FALSE option
>
```

Turning to our binary logistic regression model, the slope coefficient ($b$) for the association between X and Y_DI3 was 0.077 (SE = .021), suggesting a 1 point change in X caused a .077 change in Y_DI3. The *z*-statistic of the estimated effects of X on Y_DI3 was 3.765, while the $x^2$-statistic was 14.174. These findings, again, suggest that the magnitude of the slope coefficient of X on Y_DI3 being different than zero is substantially attenuated when compared to the magnitude of the slope coefficient of X on Y being different than zero. Moreover, recoding Y into Y_DI3 resulted in a meaningful loss of variation.

```
> ### Appropriate Model
> M4<-glm(Y_DI3~X,data = DF, family = binomial(link="logit"))
>
> # Model Results
> summary(M4)

Call:
glm(formula = Y_DI3 ~ X, family = binomial(link = "logit"), data = DF)

Deviance Residuals:
      Min          1Q      Median          3Q         Max
-0.271194656 -0.077576747 -0.059253833 -0.045740183   3.850366527

Coefficients:
              Estimate    Std. Error   z value             Pr(>|z|)
(Intercept) -6.3398628579  0.2605147236 -24.33591 < 0.000000000000000222 ***
X            0.0776177392  0.0206159534   3.76494            0.00016659 ***
---
Signif. codes:  0 '***' 0.001 '**' 0.01 '*' 0.05 '.' 0.1 ' ' 1

(Dispersion parameter for binomial family taken to be 1)

    Null deviance: 337.4921079  on 9999  degrees of freedom
Residual deviance: 323.1462601  on 9998  degrees of freedom
AIC: 327.1462601

Number of Fisher Scoring iterations: 9

> # Chi-Square Test
> linearHypothesis(M4,c("X = 0"))
Linear hypothesis test

Hypothesis:
X = 0

Model 1: restricted model
Model 2: Y_DI3 ~ X

  Res.Df Df    Chisq Pr(>Chisq)
1   9999
2   9998  1 14.17474 0.00016659 ***
---
Signif. codes:  0 '***' 0.001 '**' 0.01 '*' 0.05 '.' 0.1 ' ' 1
>
```

## 2.7. Ordered Y: Version 1

For the first ordered operationalization of Y, we split the distribution into four quartiles providing a value of "1" to cases that scored between the minimum value and the 25[th] percentile, a value of "2" to cases that score between the 25[th] percentile and the median, a value of "3" to cases that scored between the median and the 75[th] percentile, and a value of "4" to cases that scored between the 75[th] percentile and the maximum value. Using this coding scheme, an equal number of cases received a "1", "2", "3", or "4."

```
> ## Ordered Recode 1 ####
> summary(DF$Y)
         Min.       1st Qu.        Median          Mean       3rd Qu.          Max.
-145.9427130333 -21.3951226685   0.0266760098  -0.0724456971 21.3101884312 122.8730438723
> DF$Y_OR1<-NA
> DF$Y_OR1[DF$Y>=quantile(DF$Y,0) & DF$Y < quantile(DF$Y,.25)]<-1
> DF$Y_OR1[DF$Y>=quantile(DF$Y,.25) & DF$Y < quantile(DF$Y,.50)]<-2
> DF$Y_OR1[DF$Y>=quantile(DF$Y,.50) & DF$Y < quantile(DF$Y,.75)]<-3
> DF$Y_OR1[DF$Y>=quantile(DF$Y,.75) & DF$Y <= quantile(DF$Y,1)]<-4
```

```
> summary(DF$Y_OR1)
   Min. 1st Qu.  Median    Mean 3rd Qu.    Max.
   1.00    1.75    2.50    2.50    3.25    4.00
> table(DF$Y_OR1)

   1    2    3    4
2500 2500 2500 2500
```

The Spearman correlation coefficient between X and Y_OR1 was equal to .30, which is approximately .01 smaller than the Spearman correlation coefficient between X and Y.

```
> corr.test(DF$X, DF$Y_OR1, method = "spearman")
Call:corr.test(x = DF$X, y = DF$Y_OR1, method = "spearman")
Correlation matrix
[1] 0.3
Sample Size
[1] 10000
These are the unadjusted probability values.
  The probability values  adjusted for multiple tests are in the p.adj object.
[1] 0

 To see confidence intervals of the correlations, print with the short=FALSE option
```

Now, let's estimate an ordered logistic regression model (polr from the *psych* package is the function to estimate this model). Looking at the model results, the slope coefficient (*b*) for the association between X and Y_OR1 was 0.058 (SE = .002), suggesting a 1 point change in X caused a .058 change in Y_OR1.

We can look at the *t*-statistic and $x^2$-statistic to understand if the magnitude of the effects of X on Y_OR1 varies between the two models. The *t*-statistic of the estimated effects of X on Y_OR1 was 30.443, while the $x^2$-statistic was 926.762. These values, again, suggest that the slope coefficient of the association between X and Y_OR1 was statistically different from zero. Nevertheless, it seems that the magnitude of the slope coefficient of X on Y_OR1 being different than zero is slightly attenuated when compared to the magnitude of the slope coefficient of X on Y, but stronger when compared to X on Y_DI1, Y_DI2, or Y_DI3. This suggests that recoding Y into Y_OR1 resulted in some loss of variation, but not to the degree of the dichotomous operationalizations of Y.

```
> ### Appropriate Model
> M5<-polr(as.factor(Y_OR1)~X,data = DF, Hess=TRUE) # Ordered Logistic Regression
> # Model Results
> summary(M5) # Ordered Logistic Regression
Call:
polr(formula = as.factor(Y_OR1) ~ X, data = DF, Hess = TRUE)

Coefficients:
       Value    Std. Error    t value
X 0.0578087369 0.00189893143 30.4427721

Intercepts:
    Value        Std. Error    t value
1|2  -1.169513429   0.023857898  -49.019969444
2|3   0.006946834   0.020594298    0.337318290
3|4   1.184907097   0.023948396   49.477513402

Residual Deviance: 26753.3373125
AIC: 26761.3373125
> # Chi-Square Test
> linearHypothesis(M5,c("X = 0"))
Linear hypothesis test

Hypothesis:
X = 0

Model 1: restricted model
Model 2: as.factor(Y_OR1) ~ X
```

```
  Res.Df Df    Chisq           Pr(>Chisq)
1   9997
2   9996  1 926.76237 < 0.0000000000000000222 ***
---
Signif. codes:  0 '***' 0.001 '**' 0.01 '*' 0.05 '.' 0.1 ' ' 1
>
```

## 2.8. Ordered Y: Version 2

For the second ordered operationalization of Y, we split the distribution into four quartiles providing a value of "1" to cases that scored between the minimum value and the 50th percentile, a value of "2" to cases that score between the 50th percentile and the 60th percentile, a value of "3" to cases that scored between the 60th percentile and the 90th percentile, and a value of "4" to cases that scored between the 90th percentile and the maximum value. Using this coding scheme, 5,000 cases received a 1, 3,000 cases received a 3, 1,000 cases received a 2, and 1,000 cases received a 4.

```
> ## Ordered Recode 2 ####
> summary(DF$Y)
           Min.        1st Qu.         Median           Mean        3rd Qu.           Max.
-145.9427130333  -21.3951226685     0.0266760098    -0.0724456971  21.3101884312  122.8730438723
> DF$Y_OR2<-NA
> DF$Y_OR2[DF$Y>=quantile(DF$Y,0) & DF$Y < quantile(DF$Y,.50)]<-1
> DF$Y_OR2[DF$Y>=quantile(DF$Y,.50) & DF$Y < quantile(DF$Y,.60)]<-2
> DF$Y_OR2[DF$Y>=quantile(DF$Y,.60) & DF$Y < quantile(DF$Y,.90)]<-3
> DF$Y_OR2[DF$Y>=quantile(DF$Y,.90) & DF$Y <= quantile(DF$Y,1)]<-4
> summary(DF$Y_OR2)
   Min. 1st Qu.  Median    Mean 3rd Qu.    Max.
    1.0     1.0     1.5     2.0     3.0     4.0
> table(DF$Y_OR2)

   1    2    3    4
5000 1000 3000 1000
>
```

The Spearman correlation coefficient between X and Y_OR2 is equal to .28, which is approximately .03 smaller than the Spearman correlation coefficient between X and Y.

```
> corr.test(DF$X, DF$Y_OR2, method = "spearman")
Call:corr.test(x = DF$X, y = DF$Y_OR2, method = "spearman")
Correlation matrix
[1] 0.28
Sample Size
[1] 10000
These are the unadjusted probability values.
  The probability values  adjusted for multiple tests are in the p.adj object.
[1] 0

 To see confidence intervals of the correlations, print with the short=FALSE option
>
```

Looking at the model results, the slope coefficient (*b*) for the association between X and Y_OR2 was 0.058 (SE = .002), suggesting a 1 point change in X caused a .058 change in Y_OR2. The *t*-statistic of the estimated effects of X on Y_OR2 was 28.697, while the $x^2$-statistic was 823.511. These values, again, suggest that recoding Y into Y_OR2 resulted in some loss of variation, but not to the degree of the dichotomous operationalizations of Y.

```
> ### Appropriate Model
> M6<-polr(as.factor(Y_OR2)~X,data = DF, Hess=TRUE) # Ordered Logistic Regression
> # Model Results
> summary(M6) # Ordered Logistic Regression
Call:
polr(formula = as.factor(Y_OR2) ~ X, data = DF, Hess = TRUE)

Coefficients:
    Value  Std. Error   t value
```

```
X 0.0584219241 0.0020358283 28.6968818

Intercepts:
    Value          Std. Error   t value
1|2  0.005488792  0.020702430  0.265127913
2|3  0.441008060  0.021172836 20.828955441
3|4  2.334060459  0.034512266 67.629882084

Residual Deviance: 22484.9999904
AIC: 22492.9999904
> # Chi-Square Test
> linearHypothesis(M6,c("X = 0"))
Linear hypothesis test

Hypothesis:
X = 0

Model 1: restricted model
Model 2: as.factor(Y_OR2) ~ X

  Res.Df Df    Chisq             Pr(>Chisq)
1   9997
2   9996  1 823.51103 < 0.0000000000000000222 ***
---
Signif. codes:  0 '***' 0.001 '**' 0.01 '*' 0.05 '.' 0.1 ' ' 1
>
```

## 2.9. Ordered Y: Version 3

For the third ordered operationalization of Y, we split the distribution into four quartiles providing a value of "1" to cases that scored between the minimum value and the 10th percentile, a value of "2" to cases that score between the 10th percentile and the 20th percentile, a value of "3" to cases that scored between the 20th percentile and the 30th percentile, and a value of "4" to cases that scored between the 30th percentile and the maximum value. Using this coding scheme, 7,000 cases received a 4, 1,000 cases received a 1, 1,000 cases received a 2, and 1,000 cases received a 3.

```
> ## Ordered Recode 3 ####
> summary(DF$Y)
          Min.         1st Qu.          Median           Mean         3rd Qu.           Max.
-145.9427130333  -21.3951226685     0.0266760098  -0.0724456971   21.3101884312  122.8730438723
> DF$Y_OR3<-NA
> DF$Y_OR3[DF$Y>=quantile(DF$Y,0) & DF$Y < quantile(DF$Y,.10)]<-1
> DF$Y_OR3[DF$Y>=quantile(DF$Y,.10) & DF$Y < quantile(DF$Y,.20)]<-2
> DF$Y_OR3[DF$Y>=quantile(DF$Y,.20) & DF$Y < quantile(DF$Y,.30)]<-3
> DF$Y_OR3[DF$Y>=quantile(DF$Y,.30) & DF$Y <= quantile(DF$Y,1)]<-4
> summary(DF$Y_OR3)
   Min. 1st Qu.  Median    Mean 3rd Qu.    Max.
    1.0     3.0     4.0     3.4     4.0     4.0
> table(DF$Y_OR3)

   1    2    3    4
1000 1000 1000 7000
>
```

Now, we can estimate the Spearman correlation coefficient between X and Y_OR3. The Spearman correlation coefficient is equal to .25, which is approximately .06 smaller than the Spearman correlation coefficient between X and Y.

```
> corr.test(DF$X, DF$Y_OR3, method = "spearman")
Call:corr.test(x = DF$X, y = DF$Y_OR3, method = "spearman")
Correlation matrix
[1] 0.25
Sample Size
[1] 10000
These are the unadjusted probability values.
  The probability values  adjusted for multiple tests are in the p.adj object.
[1] 0

 To see confidence intervals of the correlations, print with the short=FALSE option
>
```

Looking at the model results, the slope coefficient (*b*) for the association between X and Y_OR3 was 0.059 (SE = .002), suggesting a 1 point change in X caused a .059 change in Y_OR3. The *t*-statistic of the estimated effects of X on Y_OR3 was 25.455, while the $x^2$-statistic was 647.965. These values, again, suggest that recoding Y into Y_OR3 resulted in some loss of variation, but not to the degree of the dichotomous operationalizations of Y.

```
> ### Appropriate Model
> M7<-polr(as.factor(Y_OR3)~X,data = DF, Hess=TRUE) # Ordered Logistic Regression
> # Model Results
> summary(M7) # Ordered Logistic Regression
Call:
polr(formula = as.factor(Y_OR3) ~ X, data = DF, Hess = TRUE)

Coefficients:
      Value    Std. Error     t value
X 0.059398525 0.00233345826 25.4551478

Intercepts:
    Value       Std. Error   t value
1|2 -2.318094970  0.034730391 -66.745433332
2|3 -1.470420539  0.026333788 -55.837791427
3|4 -0.900170441  0.023020774 -39.102527996

Residual Deviance: 18109.1629174
AIC: 18117.1629174
> # Chi-Square Test
> linearHypothesis(M7,c("X = 0"))
Linear hypothesis test

Hypothesis:
X = 0

Model 1: restricted model
Model 2: as.factor(Y_OR3) ~ X

  Res.Df Df    Chisq          Pr(>Chisq)
1   9997
2   9996  1 647.96455 < 0.000000000000000222 ***
---
Signif. codes:  0 '***' 0.001 '**' 0.01 '*' 0.05 '.' 0.1 ' ' 1
>
```

## *2.10. Summary of Results*

As introduced above, recoding a construct from a continuous distribution to an ordered distribution or a binomial distribution results in the reduction in the variation in the frequency of scores on Y across the expected values of Y. This is evident in the simulated results, as the *t*-statistics, *z*-statistics, and $x^2$-statistics appear to be attenuated when moving from a continuous operationalization of the dependent variable to an ordered or dichotomous operationalization. When implementing a statistical model with appropriate assumptions about the distribution of scores on the dependent variable, this loss of variation has limited impact on the slope coefficient but could result in an increased likelihood of committing a Type 1 or Type 2 error. Specifically, moving across levels of measurement for the dependent variable directly impacts our ability to conduct an accurate hypothesis test. While the simulated demonstrations presented above did not result in a Type 1 or Type 2 error, a variety of recoding processes can result in the hypothesis test suggesting an association opposite of the causal association in the population. This is most concerning when a causal association between two continuous constructs does not exist within the population, but regressing a non-continuous operationalization of the dependent variable on the independent variable produces a hypothesis test suggesting an association exists within the population (i.e., a Type 1 error). Overall, the results of the first set of simulations suggest that we should be cautious when and/or actively try not to recode our dependent variable from a continuous operationalization to an ordered or dichotomous operationalization unless theoretical or empirical justification exists.

## 3. Data Transformations

Now that we have completed our discussion about altering the level of measurement of the dependent variable, let's discuss data transformations. Okay, so in my opinion, a data transformation refers to the process of recoding a construct to alter the structure or scale of the distribution, while maintaining a continuous operationalization of the construct. That is, a data transformation is a *subtype* of recoding process used when cleaning data. Unlike the recodes performed in the previous section, all data transformations will maintain variation in the frequency of scores on Y across the expected values of Y. Nevertheless, data transformations alter the distribution of scores by either changing the scale or structure of the distribution. As such, these techniques are commonly used to satisfy the assumptions of statistical models.

While the largest concern with most data transformations is the inability to interpret the slope coefficient on the raw scale of the dependent variable, some data transformations can change the rank order of a distribution. That is, for example, a case with the lowest score on a test could become ranked higher than the lowest score after the implementation of certain data transformations. Altering the rank order of a distribution could generate the inability to observe the causal association between X and Y within the population. A similar problem arises when we restrict the distribution of scores to minimum and maximum values not representative of the distribution in the population. That is, if we exclude the case with the lowest score on a test from an analysis solely because we restrict the possible scores that we want to examine. Conducting this data transformation, similarly, can generate the inability to observe the causal association between X and Y within the population.

As a reminder, the Spearman correlation coefficient between X and Y was .31, the slope coefficient (*b*) for the association between X and Y was 1.054 (SE = .030), the *t*-statistic of the estimated effects of X on Y was 35.143, and the $x^2$-statistic of the estimated effects of X on Y was 1235.034.

### *3.1. Multiplying by A Constant*

To be honest, this is the data transformation I frequently use when estimating a structural equation model (SEM). This technique is most beneficial when the data being analyzed possesses an ill-scaled covariance matrix (i.e., some variances/covariances are substantially larger or smaller than other variances/covariances). We can go into more detail about how this works when discussing multilevel modeling or SEM but, briefly, scaling the covariance matrix properly increases the probability of a multilevel model or SEM model converging upon a single solution.

Multiplying a continuous distribution by any constant simply shifts the distribution up or down the x-axis. When the constant is a decimal smaller than one the distribution will shift down the x-axis and the variance will become smaller. When the constant is a number larger than one the distribution will shift up the x-axis and the variance will become larger. Generally, in my research, I stick to constants that start with a 1 and are all zeros (e.g., 10, 100, 1000), constants that are all zeros and end in 1 (e.g., .01, .001, .0001), or .1 to maintain the interpretability of the slope coefficient within the scale of the original measure. Either way, while the slope coefficient (*b*) and standard error (SE) will change, the correlation coefficient, the standardized effects of X on Y ($\beta$), the $t$- statistic, and the $x^2$- statistic will remain constant using this type of transformation.

### 3.1.1. Example 1: Y*.2

In this first example, we multiply Y by .2 to create Y_Re.2. Evident by the results, the Spearman correlation coefficient between X and Y_Re.2 was .31, the slope coefficient ($b$) was .211 (SE = .006), the $t$- statistic was 35.143, and the $x^2$- statistic was 1235.034. As expected, the Spearman correlation coefficient, the $t$- statistic, and the $x^2$- statistic are identical to the values observed when estimating the association between X and Y.

```
> ### Example 1: Rescaled by .2
> DF$Y_Re.2<-Y*.2
> summary(DF$Y_Re.2)
          Min.       1st Qu.        Median          Mean       3rd Qu.          Max.
-29.18854260666  -4.27902453369   0.00533520196  -0.01448913941   4.26203768624  24.57460877446
>
>
> corr.test(DF$X,DF$Y_Re.2, method = "spearman")
Call:corr.test(x = DF$X, y = DF$Y_Re.2, method = "spearman")
Correlation matrix
[1] 0.31
Sample Size
[1] 10000
These are the unadjusted probability values.
  The probability values  adjusted for multiple tests are in the p.adj object.
[1] 0

 To see confidence intervals of the correlations, print with the short=FALSE option
>
> M2<-glm(Y_Re.2~X, data = DF, family = gaussian(link = "identity"))
>
> # Model Results
> summary(M2)

Call:
glm(formula = Y_Re.2 ~ X, family = gaussian(link = "identity"),
    data = DF)

Deviance Residuals:
         Min            1Q         Median            3Q            Max
-25.951487265   -4.003581759    0.014130651    4.052959907    23.026452047

Coefficients:
                  Estimate    Std. Error  t value            Pr(>|t|)
(Intercept) -0.0427683794  0.0600135198 -0.71265             0.47608
X            0.2108272836  0.0059991162 35.14306 < 0.0000000000000002 ***
---
Signif. codes:  0 '***' 0.001 '**' 0.01 '*' 0.05 '.' 0.1 ' ' 1

(Dispersion parameter for gaussian family taken to be 36.009750275)

    Null deviance: 404498.7660  on 9999  degrees of freedom
Residual deviance: 360025.4832  on 9998  degrees of freedom
AIC: 64220.66789

Number of Fisher Scoring iterations: 2

> # Chi-Square Test
> linearHypothesis(M2,c("X = 0"))
Linear hypothesis test

Hypothesis:
X = 0

Model 1: restricted model
Model 2: Y_Re.2 ~ X

  Res.Df Df    Chisq         Pr(>Chisq)
1   9999
2   9998  1 1235.03447 < 0.000000000000000222 ***
---
Signif. codes:  0 '***' 0.001 '**' 0.01 '*' 0.05 '.' 0.1 ' ' 1
>
```

*3.1.2. Example 1: Y*20*

In this second example, we multiply Y by 20 to create Y_Re2. Evident by the results, the Spearman correlation coefficient between X and Y_Re2 was .31, the slope coefficient (*b*) was 21.083 (SE = .600), the *t*- statistic was 35.143, and the $x^2$- statistic was 1235.034. Similar to the previous example, the Spearman correlation coefficient, the *t*- statistic, and the $x^2$- statistic are identical to the values observed when estimating the association between X and Y.

```
> ### Example 2: Rescaled by 20
> DF$Y_Re2<-Y*20
> summary(DF$Y_Re2)
         Min.         1st Qu.          Median             Mean         3rd Qu.             Max.
-2918.854260666  -427.902453369     0.533520196    -1.448913941   426.203768624   2457.460877446
>
>
> corr.test(DF$X,DF$Y_Re2, method = "spearman")
Call:corr.test(x = DF$X, y = DF$Y_Re2, method = "spearman")
Correlation matrix
[1] 0.31
Sample Size
[1] 10000
These are the unadjusted probability values.
  The probability values  adjusted for multiple tests are in the p.adj object.
[1] 0

 To see confidence intervals of the correlations, print with the short=FALSE option
>
> M3<-glm(Y_Re2~X, data = DF, family = gaussian(link = "identity"))
>
> # Model Results
> summary(M3)

Call:
glm(formula = Y_Re2 ~ X, family = gaussian(link = "identity"),
    data = DF)

Deviance Residuals:
         Min              1Q          Median              3Q             Max
-2595.1487265    -400.3581759       1.4130651     405.2959907    2302.6452047

Coefficients:
              Estimate  Std. Error  t value              Pr(>|t|)
(Intercept) -4.27683794  6.00135198 -0.71265                0.47608
X           21.08272836  0.59991162 35.14306 < 0.0000000000000002 ***
---
Signif. codes:  0 '***' 0.001 '**' 0.01 '*' 0.05 '.' 0.1 ' ' 1

(Dispersion parameter for gaussian family taken to be 360097.50275)

    Null deviance: 4044987660  on 9999  degrees of freedom
Residual deviance: 3600254832  on 9998  degrees of freedom
AIC: 156324.0716

Number of Fisher Scoring iterations: 2

> # Chi-Square Test
> linearHypothesis(M3,c("X = 0"))
Linear hypothesis test

Hypothesis:
X = 0

Model 1: restricted model
Model 2: Y_Re2 ~ X

  Res.Df Df       Chisq            Pr(>Chisq)
1   9999
2   9998  1  1235.03447 < 0.000000000000000222 ***
---
Signif. codes:  0 '***' 0.001 '**' 0.01 '*' 0.05 '.' 0.1 ' ' 1
>
```

## 3.2. Standardizing Scores on Y

Okay, so this example is a little unnecessary because the distribution of Y is already normal, but a common continuous data transformation is *z*-scoring the distribution for the dependent variable – "*z*-scoring" sounds weird, but we will roll with it! This transformation generates a continuous normal distribution from the variable being transformed, were a 1 unit change in the variable now corresponds to a 1 standard deviation change in the variable. *R* makes it easy to conduct this data transformation with the scale() function.

Given that Y is already normal, the slope coefficient (*b*) and standard error (SE) will change, but the correlation coefficient, the standardized effects of X on Y (*β*), the *t*- statistic, and the $x^2$- statistic will remain constant. That said, if you *z*-score a non-normal distribution it is possible for all of the coefficients, including the correlation coefficient, the standardized effects of X on Y (*β*), the *t*-statistic, and the $x^2$- statistic, to change. Z-scores are invasive data transformations when working with non-normal distributions as they alter the shape or rank order of the distribution.

In this example, we *z*-scored Y using the scale() function and created Y_z. Evident by the results, the Spearman correlation coefficient between X and Y_z was .31, the slope coefficient (*b*) for the association between X and Y_z was .033 (SE = .001), the *t*- statistic was 35.143, and the $x^2$- statistic was 1235.034. As expected, the Spearman correlation coefficient, the *t*- statistic, and the $x^2$-statistic are identical to the values observed when estimating the association between X and Y. When using this transformation, the slope coefficient is interpreted as a 1 unit change in X results in a .033 standard deviation change in Y.

```
> ## Standardizing Data ####
> DF$Y_z<-scale(DF$Y)
> summary(DF$Y_z)
       V1
 Min.   :-4.58687028093
 1st Qu.:-0.67048861359
 Median : 0.00311686829
 Mean   : 0.00000000000
 3rd Qu.: 0.67237395805
 Max.   : 3.86600383051
>
> corr.test(DF$X,DF$Y_z, method = "spearman")
Call:corr.test(x = DF$X, y = DF$Y_z, method = "spearman")
Correlation matrix
     [,1]
[1,] 0.31
Sample Size
[1] 10000
These are the unadjusted probability values.
  The probability values  adjusted for multiple tests are in the p.adj object.
     [,1]
[1,]    0

 To see confidence intervals of the correlations, print with the short=FALSE option
>
> M4<-glm(Y_z~X, data = DF, family = gaussian(link = "identity"))
>
> # Model Results
> summary(M4)

Call:
glm(formula = Y_z ~ X, family = gaussian(link = "identity"),
    data = DF)

Deviance Residuals:
       Min           1Q        Median           3Q          Max
-4.080204549   -0.629460360   0.002221682   0.637223805   3.620317920

Coefficients:
                 Estimate       Std. Error    t value       Pr(>|t|)
(Intercept) -0.004446183860  0.009435583934  -0.47121        0.6375
```

```
X             0.033147173144  0.000943206876 35.14306 <0.0000000000000002 ***
---
Signif. codes:  0 '***' 0.001 '**' 0.01 '*' 0.05 '.' 0.1 ' ' 1

(Dispersion parameter for gaussian family taken to be 0.890142376925)

    Null deviance: 9999.000000  on 9999  degrees of freedom
Residual deviance: 8899.643484  on 9998  degrees of freedom
AIC: 27219.03191

Number of Fisher Scoring iterations: 2

> # Chi-Square Test
> linearHypothesis(M4,c("X = 0"))
Linear hypothesis test

Hypothesis:
X = 0

Model 1: restricted model
Model 2: Y_z ~ X

  Res.Df Df      Chisq              Pr(>Chisq)
1   9999
2   9998  1 1235.03447 < 0.000000000000000222 ***
---
Signif. codes:  0 '***' 0.001 '**' 0.01 '*' 0.05 '.' 0.1 ' ' 1
>
```

### *3.3. Normalizing Scores on Y*

Normalizing a distribution means that we are going to take the structure of the distribution – as is – but require the scores to range between zero and one. To normalize a variable, we take each case's score on the variable and subtract the minimum value of the distribution and divide by the difference between the maximum and minimum values of the distribution (Equation 1).

[Equation 1]

$$\frac{y_i - \min_y}{\max_y - \min_y}$$

There aren't a large number of cases where one would only normalize the distribution of a variable, as multiplying by a constant is easier when rescaling a distribution. That said, however, it is extremely important to normalize a distribution when implementing certain data transformations.

In this example, we normalized Y using equation 1 and created Y_n. Evident by the results, the Spearman correlation coefficient between X and Y_n was .31, the slope coefficient (*b*) for the association between X and Y_n was .004 (SE = .0001), the *t*- statistic was 35.143, and the $x^2$- statistic was 1235.034. As expected, the Spearman correlation coefficient, the *t*- statistic, and the $x^2$- statistic are identical to the values observed when estimating the association between X and Y.

```
> ## Normalizing Data ####
>
> DF$Y_n<-((DF$Y)-min(DF$Y))/((max(DF$Y))-min(DF$Y))
> summary(DF$Y_n)
      Min.      1st Qu.      Median        Mean      3rd Qu.         Max.
0.000000000 0.463319531 0.543009051 0.542640316 0.622184143 1.000000000
>
>
> corr.test(DF$X,DF$Y_n, method = "spearman")
Call:corr.test(x = DF$X, y = DF$Y_n, method = "spearman")
Correlation matrix
[1] 0.31
Sample Size
[1] 10000
These are the unadjusted probability values.
```

```
   The probability values  adjusted for multiple tests are in the p.adj object.
[1] 0

 To see confidence intervals of the correlations, print with the short=FALSE option
>
> M5<-glm(Y_n~X, data = DF, family = gaussian(link = "identity"))
>
> # Model Results
> summary(M5)

Call:
glm(formula = Y_n ~ X, family = gaussian(link = "identity"),
    data = DF)

Deviance Residuals:
          Min            1Q         Median             3Q            Max
-0.4827002621  -0.0744670217   0.0002628315   0.0753854602   0.4282943141

Coefficients:
                  Estimate     Std. Error    t value               Pr(>|t|)
(Intercept) 0.542114319538 0.001116257478 485.65347 < 0.000000000000000222 ***
X           0.003921408589 0.000111584162  35.14306 < 0.000000000000000222 ***
---
Signif. codes:  0 '***' 0.001 '**' 0.01 '*' 0.05 '.' 0.1 ' ' 1

(Dispersion parameter for gaussian family taken to be 0.012458067361)

    Null deviance: 139.9419001  on 9999  degrees of freedom
Residual deviance: 124.5557575  on 9998  degrees of freedom
AIC: -15471.09839

Number of Fisher Scoring iterations: 2

> # Chi-Square Test
> linearHypothesis(M5,c("X = 0"))
Linear hypothesis test

Hypothesis:
X = 0

Model 1: restricted model
Model 2: Y_n ~ X

  Res.Df Df      Chisq              Pr(>Chisq)
1   9999
2   9998  1 1235.03447 < 0.000000000000000222 ***
---
Signif. codes:  0 '***' 0.001 '**' 0.01 '*' 0.05 '.' 0.1 ' ' 1
>
```

### *3.4. Log Transformations of Y*

Logarithmic data transformations are pretty invasive and require strong justification to implement. In the current context, invasive refers to data transformations that can alter the shape or rank order of a distribution. They tend to be most commonly implemented when trying to satisfy the normality assumption with a variable that is known to be positively skewed within the population. Specifically, calculating the logarithm of a variable reduces the skew within the distribution because the difference between the raw score and logarithm of the raw score become larger when the raw score is larger. To provide an example, the natural log of 10 is 2.303 and the natural log of 100,000,000 is 18.421.

This subset of data transformations is most commonly implemented when working with income. Specifically, income has an extreme positive skew within the population and can not satisfy the normality assumption. As such, we might take the log of the observed distribution to hopefully satisfy the normality assumption. Personally, I have not seen many people calculate the log of a distribution besides income, so if you have other examples please share them! That said, logarithmic data transformations are extremely finicky and might not achieve the desired goal. Moreover, as a reminder, you can not calculate the log of 0 or of negative values.

### 3.4.1. Base Log

In this example, we want to calculate the base log of Y. Y, however, is a normal distribution with positive and negative values. As such, there are two ways we can make the Y distribution appropriate for the base log transformation. We can 1) add a positive constant greater than the largest negative value to Y or 2) normalize the distribution on T.

Given that Y_n already existed in the dataframe, we can calculate the base log of the normalized Y distribution using log(Y_n) creating Y_log. Evident by the results, the spearman correlation between X and Y_log was .31, the slope coefficient ($b$) was .006 (SE = .0001), the $t$- statistic was 34.701, and the $x^2$- statistic was 1204.149. While the amount of variance in the distribution was slightly reduced by calculating the log of the normalized Y distribution, the reduction in variance was nominal and the results remained largely the same. Importantly, when interpreting the slope coefficient between X and Y_log, a 1 point change in X resulted in a .006 base log change in the normalized distribution of Y.

```
> ### Base Log
> DF$Y_n<-((DF$Y)-min(DF$Y-25))/((max(DF$Y+25))-min(DF$Y-25))
>
> DF$Y_log<-log(DF$Y_n)
> summary(DF$Y_log)
        Min.       1st Qu.        Median          Mean       3rd Qu.          Max.
-2.5457375465 -0.7569986990 -0.6231288423 -0.6422487284 -0.5058016710 -0.0816604777
>
>
> corr.test(DF$X,DF$Y_log, method = "spearman")
Call:corr.test(x = DF$X, y = DF$Y_log, method = "spearman")
Correlation matrix
[1] 0.31
Sample Size
[1] 10000
These are the unadjusted probability values.
  The probability values  adjusted for multiple tests are in the p.adj object.
[1] 0

 To see confidence intervals of the correlations, print with the short=FALSE option
>
> M6<-glm(Y_log~X, data = DF, family = gaussian(link = "identity"))
>
> # Model Results
> summary(M6)

Call:
glm(formula = Y_log ~ X, family = gaussian(link = "identity"),
    data = DF)

Deviance Residuals:
         Min            1Q         Median            3Q           Max
-1.7234362625  -0.1091117847    0.0169627735    0.1287641726    0.5672971592

Coefficients:
                 Estimate      Std. Error    t value              Pr(>|t|)
(Intercept) -0.643116285983  0.001864571499 -344.91372 < 0.000000000000000222 ***
X            0.006467811822  0.000186387686   34.70085 < 0.000000000000000222 ***
---
Signif. codes:  0 '***' 0.001 '**' 0.01 '*' 0.05 '.' 0.1 ' ' 1

(Dispersion parameter for gaussian family taken to be 0.0347600182253)

    Null deviance: 389.3869119  on 9999  degrees of freedom
Residual deviance: 347.5306622  on 9998  degrees of freedom
AIC: -5210.104079

Number of Fisher Scoring iterations: 2

> # Chi-Square Test
```

```
> linearHypothesis(M6,c("X = 0"))
Linear hypothesis test

Hypothesis:
X = 0

Model 1: restricted model
Model 2: Y_log ~ X

  Res.Df Df     Chisq             Pr(>Chisq)
1   9999
2   9998  1 1204.14924 < 0.0000000000000000222 ***
---
Signif. codes:  0 '***' 0.001 '**' 0.01 '*' 0.05 '.' 0.1 ' ' 1
>
```

### 3.4.2. Log 10

We can calculate the $\log_{10}$ of the normalized Y distribution using log10(Y_n) creating Y_log10. Evident by the results, the Spearman correlation coefficient between X and Y_log10 was .31, the slope coefficient (*b*) was .006 (SE = .0001), the *t*- statistic was 34.701, and the $x^2$- statistic was 1204.149. Again, the reduction in variance in the Y distribution was nominal and the Spearman correlation coefficient, the *t*- statistic, and the $x^2$- statistic were almost identical to the values observed when estimating the association between X and Y.

```
> ###  Log10
> DF$Y_n<-((DF$Y)-min(DF$Y-25))/((max(DF$Y+25))-min(DF$Y-25))
>
> DF$Y_log10<-log10(DF$Y_n)
> summary(DF$Y_log10)
         Min.       1st Qu.        Median          Mean       3rd Qu.          Max.
-1.1055997688 -0.3287603578 -0.2706214177 -0.2789250788 -0.2196668747 -0.0354646948
>
> corr.test(DF$X,DF$Y_log10, method = "spearman")
Call:corr.test(x = DF$X, y = DF$Y_log10, method = "spearman")
Correlation matrix
[1] 0.31
Sample Size
[1] 10000
These are the unadjusted probability values.
  The probability values  adjusted for multiple tests are in the p.adj object.
[1] 0

 To see confidence intervals of the correlations, print with the short=FALSE option
>
> M7<-glm(Y_log10~X, data = DF, family = gaussian(link = "identity"))
>
> # Model Results
> summary(M7)

Call:
glm(formula = Y_log10 ~ X, family = gaussian(link = "identity"),
    data = DF)

Deviance Residuals:
          Min             1Q         Median             3Q            Max
-0.7484788587  -0.0473866460   0.0073668389   0.0559215696   0.2463740259

Coefficients:
                  Estimate     Std. Error    t value             Pr(>|t|)
(Intercept) -0.2793018542246  0.0008097731131 -344.91372 < 0.0000000000000000222 ***
X            0.0028089349844  0.0000809471436   34.70085 < 0.0000000000000000222 ***
---
Signif. codes:  0 '***' 0.001 '**' 0.01 '*' 0.05 '.' 0.1 ' ' 1

(Dispersion parameter for gaussian family taken to be 0.00655614602563)

    Null deviance: 73.44292624  on 9999  degrees of freedom
Residual deviance: 65.54834796  on 9998  degrees of freedom
AIC: -21890.75298

Number of Fisher Scoring iterations: 2

> # Chi-Square Test
> linearHypothesis(M7,c("X = 0"))
Linear hypothesis test
```

```
Hypothesis:
X = 0

Model 1: restricted model
Model 2: Y_log10 ~ X

  Res.Df Df      Chisq            Pr(>Chisq)
1   9999
2   9998  1 1204.14924 < 0.000000000000000222 ***
---
Signif. codes:  0 '***' 0.001 '**' 0.01 '*' 0.05 '.' 0.1 ' ' 1
>
```

## 3.5. Power Transformations—Raw Scale

Power transformations refers to the subset of data transformations calculated by raising the scores on a distribution to the $n^{th}$ power. Power transformations, similar to logarithm transformations, are generally invasive and could result in some unintended consequences. An example of a power transformation is $Y^2$, where Y is raised to the second power or (Y*Y). At the forefront of issues associated with power transformations, is the potential alteration of the rank order of the distribution on Y when calculating $Y^2$. The rank order is most commonly affected when negative values exist within the Y distribution and the variable is raised to an even power (e.g., 2, 4, 6, 8). Remember, a negative number raised to an even power equals a positive value! This is demonstrated in the simulations below, were we subject the raw scale of Y to power transformations.

### 3.5.1. Y Squared

In this example, we calculated Y^2 in R to create Y2. Evident by the results, the Spearman correlation coefficient between X and Y2 was 0, the slope coefficient (*b*) was -1.722 (SE = 1.477), the *t*- statistic was -1.166, and the $x^2$- statistic was 1.359. These results are a large departure from reality and suggest that X does not have any statistical association with Y2. These results occurred because we altered the rank order in Y. For instance, the minimum value for Y – the largest negative value –has the maximum value on the distribution of Y2.

```
> ### Squared
> DF$Y2<-DF$Y^2
> summary(DF$Y2)
        Min.        1st Qu.         Median            Mean         3rd Qu.            Max.
    0.00002969    104.06660167    456.43430867    1011.25216329    1308.79640623 21299.27548752
>
> corr.test(DF$X,DF$Y2, method = "spearman")
Call:corr.test(x = DF$X, y = DF$Y2, method = "spearman")
Correlation matrix
[1] 0
Sample Size
[1] 10000
These are the unadjusted probability values.
  The probability values  adjusted for multiple tests are in the p.adj object.
[1] 0.91

 To see confidence intervals of the correlations, print with the short=FALSE option
>
> M8<-glm(Y2~X, data = DF, family = gaussian(link = "identity"))
>
> # Model Results
> summary(M8)

Call:
glm(formula = Y2 ~ X, family = gaussian(link = "identity"), data = DF)

Deviance Residuals:
       Min            1Q         Median            3Q            Max
 -1063.781905    -908.209722    -555.471354     301.302216    20240.074412

Coefficients:
             Estimate    Std. Error   t value              Pr(>|t|)
```

```
(Intercept) 1011.48319829   14.77947982 68.43835 < 0.0000000000000002 ***
X            -1.72241123    1.47739738 -1.16584               0.24371
---
Signif. codes:  0 '***' 0.001 '**' 0.01 '*' 0.05 '.' 0.1 ' ' 1

(Dispersion parameter for gaussian family taken to be 2183937.5233)

    Null deviance: 21837975736  on 9999  degrees of freedom
Residual deviance: 21835007358  on 9998  degrees of freedom
AIC: 174349.1706

Number of Fisher Scoring iterations: 2

> # Chi-Square Test
> linearHypothesis(M8,c("X = 0"))
Linear hypothesis test

Hypothesis:
X = 0

Model 1: restricted model
Model 2: Y2 ~ X

  Res.Df Df  Chisq Pr(>Chisq)
1   9999
2   9998  1 1.35919    0.24368
>
```

### 3.5.2. Y Raised to the .2 Power

A similar result is observed when we calculate Y^.2 in R to create Y.2. The source of the problem in this condition is that we can not raise negative values to decimal powers (e.g., -2^.2). Instead, R provides an NA for these cases. This has important implications, as the Spearman correlation coefficient between X and Y.2 was .19, the slope coefficient (*b*) was -.006 (SE = .0005), the *t*-statistic was 12.827, and the $x^2$- statistic was 164.531. These results, while better than the previous transformation, are a large departure from reality.

```
> ### raised to the .2 power
> DF$Y.2<-DF$Y^.2
> summary(DF$Y.2)
      Min.    1st Qu.     Median       Mean    3rd Qu.       Max.       NA's
0.35258635 1.59715625 1.84374283 1.79812642 2.04865388 2.61752792       4997
>
> corr.test(DF$X,DF$Y.2, method = "spearman")
Call:corr.test(x = DF$X, y = DF$Y.2, method = "spearman")
Correlation matrix
[1] 0.19
Sample Size
[1] 5003
These are the unadjusted probability values.
  The probability values  adjusted for multiple tests are in the p.adj object.
[1] 0

 To see confidence intervals of the correlations, print with the short=FALSE option
>
> M9<-glm(Y.2~X, data = DF, family = gaussian(link = "identity"))
>
> # Model Results
> summary(M9)

Call:
glm(formula = Y.2 ~ X, family = gaussian(link = "identity"),
    data = DF)

Deviance Residuals:
         Min             1Q          Median             3Q             Max
-1.4992221645  -0.1967287252   0.0469582285   0.2482599743   0.8157676896

Coefficients:
              Estimate    Std. Error    t value              Pr(>|t|)
(Intercept) 1.780398448541 0.005010141910 355.35889 < 0.000000000000000222 ***
X           0.006343692745 0.000494558503  12.82698 < 0.000000000000000222 ***
---
Signif. codes:  0 '***' 0.001 '**' 0.01 '*' 0.05 '.' 0.1 ' ' 1

(Dispersion parameter for gaussian family taken to be 0.116026399605)
```

```
    Null deviance: 599.3380165  on 5002  degrees of freedom
Residual deviance: 580.2480244  on 5001  degrees of freedom
  (4997 observations deleted due to missingness)
AIC: 3425.749096

Number of Fisher Scoring iterations: 2

> # Chi-Square Test
> linearHypothesis(M9,c("X = 0"))
Linear hypothesis test

Hypothesis:
X = 0

Model 1: restricted model
Model 2: Y.2 ~ X

  Res.Df Df    Chisq             Pr(>Chisq)
1   5002
2   5001  1 164.53145 < 0.000000000000000222 ***
---
Signif. codes:  0 '***' 0.001 '**' 0.01 '*' 0.05 '.' 0.1 ' ' 1
>
```

## 3.6. Power Transformations—Normalized Scale

To demonstrate the results when the scale is normalized, we replicated the power transformations performed above on Y_n.

### 3.6.1. Y Squared

In this example, we raised Y_n^2 and created Y2_n. Evident by the results, the Spearman correlation coefficient between X and Y2_n was .31, the slope coefficient (*b*) was .004 (SE = .0001), the *t*- statistic was 34.750, and the $x^2$- statistic was 1207.574.

```
> ### Squared
> DF$Y_n<-((DF$Y)-min(DF$Y-25))/((max(DF$Y+25))-min(DF$Y-25))
> DF$Y2_n<-DF$Y_n^2
> summary(DF$Y2_n)
        Min.         1st Qu.         Median           Mean         3rd Qu.           Max.
0.00614894293 0.22002867549 0.28757899690 0.29719459761 0.36363547996 0.84931855134
>
> corr.test(DF$X,DF$Y2_n, method = "spearman")
Call:corr.test(x = DF$X, y = DF$Y2_n, method = "spearman")
Correlation matrix
[1] 0.31
Sample Size
[1] 10000
These are the unadjusted probability values.
  The probability values  adjusted for multiple tests are in the p.adj object.
[1] 0

 To see confidence intervals of the correlations, print with the short=FALSE option
>
> M10<-glm(Y2_n~X, data = DF, family = gaussian(link = "identity"))
>
> # Model Results
> summary(M10)

Call:
glm(formula = Y2_n ~ X, family = gaussian(link = "identity"),
    data = DF)

Deviance Residuals:
         Min             1Q          Median             3Q            Max
-0.3214133267  -0.0721628155  -0.0082762638    0.0627939165    0.5195618085

Coefficients:
               Estimate     Std. Error    t value              Pr(>|t|)
(Intercept) 0.296721273971 0.001015832865 292.09655 < 0.000000000000000222 ***
```

```
X            0.003528720613 0.000101545442   34.75016 < 0.000000000000000222 ***
---
Signif. codes:  0 '***' 0.001 '**' 0.01 '*' 0.05 '.' 0.1 ' ' 1

(Dispersion parameter for gaussian family taken to be 0.0103173088415)

    Null deviance: 115.6113655  on 9999  degrees of freedom
Residual deviance: 103.1524538  on 9998  degrees of freedom
AIC: -17356.55278

Number of Fisher Scoring iterations: 2

> # Chi-Square Test
> linearHypothesis(M10,c("X = 0"))
Linear hypothesis test

Hypothesis:
X = 0

Model 1: restricted model
Model 2: Y2_n ~ X

  Res.Df Df      Chisq              Pr(>Chisq)
1   9999
2   9998  1 1207.57379 < 0.000000000000000222 ***
---
Signif. codes:  0 '***' 0.001 '**' 0.01 '*' 0.05 '.' 0.1 ' ' 1
>
```

*3.6.2. Y Raised to the .2 Power*

In this example, we raised Y_n^.2 and created Y.2_n. Evident by the results, the Spearman correlation coefficient between X and Y.2_n was .31, the slope coefficient (*b*) was .001 (SE = .000), the *t*- statistic was 34.881, and the $x^2$- statistic was 1216.697.

```
> ### raised to the .2 power
> DF$Y.2_n<-DF$Y_n^.2
> summary(DF$Y.2_n)
      Min.    1st Qu.     Median       Mean    3rd Qu.       Max.
0.601007714 0.859504052 0.882827223 0.880135741 0.903788113 0.983800550
>
> corr.test(DF$X,DF$Y.2_n, method = "spearman")
Call:corr.test(x = DF$X, y = DF$Y.2_n, method = "spearman")
Correlation matrix
[1] 0.31
Sample Size
[1] 10000
These are the unadjusted probability values.
  The probability values  adjusted for multiple tests are in the p.adj object.
[1] 0

 To see confidence intervals of the correlations, print with the short=FALSE option
>
> M11<-glm(Y.2_n~X, data = DF, family = gaussian(link = "identity"))
>
> # Model Results
> summary(M11)

Call:
glm(formula = Y.2_n ~ X, family = gaussian(link = "identity"),
    data = DF)

Deviance Residuals:
         Min              1Q          Median              3Q             Max
-0.24776294359  -0.01952569669   0.00239017516    0.02227604032    0.10308605172

Coefficients:
                 Estimate     Std. Error    t value               Pr(>|t|)
(Intercept) 0.8799846131849 0.0003231284086 2723.32791 < 0.000000000000000222 ***
X           0.0011266902551 0.0000323008029   34.88118 < 0.000000000000000222 ***
---
Signif. codes:  0 '***' 0.001 '**' 0.01 '*' 0.05 '.' 0.1 ' ' 1

(Dispersion parameter for gaussian family taken to be 0.00104393196527)

    Null deviance: 11.70738069  on 9999  degrees of freedom
```

```
Residual deviance: 10.43723179  on 9998  degrees of freedom
AIC: -40264.83913

Number of Fisher Scoring iterations: 2

> # Chi-Square Test
> linearHypothesis(M11,c("X = 0"))
Linear hypothesis test

Hypothesis:
X = 0

Model 1: restricted model
Model 2: Y.2_n ~ X

  Res.Df Df      Chisq              Pr(>Chisq)
1   9999
2   9998  1 1216.69701 < 0.000000000000000222 ***
---
Signif. codes:  0 '***' 0.001 '**' 0.01 '*' 0.05 '.' 0.1 ' ' 1
>
```

## 3.7. Rounding Raw Scores

Just for fun, let's see what happens if we round the scores on Y to the nearest whole digit! In this example, we used round(Y) and created Y_R. Evident by the results, the Spearman correlation coefficient between X and Y_R was .31, the slope coefficient (*b*) was 1.054 (SE = .030), the *t*-statistic was 35.145, and the $x^2$- statistic was 1235.153. These results are identical to the results of Y regressed on X, because simply rounding Y to the nearest whole number did not reduce the variance of the distribution. That is, the distribution of Y is continuous enough to maintain its variance even when rounded to the nearest whole number.

```
> ## Rounding Raw Scores ####
> DF$Y_R<-round(DF$Y)
> summary(DF$Y_R)
      Min.   1st Qu.    Median      Mean   3rd Qu.      Max.
  -146.0000  -21.0000    0.0000   -0.0684   21.0000  123.0000
>
> corr.test(DF$X,DF$Y_R, method = "spearman")
Call:corr.test(x = DF$X, y = DF$Y_R, method = "spearman")
Correlation matrix
[1] 0.31
Sample Size
[1] 10000
These are the unadjusted probability values.
  The probability values  adjusted for multiple tests are in the p.adj object.
[1] 0

 To see confidence intervals of the correlations, print with the short=FALSE option
>
> M12<-glm(Y_R~X, data = DF, family = gaussian(link = "identity"))
>
> # Model Results
> summary(M12)

Call:
glm(formula = Y_R ~ X, family = gaussian(link = "identity"),
    data = DF)

Deviance Residuals:
          Min              1Q           Median               3Q              Max
  -130.17651665    -20.02955118       0.18207685      20.27263931     114.91605087

Coefficients:
               Estimate    Std. Error   t value           Pr(>|t|)
(Intercept) -0.2098074058  0.3000769972  -0.69918            0.48446
X            1.0542199587  0.0299965205  35.14474 < 0.0000000000000002 ***
---
Signif. codes:  0 '***' 0.001 '**' 0.01 '*' 0.05 '.' 0.1 ' ' 1

(Dispersion parameter for gaussian family taken to be 900.300151363)

    Null deviance: 10113209.214  on 9999  degrees of freedom
```

```
Residual deviance:  9001200.913  on 9998  degrees of freedom
AIC: 96410.05256

Number of Fisher Scoring iterations: 2

> # Chi-Square Test
> linearHypothesis(M12,c("X = 0"))
Linear hypothesis test

Hypothesis:
X = 0

Model 1: restricted model
Model 2: Y_R ~ X

  Res.Df Df     Chisq              Pr(>Chisq)
1   9999
2   9998  1 1235.15285 < 0.000000000000000222 ***
---
Signif. codes:  0 '***' 0.001 '**' 0.01 '*' 0.05 '.' 0.1 ' ' 1
>
```

## 3.8. Excluding Part of the Distribution

Finally, let's remove cases on Y from the distribution. In this example, we exclude cases above 5 and below -5 from the distribution of Y to create Y_C. Evident by the results, the Spearman correlation coefficient between X and Y_R was .03, the slope coefficient (*b*) was .012 (SE = .009), the *t*- statistic was 1.307, and the $x^2$- statistic was 1.708. These results are a large departure from reality, suggesting that X does not have any statistical association with Y_C. These coefficients were produced because Y_C does not approximate the variance of Y. In particular, the distribution of Y_C is so distinct from Y, that the statistical association between X and Y_C can not be used to approximate the statistical association between X and Y. Nevertheless, I do not want to go into more detail at this moment because we will talk more about issues related to excluding cases from the distribution when we cover measurement error.

```
> ## Removing Part of the Distribution  ####
> summary(DF$Y)
          Min.         1st Qu.          Median            Mean         3rd Qu.            Max.
-145.9427130333  -21.3951226685    0.0266760098   -0.0724456971   21.3101884312   122.8730438723
>
> DF$Y_C<-DF$Y
> DF$Y_C[DF$Y_C>=5|DF$Y_C<=-5]<-NA
> summary(DF$Y)
          Min.         1st Qu.          Median            Mean         3rd Qu.            Max.
-145.9427130333  -21.3951226685    0.0266760098   -0.0724456971   21.3101884312   122.8730438723
> summary(DF$Y_C)
       Min.     1st Qu.      Median        Mean     3rd Qu.        Max.        NA's
-4.99917286 -2.58365734 -0.01183147 -0.06714816  2.39161687  4.99348388        8790
>
> corr.test(DF$X,DF$Y_C, method = "spearman")
Call:corr.test(x = DF$X, y = DF$Y_C, method = "spearman")
Correlation matrix
[1] 0.03
Sample Size
[1] 1210
These are the unadjusted probability values.
  The probability values  adjusted for multiple tests are in the p.adj object.
[1] 0.25

 To see confidence intervals of the correlations, print with the short=FALSE option
>
> M13<-glm(Y_C~X, data = DF, family = gaussian(link = "identity"))
>
> # Model Results
> summary(M13)

Call:
glm(formula = Y_C ~ X, family = gaussian(link = "identity"),
    data = DF)

Deviance Residuals:
```

```
         Min          1Q      Median          3Q         Max
-5.135155750  -2.514051022  0.056991025  2.498079790  5.153721996

Coefficients:
                Estimate     Std. Error  t value Pr(>|t|)
(Intercept) -0.06655192217  0.08372769544 -0.79486  0.42685
X            0.01190733748  0.00911156457  1.30684  0.19152

(Dispersion parameter for gaussian family taken to be 8.48224377965)

    Null deviance: 10261.03668  on 1209  degrees of freedom
Residual deviance: 10246.55049  on 1208  degrees of freedom
  (8790 observations deleted due to missingness)
AIC: 6024.77936

Number of Fisher Scoring iterations: 2

> # Chi-Square Test
> linearHypothesis(M13,c("X = 0"))
Linear hypothesis test

Hypothesis:
X = 0

Model 1: restricted model
Model 2: Y_C ~ X

  Res.Df Df   Chisq Pr(>Chisq)
1   1209
2   1208  1 1.70783     0.19127
>
```

### 3.9. Summary of Results

A variety of different continuous data transformations were discussed within the current section. The results of the simulations can be summarized in four-points:

1. Multiplying by a constant – or adding a constant (which we did not review) – or normalizing a construct rescales the distribution of the construct. These data transformation techniques are often considered not invasive as they can not alter the shape or rank order associated with the distribution of a construct.

2. Standardizing the distribution of a construct will have limited impact on the shape, rank order, or variation *if* the construct is normally distributed within the sample and in the population. However, it is important to remember that standardizing a distribution can alter the shape or rank order of a construct if the distribution of the construct in the sample and population is not normal.

3. Invasive data transformations, such as log or power transformations, commonly require multiple steps to ensure that the shape and rank order in the transformed distribution is not substantively altered by the implementation of the technique. These transformations are finicky and can result in biased statistical estimates if they are not properly implemented.

4. Removing cases from the distribution of a construct without theoretical or empirical justification – e.g., the desire to generalize to a subset of the population – could result in estimates that are not representative of the causal association within the population.

In summary, when implementing continuous data transformations, it is always important to perform multiple checks to ensure that the transformation did not alter our ability to make inferences about the causal association in the population. Moreover, when possible try to avoid implementing invasive data transformations as they could have important ramifications for the association of interest.

**4. Conclusion**
I am happy that you made it this far and we are almost done talking about the measurement of the dependent variable, so let's make this conclusion quick. As I have stated multiple times throughout this entry, **when possible operationalize all of your measures as continuous constructs without the implementation of a continuous data transformation**! By doing this, you will ensure that the maximum amount of variation in the dependent variable exists when estimating your statistical model, as well as not run the risk of altering the shape or rank order of the distribution of a construct. You might be saying that you need your dependent variable to be normal to estimate a linear regression model, but this is simply not true. A variety of statistical techniques and estimators can be implemented to evaluate the effects of a predictor on a non-normally distributed dependent variable. Personally, a large amount of my work employs the gambit of weighted least squares estimators to handle non-normally distributed endogenous and lagged-endogenous variables — SEM terms for dependent variables. All of this stated, if it is theoretically or empirically relevant to change the level of measurement or implement a continuous data transformation, please do so. Just take some time to recognize the loss of information or potential problems that could occur when implementing these techniques.

***Ending Note**: I had to end this entry somewhere, but I am not happy that I did not get to everything. I could write a book about the effects of data transformations, let alone the effects of different operationalizations. If I do happen to get a book contract maybe I will focus on biases related to measurement, but in the meantime a 30-page description of the effects will do. Moreover, some of the following entries will give me the opportunity to discuss some residual issues related to the measurement of the dependent variable. I do, however, hope you enjoyed this entry.*

---

[i] Although perfectly fine for continuous-continuous, ordered-continuous, and ordered-ordered constructs, this coefficient is not preferred when a dichotomous variable is included (Point-biserial is preferred: https://www.r-bloggers.com/2021/07/point-biserial-correlation-in-r-quick-guide/). Nevertheless, for the sake of the demonstration it will do!

[ii] A replication of these simulations is provided in the R-script. Rather than specifying that X has a linear causal influence on Y, X is specified to have a curvilinear causal influence on Y.

[iii] Truthfully, I don't know if this math is correct, but it looks correct. 5 primary digits with a possible of 6 digits additional digits per number which each have 10 possible options…

**Entry 14: Level of Measurement of The Independent Variable**

**1. Introduction**

The independent variable – other terms include the treatment (0,1), predictor, or exogenous variable – is the mechanism in the population that we hypothesize causes changes in the dependent variable. That is, we believe that when a cases' score on the independent variable increases or decreases, a cases' score on the dependent variable will vary to some degree. In line with this postulation, it is extremely important for us to operationalize the independent variable on the scale that exists in the population. If we do not, we run the risk of reducing the magnitude of the association between the independent variable and dependent variable, potentially leading to interpretations and conclusions not representative of the association in the population. Let's get right into our simulations, as a lot of the discussions important to altering the level of measurement for a construct were included in Entry 13.

***Brief Recommendation:*** *I suggest reading Entry 13 – The 30-page summary, I know – as it is important to the discussions and analyses presented in this entry.*

**2. Continuous Operationalization of X**

To establish a baseline, we can begin by simulating our independent variable (X) by drawing scores for 10000 cases from a continuous normal distribution with a mean of 0 and a standard deviation of 10. Afterwards, we can specify scores on the dependent variable (Y) to be equal to – or causally influenced by – the scores on the independent variable (X) plus a random draw from a normal distribution with a mean of 0 and a standard deviation of 1. Through this specification process, a 1 point increase in X will be associated with a 1 point increase in Y.

```
n<-10000

set.seed(1992)

X<-1*rnorm(n,0,10)
Y<-1*X+1*rnorm(n,0,30)

DF<-data.frame(X,Y)
```

Following the protocol described in Entry 13, we estimate 1) a spearman correlation, 2) a bivariate linear regression model, and 3) a linear hypothesis test to observe the test statistics for the association. The spearman correlation between X and Y was .31, the *t*-score was 35.143, and the $X^2$-square score was 1235.034. The hypothesis tests provide evidence that the slope coefficient of the association between X and Y was statistically different from zero.

```
> corr.test(DF$X,DF$Y, method = "spearman")
Call:corr.test(x = DF$X, y = DF$Y, method = "spearman")
Correlation matrix
[1] 0.31
Sample Size
[1] 10000
These are the unadjusted probability values.
  The probability values  adjusted for multiple tests are in the p.adj object.
[1] 0

 To see confidence intervals of the correlations, print with the short=FALSE option
>
> M1<-glm(Y~X, data = DF, family = gaussian(link = "identity"))
>
> # Model Results
> summary(M1)
```

```
Call:
glm(formula = Y ~ X, family = gaussian(link = "identity"), data = DF)

Deviance Residuals:
         Min           1Q        Median           3Q          Max
-129.75743633  -20.01790880    0.07065325   20.26479953  115.13226023

Coefficients:
              Estimate   Std. Error   t value              Pr(>|t|)
(Intercept) -0.213841897  0.300067599  -0.71265              0.47608
X            1.054136418  0.029995581  35.14306 < 0.0000000000000002 ***
---
Signif. codes:  0 '***' 0.001 '**' 0.01 '*' 0.05 '.' 0.1 ' ' 1

(Dispersion parameter for gaussian family taken to be 900.243756876)

    Null deviance: 10112469.149  on 9999  degrees of freedom
Residual deviance:  9000637.081  on 9998  degrees of freedom
AIC: 96409.42614

Number of Fisher Scoring iterations: 2

>
> # Chi-Square Test
> linearHypothesis(M1,c("X = 0"))
Linear hypothesis test

Hypothesis:
X = 0

Model 1: restricted model
Model 2: Y ~ X

  Res.Df Df      Chisq           Pr(>Chisq)
1   9999
2   9998  1 1235.03447 < 0.000000000000000222 ***
---
Signif. codes:  0 '***' 0.001 '**' 0.01 '*' 0.05 '.' 0.1 ' ' 1
>
```

The statistical association between X and Y is further demonstrated in Figure 1, where a clear linear trend exists between the constructs. Noting the scales of the X and Y-axes, it can be deduced that the regression line suggests that approximately a 1 point increase in X is associated with a 1 point increase in Y.

[Figure 1]

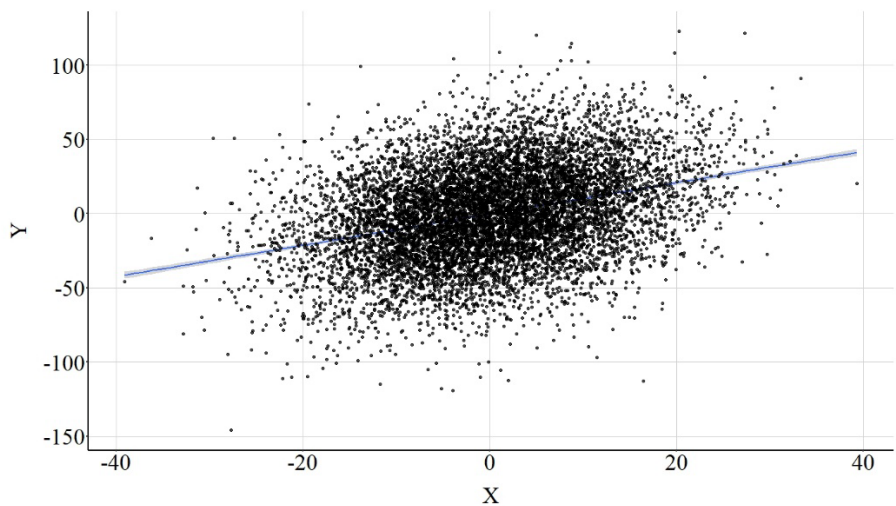

## 3. Altering the Measurement of X

For the sake of brevity – and the relative similarity to Entry 13 – we will primarily review the effects of altering the measurement of X through the employment of figures. While the coefficients associated with the spearman correlation, bivariate linear regression model, and linear hypothesis test are provided in the text, limited syntax is provided throughout the current entry. That said, you can find the *R*-syntax used to conduct all of the simulations and estimate all of the models at www.ianasilver.com, as well as the *R*-syntax used to produce the figures.

### 3.1. Dichotomous Recode 1

For our first recode of the independent variable – or X – we can split the distribution of X at the median, providing a value of 0 to cases that scored below the median and a value of 1 that scored above the median.

```
## Dichotomous Recode 1 ####
> summary(DF$X)
         Min.      1st Qu.       Median         Mean      3rd Qu.
-39.0733074883  -6.6657355322  0.0074956162  0.1341346317  6.9266877113
         Max.
 39.3637051101
> DF$X_DI1<-NA
> DF$X_DI1[DF$X<=median(DF$X)]<-0
> DF$X_DI1[DF$X>median(DF$X)]<-1
> summary(DF$X_DI1)
   Min. 1st Qu.  Median    Mean 3rd Qu.    Max.
    0.0     0.0     0.5     0.5     1.0     1.0
> table(DF$X_DI1)

   0    1
5000 5000
```

Through this recode process, the linear association in Figure 2 is produced. Regarding the test statistics, the spearman correlation between the dichotomous version of X and Y was .26, the *t*-score was 27.783, and the $X^2$-square score was 771.918. Evident by the attenuation in the key test statistics, dichotomizing the independent variable reduced the amount of variation in Y explained by X.

[Figure 2]

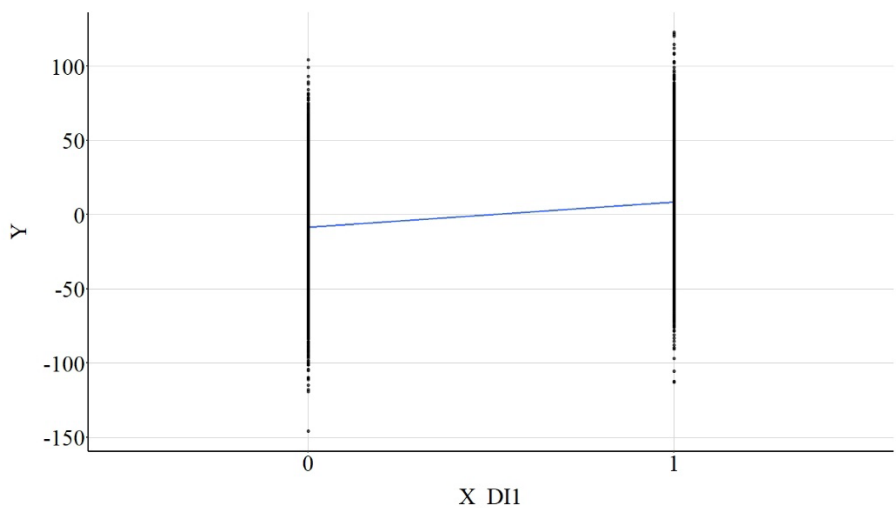

### 3.2. Dichotomous Recode 2

We can implement an alternative strategy and dichotomize the scores on X by splitting the distribution at the first quartile. That is, scores above the first quartile will receive a value of 1, while scores below the first quartile will receive a value of 0. Distinct from the previous strategy, 75% of the cases received a value of 1 and 25% of the cases received a value of 0.

```
## Dichotomous Recode 2 ####
> summary(DF$X)
        Min.       1st Qu.        Median          Mean       3rd Qu.
-39.0733074883  -6.6657355322  0.0074956162  0.1341346317  6.9266877113
        Max.
 39.3637051101
> DF$X_DI2<-NA
> DF$X_DI2[DF$X<=quantile(DF$X,.25)]<-0
> DF$X_DI2[DF$X>quantile(DF$X,.25)]<-1
> summary(DF$X_DI2)
   Min. 1st Qu.  Median    Mean 3rd Qu.    Max.
   0.00    0.75    1.00    0.75    1.00    1.00
> table(DF$X_DI2)

   0    1
2500 7500
>
```

Recoding X into a dichotomy by splitting the distribution at the first quartile produced test statistics that were attenuated when compared to the test statistics when X was coded continuously. Importantly, the test statistics were also further attenuated when compared to the test statistics produced when X was dichotomized using the median value. That is, the spearman correlation between second dichotomous version of X and Y was .23, the *t*-score was 23.804, and the $X^2$-square score was 566.616. Nevertheless, Figure 3 highlights that the linear regression line between the second dichotomous version of X and Y is almost identical to the linear regression line between first dichotomous version of X and Y.

[Figure 3]

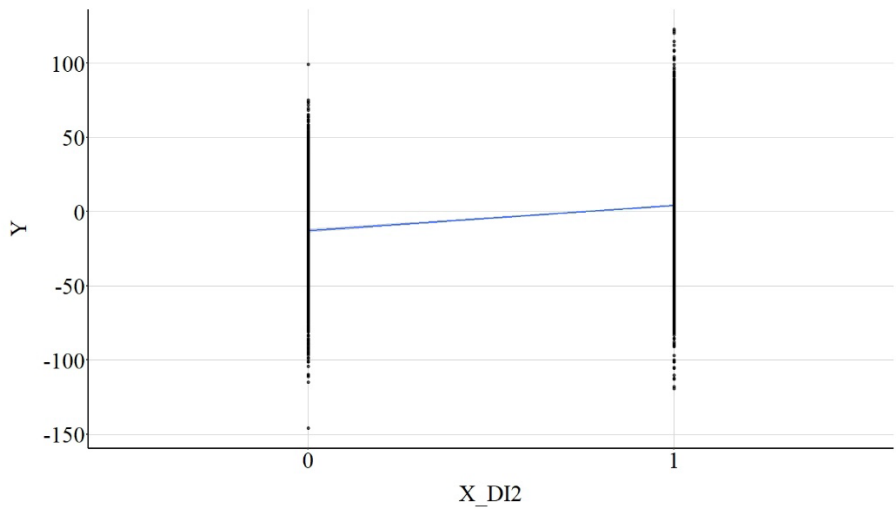

### 3.3. Dichotomous Recode 3
We can also dichotomize the scores on X by splitting the distribution at a more extreme value, such as 30. Splitting the distribution at 30, means that the 12 cases that score higher than 30 received a value of 1 on the third dichotomous version of X and the 9,988 cases that scored below 30 received a value of 0.

```
## Dichotomous Recode 3 ####
> summary(DF$X)
         Min.     1st Qu.      Median        Mean     3rd Qu.
 -39.0733074883  -6.6657355322  0.0074956162  0.1341346317  6.9266877113
          Max.
  39.3637051101
> DF$X_DI3<-NA
> DF$X_DI3[DF$X<=30]<-0
> DF$X_DI3[DF$X>30]<-1
> summary(DF$X_DI3)
   Min. 1st Qu.  Median    Mean 3rd Qu.    Max.
 0.0000  0.0000  0.0000  0.0012  0.0000  1.0000
> table(DF$X_DI3)

   0    1
9988   12
>
```

Using this dichotomous version of X, we can estimate the association with Y and identify that the association is even further attenuated. In particular, the spearman correlation between the third dichotomous version of X and Y was .04, the *t*-score was 4.320, and the $X^2$-square score was 18.660. The error in the statistical association between the third dichotomous version of X and Y is evident in Figure 4, where the 95% confidence interval band (gray area) spreads out substantially as we move from 0 to 1 on X.

[Figure 4]

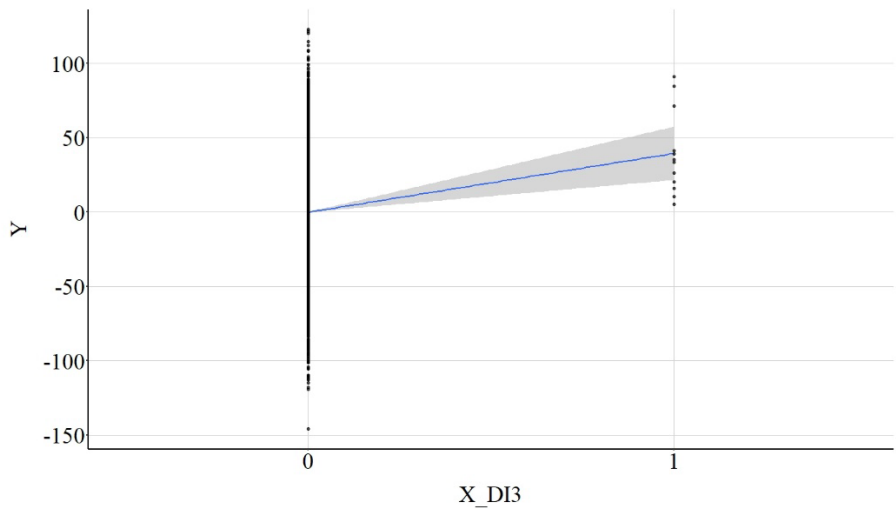

### 3.4. Ordered Recode 1
Following suit, we can create an ordered measure for the independent variable by splitting the distribution of X at the quartiles and assigning cases a value of 1, 2, 3, or 4. This process is used in the syntax below to create X_OR1.

```
> summary(DF$X)
         Min.     1st Qu.      Median        Mean     3rd Qu.
 -39.0733074883  -6.6657355322  0.0074956162  0.1341346317  6.9266877113
          Max.
  39.3637051101
> DF$X_OR1<-NA
```

```
> DF$X_OR1[DF$X>=quantile(DF$X,0) & DF$X < quantile(DF$X,.25)]<-1
> DF$X_OR1[DF$X>=quantile(DF$X,.25) & DF$X < quantile(DF$X,.50)]<-2
> DF$X_OR1[DF$X>=quantile(DF$X,.50) & DF$X < quantile(DF$X,.75)]<-3
> DF$X_OR1[DF$X>=quantile(DF$X,.75) & DF$X <= quantile(DF$X,1)]<-4
> summary(DF$X_OR1)
   Min. 1st Qu.  Median    Mean 3rd Qu.    Max.
   1.00    1.75    2.50    2.50    3.25    4.00
> table(DF$X_OR1)

   1    2    3    4
2500 2500 2500 2500
>
```

After creating X_OR1, we can regress Y on X_OR1 and estimate the linear association. The spearman correlation between the first ordered version of X and Y was .30, the *t*-score was 31.630, and the $X^2$-square score was 1000.445. These test-statistics are extremely close to the values produced when estimating the linear association between the continuous version of X and Y. Figure 5 highlights that the linear regression line between the first ordered version of X and Y is almost identical to the linear regression line between the continuous version of X and Y.

[Figure 5]

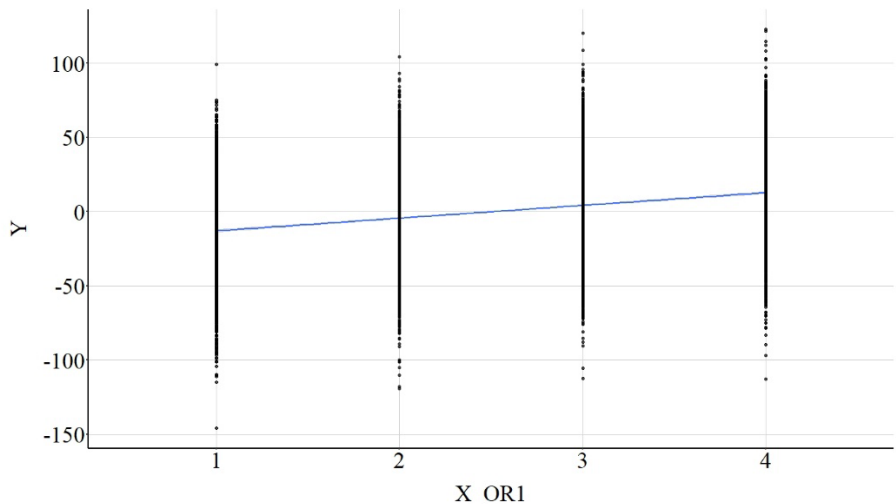

### 3.5. Ordered Recode 2
We can also create an ordered measure for X by splitting the distribution at more extreme values. Here we split the distribution into four quartiles providing a value of "1" to cases that scored between the minimum value and the 50[th] percentile, a value of "2" to cases that score between the 50[th] percentile and the 60[th] percentile, a value of "3" to cases that scored between the 60[th] percentile and the 90[th] percentile, and a value of "4" to cases that scored between the 90[th] percentile and the maximum value.

```
> summary(DF$X)
          Min.          1st Qu.           Median             Mean          3rd Qu.
 -39.0733074883   -6.6657355322     0.0074956162     0.1341346317     6.9266877113
          Max.
  39.3637051101
> DF$X_OR2<-NA
> DF$X_OR2[DF$X>=quantile(DF$X,0) & DF$X < quantile(DF$X,.50)]<-1
> DF$X_OR2[DF$X>=quantile(DF$X,.50) & DF$X < quantile(DF$X,.60)]<-2
```

```
> DF$X_OR2[DF$X>=quantile(DF$X,.60) & DF$X < quantile(DF$X,.90)]<-3
> DF$X_OR2[DF$X>=quantile(DF$X,.90) & DF$X <= quantile(DF$X,1)]<-4
> summary(DF$X_OR2)
   Min. 1st Qu. Median   Mean 3rd Qu.   Max.
    1.0     1.0    1.5    2.0     3.0    4.0
> table(DF$X_OR2)

   1    2    3    4
5000 1000 3000 1000
```

Similar to X_OR1, we can regress Y on X_OR2 and estimate the linear association. The spearman correlation between this ordered version of X and Y was .29, the *t*-score was 30.709, and the $X^2$-square score was 943.050. Figure 6 highlights that the linear regression line between second ordered version of X and Y is almost identical to the linear regression line between the first ordered version of X and Y, as well as the association between the continuous version of X and Y.

[Figure 6]

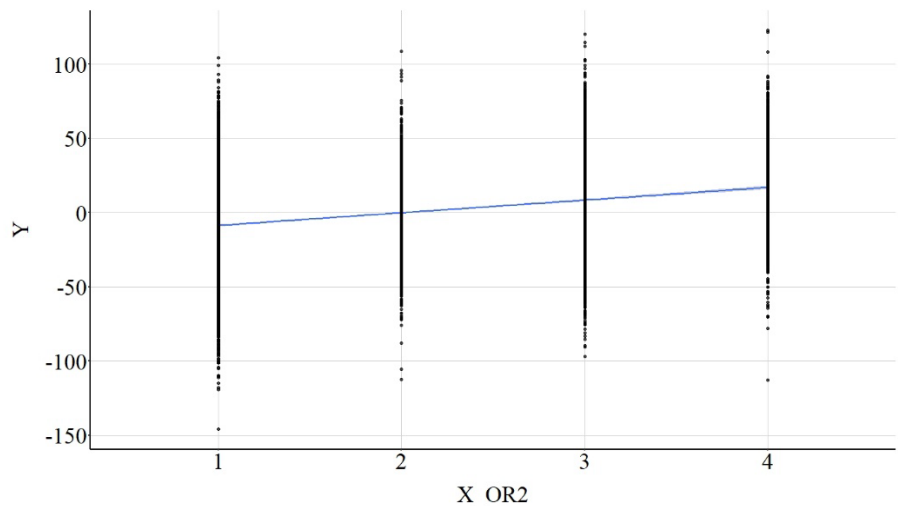

### 3.6. Ordered Recode 3
Finally, we can make the split even more extreme! For X_OR3 a value of "1" was assigned to cases that scored between the minimum value and the 10th percentile, a value of "2" was assigned to cases that score between the 10th percentile and the 20th percentile, a value of "3" was assigned to cases that scored between the 20th percentile and the 30th percentile, and a value of "4" was assigned to cases that scored between the 30th percentile and the maximum value.

```
> summary(DF$X)
          Min.        1st Qu.         Median           Mean        3rd Qu.
-39.0733074883  -6.6657355322   0.0074956162   0.1341346317   6.9266877113
          Max.
  39.3637051101
> DF$X_OR3<-NA
> DF$X_OR3[DF$X>=quantile(DF$X,0) & DF$X < quantile(DF$X,.10)]<-1
> DF$X_OR3[DF$X>=quantile(DF$X,.10) & DF$X < quantile(DF$X,.20)]<-2
> DF$X_OR3[DF$X>=quantile(DF$X,.20) & DF$X < quantile(DF$X,.30)]<-3
> DF$X_OR3[DF$X>=quantile(DF$X,.30) & DF$X <= quantile(DF$X,1)]<-4
> summary(DF$X_OR3)
   Min. 1st Qu. Median   Mean 3rd Qu.   Max.
    1.0     3.0    4.0    3.4     4.0    4.0
> table(DF$X_OR3)
```

```
   1    2    3    4
1000 1000 1000 7000
```

When the linear association between X_OR3 and Y was estimated, the spearman correlation was .25, the *t*-score was 26.559, and the $X^2$-square score was 705.404. Figure 7 highlights that the linear regression line between the third ordered version of X and Y is attenuated when compared to the linear regression line between the previous ordered versions of X and Y, as well as the association between the continuous version of X and Y.

[Figure 7]

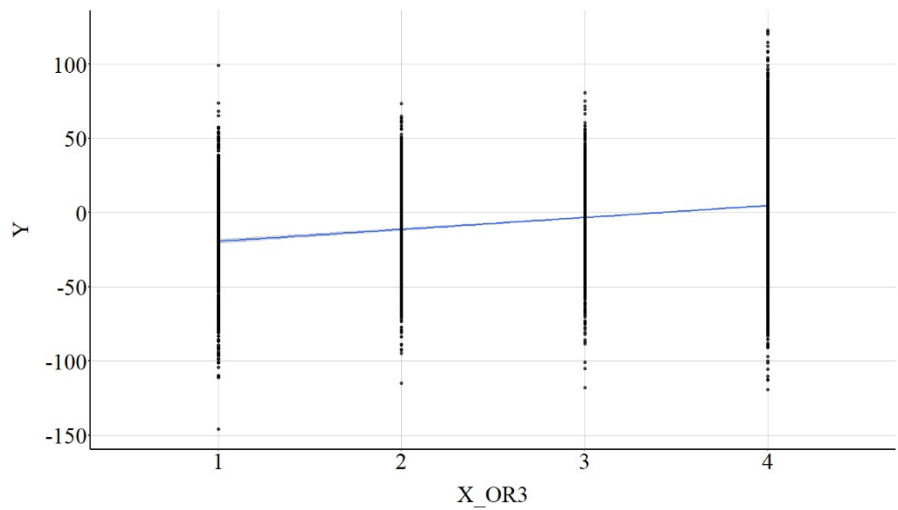

## 4. Data Transformations
Below we will speed up our review of the materials as it is extremely similar to the results presented in Entry 13.

### *4.1. Multiplying By a Constant*
*4.1.1. Example 1: X*.2*

```
> DF$X_Re.2<-X*.2
> summary(DF$X_Re.2)
         Min.         1st Qu.          Median            Mean         3rd Qu.
-7.81466149766 -1.33314710645  0.00149912325  0.02682692633  1.38533754226
         Max.
 7.87274102203
>
```

For this example, we can multiply the cases' scores on X by .2 – a constant – and shift the distribution of scores on X closer to 0. Visually, the distribution of X and X_Re.2 are identical excluding the magnitude of a unit change on the distribution. Consistent with the identical distributions, the magnitude of the association between X_Re.2 and Y is identical to the association between X and Y, which is demonstrated in both the test statistics and Figure 8. Specifically, the spearman correlation was .31, the *t*-score was 35.143, and the $X^2$-square score was 1235.034.

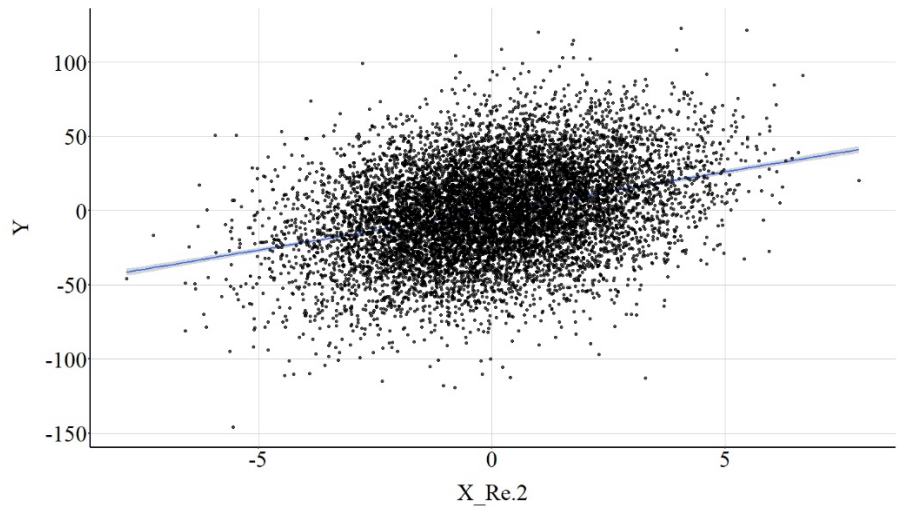

*4.1.2. Example 1: X*20*

```
> DF$X_Re2<-X*20
> summary(DF$X_Re2)
         Min.         1st Qu.          Median            Mean         3rd Qu.
-781.466149766 -133.314710645     0.149912325     2.682692633   138.533754226
         Max.
 787.274102203
>
```

Similarly, we can multiply X by 20 to observe the effects on the association between X_Re2 and Y. Similar to the first example, the magnitude of the association between X_Re2 and Y is identical to the association between X and Y, which is demonstrated in both the test statistics and Figure 9. Specifically, the spearman correlation was .31, the *t*-score was 35.143, and the $X^2$-square score was 1235.034.

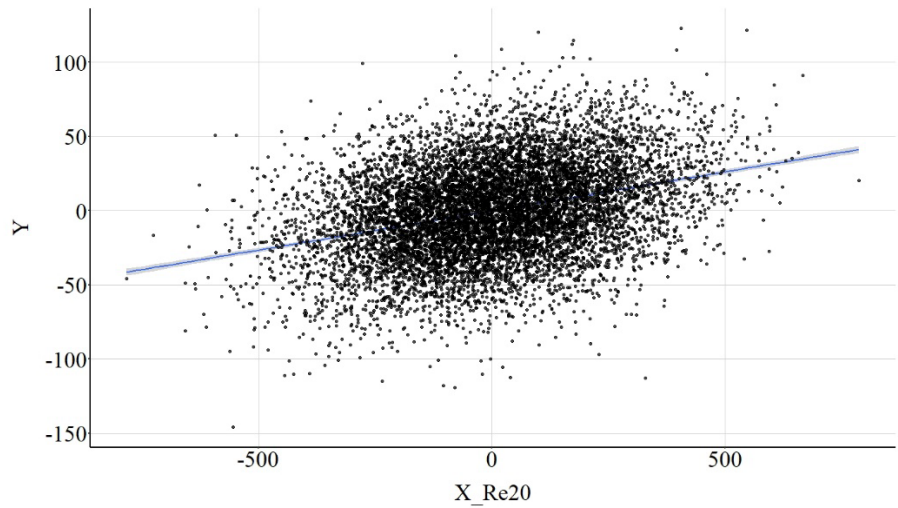

## *4.2. Standardizing Scores on X*

```
> DF$X_z<-scale(DF$X)
> summary(DF$X_z)
       V1
 Min.   :-3.9194399384
 1st Qu.:-0.6797608121
 Median :-0.0126596888
 Mean   : 0.0000000000
 3rd Qu.: 0.6790293471
 Max.   : 3.9216520381
>
```

Using the scale function in *R*, we can standardize the distribution of X to possess a mean of 0 and a standard deviation of 1. Although the scale of X has changed, the magnitude of the association between X_Z and Y is identical to the association between X and Y (Figure 10). In particular, the spearman correlation was .31, the *t*-score was 35.143, and the $X^2$-square score was 1235.034.

[Figure 10]

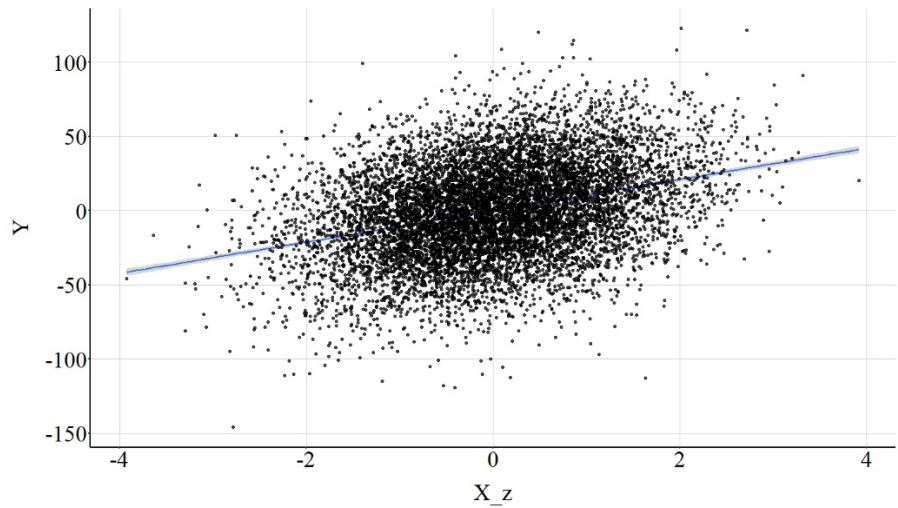

### 4.3. Normalizing Scores on X

```
> DF$X_n<-((DF$X)-min(DF$X))/((max(DF$X))-min(DF$X))
> summary(DF$X_n)
       Min.     1st Qu.      Median        Mean     3rd Qu.        Max.
0.000000000 0.413166831 0.498244411 0.499858942 0.586457766 1.000000000
>
```

We can also normalize the distribution of X to ensure that all cases receive a score between 0 and 1. Similar to the standardized distribution, the magnitude of the association between X_n and Y is identical to the association between X and Y (Figure 11). Specifically, the spearman correlation was .31, the *t*-score was 35.143, and the $X^2$-square score was 1235.034.

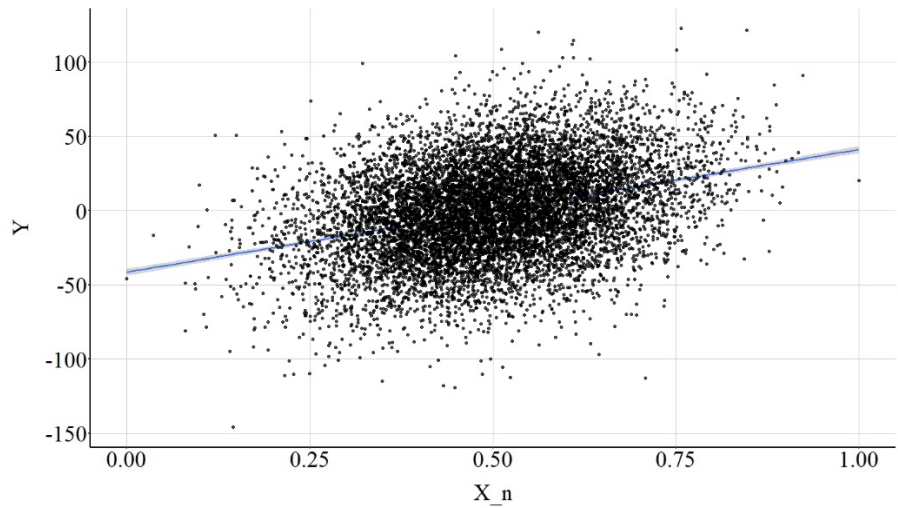

## *4.4. Log Transformation of X*
### *4.4.1. Base Log*

```
> DF$X_n<-((DF$X)-min(DF$X-25))/((max(DF$X+25))-min(DF$X-25))
>
> DF$X_log<-log(DF$X_n)
> summary(DF$X_log)
       Min.     1st Qu.      Median        Mean     3rd Qu.        Max.
-1.636562785 -0.805262400 -0.695293774 -0.705913906 -0.592758801 -0.216475756
>
```

After normalizing X, we can implement a base log transformation and examine the effects of X_log on Y. The magnitude of the association between X_log and Y is slightly attenuated when compared to the association between X and Y, but the effects are largely similar (Figure 12). Specifically, the spearman correlation was .31, the *t*-score was 34.888, and the $X^2$-square score was 1217.141.

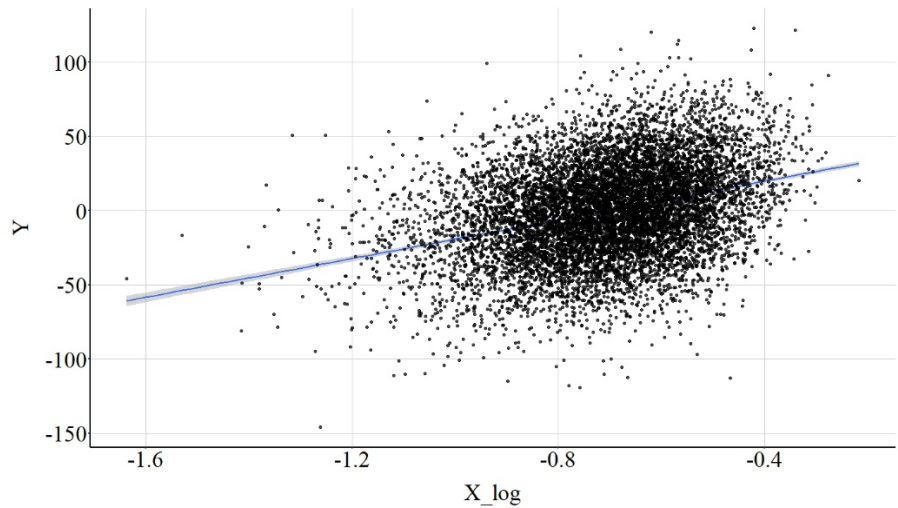[Figure 12]

### 4.4.2. Log 10

```
> DF$X_n<-((DF$X)-min(DF$X-25))/((max(DF$X+25))-min(DF$X-25))
> 
> DF$X_log10<-log10(DF$X_n)
> summary(DF$X_log10)
       Min.        1st Qu.         Median           Mean        3rd Qu. 
-0.7107501868 -0.3497210166 -0.3019622495 -0.3065745140 -0.2574318762 
       Max. 
-0.0940142264 
> 
```

We can also implement a log 10 transformation and examine the effects of X_log10 on Y. The magnitude of the association between X_log10 and Y was identical to the magnitude of the association between X_log and Y (Figure 13). Specifically, the spearman correlation was .31, the $t$-score was 34.888, and the $X^2$-square score was 1217.141.

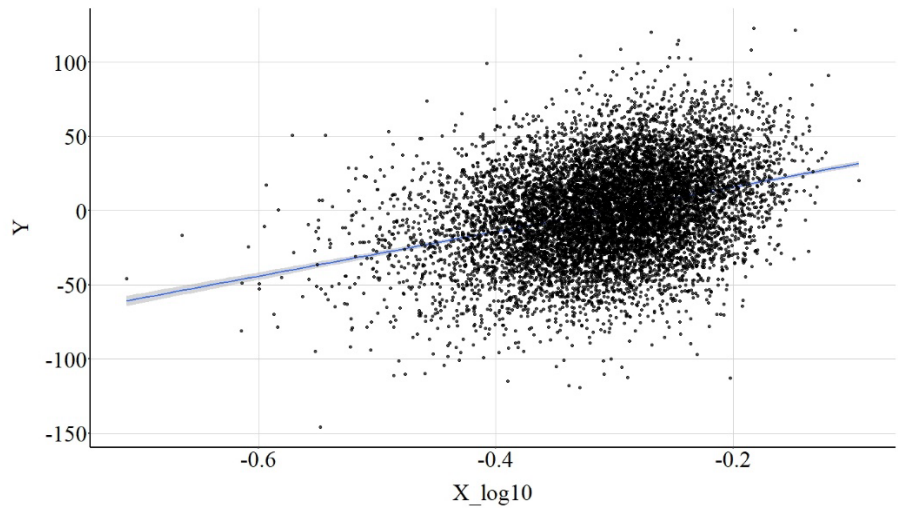

## *4.5. Power Transformations-Raw Scale*
### *4.5.1. X Squared*

```
> DF$X2<-DF$X^2
> summary(DF$X2)
         Min.        1st Qu.         Median           Mean        3rd Qu.
   0.000000002   10.480235412   46.342638585  100.074550598  132.529852859
         Max.
 1549.501279997
>
```

As demonstrated in the current example (X raised to the 2$^{nd}$ power), subjecting the raw scale of X to a power transformation, however, has undesirable consequences. In particular, the association between X2 and Y is null, which is demonstrated by a spearman correlation of .01, a *t*-score of 1.044, and a $X^2$-square score of 1.091 and further supported by Figure 14.

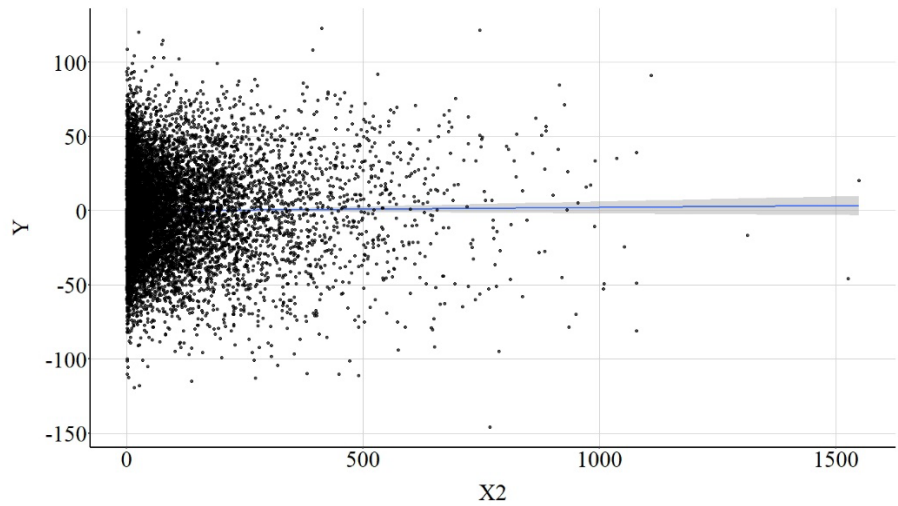

[Figure 14]

### 4.5.2. X Raised to the .2 Power

```
> DF$X.2<-DF$X^.2
> summary(DF$X.2)
     Min.    1st Qu.    Median      Mean    3rd Qu.      Max.       NA's
0.26917726 1.26999736 1.47234769 1.43429322 1.63602339 2.08458301    4997
>
```

Raising the raw scale of X to the power of .2, does not have as drastic of an impact on the association. Nevertheless, the linear association between X.2 and Y is still attenuated. The spearman correlation was .18, the *t*-score was 12.838, and the $X^2$-square score was 164.813 (Figure 15).

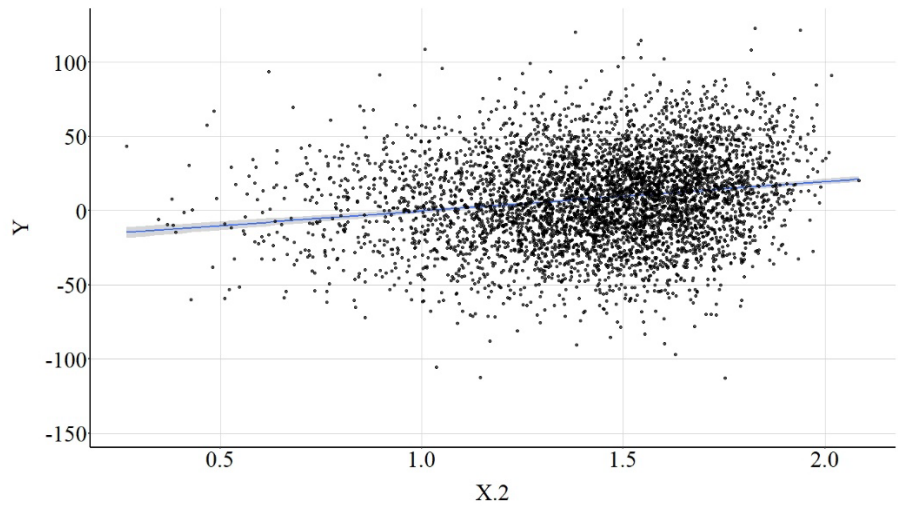

[Figure 15]

## 4.6 Power Transformations-Normalized Scale
### 4.6.1. X Squared

```
> DF$X_n<-((DF$X)-min(DF$X-25))/((max(DF$X+25))-min(DF$X-25))
>
> DF$X2_n<-DF$X_n^2
> summary(DF$X2_n)
       Min.     1st Qu.      Median        Mean     3rd Qu.        Max.
0.0378878208 0.1997827407 0.2489290038 0.2559793426 0.3055879661 0.6485919396
>
```

Raising the normalize scores on X to the 2nd power has limited impact on the magnitude of the association between X2_n and Y. Specifically, the spearman correlation was .31, the *t*-score was 35.143, and the $X^2$-square score was 1235.034. (Figure 16).

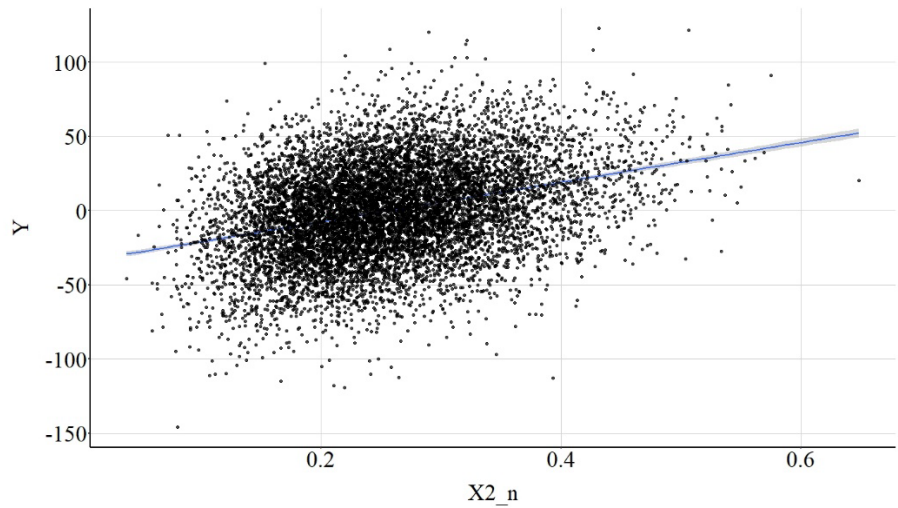

[Figure 16]

*4.6.2. X Raised to the .2 Power*

```
> DF$X.2_n<-DF$X_n^.2
> summary(DF$X.2_n)
      Min.    1st Qu.     Median       Mean    3rd Qu.       Max.
0.720858398 0.851247397 0.870176900 0.868777189 0.888205841 0.957628703
>
```

Similarly, raising the normalize scores on X to the .2 power produces results identical to the association between X and Y. Specifically, the spearman correlation was .31, the *t*-score was 35.143, and the $X^2$-square score was 1235.034. (Figure 17).

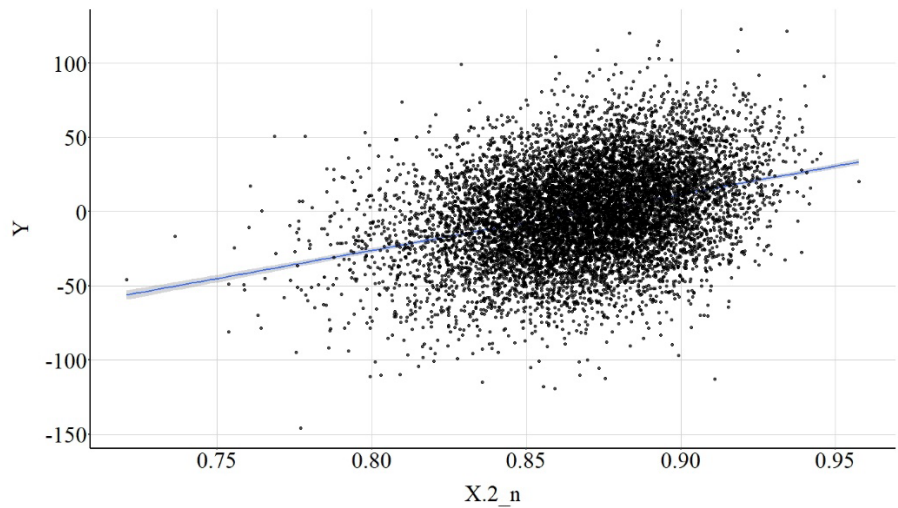

## *4.7. Rounding Raw Scores*

```
> DF$X_R<-round(DF$X)
> summary(DF$X_R)
    Min.  1st Qu.   Median     Mean  3rd Qu.     Max.
-39.0000  -7.0000   0.0000   0.1344   7.0000  39.0000
>
```

Concerning rounding the distribution of X to the nearest whole number, the statistical tests suggest that the association between X_R and Y is largely identical to the association between X and Y. Specifically, the spearman correlation was .31, the *t*-score was 35.121, and the $X^2$-square score was 1233.511. (Figure 18).

[Figure 18]

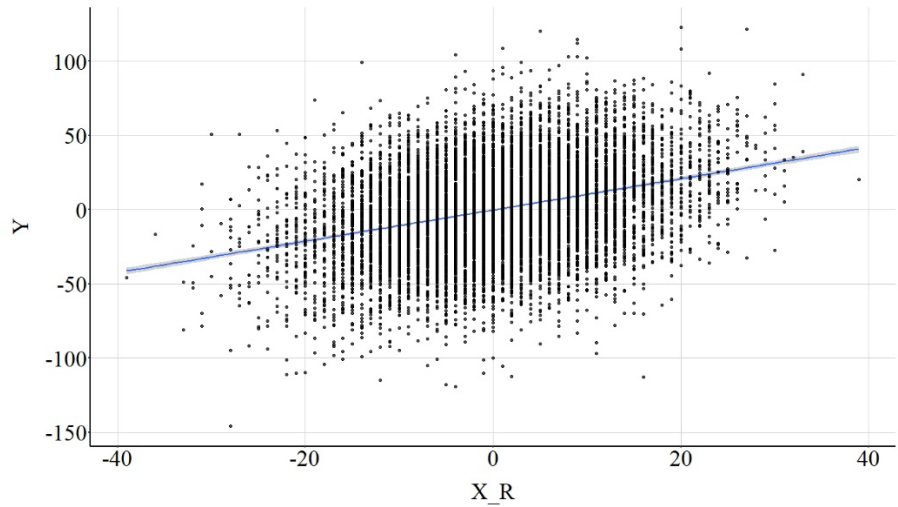

## 4.8. Excluding Part of the Distribution

```
> DF$X_C<-DF$X
> DF$X_C[DF$X_C>=5|DF$X_C<=-5]<-NA
> summary(DF$X)
          Min.       1st Qu.        Median          Mean       3rd Qu.
 -39.0733074883  -6.6657355322   0.0074956162   0.1341346317   6.9266877113
          Max.
  39.3637051101
> summary(DF$X_C)
       Min.    1st Qu.     Median       Mean    3rd Qu.       Max.
 -4.99730075 -2.50686640 -0.12705880 -0.04453803 2.44448142 4.99699775
       NA's
       6197
>
```

Finally, excluding part of the distribution of X can diminish the magnitude of the association between X_C and Y. That is, the test statistics suggest that X_C has a weaker influence on Y when compared to the association between X and Y. Specifically, the spearman correlation was .11, the *t*-score was 6.998, and the $X^2$-square score was 48.975. (Figure 19).

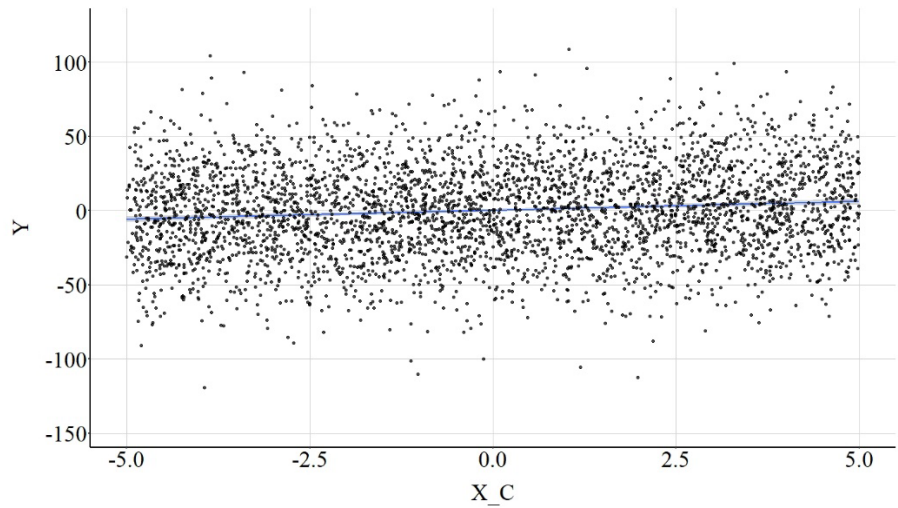

[Figure 19]

## 5. Conclusion

For the sake of clarity, I am going to restate the main conclusion of Entry 13: ***when possible operationalize all of your measures as continuous constructs without the implementation of a continuous data transformation***! I previously, however, forgot to mention that the majority of common data transformations can introduce bias into our statistical models if the variable was not normally distributed prior to the data transformation. While we will address this in a future entry, please know that the results reviewed here do not provide justification for or refute the implementation of strategies used to alter the level of measurement or transform the distribution of continuous constructs. If it is theoretically or empirically relevant to change the level of measurement or implement a continuous data transformation, please do so. Just take some time to recognize the loss of information or potential problems that could occur when implementing these techniques.

**Entry 15: Level of Measurement Confounders, Colliders, Mediators, and Moderators**

**1. Introduction**
The prior entries in the current section reviewed the impact of altering the level of measurement of the dependent variable (Entry 13) and independent variable (Entry 14) on the magnitude of the estimated association. These entries demonstrated that the level of measurement reduced the magnitude of the estimated association between independent and dependent variable – demonstrated using the hypothesis testing statistics. Specifically, the simulations illustrated that the employment of a continuous coding scheme for both the independent and dependent variable produced estimates demonstrating the strongest statistical association, while a dichotomous coding scheme produced estimates demonstrating the weakest statistical association. Overall, altering the operationalization of the dependent variable and the independent variable impacts the estimated association. In practice, however, the dependent variable and the independent variable will likely be operationalized in a manner that maximizes the information captured across the population. As such, it can be argued that while the operationalization of the constructs of interest is important, the operationalization of covariates we introduce into multivariable models could be more important!

We tend to care less about maximizing the information captured by covariates included in our statistical models. And, on some occasions, scholars are outright encouraged to simplify multivariable models by only including dichotomous covariates.[i] To provide an example in the literature, it is common for scholars in sociology, psychology, and criminology to adjust for an individual's level of educational attainment when estimating the effects of an independent variable on a dependent variable. There are multiple ways to operationalize educational attainment – e.g., years of schooling (continuous) or highest degree earned (ordinal) –, of which the dichotomous operationalization of *Completed High School* ("1" = Yes) or *Completed College* ("1" = Yes) are commonly included in multivariable models. As it is often perceived, including either *Completed High School* or *Completed College* in a multivariable model adjusts for the confounding effects of educational attainment on the association being estimated (for example, when estimating the effects of criminal behavior on employment). Although beneficial if educational attainment does confound the association, operationalizing educational attainment as a dichotomous construct (e.g*., completed high school*) limits the amount of variation adjusted for when including the covariate in a multivariable model. In this sense, the estimated association between the independent and dependent variable will vary – or become biased – conditional upon the level of measurement of the confounder.[ii]

Similarly, the bias associated with the inclusion of a collider in a statistical model will vary conditional upon the operationalization of a collider. For example, opposite of a confounder, operationalizing a collider as an ordinal or dichotomous construct– when compared to a continuous construct – can reduce the amount of bias observed when estimating the association between the independent and dependent variable of interest. On an analogous note, the level of measurement of a mediator when included as covariates influences the estimates corresponding to the direct effects of the independent variable on the dependent variable, while the level of measurement of a moderator influences the estimates corresponding to an interaction effect. For instance, including a mediator measured dichotomously in a multivariable model will produce estimates for the direct effects of the independent variable on the dependent variable distinct from the estimates produced if the mediator was measured ordinally.

More broadly, operationalizing a covariate as a continuous construct adjusts for the maximum amount of variation shared between the covariate, independent variable, and dependent variable. However, when the covariate is operationalized using an ordered scale – or as a dichotomy – the amount of shared variation adjusted for through the estimation of a multivariable model is reduced. The reduction in the shared variation, in turn impacts the estimates produced by the multivariable model. And, in particular, can introduce – or reduce – bias when estimating the association between the covariate and the dependent variable, as well as when estimating the association between the independent variable and the dependent variable. As demonstrated in the simulations below, these biases can substantively alter our interpretation of the association between the independent and dependent variable of interest.

## 2. Confounder

To briefly review, a confounder is a construct that causes variation in both the independent and dependent variable. Not adjusting for a confounder in a statistical model will result in a biased estimate of the effects of the independent variable on the dependent variable. The direction and magnitude of that bias, however, is conditional upon the direction and magnitude of the causal effects of the confounder on the independent variable and the dependent variable. Entry 7 provides a detailed review of confounder bias.

To begin our simulation, we first must specify the confounder ("Con" in the R-code). 10,000 scores for Con were drawn from a normal distribution with a mean of 0 and a standard deviation of 10. After simulating a vector of scores to represent the confounder, scores on X were specified to be causally influenced by scores on Con plus a random value drawn from a normal distribution with a mean of 0 and a standard deviation of 10. Concerning the dependent variable, scores on Y were specified to be causally influenced by scores on X and scores on Con plus a random value drawn from a normal distribution with a mean of 0 and a standard deviation of 30. Using the specification below, a 1-point increase in X was specified to cause a 1-point increase in Y, while a 1-point increase in Con was specified to cause a 4-point increase in Y. A 1-point increase in Con was also specified to cause a 4-point increase in X.

To ensure that the simulation specification produced a confounded association between X and Y, a bivariate linear regression model was estimated. Evident by the findings, the estimates suggested that a 1-point increase in X corresponded to approximately a 2-point increase in Y, highlighting that our model needs to adjust for the variation in Con to produce an estimate representative of the true causal association between X and Y (i.e., $b = 1$).

```
> # Confounders ####
> n<-10000
>
> set.seed(1992)
>
> Con<-1*rnorm(n,0,10)
> X<-1*rnorm(n,0,10)+4*Con
> Y<-1*X+1*rnorm(n,0,30)+4*Con
>
> DF<-data.frame(Con,X,Y)
>
> M1<-glm(Y~X, data = DF, family = gaussian(link = "identity"))
> summary(M1)
```

```
Call:
glm(formula = Y ~ X, family = gaussian(link = "identity"), data = DF)

Deviance Residuals:
    Min       1Q   Median       3Q      Max
-109.617  -21.412    0.056   21.120  117.419

Coefficients:
             Estimate Std. Error t value Pr(>|t|)
(Intercept) -0.663385   0.312699  -2.121   0.0339 *
X            1.929342   0.007549 255.559   <2e-16 ***
---
Signif. codes:  0 '***' 0.001 '**' 0.01 '*' 0.05 '.' 0.1 ' ' 1

(Dispersion parameter for gaussian family taken to be 977.681)

    Null deviance: 73627829  on 9999  degrees of freedom
Residual deviance:  9774855  on 9998  degrees of freedom
AIC: 97235

Number of Fisher Scoring iterations: 2

>
```

## 2.1. Continuous Confounder

Now let's estimate a multivariable model adjusting for the variation in Con. In the model presented below, Con is coded as a continuous construct. Adjusting for the variation in Con, evident by the results, permits us to produce an estimate representative of the true causal association between X and Y. In particular, the slope of the association between X and Y when adjusting for Con is .971, which is substantively identical to the true causal association between X and Y (1.00). The similarity between the estimated association and the specified association is further highlighted in Figure 1, where the estimated linear regression line is visually identical to a slope of 1.00.

```
> ## Adjusting for Continuous Confounder ####
>
> M2<-glm(Y~X+Con, data = DF, family = gaussian(link = "identity"))
> summary(M2)

Call:
glm(formula = Y ~ X + Con, family = gaussian(link = "identity"),
    data = DF)

Deviance Residuals:
    Min       1Q   Median       3Q      Max
-110.923  -19.962    0.249   20.273  126.040

Coefficients:
             Estimate Std. Error t value Pr(>|t|)
(Intercept) -0.76368    0.29672  -2.574   0.0101 *
X            0.97098    0.02967  32.729   <2e-16 ***
Con          4.08914    0.12284  33.289   <2e-16 ***
---
Signif. codes:  0 '***' 0.001 '**' 0.01 '*' 0.05 '.' 0.1 ' ' 1

(Dispersion parameter for gaussian family taken to be 880.2107)

    Null deviance: 73627829  on 9999  degrees of freedom
Residual deviance:  8799466  on 9997  degrees of freedom
AIC: 96185

Number of Fisher Scoring iterations: 2

>
```

[Figure 1]

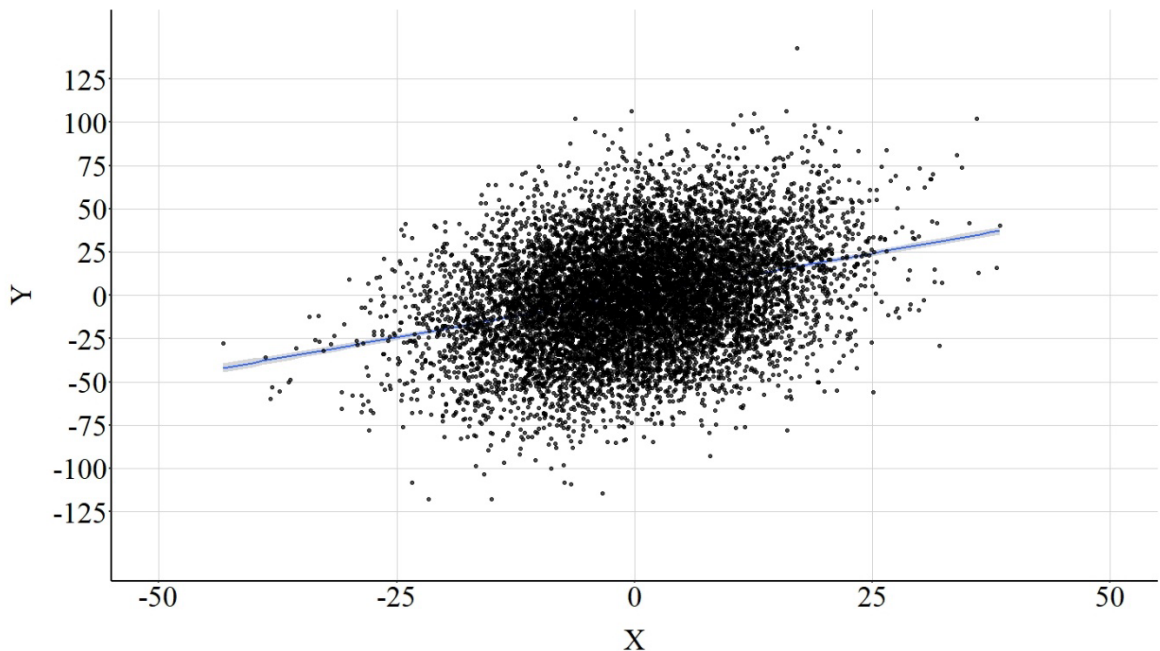

*2.2. Ordered Confounder*

Now let's recode Con as an ordered construct and include it as a covariate in a multivariable model. To recode Con, we can split the distribution at the quartiles. That is, cases that initially scored between 0 and the 25$^{th}$ percentile received a value of "1", cases that scored between the 25$^{th}$ and the 50$^{th}$ percentile received a value of "2", cases that scored between the 50$^{th}$ and the 75$^{th}$ percentile received a value of "3", and cases that scored between the 75$^{th}$ and the 100$^{th}$ percentile received a value of "4." This coding scheme generated a distribution where an equal number of cases – 2,500 – received a 1, 2, 3, or 4. This measure was re-labeled as Con_OR1, and included as a covariate when regressing Y on X. The results of the regression model adjusting for Con_OR1 are presented below. Focusing on the association between X and Y, the estimated slope coefficient was 1.686 suggesting that a 1-point increase in X corresponded to a 1.686-point increase in Y. Substantively, the estimates for the association between X and Y when adjusting for Con_OR1 are distinct from the estimates produced by the model adjusting for the continuous version of Con. In particular, adjusting for Con_OR1 produced a slope coefficient closer to the confounded slope coefficient ($b = 1.929$) than the unconfounded slope coefficient ($b = .971$). The distinction between adjusting for the continuous version of Con and the ordered version of Con is further highlighted by Figure 2 (pay close attention to the scale of the X and Y axes).

```
> ## Adjusting for Ordered Confounder ####
>
> summary(DF$Con)
    Min.  1st Qu.   Median     Mean  3rd Qu.     Max.
-39.0733  -6.6657   0.0075   0.1341   6.9267  39.3637
> DF$Con_OR1<-NA
> DF$Con_OR1[DF$Con>=quantile(DF$Con,0) & DF$Con < quantile(DF$Con,.25)]<-1
> DF$Con_OR1[DF$Con>=quantile(DF$Con,.25) & DF$Con < quantile(DF$Con,.50)]<-2
> DF$Con_OR1[DF$Con>=quantile(DF$Con,.50) & DF$Con < quantile(DF$Con,.75)]<-3
```

```
> DF$Con_OR1[DF$Con>=quantile(DF$Con,.75) & DF$Con <= quantile(DF$Con,1)]<-4
> summary(DF$Con_OR1)
   Min. 1st Qu.  Median    Mean 3rd Qu.    Max.
   1.00    1.75    2.50    2.50    3.25    4.00
> table(DF$Con_OR1)

   1    2    3    4
2500 2500 2500 2500
>
> M2<-glm(Y~X+Con_OR1, data = DF, family = gaussian(link = "identity"))
> summary(M2)

Call:
glm(formula = Y ~ X + Con_OR1, family = gaussian(link = "identity"),
    data = DF)

Deviance Residuals:
     Min        1Q    Median        3Q       Max
-107.874   -21.115     0.107    20.940   114.571

Coefficients:
            Estimate Std. Error t value Pr(>|t|)
(Intercept) -25.6402     1.5888  -16.14   <2e-16 ***
X             1.6862     0.0169   99.75   <2e-16 ***
Con_OR1      10.0362     0.6262   16.03   <2e-16 ***
---
Signif. codes:  0 '***' 0.001 '**' 0.01 '*' 0.05 '.' 0.1 ' ' 1

(Dispersion parameter for gaussian family taken to be 953.2869)

    Null deviance: 73627829  on 9999  degrees of freedom
Residual deviance:  9530009  on 9997  degrees of freedom
AIC: 96983

Number of Fisher Scoring iterations: 2

>
```

[Figure 2]

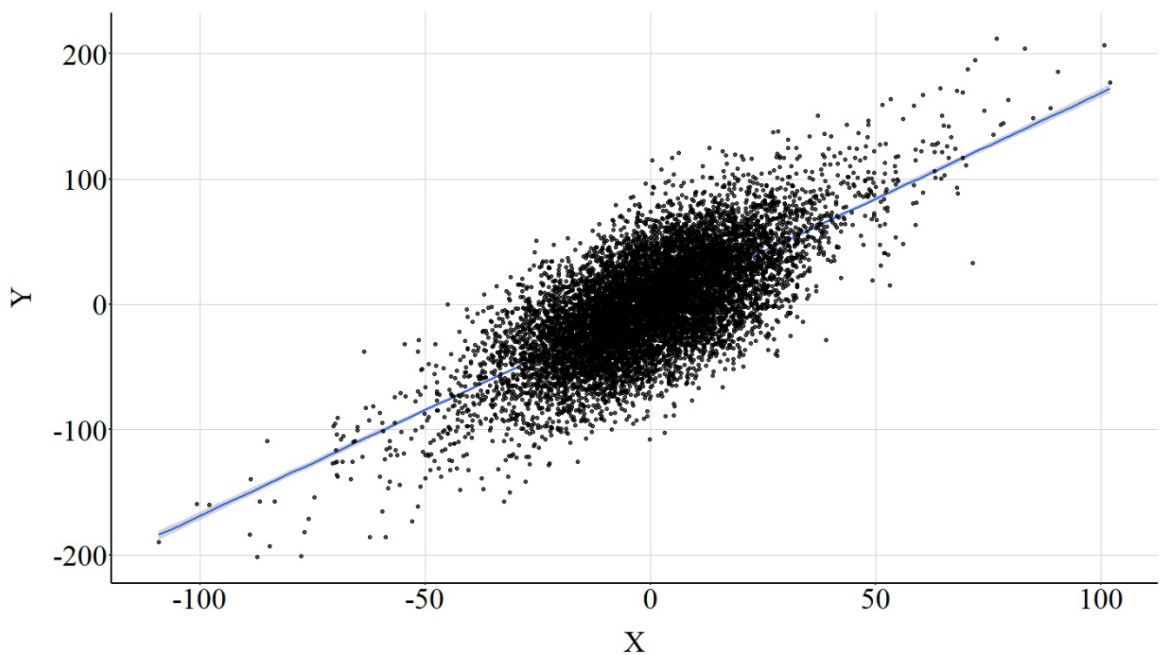

## 2.3. Dichotomous Confounder

Following the trend, now let's dichotomize Con by splitting the distribution at the median. Specifically, cases with a score on Con equal to or below the mediate received a score of "0" on Con_DI1, while cases with a score on Con greater than the median received a score "1" on Con_DI1. Including this as a covariate in the model produced estimates almost identical to the confounder regression model (i.e., the bivariate linear regression model of X on Y). Specifically, the estimated association suggests that a 1-point change in X corresponded to a 1.842-point change in Y after adjusting for the variation in Con_DI1. Of particular interest, these findings suggest that adjusting for Con_DI1 does not appear to produce estimates substantively distinct from the estimates produced when we did not adjust for the variation in Con. This interpretation is further supported by Figure 3.

```
> ## Adjusting for Dichotomous Confounder ####
> 
> summary(DF$Con)
    Min.  1st Qu.   Median     Mean  3rd Qu.     Max.
-39.0733  -6.6657   0.0075   0.1341   6.9267  39.3637
> DF$Con_DI1<-NA
> DF$Con_DI1[DF$Con<=median(DF$Con)]<-0
> DF$Con_DI1[DF$Con>median(DF$Con)]<-1
> summary(DF$Con_DI1)
   Min. 1st Qu.  Median    Mean 3rd Qu.    Max.
    0.0     0.0     0.5     0.5     1.0     1.0
> table(DF$Con_DI1)

   0    1
5000 5000
> 
> M2<-glm(Y~X+Con_DI1, data = DF, family = gaussian(link = "identity"))
> summary(M2)

Call:
glm(formula = Y ~ X + Con_DI1, family = gaussian(link = "identity"),
    data = DF)

Deviance Residuals:
    Min       1Q   Median       3Q      Max
-109.69   -21.22     0.26    21.11   117.15

Coefficients:
             Estimate Std. Error t value Pr(>|t|)
(Intercept) -5.25427    0.58067  -9.049   <2e-16 ***
X            1.84246    0.01194 154.323   <2e-16 ***
Con_DI1      9.26304    0.98896   9.366   <2e-16 ***
---
Signif. codes:  0 '***' 0.001 '**' 0.01 '*' 0.05 '.' 0.1 ' ' 1

(Dispersion parameter for gaussian family taken to be 969.2729)

    Null deviance: 73627829  on 9999  degrees of freedom
Residual deviance:  9689821  on 9997  degrees of freedom
AIC: 97149

Number of Fisher Scoring iterations: 2

> 
```

[Figure 3]

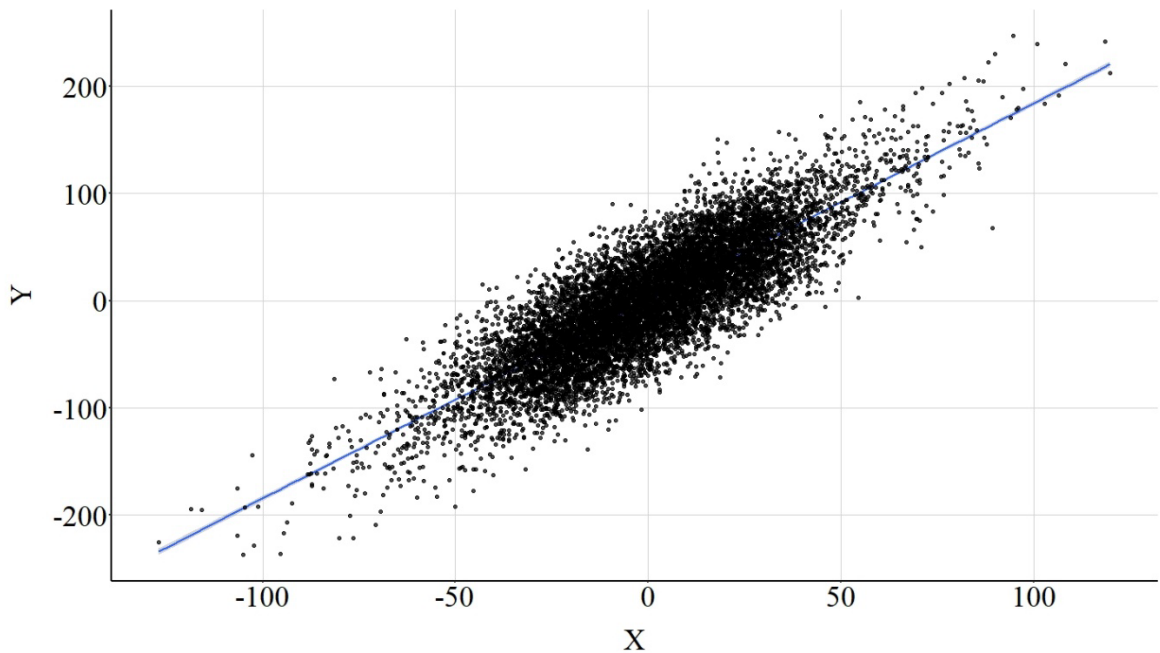

*2.4. Summary of Results*

As demonstrated, the estimated effects of X on Y produced by the multivariable model varied conditional upon the level of measurement for the confounder. When measured continuously, adjusting for the confounder in a multivariable model produced estimates for the association between X and Y that emulated the specification of the simulated data. When measured ordinally, adjusting for the confounder in a multivariable model produced estimates for the association between X and Y closer to the bivariate model than the model adjusting for the continuous version of the confounder. When measured dichotomously, adjusting for the confounder in a multivariable model produced estimates for the association between X and Y extremely similar to the bivariate model. Overall, when a construct is postulated to confound an association of interest, operationalizing the construct continuously and including it as a covariate in a multivariable model will provide the best opportunity – when compared to an ordered or dichotomous operationalization – to observe the causal association between the variables of interest.

**3. Collider**

Collider variables exist when two or more constructs cause variation in a single construct. In this simulation, our independent and dependent variables (X and Y) will be specified to cause variation in a single construct (Col) and generate a collision. Distinct from a confounder, the inclusion of a collider in a multivariable model will result in the estimation of a biased association between the independent variable and the dependent variable (Reviewed in Entry 8). The direction and magnitude of the bias is conditional upon the direction and magnitude of the causal influence of the independent variable and the dependent variable on the collider.

To simulate a collision, or a collider variable, we first simulate scores drawn from a normal distribution with a mean of 0 and a standard deviation of 10 to represent our independent variable (X). After which, we specify that scores on Y are causally influenced by scores on X – where a 1-point increase in X corresponds to a 1-point increase in Y – plus a random value drawn from a normal distribution with a mean of 0 and a standard deviation of 30. Finally, we simulate the collider (Col) by specifying that scores are causally influenced by X and Y, plus a value randomly drawn from a normal distribution with a mean of 0 and a standard deviation of 10. In this simulation, a 1-point increase in X or Y was specified to result in a 4-point increase in the collider.

To begin our analysis, let's estimate the bivariate association between X and Y. The results of the model – presented below – emulated the simulation specification, producing evidence that a 1-point increase in X corresponded to a 1.054-point increase in Y. The difference ($\Delta b = .054$) corresponds to the random variation specified to exist in X and Y. Importantly, in this simulation, association between X and Y will only become biased when Col is included as a covariate in the regression equation. So, let's do that.

```
> # Colliders ####
> n<-10000
> 
> set.seed(1992)
> 
> 
> X<-1*rnorm(n,0,10)
> Y<-1*X+1*rnorm(n,0,30)
> Col<-1*rnorm(n,0,10)+ 4*X+4*Y
> 
> DF<-data.frame(Col,X,Y)
> 
> M1<-glm(Y~X, data = DF, family = gaussian(link = "identity"))
> summary(M1)

Call:
glm(formula = Y ~ X, family = gaussian(link = "identity"), data = DF)

Deviance Residuals:
     Min        1Q    Median        3Q       Max
-129.757   -20.018     0.071    20.265   115.132

Coefficients:
            Estimate Std. Error t value Pr(>|t|)
(Intercept) -0.2138     0.3001  -0.713    0.476
X            1.0541     0.0300  35.143   <2e-16 ***
---
Signif. codes:  0 '***' 0.001 '**' 0.01 '*' 0.05 '.' 0.1 ' ' 1

(Dispersion parameter for gaussian family taken to be 900.2438)

    Null deviance: 10112469  on 9999  degrees of freedom
Residual deviance:  9000637  on 9998  degrees of freedom
AIC: 96409

Number of Fisher Scoring iterations: 2

>
```

### 3.1. Continuous Collider

Here, we estimate a multivariable regression model evaluating the effects of X on Y adjusting for the variation in Col. The estimates of this regression model suggest that X has a negative influence on Y, where a 1-point increase in X corresponds to a -.985 reduction in Y. This substantive departure from the bivariate association is observed because Col was included in the model as a covariate, highlighting the effects of collider bias on the interpretation corresponding to the association of interest. Figure 4 highlights the negative association between X and Y that is observed when the multivariable model adjusts for Col.

```
> ## Adjusting for Continuous Collider ####
>
> M2<-glm(Y~X+Col, data = DF, family = gaussian(link = "identity"))
> summary(M2)

Call:
glm(formula = Y ~ X + Col, family = gaussian(link = "identity"),
    data = DF)

Deviance Residuals:
     Min        1Q    Median        3Q       Max
-10.0943   -1.6864   -0.0158    1.6730    9.1198

Coefficients:
             Estimate Std. Error  t value Pr(>|t|)
(Intercept)  0.061816   0.024663    2.506   0.0122 *
X           -0.985497   0.002985 -330.201   <2e-16 ***
Col          0.248512   0.000205 1212.479   <2e-16 ***
---
Signif. codes:  0 '***' 0.001 '**' 0.01 '*' 0.05 '.' 0.1 ' ' 1

(Dispersion parameter for gaussian family taken to be 6.081088)

    Null deviance: 10112469  on 9999  degrees of freedom
Residual deviance:    60793  on 9997  degrees of freedom
AIC: 46436

Number of Fisher Scoring iterations: 2

>
```

[Figure 4]

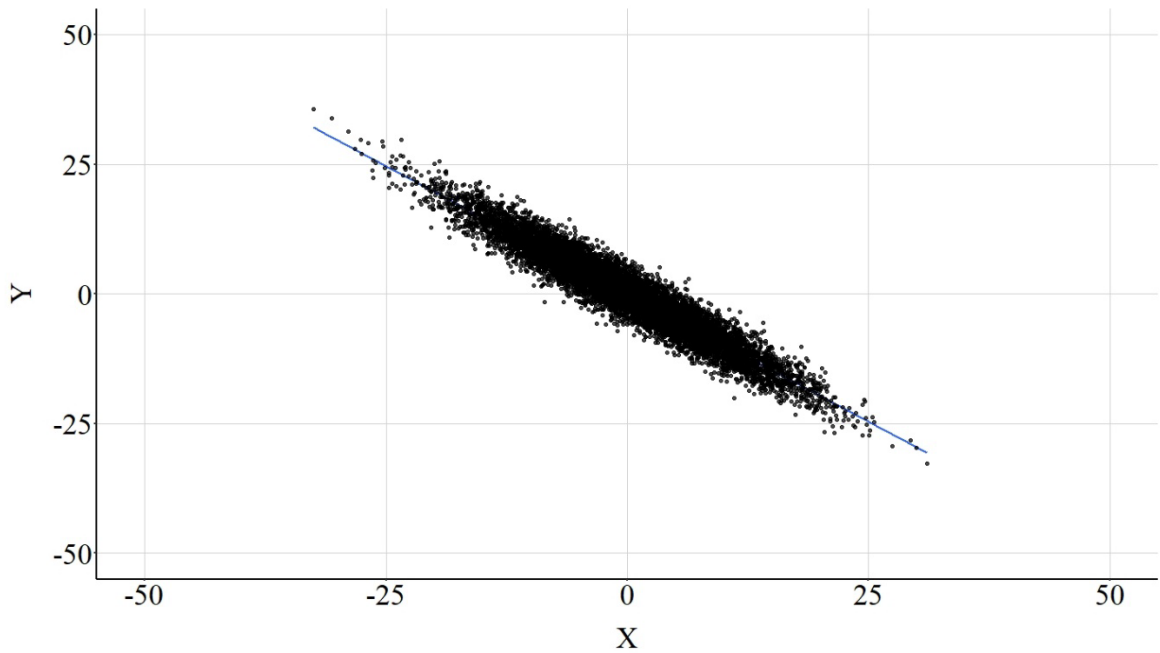

### 3.2. Ordered Collider

Now let's recode Col into an ordinal variable. Here, following the steps discussed when recoding Con to Con_OR1, we can create Col_OR1. The cases received a 1, 2, 3, or 4 on Col_OR1 conditional upon their ranked position in the Col distribution. While the estimated effects of X on Y remained negative when adjusting for Col_OR1, the slope coefficient became attenuated (closer to zero) when compared to the model adjusting for the continuous version of Col. Specifically, the results of this model suggest that a 1-point increase in X corresponded to a .563 reduction in Y. Figure 5 further highlights the negative association between X and Y that is observed when the multivariable model adjusts for Col_OR1, while also illustrating the distinct effects of the continuous and ordered versions of Col when estimating the association between X and Y.

```
> ## Adjusting for Ordered Collider ####
>
> summary(DF$Col)
     Min.  1st Qu.   Median     Mean  3rd Qu.     Max.
 -693.6215 -97.7149   1.2271  -0.0083  98.2333 608.0042
> DF$Col_OR1<-NA
> DF$Col_OR1[DF$Col>=quantile(DF$Col,0) & DF$Col < quantile(DF$Col,.25)]<-1
> DF$Col_OR1[DF$Col>=quantile(DF$Col,.25) & DF$Col < quantile(DF$Col,.50)]<-2
> DF$Col_OR1[DF$Col>=quantile(DF$Col,.50) & DF$Col < quantile(DF$Col,.75)]<-3
> DF$Col_OR1[DF$Col>=quantile(DF$Col,.75) & DF$Col <= quantile(DF$Col,1)]<-4
> summary(DF$Col_OR1)
   Min. 1st Qu.  Median    Mean 3rd Qu.    Max.
   1.00    1.75    2.50    2.50    3.25    4.00
> table(DF$Col_OR1)

   1    2    3    4
2500 2500 2500 2500
>
> M2<-glm(Y~X+Col_OR1, data = DF, family = gaussian(link = "identity"))
> summary(M2)

Call:
glm(formula = Y ~ X + Col_OR1, family = gaussian(link = "identity"),
    data = DF)

Deviance Residuals:
     Min        1Q    Median        3Q       Max
 -119.692    -8.028     0.146     8.188    95.315

Coefficients:
             Estimate Std. Error  t value Pr(>|t|)
(Intercept) -69.74158    0.38373 -181.74   <2e-16 ***
X            -0.56283    0.01607  -35.03   <2e-16 ***
Col_OR1      27.89785    0.14376  194.06   <2e-16 ***
---
Signif. codes:  0 '***' 0.001 '**' 0.01 '*' 0.05 '.' 0.1 ' ' 1

(Dispersion parameter for gaussian family taken to be 188.8634)

    Null deviance: 10112469  on 9999  degrees of freedom
Residual deviance:  1888068  on 9997  degrees of freedom
AIC: 80794

Number of Fisher Scoring iterations: 2

>
```

[Figure 5]

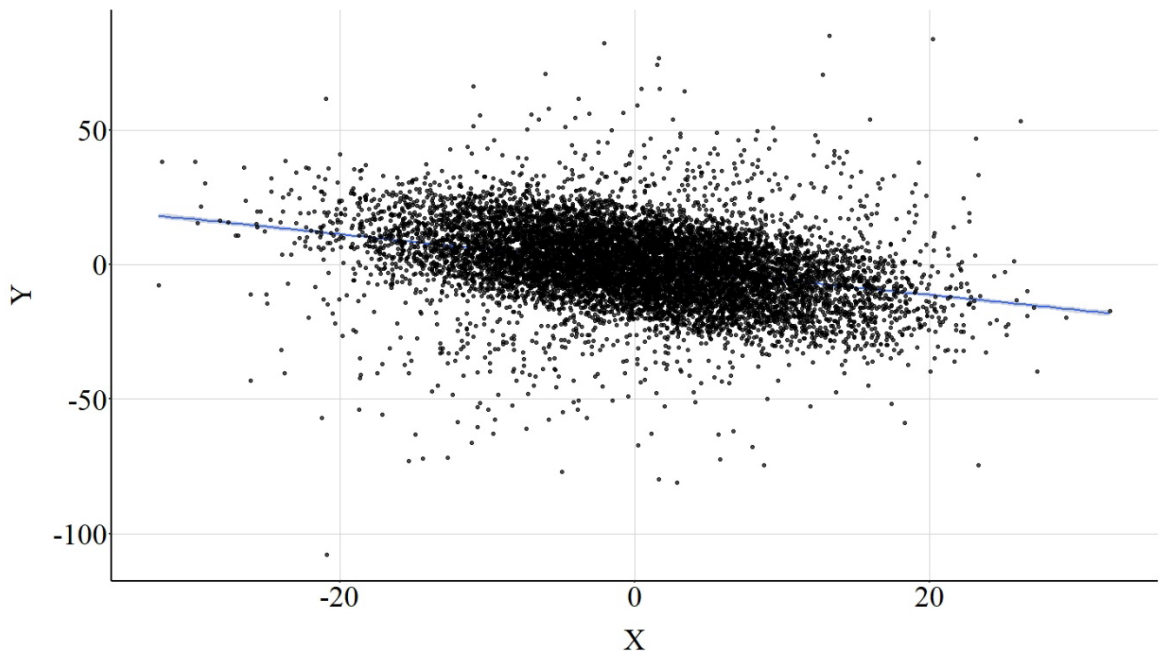

### 3.3. Dichotomous Collider

Similar to the confounder demonstration, let's recode our collider into a dichotomous construct and introduce it into our multivariable model. Cases that scored below or equal to the median of Col received a "0" on Col_DI1, while cases that scored above the median received a "1" on Col_DI1. When adjusting for Col_DI1, the results of the multivariable regression model suggested that a 1-point increase in X corresponded to a .02-point reduction in Y. The magnitude of this point estimate is extremely close to zero, suggesting that the variation in X does not predict the variation in Y when adjusting for Col_DI1. The interpretation associated with these estimates is distinct from the preceding models, which suggested that X has a notable negative influence on Y. The horizontal regression line presented in Figure 6 further highlights the distinction between the estimates produced by the model below and the estimates produced by the prior models.

```
> ## Adjusting for Dichotomous Collider ####
>
> summary(DF$Col)
     Min.   1st Qu.    Median      Mean   3rd Qu.      Max.
-693.6215  -97.7149    1.2271   -0.0083   98.2333  608.0042
> DF$Col_DI1<-NA
> DF$Col_DI1[DF$Col<=median(DF$Col)]<-0
> DF$Col_DI1[DF$Col>median(DF$Col)]<-1
> summary(DF$Col_DI1)
   Min. 1st Qu.  Median    Mean 3rd Qu.    Max.
    0.0     0.0     0.5     0.5     1.0     1.0
> table(DF$Col_DI1)

   0    1
5000 5000
>
> M2<-glm(Y~X+Col_DI1, data = DF, family = gaussian(link = "identity"))
> summary(M2)

Call:
glm(formula = Y ~ X + Col_DI1, family = gaussian(link = "identity"),
```

```
         data = DF)

Deviance Residuals:
     Min        1Q    Median        3Q       Max
-121.920   -13.379    -0.124    13.835    98.838

Coefficients:
             Estimate Std. Error  t value Pr(>|t|)
(Intercept) -24.59034    0.30358  -81.001   <2e-16 ***
X            -0.02049    0.02264   -0.905    0.366
Col_DI1      49.04129    0.45302  108.254   <2e-16 ***
---
Signif. codes:  0 '***' 0.001 '**' 0.01 '*' 0.05 '.' 0.1 ' ' 1

(Dispersion parameter for gaussian family taken to be 414.4736)

    Null deviance: 10112469  on 9999  degrees of freedom
Residual deviance:  4143493  on 9997  degrees of freedom
AIC: 88654

Number of Fisher Scoring iterations: 2

>
```

[Figure 6]

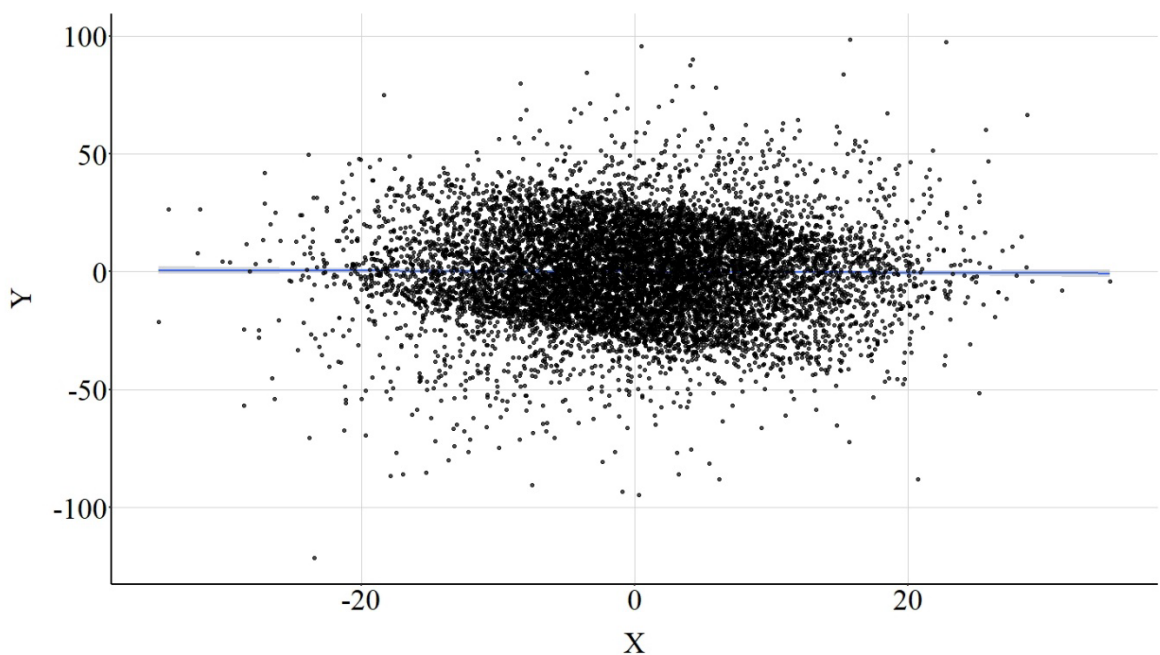

### 3.4. Summary of Results

The results of the three models illustrate the importance of the level of measurement for a collider when estimating an association of interest. Introducing a continuous collider into a multivariable model will increase – comparatively – the bias in the estimated association between the independent and dependent variable. Introducing a dichotomous collider into a multivariable model will decrease – comparatively – the bias in the estimated association between the independent and dependent variable. Prominently, however, it is important to consider that the level of measurement for a collider can result in distinct interpretations for the association of interest. As illustrated, including the continuous and ordinal versions of the collider in a multivariable model resulted in estimates suggesting a negative association between X and Y, while including the dichotomous version of the collider in a multivariable model resulted in estimates suggesting no association between X and Y.

## 4. Mediators

Mediators represent constructs where the variation is caused by the independent variable of interest, and cause variation in the dependent variable of interest. As discussed in Entry 9, the effects of X on Y can be demarcated into three categories: total effects, direct effects, and indirect effects. When a mediator does exist, the estimates produced when modeling the bivariate association between X and Y represents the total effect of X on Y. The estimates corresponding to the association between X and Y when a mediator is included as a covariate in a regression model represents the direct effect of X on Y. The indirect effects can not be illuminated unless multiple regression models or a path model is estimated. In this section, we will focus on estimating the direct effect of X on Y when adjusting for a continuous, ordered, or dichotomous mediator.

To simulate the data, we can first begin by specifying for X to be a vector of scores randomly drawn from a normal distribution with a mean of 0 and a standard deviation of 10. The Mediator (Med in the code) is specified to be causally influenced by the cases' score on X plus a random value drawn from a normal distribution with a mean of 0 and a standard deviation of 10. A 1-point increase in X was specified to be associated with a 4-point increase in Med. Y was specified to be causally influenced by X (i.e., the direct pathway) and Med (i.e., the indirect pathway) plus a random value drawn from a normal distribution with a mean of 0 and a standard deviation of 30. The direct effects of X on Y were specified to equal 1, where a 1-point increase in X would correspond with a 1-point increase in Y, while a 1-point increase in Med was specified to cause a 4-point increase in Y.

Let's start by estimating the total effect of X on Y using a bivariate regression model. The results of this bivariate regression model suggest that a 1-point increase in X results in a 17-point increase in Y. This estimate is *correct*, as the total effect of X on Y would be calculated as the multiplication of the indirect pathway plus the direct pathway, which is (4*4) + 1 ≈ 17.

```
> # Mediators ####
> n<-10000
> 
> set.seed(1992)
> 
> 
> X<-1*rnorm(n,0,10)
> Med<-1*rnorm(n,0,10)+ 4*X
> Y<-1*X+4*Med+1*rnorm(n,0,30)
> 
> 
> DF<-data.frame(Med,X,Y)
> 
> M1<-glm(Y~X, data = DF, family = gaussian(link = "identity"))
> summary(M1)

Call:
glm(formula = Y ~ X, family = gaussian(link = "identity"), data = DF)

Deviance Residuals:
    Min       1Q   Median       3Q      Max
-182.689  -33.573    0.216   32.721  209.901

Coefficients:
            Estimate Std. Error t value Pr(>|t|)
(Intercept) -1.04673    0.49577  -2.111   0.0348 *
```

```
X           17.04472    0.04956 343.932   <2e-16 ***
---
Signif. codes:  0 '***' 0.001 '**' 0.01 '*' 0.05 '.' 0.1 ' ' 1

(Dispersion parameter for gaussian family taken to be 2457.418)

    Null deviance: 315255964  on 9999  degrees of freedom
Residual deviance:  24569262  on 9998  degrees of freedom
AIC: 106451

Number of Fisher Scoring iterations: 2

>
```

*4.1. Continuous Mediator*

To estimate the direct effect of X on Y, we can include the continuous mediator – Med – as a covariate in the model. The results of this multivariable model suggest, consistent with the simulation specification, that a 1-point increase in X corresponds to a 1.089-point increase in Y. The direct effect of X on Y when adjusting for the continuous mediator is illustrated in Figure 7.

```
> ## Adjusting for Continuous Mediator ####
>
> M2<-glm(Y~X+Med, data = DF, family = gaussian(link = "identity"))
> summary(M2)

Call:
glm(formula = Y ~ X + Med, family = gaussian(link = "identity"),
    data = DF)

Deviance Residuals:
     Min        1Q    Median        3Q       Max
-110.923   -19.962     0.249    20.273   126.040

Coefficients:
            Estimate Std. Error t value Pr(>|t|)
(Intercept) -0.76368    0.29672  -2.574   0.0101 *
X            1.08914    0.12284   8.866   <2e-16 ***
Med          3.97098    0.02967 133.850   <2e-16 ***
---
Signif. codes:  0 '***' 0.001 '**' 0.01 '*' 0.05 '.' 0.1 ' ' 1

(Dispersion parameter for gaussian family taken to be 880.2107)

    Null deviance: 315255964  on 9999  degrees of freedom
Residual deviance:   8799466  on 9997  degrees of freedom
AIC: 96185

Number of Fisher Scoring iterations: 2

>
```

[Figure 7]

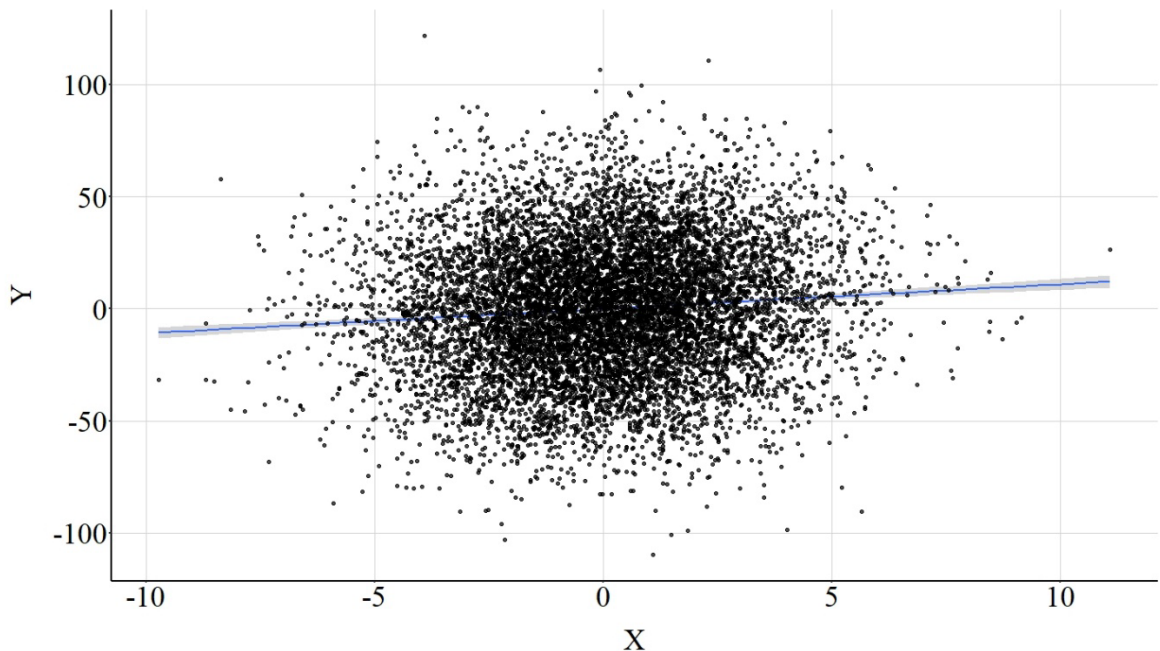

## 4.2. Ordered Mediator

Now let's focus on our ordered version of the mediator (Med_OR1). Med_OR1 was created using the same recoding scheme as previously discussed. When Med_OR1 was included as a covariate in the multivariable model, the magnitude of the direct effect of X on Y was estimated to be 13.007. That is, a 1-point increase in X corresponds to a 13.006-point increase in Y. Substantively, this estimate is distinct from the true direct effect ($b = 1.000$) and closer to the total effect of X on Y (17.000). The change in the magnitude of the estimated direct effect can be observed by comparing the regression line presented in Figure 7 to the regression line presented in Figure 8.

```
> ## Adjusting for Ordered Mediator ####
> 
> summary(DF$Med)
    Min.   1st Qu.    Median      Mean   3rd Qu.      Max.
-158.5572  -27.8877    0.2188    0.4677   28.5786  152.4684
> DF$Med_OR1<-NA
> DF$Med_OR1[DF$Med>=quantile(DF$Med,0)   & DF$Med <  quantile(DF$Med,.25)]<-1
> DF$Med_OR1[DF$Med>=quantile(DF$Med,.25) & DF$Med <  quantile(DF$Med,.50)]<-2
> DF$Med_OR1[DF$Med>=quantile(DF$Med,.50) & DF$Med <  quantile(DF$Med,.75)]<-3
> DF$Med_OR1[DF$Med>=quantile(DF$Med,.75) & DF$Med <= quantile(DF$Med,1)]<-4
> summary(DF$Med_OR1)
   Min. 1st Qu.  Median    Mean 3rd Qu.    Max.
   1.00    1.75    2.50    2.50    3.25    4.00
> table(DF$Med_OR1)

   1    2    3    4
2500 2500 2500 2500
> 
> M2<-glm(Y~X+Med_OR1, data = DF, family = gaussian(link = "identity"))
> summary(M2)

Call:
glm(formula = Y ~ X + Med_OR1, family = gaussian(link = "identity"),
    data = DF)

Deviance Residuals:
```

```
     Min       1Q   Median       3Q      Max
 -204.57    -29.45    -0.24    29.63   209.92

Coefficients:
             Estimate Std. Error  t value Pr(>|t|)
(Intercept) -101.0341     2.3482   -43.03   <2e-16 ***
X             13.0065     0.1036   125.60   <2e-16 ***
Med_OR1       40.2116     0.9265    43.40   <2e-16 ***
---
Signif. codes:  0 '***' 0.001 '**' 0.01 '*' 0.05 '.' 0.1 ' ' 1

(Dispersion parameter for gaussian family taken to be 2067.981)

    Null deviance: 315255964  on 9999  degrees of freedom
Residual deviance:  20673608  on 9997  degrees of freedom
AIC: 104727

Number of Fisher Scoring iterations: 2

>
```

[Figure 8]

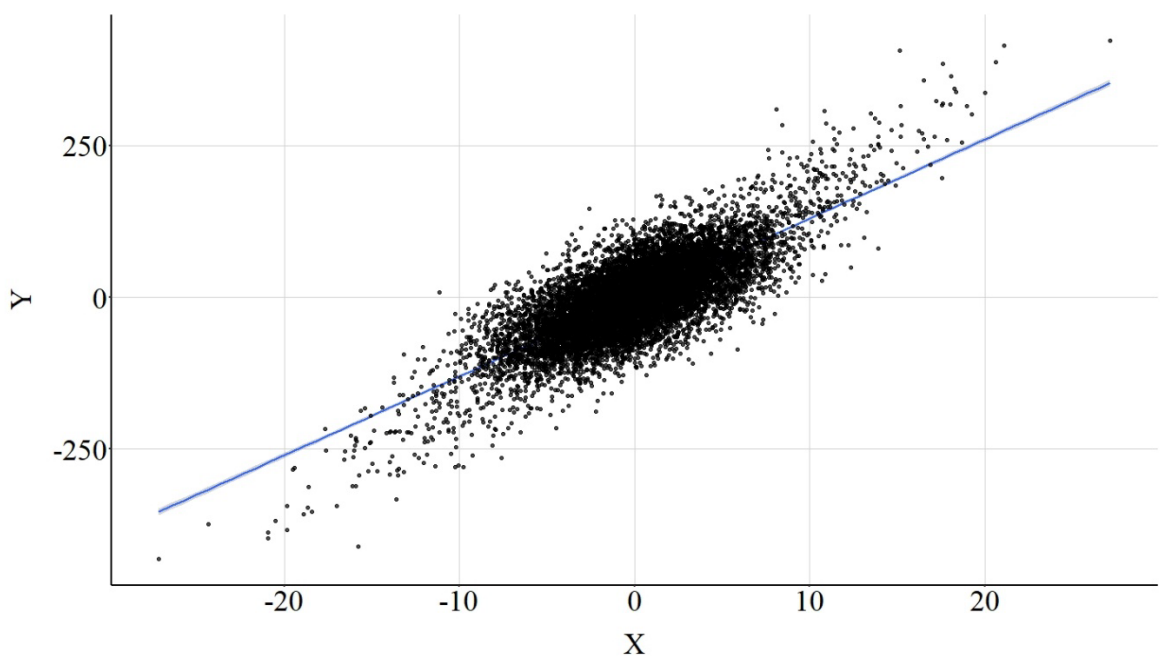

### 4.3. Dichotomous Mediator

Similar to Med_OR1, the multivariable regression model with Med_DI1 included as a covariate produced results suggesting that a 1-point increase in X corresponds to a 15.641-point increase in Y. Again, the estimated direct effect of X on Y when adjusting for Med_DI1 is distinct from the true direct effect ($b = 1.000$) and closer to the total effect of X on Y (17.000; see Figure 9).

```
> ## Adjusting for Dichotomous Mediator ####
>
> summary(DF$Med)
     Min.   1st Qu.    Median      Mean   3rd Qu.      Max.
 -158.5572  -27.8877    0.2188    0.4677   28.5786  152.4684
> DF$Med_DI1<-NA
```

```
> DF$Med_DI1[DF$Med<=median(DF$Med)]<-0
> DF$Med_DI1[DF$Med>median(DF$Med)]<-1
> summary(DF$Med_DI1)
   Min. 1st Qu.  Median    Mean 3rd Qu.    Max.
    0.0     0.0     0.5     0.5     1.0     1.0
> table(DF$Med_DI1)

   0    1
5000 5000
>
> M2<-glm(Y~X+Med_DI1, data = DF, family = gaussian(link = "identity"))
> summary(M2)

Call:
glm(formula = Y ~ X + Med_DI1, family = gaussian(link = "identity"),
    data = DF)

Deviance Residuals:
     Min       1Q    Median       3Q      Max
-186.806  -31.904     0.121    32.072  193.137

Coefficients:
            Estimate Std. Error t value Pr(>|t|)
(Intercept) -18.8926     0.9013  -20.96   <2e-16 ***
X            15.6409     0.0769  203.39   <2e-16 ***
Med_DI1      36.0684     1.5385   23.45   <2e-16 ***
---
Signif. codes:  0 '***' 0.001 '**' 0.01 '*' 0.05 '.' 0.1 ' ' 1

(Dispersion parameter for gaussian family taken to be 2329.579)

    Null deviance: 315255964  on 9999  degrees of freedom
Residual deviance:  23288800  on 9997  degrees of freedom
AIC: 105918

Number of Fisher Scoring iterations: 2

>
```

[Figure 9]

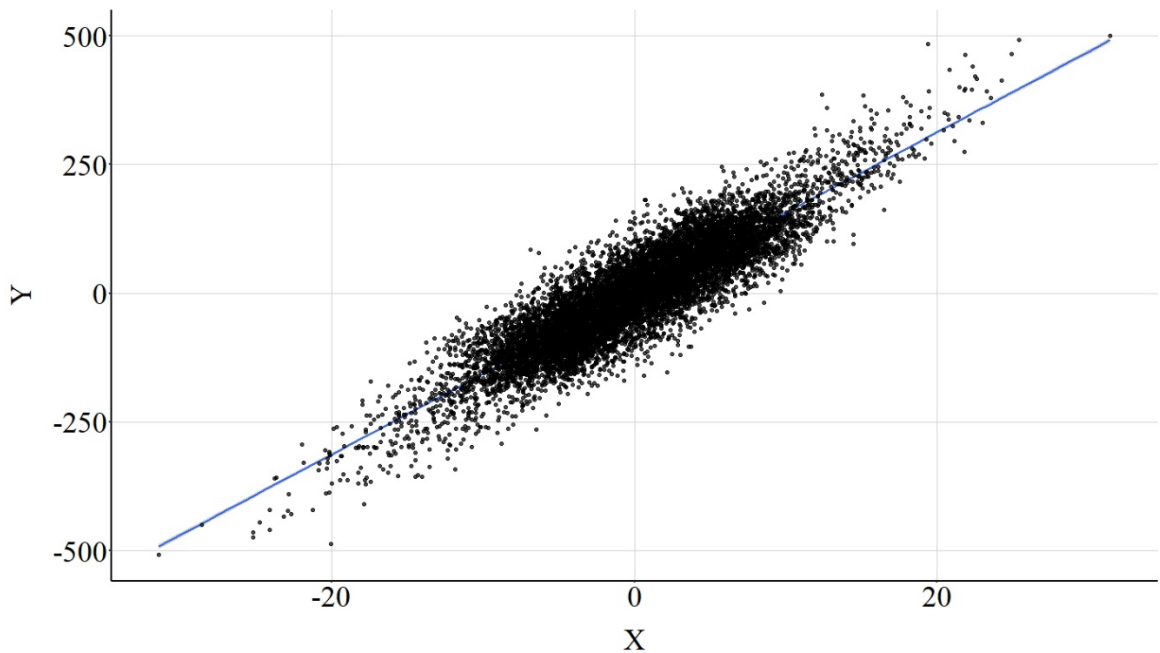

*4.4. Summary of Results*
The results, similar to the previous simulations, highlight how important the level of measurement of a mediator is to the estimation of the direct effect of an independent variable on a dependent variable. As highlighted by the findings, only the continuous version of the mediator provided the ability to observe the direct effect of X on Y. Moreover, the ordered and dichotomous versions of the mediator produced estimates for the direct effect of X on Y closer to the total effect than the true direct effect. Overall, when estimating the direct and indirect pathways, the results appear to suggest that the level of measurement of the mediator is fundamental to deriving unbiased estimates.

**5. Moderators**
Lastly, let's observe how the level of measurement influences the estimation of moderating effects. Briefly, a moderator is a construct that influences the magnitude of the causal effects of an independent variable on a dependent variable. Specifically, the magnitude, direction, or existence of a direct causal pathway between an independent variable and a dependent variable differs across the distribution of the moderator, generating a condition where the variation in the moderator possesses an indirect causal influence on a dependent variable. For a complete discussion of moderators and the existence of moderators, see Entry 9.

In this simulation, we begin by specifying for X to be a vector of scores randomly drawn from a normal distribution with a mean of 0 and a standard deviation of 10. Similarly, we specify Mod to be a vector of scores randomly drawn from a normal distribution with a mean of 0 and a standard deviation of 10. Finally, we specify that scores on Y are causally influenced by X plus a random value drawn from a normal distribution with a mean of 0 and a standard deviation of 30, but include an interaction between X and the moderator. The inclusion of this interaction term in the simulation creates a situation where the moderated effects of X on Y can only be observed after specifying a multivariable model including the interaction between X and Mod. The bivariate model presented below provides the unmoderated direct causal effects of X on Y, which suggests that a 1-point increase in X is associated with a .777 increase in Y. Now let's observe how the level of measurement of the moderator influences the estimation of the moderated effects.

```
> n<-10000
>
> set.seed(1992)
>
>
> X<-1*rnorm(n,0,10)
> Mod<-1*rnorm(n,0,10)
> Y<-1*X+4*(X*Mod)+1*rnorm(n,0,30)
>
>
> DF<-data.frame(Mod,X,Y)
>
> M1<-glm(Y~X, data = DF, family = gaussian(link = "identity"))
> summary(M1)

Call:
glm(formula = Y ~ X, family = gaussian(link = "identity"), data = DF)

Deviance Residuals:
        Min           1Q       Median           3Q          Max
-3223.773839   -149.470470    -6.570033    145.430447   4357.864461

Coefficients:
              Estimate  Std. Error  t value  Pr(>|t|)
(Intercept) 6.449943835 4.062862035  1.58754  0.112423
X           0.777023852 0.406134844  1.91322  0.055749 .
---
Signif. codes:  0 '***' 0.001 '**' 0.01 '*' 0.05 '.' 0.1 ' ' 1

(Dispersion parameter for gaussian family taken to be 165038.802029)

    Null deviance: 1650662050  on 9999  degrees of freedom
Residual deviance: 1650057943  on 9998  degrees of freedom
AIC: 148522.1294

Number of Fisher Scoring iterations: 2

>
```

## 5.1. Continuous Moderator

The model presented below estimates the effects of X on Y while adjusting for the interaction between X and the moderator. Evident by the estimates corresponding to X:Mod, the causal effects of X on Y vary across levels of Mod. These estimates suggest that as Mod increases and as X increases, the magnitude of the association between X and Y increases by 4. That is, if Mod increases by 1-point and X increases by 1-point, the slope of the effects of X on Y increases by 4-points.

```
> ## Adjusting for Continuous Moderator ####
>
> M2<-glm(Y~X+(X*Mod), data = DF, family = gaussian(link = "identity"))
> summary(M2)

Call:
glm(formula = Y ~ X + (X * Mod), family = gaussian(link = "identity"),
    data = DF)

Deviance Residuals:
         Min            1Q        Median            3Q           Max
-110.93669769  -19.96835512    0.21990476    20.27815372   126.05903138

Coefficients:
```

```
                Estimate    Std. Error      t value              Pr(>|t|)
(Intercept) -0.76240085432  0.29677836552   -2.56892              0.010216 *
X            0.97302528493  0.02966651509   32.79877 < 0.0000000000000002 ***
Mod         -0.02906768277  0.02966933741   -0.97972              0.327247
X:Mod        3.99929135359  0.00292900378 1365.41010 < 0.0000000000000002 ***
---
Signif. codes:  0 '***' 0.001 '**' 0.01 '*' 0.05 '.' 0.1 ' ' 1

(Dispersion parameter for gaussian family taken to be 880.293582032)

    Null deviance: 1650662050.236  on 9999  degrees of freedom
Residual deviance:    8799414.646  on 9996  degrees of freedom
AIC: 96187.32454

Number of Fisher Scoring iterations: 2

>
```

## 5.2. Ordered Moderator

Now let's observe the interaction effects when Mod_OR1 is included instead of the continuous version of the moderator. Mod_OR1 was created using a coding scheme identical to the previous examples. Evident by the regression model, the interaction between X and the ordered version of the moderator (X:Mod_OR1) produced an estimate suggesting that the slope of the effects of X on Y increased by 33-points for every 1-point increase in X and 1-point increase in Mod_OR1. The magnitude of this interaction effect is substantively larger than the magnitude of the interaction between X and the continuous version of the moderator.

```
> ## Adjusting for Ordered Moderator ####
>
> summary(DF$Mod)
     Min.   1st Qu.    Median      Mean   3rd Qu.      Max.
-43.02615  -6.76873   0.01920  -0.06886   6.66684  38.39675
> DF$Mod_OR1<-NA
> DF$Mod_OR1[DF$Mod>=quantile(DF$Mod,0) & DF$Mod < quantile(DF$Mod,.25)]<-1
> DF$Mod_OR1[DF$Mod>=quantile(DF$Mod,.25) & DF$Mod < quantile(DF$Mod,.50)]<-2
> DF$Mod_OR1[DF$Mod>=quantile(DF$Mod,.50) & DF$Mod < quantile(DF$Mod,.75)]<-3
> DF$Mod_OR1[DF$Mod>=quantile(DF$Mod,.75) & DF$Mod <= quantile(DF$Mod,1)]<-4
> summary(DF$Mod_OR1)
   Min. 1st Qu.  Median    Mean 3rd Qu.    Max.
   1.00    1.75    2.50    2.50    3.25    4.00
> table(DF$Mod_OR1)

   1    2    3    4
2500 2500 2500 2500
>
> M2<-glm(Y~X+(X*Mod_OR1), data = DF, family = gaussian(link = "identity"))
> summary(M2)

Call:
glm(formula = Y ~ X + (X * Mod_OR1), family = gaussian(link = "identity"),
    data = DF)

Deviance Residuals:
     Min        1Q    Median        3Q       Max
-1986.18    -58.22     -2.17     55.67   2959.57

Coefficients:
             Estimate Std. Error  t value Pr(>|t|)
(Intercept)    2.2615     3.9383    0.574    0.566
X            -83.3053     0.3964 -210.146   <2e-16 ***
Mod_OR1       -0.9027     1.4382   -0.628    0.530
X:Mod_OR1     33.4260     0.1440  232.045   <2e-16 ***
---
Signif. codes:  0 '***' 0.001 '**' 0.01 '*' 0.05 '.' 0.1 ' ' 1

(Dispersion parameter for gaussian family taken to be 25846.16)

    Null deviance: 1650662050  on 9999  degrees of freedom
Residual deviance:  258358255  on 9996  degrees of freedom
AIC: 129984

Number of Fisher Scoring iterations: 2
```

## 5.3. Dichotomous Moderator

Similarly, the interaction between X and the dichotomous version of the moderator (X:Mod_DI1) produced an estimate suggesting that the slope of the effects of X on Y increased by 64-points for every 1-point increase in X and 1-point increase in Mod_DI1. Again, the magnitude of this interaction effect is substantively larger than the magnitude of the interaction between X and the continuous version of the moderator, as well as the magnitude of the interaction between X and the ordered version of the moderator.

```
> ## Adjusting for Dichotomous Moderator ####
> 
> summary(DF$Mod)
         Min.       1st Qu.        Median          Mean       3rd Qu.          Max.
-43.0261512727  -6.7687334100   0.0192011481  -0.0688601096   6.6668399946  38.3967546327
> DF$Mod_DI1<-NA
> DF$Mod_DI1[DF$Mod<=median(DF$Mod)]<-0
> DF$Mod_DI1[DF$Mod>median(DF$Mod)]<-1
> summary(DF$Mod_DI1)
   Min. 1st Qu.  Median    Mean 3rd Qu.    Max.
    0.0     0.0     0.5     0.5     1.0     1.0
> table(DF$Mod_DI1)

   0    1
5000 5000
> 
> M2<-glm(Y~X+(X*Mod_DI1), data = DF, family = gaussian(link = "identity"))
> summary(M2)

Call:
glm(formula = Y ~ X + (X * Mod_DI1), family = gaussian(link = "identity"),
    data = DF)

Deviance Residuals:
        Min            1Q         Median            3Q           Max
-2287.190370    -98.823187      -4.062528      93.622929    3452.845917

Coefficients:
                Estimate    Std. Error   t value              Pr(>|t|)
(Intercept)    5.971904423   3.521726935   1.69573              0.089968 .
X            -31.907528830   0.355377521 -89.78488 < 0.00000000000000002 ***
Mod_DI1       -7.552764240   4.981359320  -1.51621              0.129499
X:Mod_DI1     64.191195581   0.498035661 128.88875 < 0.00000000000000002 ***
---
Signif. codes:  0 '***' 0.001 '**' 0.01 '*' 0.05 '.' 0.1 ' ' 1

(Dispersion parameter for gaussian family taken to be 62012.8025413)

    Null deviance: 1650662050.2  on 9999  degrees of freedom
Residual deviance:  619879974.2  on 9996  degrees of freedom
AIC: 138735.7312

Number of Fisher Scoring iterations: 2

>
```

## 5.4. Summary of Results

The results suggest that the estimates corresponding to an interaction term vary substantively conditional upon the level of measurement of the moderating construct. Under the conditions of this simulation, the magnitude of the interaction effects was substantively smaller when the moderator was measured continuously. On the other side of the coin, the interaction effects were substantively larger when the moderator was measured as an ordered construct and even larger when the moderator was measured dichotomously. These results suggest that the moderating effects of a construct vary across the level of measurement of the moderator.

## 6. Conclusion

The current entry highlights the importance of the level of measurement for covariates included in our statistical models. The simulations, largely, demonstrate that the level of measurement for covariates – whether it be a confounder, collider, mediator, or moderator – will influence the estimated effects of the independent variable on the dependent variable and result in meaningfully distinct interpretations of the association of interest. Of particular importance, measuring a confounder, mediator, or moderator as an ordered or dichotomous construct will likely increase the bias in the estimated effects of the independent variable on the dependent variable, when compared to a continuous measure. However, measuring a collider as an ordered or dichotomous construct will likely decrease the bias in the estimated effects of the independent variable on the dependent variable, when compared to a continuous measure. The takeaway from this entry should be: ***the level of measurement of the covariates included in our statistical models does influence the estimation of the association of interest and should be considered when specifying a multivariable model.*** Personally, given these simulations and my knowledge of statistics, I believe that operationalizing covariates as continuous constructs is ***more important*** to the production of unbiased statistical estimates than operationalizing the independent variable and the dependent variable as continuous constructs. Nevertheless, it is not always possible to operationalize all of our constructs as continuous variables and, as such, we should always strive to operationalize all of our measures as continuous when possible.

---

[i] To provide an example, it is commonly encouraged to dichotomize constructs when conducting *propensity score matching*. By dichotomizing a construct, the likelihood of achieving post-matching balance (i.e., the distribution is identical across the treatment and control groups) is increased substantially increased. However, while I understand the downside of poor post-matching balance, is losing information associated with the matching covariates worth the increased balance by matching respondents on only dichotomous constructs? Of course, there is no right answer to this question, as it is open to debate.

[ii] The direction of the bias is conditional upon the direction of the effects of the confounder on the independent and dependent variables of interest and the true association between the independent and dependent variable.

# Resources, References, R-Packages

## Resources

Peng, R. D. (2016). *R programming for data science*. Leanpub.
https://bookdown.org/rdpeng/rprogdatascience/

## References

***Exploring the Effects of Violating the Fundamental Assumptions of Regression Models***
1. Draper, N. R., & Smith, H. (1998). *Applied regression analysis*. New York, NY: John Wiley & Sons.
2. Fox, J. (2015). *Applied regression analysis and generalized linear models*. Thousand Oaks, CA: Sage Publications.

***Assumptions When Examining Causal Associations***
1. Rosenbaum, P. R. (2004). Randomized experiments and observational studies: causal inference in statistics. Retrieved from http://citeseerx.ist.psu.edu/viewdoc/download?doi=10.1.1.586.3844&rep=rep1&type=pdf
2. Fisher, R. A. (1936). Design of experiments. *Br Med J*, *1*(3923), 554-554.
3. Neyman, J. (1942) Basic ideas and some recent results of the theory of testing statistical hypotheses. Journal of the Royal Statistical Society, 105, 292-327.
4. Rosenbaum P. R. (1984) From association to causation in observational studies. Journal of the American Statistical Association, 79, 41-48.
5. Rosenbaum, P. R. (1984). Conditional permutation tests and the propensity score in observational studies. Journal of the American Statistical Association 79, 565-574.
6. Rosenbaum, P. R. (1987) Sensitivity analysis for certain permutation inferences in matched observational studies. Biometrika, 74, 13-26.
7. Rosenbaum, P. R. (1987). Model-based direct adjustment. Journal of the American Statistical Association 82, 387-394.
8. Rosenbaum, P. R. & Rubin, D. B. (1983) The central role of the propensity score in observational studies for causal effects. Biometrika, 70, 41-55.
9. Rosenbaum, P. & Rubin, D. (1984) Reducing bias in observational studies using subclassification on the propensity score. Journal of the American Statistical Association, 79, 516-524.
10. Pearl, J. (2000). Causal Inference: Models, reasoning and inference. *Cambridge, UK: Cambridge University Press*.
11. Pearl, J. (2009). Causal inference in statistics: An overview. *Statistics surveys*, *3*, 96-146.
12. Pearl, J. (2014). *Probabilistic reasoning in intelligent systems: networks of plausible inference*. Elsevier.
13. Pearl, J. (2009). *Causality*. Cambridge university press.
14. Pearl, J. (1995). Causal diagrams for empirical research. *Biometrika*, *82*(4), 669-688.
15. Greenland, S., Pearl, J., & Robins, J. M. (1999). Causal diagrams for epidemiologic research. *Epidemiology*, 37-48.
16. Efird, J. (2011). Blocked randomization with randomly selected block sizes. *International journal of environmental research and public health*, *8*(1), 15-20.
17. Lachin, J. M., Matts, J. P., & Wei, L. J. (1988). Randomization in clinical trials: conclusions and recommendations. *Controlled clinical trials*, *9*(4), 365-374.

18. Neubauer, P. B., & Neubauer, A. (1996). Nature's thumbprint: The new genetics of personality. Columbia University Press.

**R-Packages**

Behrendt, S. (2014). lm.beta: Add Standardized Regression Coefficients to lm-Objects. R package version 1.5-1. https://CRAN.R-project.org/package=lm.beta

Chang, W. (2014). extrafont: Tools for using fonts. R package version 0.17. https://CRAN.R-project.org/package=extrafont

Daniel E. Ho, Kosuke Imai, Gary King, Elizabeth A. Stuart (2011). MatchIt: Nonparametric Preprocessing for Parametric Causal Inference. Journal of Statistical Software, Vol. 42, No. 8, pp. 1-28. https://doi.org/10.18637/jss.v042.i08

Gaujoux, R. (2020). doRNG: Generic Reproducible Parallel Backend for 'foreach' Loops. R package version 1.8.2. https://CRAN.R-project.org/package=doRNG

Grosjean, P. and Ibanez, F. (2018). pastecs: Package for Analysis of Space-Time Ecological Series. R package version 1.3.21. https://CRAN.R-project.org/package=pastecs

Alboukadel Kassambara (2020). ggpubr: 'ggplot2' Based Publication Ready Plots. R package version 0.4.0. https://CRAN.R-project.org/package=ggpubr

Microsoft Corporation and Steve Weston (2020). doParallel: Foreach Parallel Adaptor for the 'parallel' Package. R package version 1.0.16. https://CRAN.R-project.org/package=doParallel

R Core Team (2017). R: A language and environment for statistical computing. R Foundation for Statistical Computing, Vienna, Austria. URL https://www.R-project.org/.

RStudio Team (2020). RStudio: Integrated Development for R. RStudio, PBC, Boston, MA URL http://www.rstudio.com/.

Revelle, W. (2021) psych: Procedures for Personality and Psychological Research, Northwestern University, Evanston, Illinois, USA, https://CRAN.R-project.org/package=psych Version = 2.1.9.

Rosseel, Y. (2012). Lavaan: An R package for structural equation modeling and more. Version 0.5–12 (BETA). *Journal of statistical software*, *48*(2), 1-36.

Schloerke, B., Crowley, J., Cook, D., Hofmann, H., Wickham, H., Briatte, F., ... & Larmarange, J. (2011). Ggally: Extension to ggplot2.

Stuart, E. A., King, G., Imai, K., & Ho, D. (2011). MatchIt: nonparametric preprocessing for parametric causal inference. *Journal of statistical software*.

Venables WN, Ripley BD (2002). Modern Applied Statistics with S, Fourth edition. Springer, New York. ISBN 0-387-95457-0, http://www.stats.ox.ac.uk/pub/MASS4/.

Wickham H (2016). ggplot2: Elegant Graphics for Data Analysis. Springer-Verlag New York. ISBN 978-3-319-24277-4, https://ggplot2.tidyverse.org.

Wickham et al., (2019). Welcome to the tidyverse. Journal of Open Source Software, 4(43), 1686, https://doi.org/10.21105/joss.01686